# The Design of the NetBSD I/O Subsystems

SungWon Chung

Pusan National University

This book is dedicated to the open-source code developers in the NetBSD community. The original copy of this publication is available in an electronic form and it can be downloaded for free from http://arXiv.org.

NetBSD is the registered trademark of The NetBSD Foundation, Inc.

# Contents

# Preface

Influenced by the documentation of Fast File System (FFS) in NetBSD Operating System Release 1.6, which is a derivative of 4.4BSD-Lite [1], this book aims to provide necessary information to design and implement a new filesystem by describing the implementation of the NetBSD FFS while trying to answer the following questions. This work contains many direct excerpts from the books and papers listed in bibliography section as well as text manual entries from the NetBSD operation system. What I did is merely connecting the concepts in books, papers, and manuals, to the assembly and C source code of the NetBSD Operating System on 64-bit UltraSPARC platform.

- How the root filesystem is mounted ?

- How the system call request by application program is executed by the virtual filesystem layer ?

- How can we add a new filesystem as another virtual filesystem layer ?

- How the UFS integrates FFS as a virtual filesystem layer ?

- How the buffer cache of the kernel works with the filesystem ?

- How the UVM works with the filesystem ?

- How to replace buffer cache ?

I am debted to my project team members, Young-Jin Shin and Woo-Young Park who generously have offered insightful comments and spiritual encouragement towards the completion of our project. I am grateful to the USENIX Association for financial support through the *USENIX Student Research Grant* program in 2002. I sincerely wish to thank my project team advisor, Professor Sang-Hwa Chung at the Division of Electrical and Computer Engineering in Pusan National University.

SungWon Chung

Busan, Korea

27 November 2002

sungwon@ieee.org

**A note added in 2016 when archiving this publication to arXiv.org:** Since many part of this book need clarification and correction, the readers' generous understanding will be greatly appreciated until a second edition is available in the future. This archival version is the same as the initial 2002 release of this publication except few typo corrections and updates here in the preface. During the past 14 years, I had switched my main field of interests from operating system software to integrated circuits design while my two project members Young-Jin Shin and Woo-Young Park have kept their endeavors and have had successful careers as well-known technical writers, entrepreneurs, and software engineers. For the second edition, we are planning to include a simple example of new virtual filesystem development. Your comments on any other suggestions will be extremely valuable to us.

# Source Code Copyright

## The NetBSD Foundation

The NetBSD specific code contains the following copyright notice.

```
/*-
 * Copyright (c) 1998, 2000 The NetBSD Foundation, Inc.
 * All rights reserved.
 *
 * This code is derived from software contributed to The NetBSD Foundation
 * by Jason R. Thorpe of the Numerical Aerospace Simulation Facility,
 * NASA Ames Research Center.
 *
 * Redistribution and use in source and binary forms, with or without
 * modification, are permitted provided that the following conditions
 * are met:
 * 1. Redistributions of source code must retain the above copyright
 *    notice, this list of conditions and the following disclaimer.
 * 2. Redistributions in binary form must reproduce the above copyright
 *    notice, this list of conditions and the following disclaimer in the
 *    documentation and/or other materials provided with the distribution.
 * 3. All advertising materials mentioning features or use of this software
 *    must display the following acknowledgement:
 *       This product includes software developed by the NetBSD
 *       Foundation, Inc. and its contributors.
 * 4. Neither the name of The NetBSD Foundation nor the names of its
 *    contributors may be used to endorse or promote products derived
 *    from this software without specific prior written permission.
 *
 * THIS SOFTWARE IS PROVIDED BY THE NETBSD FOUNDATION, INC. AND CONTRIBUTORS
 * ``AS IS'' AND ANY EXPRESS OR IMPLIED WARRANTIES, INCLUDING, BUT NOT LIMITED
 * TO, THE IMPLIED WARRANTIES OF MERCHANTABILITY AND FITNESS FOR A PARTICULAR
 * PURPOSE ARE DISCLAIMED.  IN NO EVENT SHALL THE FOUNDATION OR CONTRIBUTORS
 * BE LIABLE FOR ANY DIRECT, INDIRECT, INCIDENTAL, SPECIAL, EXEMPLARY, OR
 * CONSEQUENTIAL DAMAGES (INCLUDING, BUT NOT LIMITED TO, PROCUREMENT OF
 * SUBSTITUTE GOODS OR SERVICES; LOSS OF USE, DATA, OR PROFITS; OR BUSINESS
 * INTERRUPTION) HOWEVER CAUSED AND ON ANY THEORY OF LIABILITY, WHETHER IN
 * CONTRACT, STRICT LIABILITY, OR TORT (INCLUDING NEGLIGENCE OR OTHERWISE)
 * ARISING IN ANY WAY OUT OF THE USE OF THIS SOFTWARE, EVEN IF ADVISED OF THE
 * POSSIBILITY OF SUCH DAMAGE.
 */
```

## University of California at Berkeley

All the source code in this book that is taken from the 4.4BSD-Lite release contains the following copyright notice.

```
/*
 * Copyright (c) 1989, 1993
 *       The Regents of the University of California.  All rights reserved.
 *
 * This code is derived from software contributed
 * to Berkeley by John Heidemann of the UCLA Ficus project.
 *
```

```
 * Source: * @(#)i405_init.c 2.10 92/04/27 UCLA Ficus project
 *
 * Redistribution and use in source and binary forms, with or without
 * modification, are permitted provided that the following conditions
 * are met:
 * 1. Redistributions of source code must retain the above copyright
 *    notice, this list of conditions and the following disclaimer.
 * 2. Redistributions in binary form must reproduce the above copyright
 *    notice, this list of conditions and the following disclaimer in the
 *    documentation and/or other materials provided with the distribution.
 * 3. All advertising materials mentioning features or use of this software
 *    must display the following acknowledgement:
 *       This product includes software developed by the University of
 *       California, Berkeley and its contributors.
 * 4. Neither the name of the University nor the names of its contributors
 *    may be used to endorse or promote products derived from this software
 *    without specific prior written permission.
 *
 * THIS SOFTWARE IS PROVIDED BY THE REGENTS AND CONTRIBUTORS ``AS IS'' AND
 * ANY EXPRESS OR IMPLIED WARRANTIES, INCLUDING, BUT NOT LIMITED TO, THE
 * IMPLIED WARRANTIES OF MERCHANTABILITY AND FITNESS FOR A PARTICULAR PURPOSE
 * ARE DISCLAIMED.  IN NO EVENT SHALL THE REGENTS OR CONTRIBUTORS BE LIABLE
 * FOR ANY DIRECT, INDIRECT, INCIDENTAL, SPECIAL, EXEMPLARY, OR CONSEQUENTIAL
 * DAMAGES (INCLUDING, BUT NOT LIMITED TO, PROCUREMENT OF SUBSTITUTE GOODS
 * OR SERVICES; LOSS OF USE, DATA, OR PROFITS; OR BUSINESS INTERRUPTION)
 * HOWEVER CAUSED AND ON ANY THEORY OF LIABILITY, WHETHER IN CONTRACT, STRICT
 * LIABILITY, OR TORT (INCLUDING NEGLIGENCE OR OTHERWISE) ARISING IN ANY WAY
 * OUT OF THE USE OF THIS SOFTWARE, EVEN IF ADVISED OF THE POSSIBILITY OF
 * SUCH DAMAGE.
 *
 *      @(#)vfs_init.c  8.5 (Berkeley) 5/11/95
 */
```

## Washington University

UVM code contains the following copyright notice.

```
/*
 *
 * Copyright (c) 1997 Charles D. Cranor and Washington University.
 * All rights reserved.
 *
 * Redistribution and use in source and binary forms, with or without
 * modification, are permitted provided that the following conditions
 * are met:
 * 1. Redistributions of source code must retain the above copyright
 *    notice, this list of conditions and the following disclaimer.
 * 2. Redistributions in binary form must reproduce the above copyright
 *    notice, this list of conditions and the following disclaimer in the
 *    documentation and/or other materials provided with the distribution.
 * 3. All advertising materials mentioning features or use of this software
 *    must display the following acknowledgement:
 *       This product includes software developed by Charles D. Cranor and
 *       Washington University.
```

```
 * 4. The name of the author may not be used to endorse or promote products
 *    derived from this software without specific prior written permission.
 *
 * THIS SOFTWARE IS PROVIDED BY THE AUTHOR ``AS IS'' AND ANY EXPRESS OR
 * IMPLIED WARRANTIES, INCLUDING, BUT NOT LIMITED TO, THE IMPLIED WARRANTIES
 * OF MERCHANTABILITY AND FITNESS FOR A PARTICULAR PURPOSE ARE DISCLAIMED.
 * IN NO EVENT SHALL THE AUTHOR BE LIABLE FOR ANY DIRECT, INDIRECT,
 * INCIDENTAL, SPECIAL, EXEMPLARY, OR CONSEQUENTIAL DAMAGES (INCLUDING, BUT
 * NOT LIMITED TO, PROCUREMENT OF SUBSTITUTE GOODS OR SERVICES; LOSS OF USE,
 * DATA, OR PROFITS; OR BUSINESS INTERRUPTION) HOWEVER CAUSED AND ON ANY
 * THEORY OF LIABILITY, WHETHER IN CONTRACT, STRICT LIABILITY, OR TORT
 * (INCLUDING NEGLIGENCE OR OTHERWISE) ARISING IN ANY WAY OUT OF THE USE OF
 * THIS SOFTWARE, EVEN IF ADVISED OF THE POSSIBILITY OF SUCH DAMAGE.
 */
```

# Part I

# Basics to Learn Filesystem

# Chapter 1

# Welcome to the World of Kernel !

In this chapter, the procedure involved in mounting root filesystem is described.

## 1.1 How Does a System Call Dive into Kernel from User Program ?

In this section, we present how a system call works with an example using a filesystem related system call.

### 1.1.1 Example: `write` system call

Let's see how a system call such as `write`, used in the below user application program, is processed.

──────────────────────────────────────── hello.c

```
main() {
    char *string = "Hello World ?";

    write(0, string, strlen(string));
}
```

──────────────────────────────────────── hello.c

`write` function is defined in `libc` library of GNU C compiler. For sparc64 platform, the source code of `write` function in `libc` library is shown below.

──────────────────────────────────────── src/lib/libc/obj/write.S

```
1 #include "SYS.h"
2 RSYSCALL(write)
```

──────────────────────────────────────── src/lib/libc/obj/write.S

RSYSCALL macro used in the above platform-independent `write.S` code is defined in a machine independent way in `libc` library source, `src/lib/libc/arch/sparc64/SYS.h`.

──────────────────────────────────────── src/lib/libc/arch/sparc64/SYS.h

```
87 /*
88  * RSYSCALL is used when the system call should just return.  Here
89  * we use the SYSCALL_G7RFLAG to put the 'success' return address in %g7
90  * and avoid a branch.
91  */
92 #define RSYSCALL(x) \
93          ENTRY(x); mov (_CAT(SYS_,x))|SYSCALL_G7RFLAG,%g1; add %o7,8,%g7; \
94          t ST_SYSCALL; ERROR()
/src/lib/libc/arch/sparc64
```

—————————————————————————— src/lib/libc/arch/sparc64/SYS.h

Thus, `write` system call used in `hello.c` executes assembly code of `line 93` where argument `x` is replaced with `write`. As a result, system call number is stored `%g1` register, and Ultra SPARC CPU Trap is occured by `t` instruction.

## 1.1.2  Ultra SPARC `0x7c` CPU Trap

Ultra SPARC CPU trap number, `ST_SYSCALL` is defined as

————————————————————————— arch/sparc64/include/trap.h

```
106 #define T_SUN_SYSCALL    0x100   /* system call */
...
150 #define ST_SYSCALL       (T_SUN_SYSCALL & 0x7f)
```

————————————————————————— arch/sparc64/include/trap.h

How Ultra SPARC CPU trap is processed is determined by CPU initialization stage during bootstrap, according to `arch/sparc64/sparc64/locore.s`. This part of the kernel source code is listed below.

————————————————————————— arch/sparc64/include/trap.h

```
636 #define SYSCALL          VTRAP(0x100, syscall_setup)
...
...
...
805          .globl  _C_LABEL(trapbase)
806 _C_LABEL(trapbase):
807          b dostart; nop; TA8     ! 000 = reserved -- Use it to boot
808          /* We should not get the next 5 traps */
809          UTRAP(0x001)            ! 001 = POR Reset -- ROM should get this
810          UTRAP(0x002)            ! 002 = WDR -- ROM should get this
811          UTRAP(0x003)            ! 003 = XIR -- ROM should get this
...
1010         UTRAP(0x0fc); TA32      ! 0x0fc fill_7_other
1011 TABLE(syscall):
1012         SYSCALL                 ! 0x100 = sun syscall
```

————————————————————————— arch/sparc64/sparc64/locore.s

Remember that `write` function defined in `libc` library requests CPU Trap, `ST_SYSCALL` that is defined as `0x7c`. So, according to **line 1012** of `arch/sparc64/include/trap.h`, jump to `syscall_setup` label is made.

Source code from the `syscall_setup` is shown below.

――――――――――――――――――――――――――― - arch/sparc64/sparc64/locore.s

```
3905 /*
3906  * syscall_setup() builds a trap frame and calls syscall().
3907  * sun_syscall is same but delivers sun system call number
3908  * XXX  should not have to save&reload ALL the registers just for
3909  * ptrace...
3910  */
3911 syscall_setup:
3912 #ifdef TRAPS_USE_IG
3913     wrpr    %g0, PSTATE_KERN|PSTATE_IG, %pstate ! DEBUG
3914 #endif
3915     TRAP_SETUP(-CC64FSZ-TF_SIZE)
3916
3917 #ifdef DEBUG
3918     rdpr    %tt, %o1    ! debug
3919     sth %o1, [%sp + CC64FSZ + STKB + TF_TT]! debug
3920 #endif
3921
3922     wrpr    %g0, PSTATE_KERN, %pstate   ! Get back to normal globals
3923     stx %g1, [%sp + CC64FSZ + STKB + TF_G + ( 1*8)]
3924     mov %g1, %o1            ! code
3925     rdpr    %tpc, %o2         ! (pc)
3926     stx %g2, [%sp + CC64FSZ + STKB + TF_G + ( 2*8)]
3927     rdpr    %tstate, %g1
3928     stx %g3, [%sp + CC64FSZ + STKB + TF_G + ( 3*8)]
3929     rdpr    %tnpc, %o3
3930     stx %g4, [%sp + CC64FSZ + STKB + TF_G + ( 4*8)]
3931     rd  %y, %o4
3932     stx %g5, [%sp + CC64FSZ + STKB + TF_G + ( 5*8)]
3933     stx %g6, [%sp + CC64FSZ + STKB + TF_G + ( 6*8)]
3934     CHKPT(%g5,%g6,0x31)
3935     wrpr    %g0, 0, %tl        ! return to tl=0
3936     stx %g7, [%sp + CC64FSZ + STKB + TF_G + ( 7*8)]
3937     add %sp, CC64FSZ + STKB, %o0    ! (&tf)
3938
3939     stx %g1, [%sp + CC64FSZ + STKB + TF_TSTATE]
3940     stx %o2, [%sp + CC64FSZ + STKB + TF_PC]
3941     stx %o3, [%sp + CC64FSZ + STKB + TF_NPC]
3942     st  %o4, [%sp + CC64FSZ + STKB + TF_Y]
3943
3944     rdpr    %pil, %g5
3945     stb %g5, [%sp + CC64FSZ + STKB + TF_PIL]
3946     stb %g5, [%sp + CC64FSZ + STKB + TF_OLDPIL]
3947
3948     !! In the EMBEDANY memory model %g4 points to the start of the data segment.
3949     !! In our case we need to clear it before calling any C-code
3950     clr %g4
3951     wr  %g0, ASI_PRIMARY_NOFAULT, %asi  ! Restore default ASI
3952
3953     call    _C_LABEL(syscall)        ! syscall(&tf, code, pc)
3954      wrpr   %g0, PSTATE_INTR, %pstate   ! turn on interrupts
3955
3956     /* see 'proc_trampoline' for the reason for this label */
```

```
3957 return_from_syscall:
3958     wrpr    %g0, PSTATE_KERN, %pstate    ! Disable intterrupts
3959     CHKPT(%o1,%o2,0x32)
3960     wrpr    %g0, 0, %tl          ! Return to tl==0
3961     CHKPT(%o1,%o2,4)
3962     ba,a,pt %icc, return_from_trap
3963      nop
3964     NOTREACHED
```

———————————————————————————————————— - arch/sparc64/sparc64/locore.s

Notice that in **line 3953**, jump to `syscall` function defined in `arch/sparc64/sparc64/trap.c`.
`trap.c` is somewhat complex since it supports system call emulation such as
Solaris or Ultra Linux. Critical part of `trap.c` managing NetBSD specific system
call is shown below.

———————————————————————————————————— - arch/sparc64/sparc64/trap.s

```
1721 void
1722 syscall(tf, code, pc)
1723     register_t code;
1724     struct trapframe64 *tf;
1725     register_t pc;
1726 {
1727     int i, nsys, nap;
1728     int64_t *ap;
1729     const struct sysent *callp;
1730     struct proc *p;
1731     int error = 0, new;
1732     union args {
1733         register32_t i[8];
1734         register64_t l[8];
1735     } args;
...
1766     p = curproc;
...
1780     callp = p->p_emul->e_sysent;
1781     nsys = p->p_emul->e_nsysent;
...
1851         callp += code;
...

1876                 error = copyin((caddr_t)(u_long)tf->tf_out[6] + BIAS +
1877                         offsetof(struct frame64, fr_argx),
1878                         &args.l[nap],
1879                         (i - nap) * sizeof(register64_t));
...
1997     error = (*callp->sy_call)(p, &args, rval);
```

———————————————————————————————————— - arch/sparc64/sparc64/trap.s

By **line 1997**, a function pointed by a function pointer is called. The function
pointer is set by **line 1780** and **line 1851**. To explain what this function pointer
means, kernel structure for a process should briefly described.

Kernel structure to describe a process is `struct proc` and it contains so called
`per-process emulation information` in `const struct emul *p_emul` structure.

——————————————————————————————— sys/proc.h

```
149 struct proc {
...
173         pid_t           p_pid;          /* Process identifier */
...
218         const struct emul *p_emul;      /* Emulation information */
219         void            *p_emuldata;    /*
...
264 };
```

——————————————————————————————— sys/proc.h

This member is used to run a SUN Solaris or Linux binary on NetBSD/sparc64.
However, for native NetBSD/sparc64 binary, this member is initialized by `kern/init_main.c`
to point a `const struct emul emul_netbsd` structure defined in `kern/kern_exec.c`.
The source for this initialization is

——————————————————————————————— sys/proc.h

```
165 /*
166  * System startup; initialize the world, create process 0, mount root
167  * filesystem, and fork to create init and pagedaemon.  Most of the
168  * hard work is done in the lower-level initialization routines including
169  * startup(), which does memory initialization and autoconfiguration.
170  */
171 void
172 main(void)
173 {
174         struct proc *p;
...
188         /*
189          * Initialize the current process pointer (curproc) before
190          * any possible traps/probes to simplify trap processing.
191          */
192         simple_lock_init(&proc0.p_raslock);
193         p = &proc0;
194         curproc = p;
...
274         p->p_emul = &emul_netbsd;
...
545         /*
546          * Okay, now we can let init(8) exec!  It's off to userland!
547          */
548         start_init_exec = 1;
549         wakeup((void *)&start_init_exec);
550
551         /* The scheduler is an infinite loop. */
552         uvm_scheduler();
553         /* NOTREACHED */
554 }
```

——————————————————————————————— sys/proc.h

and,

———————————————————————————————— kern/kern_exec.h

```
128 const struct emul emul_netbsd = {
129         "netbsd",
130         NULL,           /* emulation path */
131 #ifndef __HAVE_MINIMAL_EMUL
132         EMUL_HAS_SYS___syscall,
133         NULL,
134         SYS_syscall,
135         SYS_NSYSENT,
136 #endif
137         sysent,
138 #ifdef SYSCALL_DEBUG
139         syscallnames,
140 #else
141         NULL,
142 #endif
143         sendsig,
144         trapsignal,
145         sigcode,
146         esigcode,
147         setregs,
148         NULL,
149         NULL,
150         NULL,
151 #ifdef __HAVE_SYSCALL_INTERN
152         syscall_intern,
153 #else
154         syscall,
155 #endif
156         NULL,
157         NULL,
158 };
```

———————————————————————————————— kern/kern_exec.h

where the definition of `struct emul` structure is

———————————————————————————————— kern/kern_exec.h

```
 94 struct emul {
 95         const char      *e_name;        /* Symbolic name */
 96         const char      *e_path;        /* Extra emulation path (NULL if none)*/
 97 #ifndef __HAVE_MINIMAL_EMUL
 98         int             e_flags;        /* Miscellaneous flags, see above */
 99                                         /* Syscall handling function */
100         const int       *e_errno;       /* Errno array */
101         int             e_nosys;        /* Offset of the nosys() syscall */
102         int             e_nsysent;      /* Number of system call entries */
103 #endif
104         const struct sysent *e_sysent;  /* System call array */
105         const char * const *e_syscallnames; /* System call name array */
106                                         /* Signal sending function */
107         void            (*e_sendsig) __P((int, sigset_t *, u_long));
108         void            (*e_trapsignal) __P((struct proc *, int, u_long));
```

```
109        char          *e_sigcode;      /* Start of sigcode */
110        char          *e_esigcode;     /* End of sigcode */
111                                        /* Set registers before execution */
112        void          (*e_setregs) __P((struct proc *, struct exec_package *,
113                           u_long));
114
115                                        /* Per-process hooks */
116        void          (*e_proc_exec) __P((struct proc *,
117                               struct exec_package *));
118        void          (*e_proc_fork) __P((struct proc *p,
119                                struct proc *parent));
120        void          (*e_proc_exit) __P((struct proc *));
121
122 #ifdef __HAVE_SYSCALL_INTERN
123        void          (*e_syscall_intern) __P((struct proc *));
124 #else
125        void          (*e_syscall) __P((void));
126 #endif
127                                        /* Emulation specific sysctl */
128        int           (*e_sysctl) __P((int *, u_int , void *, size_t *,
129                           void *, size_t, struct proc *p));
130        int           (*e_fault) __P((struct proc *, vaddr_t, int, int));
131 };
```

———————————————————————————————— kern/kern_exec.h

The `emul_netbsd` structure has a member whose definition is `const struct sysent *e_sysent` and this member points, by the initialization of `kern/kern_exec.c`, to `struct sysent sysent[]` array of structure which is defined in `init_sysent.c` as

———————————————————————————————— kern/init_sysent.h

```
 71 struct sysent sysent[] = {
 72        { 0, 0, 0,
 73            sys_nosys },                 /* 0 = syscall (indir) */
 74        { 1, s(struct sys_exit_args), 0,
 75            sys_exit },                  /* 1 = exit */
 76        { 0, 0, 0,
 77            sys_fork },                  /* 2 = fork */
 78        { 3, s(struct sys_read_args), 0,
 79            sys_read },                  /* 3 = read */
 80        { 3, s(struct sys_write_args), 0,
 81            sys_write },                 /* 4 = write */
 82        { 3, s(struct sys_open_args), 0,
 83            sys_open },                  /* 5 = open */
...
1227           sys_nosys },                 /* 511 = filler */
1228 };
```

———————————————————————————————— kern/init_sysent.h

where `struct sysent` is defined as

———————————————————————————————— kern/init_sysent.h

```
131 extern struct sysent {        /* system call table */
132        short   sy_narg;       /* number of args */
133        short   sy_argsize;    /* total size of arguments */
134        int     sy_flags;      /* flags. see below */
135        sy_call_t *sy_call;    /* implementing function */
136 } sysent[];
```

——————————————————————————————— kern/init_sysent.h

Now, based on the description up to now, we can exactly understand what the **line 1997** means. Actually, It means that jump to the `sys_write` function.

### 1.1.3  Jump to the File Descriptor Layer

`sys_write` function is defined in `sys_generic.c` as

——————————————————————————————— kern/sys_generic.c

```
278 /*
279  * Write system call
280  */
281 int
282 sys_write(struct proc *p, void *v, register_t *retval)
283 {
284        struct sys_write_args /* {
285                syscallarg(int)                fd;
286                syscallarg(const void *)       buf;
287                syscallarg(size_t)             nbyte;
288        } */ *uap = v;
289        int             fd;
290        struct file     *fp;
291        struct filedesc *fdp;
292
293        fd = SCARG(uap, fd);
294        fdp = p->p_fd;
295
296        if ((fp = fd_getfile(fdp, fd)) == NULL)
297                return (EBADF);
298
299        if ((fp->f_flag & FWRITE) == 0)
300                return (EBADF);
301
302        FILE_USE(fp);
303
304        /* dofilewrite() will unuse the descriptor for us */
305        return (dofilewrite(p, fd, fp, SCARG(uap, buf), SCARG(uap, nbyte),
306            &fp->f_offset, FOF_UPDATE_OFFSET, retval));
307 }
308
309 int
310 dofilewrite(struct proc *p, int fd, struct file *fp, const void *buf,
311        size_t nbyte, off_t *offset, int flags, register_t *retval)
312 {
313        struct uio      auio;
314        struct iovec    aiov;
```

```
315        size_t          cnt;
316        int             error;
317 #ifdef KTRACE
318        struct iovec    ktriov;
319 #endif
320
321        error = 0;
322        aiov.iov_base = (caddr_t)buf;          /* XXX kills const */
323        aiov.iov_len = nbyte;
324        auio.uio_iov = &aiov;
325        auio.uio_iovcnt = 1;
326        auio.uio_resid = nbyte;
327        auio.uio_rw = UIO_WRITE;
328        auio.uio_segflg = UIO_USERSPACE;
329        auio.uio_procp = p;
330
331        /*
332         * Writes return ssize_t because -1 is returned on error.  Therefore
333         * we must restrict the length to SSIZE_MAX to avoid garbage return
334         * values.
335         */
336        if (auio.uio_resid > SSIZE_MAX) {
337                error = EINVAL;
338                goto out;
339        }
340
341 #ifdef KTRACE
342        /*
343         * if tracing, save a copy of iovec
344         */
345        if (KTRPOINT(p, KTR_GENIO))
346                ktriov = aiov;
347 #endif
348        cnt = auio.uio_resid;
349        error = (*fp->f_ops->fo_write)(fp, offset, &auio, fp->f_cred, flags);
350        if (error) {
351                if (auio.uio_resid != cnt && (error == ERESTART ||
352                    error == EINTR || error == EWOULDBLOCK))
353                        error = 0;
354                if (error == EPIPE)
355                        psignal(p, SIGPIPE);
356        }
357        cnt -= auio.uio_resid;
358 #ifdef KTRACE
359        if (KTRPOINT(p, KTR_GENIO) && error == 0)
360                ktrgenio(p, fd, UIO_WRITE, &ktriov, cnt, error);
361 #endif
362        *retval = cnt;
363 out:
364        FILE_UNUSE(fp, p);
365        return (error);
366 }
```

Do you think it is the whole kernel source code to process `write` system call ? Unfortunately, there remains somewhat long way for us to walk before we reach the realm of the fast filesystem code. / :)

See the **line 349** of `kern/sys/generic.c`. You may wonder how the `f_ops` member of `fp` structure is set. It is initialized when `open` system call is executed as,

──────────────────────────────────────────────────── kern/vfs_syscalls.c

```
986 /*
987  * Check permissions, allocate an open file structure,
988  * and call the device open routine if any.
989  */
990 int
991 sys_open(p, v, retval)
992         struct proc *p;
993         void *v;
994         register_t *retval;
995 {
996         struct sys_open_args /* {
997                 syscallarg(const char *) path;
998                 syscallarg(int) flags;
999                 syscallarg(int) mode;
1000        } */ *uap = v;
1001        struct cwdinfo *cwdi = p->p_cwdi;
1002        struct filedesc *fdp = p->p_fd;
1003        struct file *fp;
1004        struct vnode *vp;
1005        int flags, cmode;
1006        int type, indx, error;
1007        struct flock lf;
1008        struct nameidata nd;
1009
1010        flags = FFLAGS(SCARG(uap, flags));
1011        if ((flags & (FREAD | FWRITE)) == 0)
1012                return (EINVAL);
1013        /* falloc() will use the file descriptor for us */
1014        if ((error = falloc(p, &fp, &indx)) != 0)
1015                return (error);
1016        cmode = ((SCARG(uap, mode) &~ cwdi->cwdi_cmask) & ALLPERMS) &~ S_ISTXT;
1017        NDINIT(&nd, LOOKUP, FOLLOW, UIO_USERSPACE, SCARG(uap, path), p);
1018        p->p_dupfd = -indx - 1;                 /* XXX check for fdopen */
1019        if ((error = vn_open(&nd, flags, cmode)) != 0) {
1020                FILE_UNUSE(fp, p);
1021                ffree(fp);
1022                if ((error == ENODEV || error == ENXIO) &&
1023                    p->p_dupfd >= 0 &&                  /* XXX from fdopen */
1024                    (error =
1025                        dupfdopen(p, indx, p->p_dupfd, flags, error)) == 0) {
1026                        *retval = indx;
1027                        return (0);
1028                }
1029                if (error == ERESTART)
1030                        error = EINTR;
```

```
1031                    fdremove(fdp, indx);
1032                    return (error);
1033            }
1034        p->p_dupfd = 0;
1035        vp = nd.ni_vp;
1036        fp->f_flag = flags & FMASK;
1037        fp->f_type = DTYPE_VNODE;
1038        fp->f_ops = &vnops;
1039        fp->f_data = (caddr_t)vp;
1040        if (flags & (O_EXLOCK | O_SHLOCK)) {
1041                lf.l_whence = SEEK_SET;
1042                lf.l_start = 0;
1043                lf.l_len = 0;
1044                if (flags & O_EXLOCK)
1045                        lf.l_type = F_WRLCK;
1046                else
1047                        lf.l_type = F_RDLCK;
1048                type = F_FLOCK;
1049                if ((flags & FNONBLOCK) == 0)
1050                        type |= F_WAIT;
1051                VOP_UNLOCK(vp, 0);
1052                error = VOP_ADVLOCK(vp, (caddr_t)fp, F_SETLK, &lf, type);
1053                if (error) {
1054                        (void) vn_close(vp, fp->f_flag, fp->f_cred, p);
1055                        FILE_UNUSE(fp, p);
1056                        ffree(fp);
1057                        fdremove(fdp, indx);
1058                        return (error);
1059                }
1060                vn_lock(vp, LK_EXCLUSIVE | LK_RETRY);
1061                fp->f_flag |= FHASLOCK;
1062        }
1063        VOP_UNLOCK(vp, 0);
1064        *retval = indx;
1065        FILE_SET_MATURE(fp);
1066        FILE_UNUSE(fp, p);
1067        return (0);
1068 }
```

———————————————————————————————————— kern/vfs_syscalls.c

You can check that this code segment is described by the page 205-207 of a book titled as '*the design and implementation of the 4.4BSD operating system*'

For more important, see **1038** of `vfs_syscalls.c`. Did you have a sense what this means ? By this code line, the `f_ops` member of `fp` structure in **line 349** of `kern/sys_generic.c` points `vnops` global variable which is defined as,

———————————————————————————————————— kern/vfs_vnops.c

```
82 struct  fileops vnops = {
83         vn_read, vn_write, vn_ioctl, vn_fcntl, vn_poll,
84         vn_statfile, vn_closefile, vn_kqfilter
85 };
```

———————————————————————————————————— kern/vfs_vnops.c

where the definition of `struct fileops` is embedded in the definition of `file` structure as,

——————————————————————————————————— sys/file.h

```
53 /*
54  * Kernel descriptor table.
55  * One entry for each open kernel vnode and socket.
56  */
57 struct file {
58         LIST_ENTRY(file) f_list;          /* list of active files */
59         int              f_flag;          /* see fcntl.h */
60         int              f_iflags;        /* internal flags */
61 #define DTYPE_VNODE    1                   /* file */
62 #define DTYPE_SOCKET   2                   /* communications endpoint */
63 #define DTYPE_PIPE     3                   /* pipe */
64 #define DTYPE_KQUEUE   4                   /* event queue */
65 #define DTYPE_MISC     5                   /* misc file descriptor type */
66         int              f_type;          /* descriptor type */
67         u_int            f_count;         /* reference count */
68         u_int            f_msgcount;      /* references from message queue */
69         int              f_usecount;      /* number active users */
70         struct ucred     *f_cred;         /* creds associated with descriptor */
71         struct fileops {
72                 int    (*fo_read)    (struct file *fp, off_t *offset,
73                                          struct uio *uio,
74                                          struct ucred *cred, int flags);
75                 int    (*fo_write)   (struct file *fp, off_t *offset,
76                                          struct uio *uio,
77                                          struct ucred *cred, int flags);
78                 int    (*fo_ioctl)   (struct file *fp, u_long com,
79                                          caddr_t data, struct proc *p);
80                 int    (*fo_fcntl)   (struct file *fp, u_int com,
81                                          caddr_t data, struct proc *p);
82                 int    (*fo_poll)    (struct file *fp, int events,
83                                          struct proc *p);
84                 int    (*fo_stat)    (struct file *fp, struct stat *sp,
85                                          struct proc *p);
86                 int    (*fo_close)   (struct file *fp, struct proc *p);
87                 int    (*fo_kqfilter) (struct file *fp, struct knote *kn);
88         } *f_ops;
89         off_t            f_offset;
90         caddr_t          f_data;          /* descriptor data, e.g. vnode/socket */
91 };
```

——————————————————————————————————— sys/file.h

Based on the above code, **line 349** of `kern_sysgeneric.c` makes a jump to `vn_write`.

## 1.1.4   Arriving at Virtual Filesystem Operations

The `vn_write` function is defined in `vfs_vnops.c` as,

——————————————————————————————————— kern/vfs_vnops.c

```
526 /*
527  * File table vnode write routine.
528  */
529 static int
530 vn_write(fp, offset, uio, cred, flags)
531         struct file *fp;
532         off_t *offset;
533         struct uio *uio;
534         struct ucred *cred;
535         int flags;
536 {
537         struct vnode *vp = (struct vnode *)fp->f_data;
538         int count, error, ioflag = IO_UNIT;
539
540         if (vp->v_type == VREG && (fp->f_flag & O_APPEND))
541                 ioflag |= IO_APPEND;
542         if (fp->f_flag & FNONBLOCK)
543                 ioflag |= IO_NDELAY;
544         if (fp->f_flag & FFSYNC ||
545             (vp->v_mount && (vp->v_mount->mnt_flag & MNT_SYNCHRONOUS)))
546                 ioflag |= IO_SYNC;
547         else if (fp->f_flag & FDSYNC)
548                 ioflag |= IO_DSYNC;
549         if (fp->f_flag & FALTIO)
550                 ioflag |= IO_ALTSEMANTICS;
551         VOP_LEASE(vp, uio->uio_procp, cred, LEASE_WRITE);
552         vn_lock(vp, LK_EXCLUSIVE | LK_RETRY);
553         uio->uio_offset = *offset;
554         count = uio->uio_resid;
555         error = VOP_WRITE(vp, uio, ioflag, cred);
556         if (flags & FOF_UPDATE_OFFSET) {
557                 if (ioflag & IO_APPEND)
558                         *offset = uio->uio_offset;
559                 else
560                         *offset += count - uio->uio_resid;
561         }
562         VOP_UNLOCK(vp, 0);
563         return (error);
564 }
```

──────────────────────────────────── kern/vfs_vnops.c

By the functions used in line **line 551, 555** — `VOP_LEASE`, `VOP_WRITE` — are calls to virtual filesystem operations. Before describing the jump to virtual file system code by this function, we should explain architecture and source code for virtual filesystem layer in NetBSD/sparc64. Therefore, we postpone further description to the next chapter telling about virtual filesystem layer implementation.

Starting from `write` system call in `hello.c`, we arrived just before the filesystem code. Isn't it interesting ?

## 1.2   General Data Structures in Kernel such as List, Hash, Queue, ...

In NetBSD, general data structure manipulation macros are provided. Conceptually, they are equivalent to templates of C++ language. To use these macro, the only thing to do is including `sys/queue.h` header file.

Those built-in macros in NetBSD supports five types of data structures: singly-linked lists, linked-lists, simple queues, tail queues, and circular queues. They are used by the various parts of kernel. For example, *buffer cache* uses lists and tail queues.

All five data structures support the following four functionality:

- Insertion of a new entry at the head of the list
- Insertion of a new entry before or after any element in the list
- Removal of any entry in the list
- Forward traversal through the list

All doubly linked types of data structures (lists, tail queues, and circle queues) additionally allow:

- Insertion of a new entry before any element in the list.
- O(1) removal of any entry in the list.

However, code size and execution time of operations (except for removal) is about twice that of the singly-linked data structures.

### 1.2.1   Linked-Lists

Linked lists are the simplest of the doubly linked data structures.

Here is an example using linked lists.

**An Example**

```
LIST_HEAD(listhead, entry) head;
struct listhead *headp;          /* List head. */
struct entry {
        ...
        LIST_ENTRY(entry) entries;     /* List. */
        ...
} *n1, *n2, *np;

LIST_INIT(&head);                       /* Initialize the list. */

n1 = malloc(sizeof(struct entry));      /* Insert at the head. */
LIST_INSERT_HEAD(&head, n1, entries);

n2 = malloc(sizeof(struct entry));      /* Insert after. */
LIST_INSERT_AFTER(n1, n2, entries);

n2 = malloc(sizeof(struct entry));      /* Insert before. */
LIST_INSERT_BEFORE(n1, n2, entries);
                                        /* Forward traversal. */
```

```
LIST_FOREACH(np, &head, entries)
        np-> ...
                                        /* Delete. */
while (LIST_FIRST(&head) != NULL)
        LIST_REMOVE(LIST_FIRST(&head), entries);
if (LIST_EMPTY(&head))                  /* Test for emptiness. */
        printf("nothing to do\n");
```

From now on, with this example, we will describe how to use built-in macros about linked-lists.

### List Definition

A list is headed by a structure defined by the `LIST_HEAD` macro. This macro is defined in `sys/queue.h` as,

———————————————————————————————————— sys/queue.h

```
87 /*
88  * List definitions.
89  */
90 #define LIST_HEAD(name, type)                               \
91 struct name {                                               \
92         struct type *lh_first;  /* first element */         \
93 }
```

———————————————————————————————————— sys/queue.h

This structure contains a single pointer to the first element on the list. The elements are doubly linked so that an arbitrary element can be removed without traversing the list. New elements can be added to the list after an existing element, before an existing element, or at the head of the list. A `LIST_HEAD` structure is declared as follows:

```
LIST_HEAD(HEADNAME, TYPE) head;
```

where `HEADNAME` is the name of the structure to be defined, and `TYPE` is the type of the elements to be linked into the list. A pointer to the head of the list can later be declared as:

```
struct HEADNAME *headp;
```

(The names head and headp are user selectable.)

**Declaring Entry**

The macro `LIST_ENTRY` declares a structure that connects the elements in the list.
This macro is defined in **sys/queue.h** as,

———————————————————————————————— sys/queue.h

```
 98 #define LIST_ENTRY(type)                                             \
 99 struct {                                                             \
100        struct type *le_next;   /* next element */                    \
101        struct type **le_prev;  /* address of previous next element */ \
102 }
```

———————————————————————————————— sys/queue.h

**List Initialization**

The macro `LIST_INIT` initializes the list referenced by head. This marco is defined
as,

———————————————————————————————— sys/queue.h

```
128 #define LIST_INIT(head) do {                                         \
129        (head)->lh_first = NULL;                                      \
130 } while (/*CONSTCOND*/0)
```

———————————————————————————————— sys/queue.h

**Entry Insertion**

   `LIST_INSERT_HEAD` macro inserts the new element `elm` at the head of
      the list.

   `LIST_INSERT_AFTER` inserts the new element `elm` after the element `listelm`.

   `LIST_INSERT_BEFORE` inserts the new element `elm` before the element
      `listelm`.

Those macros are defined in **sys/queue.h** as,

———————————————————————————————— sys/queue.h

```
132 #define LIST_INSERT_AFTER(listelm, elm, field) do {                  \
133        QUEUEDEBUG_LIST_OP((listelm), field)                          \
134        if (((elm)->field.le_next = (listelm)->field.le_next) != NULL) \
135                (listelm)->field.le_next->field.le_prev =             \
136                        &(elm)->field.le_next;                        \
137        (listelm)->field.le_next = (elm);                             \
138        (elm)->field.le_prev = &(listelm)->field.le_next;             \
139 } while (/*CONSTCOND*/0)
140
141 #define LIST_INSERT_BEFORE(listelm, elm, field) do {                 \
142        QUEUEDEBUG_LIST_OP((listelm), field)                          \
143        (elm)->field.le_prev = (listelm)->field.le_prev;              \
144        (elm)->field.le_next = (listelm);                             \
145        *(listelm)->field.le_prev = (elm);                            \
146        (listelm)->field.le_prev = &(elm)->field.le_next;             \
147 } while (/*CONSTCOND*/0)
148
```

```
149 #define LIST_INSERT_HEAD(head, elm, field) do {               \
150         QUEUEDEBUG_LIST_INSERT_HEAD((head), (elm), field)       \
151         if (((elm)->field.le_next = (head)->lh_first) != NULL)  \
152                 (head)->lh_first->field.le_prev = &(elm)->field.le_next;\
153         (head)->lh_first = (elm);                               \
154         (elm)->field.le_prev = &(head)->lh_first;               \
155 } while (/*CONSTCOND*/0)
```

———————————————————————————————— sys/queue.h

From the definition, macros beginning with `QUEUEDEBUG` is assertion macros. They
are meaningful only if `QUEUEDEBUG` macro is defined. These macros are defined as

———————————————————————————————— sys/queue.h

```
107 #if defined(_KERNEL) && defined(QUEUEDEBUG)
108 #define QUEUEDEBUG_LIST_INSERT_HEAD(head, elm, field)          \
109         if ((head)->lh_first &&                                 \
110             (head)->lh_first->field.le_prev != &(head)->lh_first)  \
111                 panic("LIST_INSERT_HEAD %p %s:%d", (head), __FILE__, __LINE__);
112 #define QUEUEDEBUG_LIST_OP(elm, field)                          \
113         if ((elm)->field.le_next &&                             \
114             (elm)->field.le_next->field.le_prev !=              \
115             &(elm)->field.le_next)                              \
116                 panic("LIST_* forw %p %s:%d", (elm), __FILE__, __LINE__);\
117         if (*(elm)->field.le_prev != (elm))                     \
118                 panic("LIST_* back %p %s:%d", (elm), __FILE__, __LINE__);
119 #define QUEUEDEBUG_LIST_POSTREMOVE(elm, field)                  \
120         (elm)->field.le_next = (void *)1L;                      \
121         (elm)->field.le_prev = (void *)1L;
122 #else
123 #define QUEUEDEBUG_LIST_INSERT_HEAD(head, elm, field)
124 #define QUEUEDEBUG_LIST_OP(elm, field)
125 #define QUEUEDEBUG_LIST_POSTREMOVE(elm, field)
126 #endif
```

———————————————————————————————— sys/queue.h

### Entry Removal

The macro LIST_REMOVE removes the element elm from the list.

This marco is defined as

———————————————————————————————— sys/queue.h

```
157 #define LIST_REMOVE(elm, field) do {                           \
158         QUEUEDEBUG_LIST_OP((elm), field)                        \
159         if ((elm)->field.le_next != NULL)                       \
160                 (elm)->field.le_next->field.le_prev =           \
161                     (elm)->field.le_prev;                       \
162         *(elm)->field.le_prev = (elm)->field.le_next;           \
163         QUEUEDEBUG_LIST_POSTREMOVE((elm), field)                \
164 } while (/*CONSTCOND*/0)
```

———————————————————————————————— sys/queue.h

**List Access**

> **LIST_EMPTY** macro return true if the list head has no elements.
>
> **LIST_FIRST** macro returns the first element of the list head.
>
> **LIST_FOREACH** macro traverses the list referenced by head in the forward direction, assigning each element in turn to var.
>
> **LIST_NEXT** macro returns the element after the element elm.

Those macros are defined as

———————————————————————————————————— sys/queue.h

```
166 #define LIST_FOREACH(var, head, field)                          \
167         for ((var) = ((head)->lh_first);                        \
168                 (var);                                          \
169                 (var) = ((var)->field.le_next))
170
171 /*
172  * List access methods.
173  */
174 #define LIST_EMPTY(head)              ((head)->lh_first == NULL)
175 #define LIST_FIRST(head)              ((head)->lh_first)
176 #define LIST_NEXT(elm, field)         ((elm)->field.le_next)
```

———————————————————————————————————— sys/queue.h

Now, if you see again the previous example, you would fully understand how it works !

## 1.2.2   Tail Queues

Tail queues add the following one additional functionality:

> • Entries can be added at the end of a list.

However,

> • All list insertions and removals, except insertion before another element, must specify the head of the list.
>
> • Code size is about 15slower than linked-lists.

**An Example**

```
TAILQ_HEAD(tailhead, entry) head;
struct tailhead *headp;         /* Tail queue head. */
struct entry {
        ...
        TAILQ_ENTRY(entry) entries;     /* Tail queue. */
        ...
} *n1, *n2, *np;

TAILQ_INIT(&head);                      /* Initialize the queue. */

n1 = malloc(sizeof(struct entry));      /* Insert at the head. */
```

```
TAILQ_INSERT_HEAD(&head, n1, entries);

n1 = malloc(sizeof(struct entry));      /* Insert at the tail. */
TAILQ_INSERT_TAIL(&head, n1, entries);

n2 = malloc(sizeof(struct entry));      /* Insert after. */
TAILQ_INSERT_AFTER(&head, n1, n2, entries);

n2 = malloc(sizeof(struct entry));      /* Insert before. */
TAILQ_INSERT_BEFORE(n1, n2, entries);
                                        /* Forward traversal. */
TAILQ_FOREACH(np, &head, entries)
        np-> ...
                                        /* Reverse traversal. */
TAILQ_FOREACH_REVERSE(np, &head, tailhead, entries)
        np-> ...
                                        /* Delete. */
while (TAILQ_FIRST(&head) != NULL)
        TAILQ_REMOVE(&head, TAILQ_FIRST(&head), entries);
if (TAILQ_EMPTY(&head))                 /* Test for emptiness. */
        printf("nothing to do\n");
```

---

If you read the previous subsection about linked-list, you will not have any problem in understanding the above example. Therefore, instead of describing usages for each macro, we show the definition of those macros.

——————————————————————————————————— sys/queue.h

```
310 /*
311  * Tail queue definitions.
312  */
313 #define TAILQ_HEAD(name, type)                                      \
314 struct name {                                                       \
315         struct type *tqh_first; /* first element */                 \
316         struct type **tqh_last; /* addr of last next element */     \
317 }
318
319 #define TAILQ_HEAD_INITIALIZER(head)                                \
320         { NULL, &(head).tqh_first }
321
322 #define TAILQ_ENTRY(type)                                           \
323 struct {                                                            \
324         struct type *tqe_next;  /* next element */                  \
325         struct type **tqe_prev; /* address of previous next element */ \
326 }
327
328 /*
329  * Tail queue functions.
330  */
331 #if defined(_KERNEL) && defined(QUEUEDEBUG)
332 #define QUEUEDEBUG_TAILQ_INSERT_HEAD(head, elm, field)              \
333         if ((head)->tqh_first &&                                    \
334             (head)->tqh_first->field.tqe_prev != &(head)->tqh_first) \
```

```
335                     panic("TAILQ_INSERT_HEAD %p %s:%d", (head), __FILE__, __LINE__);
336 #define QUEUEDEBUG_TAILQ_INSERT_TAIL(head, elm, field)                      \
337             if (*(head)->tqh_last != NULL)                                  \
338                     panic("TAILQ_INSERT_TAIL %p %s:%d", (head), __FILE__, __LINE__);
339 #define QUEUEDEBUG_TAILQ_OP(elm, field)                                     \
340             if ((elm)->field.tqe_next &&                                    \
341                 (elm)->field.tqe_next->field.tqe_prev !=                    \
342                 &(elm)->field.tqe_next)                                     \
343                     panic("TAILQ_* forw %p %s:%d", (elm), __FILE__, __LINE__);\
344             if (*(elm)->field.tqe_prev != (elm))                           \
345                     panic("TAILQ_* back %p %s:%d", (elm), __FILE__, __LINE__);
346 #define QUEUEDEBUG_TAILQ_PREREMOVE(head, elm, field)                        \
347             if ((elm)->field.tqe_next == NULL &&                           \
348                 (head)->tqh_last != &(elm)->field.tqe_next)                \
349                     panic("TAILQ_PREREMOVE head %p elm %p %s:%d",          \
350                         (head), (elm), __FILE__, __LINE__);
351 #define QUEUEDEBUG_TAILQ_POSTREMOVE(elm, field)                             \
352             (elm)->field.tqe_next = (void *)1L;                            \
353             (elm)->field.tqe_prev = (void *)1L;
354 #else
355 #define QUEUEDEBUG_TAILQ_INSERT_HEAD(head, elm, field)
356 #define QUEUEDEBUG_TAILQ_INSERT_TAIL(head, elm, field)
357 #define QUEUEDEBUG_TAILQ_OP(elm, field)
358 #define QUEUEDEBUG_TAILQ_PREREMOVE(head, elm, field)
359 #define QUEUEDEBUG_TAILQ_POSTREMOVE(elm, field)
360 #endif
361
362 #define TAILQ_INIT(head) do {                                               \
363             (head)->tqh_first = NULL;                                       \
364             (head)->tqh_last = &(head)->tqh_first;                          \
365 } while (/*CONSTCOND*/0)
366
367 #define TAILQ_INSERT_HEAD(head, elm, field) do {                            \
368             QUEUEDEBUG_TAILQ_INSERT_HEAD((head), (elm), field)             \
369             if (((elm)->field.tqe_next = (head)->tqh_first) != NULL)       \
370                     (head)->tqh_first->field.tqe_prev =                    \
371                         &(elm)->field.tqe_next;                            \
372             else                                                           \
373                     (head)->tqh_last = &(elm)->field.tqe_next;             \
374             (head)->tqh_first = (elm);                                     \
375             (elm)->field.tqe_prev = &(head)->tqh_first;                    \
376 } while (/*CONSTCOND*/0)
377
378 #define TAILQ_INSERT_TAIL(head, elm, field) do {                            \
379             QUEUEDEBUG_TAILQ_INSERT_TAIL((head), (elm), field)             \
380             (elm)->field.tqe_next = NULL;                                  \
381             (elm)->field.tqe_prev = (head)->tqh_last;                      \
382             *(head)->tqh_last = (elm);                                     \
383             (head)->tqh_last = &(elm)->field.tqe_next;                     \
384 } while (/*CONSTCOND*/0)
385
386 #define TAILQ_INSERT_AFTER(head, listelm, elm, field) do {                  \
387             QUEUEDEBUG_TAILQ_OP((listelm), field)                          \
388             if (((elm)->field.tqe_next = (listelm)->field.tqe_next) != NULL)\
```

```
389                         (elm)->field.tqe_next->field.tqe_prev =         \
390                                 &(elm)->field.tqe_next;                  \
391             else                                                        \
392                         (head)->tqh_last = &(elm)->field.tqe_next;       \
393             (listelm)->field.tqe_next = (elm);                          \
394             (elm)->field.tqe_prev = &(listelm)->field.tqe_next;         \
395 } while (/*CONSTCOND*/0)
396
397 #define TAILQ_INSERT_BEFORE(listelm, elm, field) do {                   \
398             QUEUEDEBUG_TAILQ_OP((listelm), field)                       \
399             (elm)->field.tqe_prev = (listelm)->field.tqe_prev;          \
400             (elm)->field.tqe_next = (listelm);                          \
401             *(listelm)->field.tqe_prev = (elm);                         \
402             (listelm)->field.tqe_prev = &(elm)->field.tqe_next;         \
403 } while (/*CONSTCOND*/0)
404
405 #define TAILQ_REMOVE(head, elm, field) do {                             \
406             QUEUEDEBUG_TAILQ_PREREMOVE((head), (elm), field)            \
407             QUEUEDEBUG_TAILQ_OP((elm), field)                           \
408             if (((elm)->field.tqe_next) != NULL)                        \
409                         (elm)->field.tqe_next->field.tqe_prev =         \
410                                 (elm)->field.tqe_prev;                  \
411             else                                                        \
412                         (head)->tqh_last = (elm)->field.tqe_prev;       \
413             *(elm)->field.tqe_prev = (elm)->field.tqe_next;             \
414             QUEUEDEBUG_TAILQ_POSTREMOVE((elm), field);                  \
415 } while (/*CONSTCOND*/0)
416
417 /*
418  * Tail queue access methods.
419  */
420 #define TAILQ_EMPTY(head)               ((head)->tqh_first == NULL)
421 #define TAILQ_FIRST(head)               ((head)->tqh_first)
422 #define TAILQ_NEXT(elm, field)          ((elm)->field.tqe_next)
423
424 #define TAILQ_LAST(head, headname) \
425             (*(((struct headname *)((head)->tqh_last))->tqh_last))
426 #define TAILQ_PREV(elm, headname, field) \
427             (*(((struct headname *)((elm)->field.tqe_prev))->tqh_last))
428
429 #define TAILQ_FOREACH(var, head, field)                                 \
430             for ((var) = ((head)->tqh_first);                           \
431                         (var);                                          \
432                         (var) = ((var)->field.tqe_next))
433
434 #define TAILQ_FOREACH_REVERSE(var, head, headname, field)               \
435             for ((var) = (*(((struct headname *)((head)->tqh_last))->tqh_last));   \
436                         (var);                                          \
437                         (var) = (*(((struct headname *)((var)->field.tqe_prev))->tqh_last)))
```

### 1.2.3   Hash

The hash implementation in NerBSD kernel is so simple that it has only two functions: `hashinit, hashfree`. Only just read the source code will be adequate description for how to use hash in kernel.

—————————————————————————————————————— kern/kern_subr.c

```
314 /*
315  * General routine to allocate a hash table.
316  * Allocate enough memory to hold at least 'elements' list-head pointers.
317  * Return a pointer to the allocated space and set *hashmask to a pattern
318  * suitable for masking a value to use as an index into the returned array.
319  */
320 void *
321 hashinit(elements, htype, mtype, mflags, hashmask)
322         u_int elements;
323         enum hashtype htype;
324         int mtype, mflags;
325         u_long *hashmask;
326 {
327         u_long hashsize, i;
328         LIST_HEAD(, generic) *hashtbl_list;
329         TAILQ_HEAD(, generic) *hashtbl_tailq;
330         size_t esize;
331         void *p;
332
333         if (elements == 0)
334                 panic("hashinit: bad cnt");
335         for (hashsize = 1; hashsize < elements; hashsize <<= 1)
336                 continue;
337
338         switch (htype) {
339         case HASH_LIST:
340                 esize = sizeof(*hashtbl_list);
341                 break;
342         case HASH_TAILQ:
343                 esize = sizeof(*hashtbl_tailq);
344                 break;
345 #ifdef DIAGNOSTIC
346         default:
347                 panic("hashinit: invalid table type");
348 #endif
349         }
350
351         if ((p = malloc(hashsize * esize, mtype, mflags)) == NULL)
352                 return (NULL);
353
354         switch (htype) {
355         case HASH_LIST:
356                 hashtbl_list = p;
357                 for (i = 0; i < hashsize; i++)
358                         LIST_INIT(&hashtbl_list[i]);
359                 break;
360         case HASH_TAILQ:
```

```
361                    hashtbl_tailq = p;
362                    for (i = 0; i < hashsize; i++)
363                            TAILQ_INIT(&hashtbl_tailq[i]);
364                    break;
365            }
366            *hashmask = hashsize - 1;
367            return (p);
368 }
369
370 /*
371  * Free memory from hash table previosly allocated via hashinit().
372  */
373 void
374 hashdone(hashtbl, mtype)
375            void *hashtbl;
376            int mtype;
377 {
378
379            free(hashtbl, mtype);
380 }
```

——————————————————————————————— kern/kern_subr.c

## 1.3   Waiting and Sleeping in Kernel

## 1.4   Kernel Lock Manager

The lock functions provide synchronisation in the kernel by preventing multiple
threads from simultaneously executing critical sections of code accessing shared
data. A number of different locks are available:

### 1.4.1   simplelock and lock

struct simplelock
        Provides a simple spinning mutex.  A processor will busy-wait
        while trying to acquire a simplelock.  The simplelock operations
        are implemented with machine-dependent locking primitives.

        Simplelocks are usually used only by the high-level lock manager
        and to protect short, critical sections of code.  Simplelocks
        are the only locks that can be used inside an interrupt han-
        dler.  For a simplelock to be used in an interrupt handler, care
        must be taken to disable the interrupt, acquire the lock, do any
        processing, release the simplelock and re-enable the interrupt.
        This procedure is necessary to avoid deadlock between the inter-
        rupt handler and other threads executing on the same processor.

struct lock
        Provides a high-level lock supporting sleeping/spinning until
        the lock can be acquired.  The lock manager supplies both exclu-
        sive-access and shared-access locks, with recursive exclusive-
        access locks within a single thread.  It also allows upgrading a
        shared-access lock to an exclusive-access lock, as well as down-

        grading an exclusive-access lock to a shared-access lock.

If the kernel option LOCKDEBUG is enabled, additional facilities are pro-
vided to record additional lock information.  These facilities are pro-
vided to assist in determining deadlock occurrences.

## 1.4.2   Simplelock Interfaces

simple_lock_init(slock)
        The simplelock slock is initialised to the unlocked state.  A
        statically allocated simplelock also can be initialised with the
        macro SIMPLELOCK_INITIALIZER.  The effect is the same as the dy-
        namic initialisation by a call to simple_lock_init.  For exam-
        ple,

        struct simplelock slock = SIMPLELOCK_INITIALIZER;

simple_lock(slock)
        The simplelock slock is locked.  If the simplelock is held then
        execution will spin until the simplelock is acquired.  Care must
        be taken that the calling thread does not already hold the sim-
        plelock.  In this case, the simplelock can never be acquired.
        If kernel option LOCKDEBUG is enabled, a "locking against my-
        self" panic will occur.

simple_lock_try(slock)
        Try to acquire the simplelock slock without spinning.  If the
        simplelock is held by another thread then the return value is 0.
        If the simplelock was acquired successfully then the return val-
        ue is 1.

simple_lock_unlock(slock)
        The simplelock slock is unlocked.  The simplelock must be locked
        and the calling thread must be the one that last acquired the
        simplelock.  If the calling thread does not hold the simplelock,
        the simplelock will be released but the kernel behaviour is un-
        defined.

simple_lock_freecheck(start, end)
        Check that all simplelocks in the address range start to end are
        not held.  If a simplelock within the range is found, the kernel
        enters the debugger.  This function is available only with ker-
        nel option LOCKDEBUG.  It provides a mechanism for basic simple-
        lock consistency checks.

simple_lock_dump(void)
        Dump the state of all simplelocks in the kernel.  This function
        is available only with kernel option LOCKDEBUG.

## 1.4.3   Lock Interfaces

lockinit(lock, prio, wmesg, timo, flags)
        The lock lock is initialised according to the parameters provid-
        ed.  Arguments are as follows:

lock    The lock.

prio    The thread priority when it is woken up after sleeping
        on the lock.

wmesg   A sleep message used when a thread goes to sleep wait-
        ing for the lock, so that the exact reason it is sleep-
        ing can easily be identified.

timo    The maximum sleep time. Used by tsleep(9).

flags   Flags to specify the lock behaviour permanently over
        the lifetime of the lock. Valid lock flags are:

        LK_NOWAIT
                Threads should not sleep when attempting to
                acquire the lock.

        LK_SLEEPFAIL
                Threads should sleep, then return failure when
                acquiring the lock.

        LK_CANRECURSE
                Threads can acquire the lock recursively.

lockmgr(lock, flags, slock)
        Set, change or release a lock according to the parameters pro-
        vided. Arguments are as follows:

lock    The lock.

slock   Simplelock interlock. If the flag LK_INTERLOCK is set
        in flags, slock is a simplelock held by the caller.
        When the lock lock is acquired, the simplelock is re-
        leased. If the flag LK_INTERLOCK is not set, slock is
        ignored.

flags   Flags to specify the lock request type. In addition to
        the flags specified above, the following flags are
        valid:

        LK_SHARED
                Get one of many possible shared-access locks.
                If a thread holding an exclusive-access lock
                requests a shared-access lock, the exclusive-
                access lock is downgraded to a shared-access
                lock.

        LK_EXCLUSIVE
                Stop further shared-access locks, when they
                are cleared, grant a pending upgrade if it ex-
                ists, then grant an exclusive-access lock.
                Only one exclusive-access lock may exist at a

time, except that a thread holding an exclu-
sive-access lock may get additional exclusive-
access locks if it explicitly sets the LK_CAN-
RECURSE flag in the lock request, or if the
LK_CANRECURSE flag was set when the lock was
initialised.

LK_UPGRADE
          The thread must hold a shared-access lock that
          it wants to have upgraded to an exclusive-ac-
          cess lock.  Other threads may get exclusive
          access to the protected resource between the
          time that the upgrade is requested and the
          time that it is granted.

LK_EXCLUPGRADE
          The thread must hold a shared-access lock that
          it wants to have upgraded to an exclusive-ac-
          cess lock.  If the request succeeds, no other
          threads will have acquired exclusive access to
          the protected resource between the time that
          the upgrade is requested and the time that it
          is granted.  However, if another thread has
          already requested an upgrade, the request will
          fail.

LK_DOWNGRADE
          The thread must hold an exclusive-access lock
          that it wants to have downgraded to a shared-
          access lock.  If the thread holds multiple
          (recursive) exclusive-access locks, they will
          all be downgraded to shared-access locks.

LK_RELEASE
          Release one instance of a lock.

LK_DRAIN
          Wait for all activity on the lock to end, then
          mark it decommissioned.  This feature is used
          before freeing a lock that is part of a piece
          of memory that is about to be freed.

LK_REENABLE
          Lock is to be re-enabled after drain.  The
          LK_REENABLE flag may be set only at the re-
          lease of a lock obtained by a drain.

LK_SETRECURSE
          Other locks while we have it OK.

LK_RECURSEFAIL
          Attempt at recursive lock fails.

LK_SPIN  Lock spins instead of sleeping.

LK_INTERLOCK
        Unlock the simplelock slock when the lock is
        acquired.

lockstatus(lock)
    Determine the status of lock lock. Returns LK_EXCLUSIVE or
    LK_SHARED for exclusive-access and shared-access locks respec-
    tively.

lockmgr_printinfo(lock)
    Print out information about state of lock lock.

spinlockinit(lock, wmesg, flags)
    The lock lock is initialised as a spinlock according to the pa-
    rameters provided. Arguments are as follows:

    lock    The lock.

    wmesg   This is a simple name for lock.
    flags   Flags to specify the lock behaviour. Valid lock flags
            are the same as outlined above.

spinlockmgr(lock, flags, slock)
    Set, change or release a lock according to the parameters pro-
    vided. Arguments are as follows:

    lock    The spin lock.

    flags   Flags to specify the lock request type. Valid lock
            flags are the same as outlined above.

    slock   Simplelock interlock. The simplelock slock is set by
            the caller. When the lock lock is acquired, the sim-
            plelock is released.

## 1.5 Kernel Memory Allocation

## 1.6 Resource Pool Manager

Before describing virtual filesystem initialization, there needs to be brief mention of
resource-pool manager which is used as base library in implementating (1) *namei
pathname buffers*, (2) *vnode management data structures*, (3) *file structures*, (4)
*current working directory structures*, (5) *file descriptor structures*, and so forth.

resource-pool manager is implemented in `kern/subr_pool.c`. Instead of in-
vestigating details of this code, we present the API of pool allocator defined in
`sys/pool.h` to minimize the basic knowledge to access to the essentials of fast
filesystem code.

Theses utility routines provide management of pools of fixed-sized areas of mem-
ory. Resource pool set aside an amount of memory for exclusive use by the resource
pool owner.

### 1.6.1   Design of Resource-Pool Manager

Memory is allocated in pages which are split into pieces according to the pool item size. Each page is kept on a list headed by 'pr_pagelist' in the pool structure. The individual pool items are on a linked list headed by 'ph_itemlist' in each page header.

———————————————————————————————————— sys/pool.h

```
208 void            pool_init(struct pool *, size_t, u_int, u_int,
209                     int, const char *, struct pool_allocator *);
210 void            pool_destroy(struct pool *);
211
212 void            pool_set_drain_hook(struct pool *,
213                     void (*)(void *, int), void *);
214
215 void            *pool_get(struct pool *, int);
216 void            pool_put(struct pool *, void *);
217 int             pool_reclaim(struct pool *);
218
...
231 int             pool_prime(struct pool *, int);
232 void            pool_setlowat(struct pool *, int);
233 void            pool_sethiwat(struct pool *, int);
234 void            pool_sethardlimit(struct pool *, int, const char *, int);
235 void            pool_drain(void *);
236
237 /*
238  * Debugging and diagnostic aides.
239  */
240 void            pool_print(struct pool *, const char *);
241 void            pool_printit(struct pool *, const char *,
242                     void (*)(const char *, ...));
243 int             pool_chk(struct pool *, const char *);
244
245 /*
246  * Pool cache routines.
247  */
248 void            pool_cache_init(struct pool_cache *, struct pool *,
249                     int (*ctor)(void *, void *, int),
250                     void (*dtor)(void *, void *),
251                     void *);
252 void            pool_cache_destroy(struct pool_cache *);
253 void            *pool_cache_get(struct pool_cache *, int);
254 void            pool_cache_put(struct pool_cache *, void *);
255 void            pool_cache_destruct_object(struct pool_cache *, void *);
256 void            pool_cache_invalidate(struct pool_cache *);
```

———————————————————————————————————— sys/pool.h

### 1.6.2   Initializing a pool

The `pool_init` function declared in **line 208** initializes a resource pool.  This function allow other kernel parts to declare static pools that must be initialized before `malloc` kernel function is available.

The arguments are, in order,

**struct pool \*pp** is the handle identifying the pool resource instance

**size_t size** specifies the size of the memory items managed by the pool

**u_int align** specifies the memory address alignment of the items returned by `pool_get` function. This argument must be a power of two. If zero, the alignment defaults to a architecture specific natural alignment.

**u_int ioff**

**int flags**

**const char \*wchan** set the *wait channel* passed on to `tsleep` function, a kernel function similar to `sleep` in C library, if `pool_get` must wait for items to be returned to the pool. If you run `top` program, you may see this string in the `state` field.

**struct pool_allocator \*palloc** is called to add additional memory if the pool is depleted.

### 1.6.3 Destroying a Pool

`pool_destroy` function declared in **line 210** destroys a resource pool. It takes a single argument identifying the pool resource instance.

### 1.6.4 Allocating Items from a Pool

`pool_get` function allocates an item from the pool and returns a pointer to it. The arguments are, in order,

**struct pool \*pp** is the handle identifying the pool resource instance

**int flags** defines behavior in case the pooled resources are depleted. If no resources are available and `PR_WAITOK` is given, this function will wait until items are returned to the pool. If both `PR_LIMITFAIL` and `PR_WAITOK` is specified, and the pool has reached its hard limit, `pool_get` function will return `NULL` without waiting, allowing the caller to do its own garbage collection.

### 1.6.5 Returning Items to a Pool

`pool_put` function declared in **line 216** returns the pool item to the resource pool. If the number of available items in the pool exceeds the maximum pool size set by `pool_sethiwat` function and there are no outstanding requests for pool items, the excess items will be returned to the system.

The arguments are, in order,

**struct pool \*pp** is the handle identifying the pool resource instance.

**void \*item** is a pointer to a pool item previously obtained by `pool_get` function.

### 1.6.6 Using Cache to Speed Up

Pool caches provide a way for constructed objects to be cached. This can lead to performance improvements by avoiding needless object construction/destruction that is deferred until absolutely necessary.

Caches are grouped into cache groups. Each cache group references up to 16 constructed objects. The pool cache is initialized by `pool_cache_init` function.

When a cache allocates an object from the pool, it calls the object's constructor and places it into a cache group. When a cache group frees an object back to the pool, it first calls the object's destructor. This allows the object to persist in constructed form while freed to the cache.

Though pool_cache is initialized by virtual filesystem, it is not used.

### 1.6.7   Other Resource-Pool Manager API

There are some resource-pool manager API that is not described such as `pool_prime`, `pool_sethiwat`, `pool_setlowat`, `pool_set_drain_hook`, `pool_reclaim`, `pool_drain`, `pool_sethardlimit`, and so forth.

Although this API is designed and implemented by the resource-pool manager designer, it is not used in filesystem code.

# Chapter 2

# I/O System

## 2.1 I/O Mapping from User to Device

There are four main kinds of I/O in 4.4BSD.

- filesystem

- character-device interface

- block-device interface

- socket interface

The character device interface provides unstructured access to the underlying hardware, whereas the block device provides structured acess to the underlying hardware.

All I/O executed by block-device interface is done to or from I/O buffers that resides in the kernel's address space, the buffer cache. This approach requires at least one memory-to-memory copy operation to satisfy a user request.

For character-device interface, I/O operations do not go through the buffer cache; instead, they are made directly between the device and buffers in the application's virtual address space.

### 2.1.1 Device Drivers

A device driver is divided into three main sections:

1. Autoconfiguration and initialization routines

2. The top half: routines for servicing I/O requests

3. The bottom half: interrupt service routines

### 2.1.2 I/O Queueing

**Queue Processing Procedure**

The I/O queues are the primary means of communication bewteen the top and bottom halves of a device dirver. When an input or output request is received by the top half of the driver,

1. it is recorded in a data structure that is placed on per-device queue for processing.

2. When an input or output operation completes, the device dirver receives an interrupt from the controller.

3. The interrupt service routine removes the appropriate request from the device's queue,

4. notifies the requester that the command has completed, and then

5. starts the next request from the queue.

**Maintaing Queue Consistency**

Because I/O queues are shared among asynchronous routines, access to the queues must be synchronized. Routine that make up the top half of a device driver must raise the processor priority level using `splbio(), spltty(),` etc. to prevent the bottom half from being entered as a result of an interrupt while a top-half routine is manipulating an I/O queue.

### 2.1.3   Interrupt Handling

The system arranges for the unit-number parameter to be passed to the interrupt service routine for each device by installing the address of an auxiliary glue routine in the interrupt-vector table. This glue routine, rather than the actual interrupt service routine, is invoked to service the interrupt.

## 2.2   Block Devices

The task of the block-device interface is to convert from the user abstraction of a disk as an array of bytes to the structure imposed by the underlying physical medium. This operation of converting random access to an array of bytes to reads and writes of disk serctors is known as *block I/O.*

### 2.2.1   Entry Points for Block-Device Drivers

`open` commonly verify the integrity of the associated medium. `open` entry point will be called for each `open` or `mount` system call on a block special device file.

`strategy` start a read or write operation, and return immediately. Block I/O routines such as `bread` or `bwrite` routines call the device's strategy routine to read or write data not in the buffer cache. If the request is synchronous, the caller must sleep until I/O completes.

`close` Disk devices have nothing to do when a device is closed.

`dump` write all physical memory to the device.

`psize` returns the size of a disk-drive partition. This entry point is used during the bootstrap procedure to calculate the location at which a crash dump should be placed and to determine the sizes of the swap devices.

### 2.2.2   Disk Labels

**What is in it ?**

Disk label contains the information about the geometry of the disk and about the layout of he partitions.

**How is it used ?**

When the machine is powered up or the reset button is pressed, the CPU executes the *hardware bootstrap code* from the ROM. The hardware bootstrap code typically reads the first few sectors on the disk into the main memory, then branches to the address of the first location that it read. The program stored in these first few sectors is the *second-level bootstrap*. Having the *disk label* stored in the part of the disk read as part of the hardware bootstrap allows the second-level bootstrap to have the disk-label information. This information gives it the ability to find the root filesystem and hence the files, such as kernel, needed to bring up 4.4BSD.

**Format of Disk Label**

The size and location of the second-level bootstrap are dependent on the requirements of the hardware bootstrap code. Since there is no standard for disk-label formats and the hardware bootstrap code usually understands only the vendor level, it is often necessary to support both the vendor and the 4.4BSD disk labels. Here, the vendor label must be placed where the hardware bootstrap ROM code expects it; the 4.4BSD label must be placed out of the way of the vendor label but within the are that is read in by the hardware bootstrap code, so that it will be available to the second-level bootstrap.

## 2.3 Character Devices

A character device ususally maps the hardware interface into a byte stream. The character interface for disks and tapes is also called the *raw device interface*. Its primary task is to arrange for direct I/O to and from the device. It also handles the asynchronous nature of I/O by maintaing and ordering an active queue of pending transfers.

### 2.3.1 Raw Devices and Physical I/O

Most raw devices differ from block devices only in the way that they do I/O. Whereas block devices read and write to and from the system buffer cache, raw device bypasses the buffer cache. This eliminates the memory-to-memory copy, but denies the benefits of data caching. To preserve consistency between data in the buffer cache and data written directly to the device via character-device interface, the raw device should be used only when the block device is idle.

**Buffers for Character-Device Interface**

Because raw devices bypass the buffer cache, they are responsible for managing their own buffer structures. The `read` and `write` entry points for raw device driver uses `physio` function to start a raw I/O operatin. The `strategy` funtion manages buffers to map the user data buffer. This buffer is completely different and separated from buffer cache used by block-device driver.

The `strategy` function of `kern/kern_physio.c` is shown below. You may understand details after you read chapters about UVM and buffer cache. Now, just see the algorithm described in comments !

———————————————————————————————— kern/kern_physio.c

```
69 /*
70  * Do "physical I/O" on behalf of a user.  "Physical I/O" is I/O directly
71  * from the raw device to user buffers, and bypasses the buffer cache.
```

```
72  *
73  * Comments in brackets are from Leffler, et al.'s pseudo-code implementation.
74  */
75 int
76 physio(strategy, bp, dev, flags, minphys, uio)
77          void (*strategy) __P((struct buf *));
78          struct buf *bp;
79          dev_t dev;
80          int flags;
81          void (*minphys) __P((struct buf *));
82          struct uio *uio;
83 {
84          struct iovec *iovp;
85          struct proc *p = curproc;
86          int error, done, i, nobuf, s;
87          long todo;
88
89          error = 0;
90          flags &= B_READ | B_WRITE;
91
92          /* Make sure we have a buffer, creating one if necessary. */
93          if ((nobuf = (bp == NULL)) != 0) {
94
95                  bp = getphysbuf();
96                  /* bp was just malloc'd so can't already be busy */
97                  bp->b_flags |= B_BUSY;
98
99          } else {
100
101                  /* [raise the processor priority level to splbio;] */
102                  s = splbio();
103
104                  /* [while the buffer is marked busy] */
105                  while (bp->b_flags & B_BUSY) {
106                          /* [mark the buffer wanted] */
107                          bp->b_flags |= B_WANTED;
108                          /* [wait until the buffer is available] */
109                          tsleep((caddr_t)bp, PRIBIO+1, "physbuf", 0);
110                  }
111
112                  /* Mark it busy, so nobody else will use it. */
113                  bp->b_flags |= B_BUSY;
114
115                  /* [lower the priority level] */
116                  splx(s);
117          }
118
119          /* [set up the fixed part of the buffer for a transfer] */
120          bp->b_dev = dev;
121          bp->b_error = 0;
122          bp->b_proc = p;
123          LIST_INIT(&bp->b_dep);
124
125          /*
```

```
126                * [while there are data to transfer and no I/O error]
127                * Note that I/O errors are handled with a 'goto' at the bottom
128                * of the 'while' loop.
129                */
130           for (i = 0; i < uio->uio_iovcnt; i++) {
131                   iovp = &uio->uio_iov[i];
132                   while (iovp->iov_len > 0) {
133
134                           /*
135                            * [mark the buffer busy for physical I/O]
136                            * (i.e. set B_PHYS (because it's an I/O to user
137                            * memory, and B_RAW, because B_RAW is to be
138                            * "Set by physio for raw transfers.", in addition
139                            * to the "busy" and read/write flag.)
140                            */
141                           bp->b_flags = B_BUSY | B_PHYS | B_RAW | flags;
142
143                           /* [set up the buffer for a maximum-sized transfer] */
144                           bp->b_blkno = btodb(uio->uio_offset);
145                           bp->b_bcount = iovp->iov_len;
146                           bp->b_data = iovp->iov_base;
147
148                           /*
149                            * [call minphys to bound the transfer size]
150                            * and remember the amount of data to transfer,
151                            * for later comparison.
152                            */
153                           (*minphys)(bp);
154                           todo = bp->b_bcount;
155 #ifdef DIAGNOSTIC
156                           if (todo <= 0)
157                                   panic("todo(%ld) <= 0; minphys broken", todo);
158                           if (todo > MAXPHYS)
159                                   panic("todo(%ld) > MAXPHYS; minphys broken",
160                                           todo);
161 #endif
162
163                           /*
164                            * [lock the part of the user address space involved
165                            *    in the transfer]
166                            * Beware vmapbuf(); it clobbers b_data and
167                            * saves it in b_saveaddr.  However, vunmapbuf()
168                            * restores it.
169                            */
170                           PHOLD(p);
171                           error = uvm_vslock(p, bp->b_data, todo,
172                                           (flags & B_READ) ?
173                                           VM_PROT_WRITE : VM_PROT_READ);
174                           if (error) {
175                                   bp->b_flags |= B_ERROR;
176                                   bp->b_error = error;
177                                   goto after_vsunlock;
178                           }
179                           vmapbuf(bp, todo);
```

```
180
181                             /* [call strategy to start the transfer] */
182                             (*strategy)(bp);
183
184                             /*
185                              * Note that the raise/wait/lower/get error
186                              * steps below would be done by biowait(), but
187                              * we want to unlock the address space before
188                              * we lower the priority.
189                              *
190                              * [raise the priority level to splbio]
191                              */
192                             s = splbio();
193
194                             /* [wait for the transfer to complete] */
195                             while ((bp->b_flags & B_DONE) == 0)
196                                     tsleep((caddr_t) bp, PRIBIO + 1, "physio", 0);
197
198                             /* Mark it busy again, so nobody else will use it. */
199                             bp->b_flags |= B_BUSY;
200
201                             /* [lower the priority level] */
202                             splx(s);
203
204                             /*
205                              * [unlock the part of the address space previously
206                              *    locked]
207                              */
208                             vunmapbuf(bp, todo);
209                             uvm_vsunlock(p, bp->b_data, todo);
210     after_vsunlock:
211                             PRELE(p);
212
213                             /* remember error value (save a splbio/splx pair) */
214                             if (bp->b_flags & B_ERROR)
215                                     error = (bp->b_error ? bp->b_error : EIO);
216
217                             /*
218                              * [deduct the transfer size from the total number
219                              *    of data to transfer]
220                              */
221                             done = bp->b_bcount - bp->b_resid;
222                             KASSERT(done >= 0);
223                             KASSERT(done <= todo);
224
225                             iovp->iov_len -= done;
226                             iovp->iov_base = (caddr_t)iovp->iov_base + done;
227                             uio->uio_offset += done;
228                             uio->uio_resid -= done;
229
230                             /*
231                              * Now, check for an error.
232                              * Also, handle weird end-of-disk semantics.
233                              */
```

```
234                              if (error || done < todo)
235                                      goto done;
236                      }
237              }
238
239 done:
240          /*
241           * [clean up the state of the buffer]
242           * Remember if somebody wants it, so we can wake them up below.
243           * Also, if we had to steal it, give it back.
244           */
245          s = splbio();
246          bp->b_flags &= ~(B_BUSY | B_PHYS | B_RAW);
247          if (nobuf)
248                  putphysbuf(bp);
249          else {
250                  /*
251                   * [if another process is waiting for the raw I/O buffer,
252                   *    wake up processes waiting to do physical I/O;
253                   */
254                  if (bp->b_flags & B_WANTED) {
255                          bp->b_flags &= ~B_WANTED;
256                          wakeup(bp);
257                  }
258          }
259          splx(s);
260
261          return (error);
262 }
263
264 /*
265  * allocate a buffer structure for use in physical I/O.
266  */
267 struct buf *
268 getphysbuf()
269 {
270          struct buf *bp;
271          int s;
272
273          s = splbio();
274          bp = pool_get(&bufpool, PR_WAITOK);
275          splx(s);
276          memset(bp, 0, sizeof(*bp));
277          return(bp);
278 }
279
280 /*
281  * get rid of a swap buffer structure which has been used in physical I/O.
282  */
283 void
284 putphysbuf(bp)
285          struct buf *bp;
286 {
287          int s;
```

```
288
289          if (__predict_false(bp->b_flags & B_WANTED))
290                  panic("putphysbuf: private buf B_WANTED");
291          s = splbio();
292          pool_put(&bufpool, bp);
293          splx(s);
294 }
295
296 /*
297  * Leffler, et al., says on p. 231:
298  * "The minphys() routine is called by physio() to adjust the
299  * size of each I/O transfer before the latter is passed to
300  * the strategy routine..."
301  *
302  * so, just adjust the buffer's count accounting to MAXPHYS here,
303  * and return the new count;
304  */
305 void
306 minphys(bp)
307          struct buf *bp;
308 {
309
310          if (bp->b_bcount > MAXPHYS)
311                  bp->b_bcount = MAXPHYS;
312 }
```

——————————————————————————————————————————————— kern/kern_physio.c

### 2.3.2   Entry Points for Character-Device Drivers

open

clode

ioctl

mmap Map a device offset into a memory address. This entry point is called by the
      virtual-memory system to convert a logical mapping to a physical address.

read

reset

select

stop

write

## 2.4   Descriptor Management

### 2.4.1   File Descriptor, Descriptor Table, and File Entry

For user process, all I/O is done through *file descriptors*. System calls that refer to
open files take a file descriptor as an argument to specify the file. The file descriptor
is used by the kernel to index into the *descriptor table* for the current process to
locate a *file entry*.

## 2.4.2   What does the File Entry Points ?

File entry can point to *vnode* structure or *socket*.

**vnode**  The *file entry* provides a file type and a pointer to an underlying object for the descriptor. For data files, the file entry points to a vnode structure.

Special files do not have data blocks allocated on the disk; they are handled by the special-device filesystem that calls appropriate drivers to haldle I/O for them.

The virtual-memory system supports the mapping of files into a process's address space. Here, the file descriptor must reference a vnode that will be partially or completely mapped into the user's address space.

**socket**  The file entry may also reference a socket. The Sockets have a different file type, and the file entry points to a system block that is used in doing interprocess communication.

## 2.4.3   Movement of Data Inside the Kernel: `uiomove` function

Within the kernel, I/O data are described by an array of vectors. Each I/O vector or `iovec` has a base address and a length. The kernel maintains another structure, called a `uio` structure. All I/O within the kernel is described with `iovec` and `uio` structures. Movement of data is processed as following steps.

1. System calls such as `read` and `write` that are not passed an `iovec` create a `uio` to describe their arguemnts.

2. The `uio` structure reaches the part of the kernel responsible for moving the data to or from the process address space: the filesystem, the network, or a device driver.

3. In general, these parts of the kernel arrange a kernel buffer to hold the data, then use `uiomove` function to copy the data to or from the buffer or buffers describved by the `uio` structure.

4. `uiomove` function is called with a pointer to kernel data area, a data count, and a `uio` structure. As it moves data, it updates the counters and pointers of the `iovec` and `uio` structures by a corresponding amount.

5. If the kernel buffer is not as large as the area described by the `uio` structure, the `uio` structure will point to the part of the process address space just beyond the location completed most recently. Thus, while servicing a request, the kernel may call `uiomove` function multiple times, each time giving a pointer to a new kernel buffer for the next block of data.

The source for the definition of `iovec`, `uio` structure is

———————————————————————————————— sys/uio.h

```
54 struct iovec {
55     void    *iov_base;  /* Base address. */
56     size_t  iov_len;    /* Length. */
57 };
58
59 #if !defined(_POSIX_C_SOURCE) && !defined(_XOPEN_SOURCE)
```

```
60 #include <sys/ansi.h>
61
62 #ifndef off_t
63 typedef __off_t      off_t;  /* file offset */
64 #define off_t        __off_t
65 #endif
66
67 enum   uio_rw { UIO_READ, UIO_WRITE };
68
69 /* Segment flag values. */
70 enum uio_seg {
71     UIO_USERSPACE,       /* from user data space */
72     UIO_SYSSPACE         /* from system space */
73 };
74
75 struct uio {
76     struct  iovec *uio_iov; /* pointer to array of iovecs */
77     int uio_iovcnt; /* number of iovecs in array */
78     off_t   uio_offset; /* offset into file this uio corresponds to */
79     size_t  uio_resid;  /* residual i/o count */
80     enum    uio_seg uio_segflg; /* see above */
81     enum    uio_rw uio_rw;  /* see above */
82     struct  proc *uio_procp;/* process if UIO_USERSPACE */
83 };
```

———————————————————————————————————— sys/uio.h

The source for `uiomove` function is

———————————————————————————————————— kern/kern_subr.c

```
140 int
141 uiomove(buf, n, uio)
142     void *buf;
143     size_t n;
144     struct uio *uio;
145 {
146     struct iovec *iov;
147     u_int cnt;
148     int error = 0;
149     char *cp = buf;
150     struct proc *p = uio->uio_procp;
151
152 #ifdef DIAGNOSTIC
153     if (uio->uio_rw != UIO_READ && uio->uio_rw != UIO_WRITE)
154         panic("uiomove: mode");
155 #endif
156     while (n > 0 && uio->uio_resid) {
157         iov = uio->uio_iov;
158         cnt = iov->iov_len;
159         if (cnt == 0) {
160             uio->uio_iov++;
161             uio->uio_iovcnt--;
162             continue;
163         }
```

```
164            if (cnt > n)
165                cnt = n;
166            switch (uio->uio_segflg) {
167
168            case UIO_USERSPACE:
169                if (curproc->p_cpu->ci_schedstate.spc_flags &
170                    SPCF_SHOULDYIELD)
171                    preempt(NULL);
172                if (__predict_true(p == curproc)) {
173                    if (uio->uio_rw == UIO_READ)
174                        error = copyout(cp, iov->iov_base, cnt);
175                    else
176                        error = copyin(iov->iov_base, cp, cnt);
177                } else {
178                    if (uio->uio_rw == UIO_READ)
179                        error = copyout_proc(p, cp,
180                            iov->iov_base, cnt);
181                    else
182                        error = copyin_proc(p, iov->iov_base,
183                            cp, cnt);
184                }
185                if (error)
186                    return (error);
187                break;
188
189            case UIO_SYSSPACE:
190                if (uio->uio_rw == UIO_READ)
191                    error = kcopy(cp, iov->iov_base, cnt);
192                else
193                    error = kcopy(iov->iov_base, cp, cnt);
194                if (error)
195                    return (error);
196                break;
197            }
198            iov->iov_base = (caddr_t)iov->iov_base + cnt;
199            iov->iov_len -= cnt;
200            uio->uio_resid -= cnt;
201            uio->uio_offset += cnt;
202            cp += cnt;
203            KDASSERT(cnt <= n);
204            n -= cnt;
205        }
206        return (error);
207 }
```

———————————————————————————————————————— kern/kern_subr.c

where

```
    int
    copyin(const void *uaddr, void *kaddr, size_t len);
```

        Copies len bytes of data from the user-space address uaddr
        to the kernel-space address kaddr.

```
int
copyout(const void *kaddr, void *uaddr, size_t len);
```

> Copies len bytes of data from the kernel-space address
> kaddr to the user-space address uaddr.

Character device drivers that do not copy data from the process generally do not interpret the `uio` structure. Instead, there is one low-level kernel routine that arranges a direct transfer to or from the address space of the process. Here, a separate I/O operation is done for each `iovec` element.

Block device drivers does not use `uio` structures. User operations on block devies are done through the buffer cache.

# Chapter 3

# Virtual File System

The virtual filesystem, *VFS*, is the kernel interface to filesystems. The interface specifies the calls for the kernel to access filesystems. It also specifies the core functionality that a filesystem must provide to the kernel. The focus of *VFS* activity is the *vnode* and is discussed in the other chapter.

## 3.1 Architecture of Virtual File System

### 3.1.1 Why VFS is needed ?

In earlier BSD, the file entries directly referenced the local filesystem *inode*. An inode is a data structure that describes the contents of a file. However, with the advent of multiple filesystem types, the architecture had to be generalized. Thus, it was easier and more logical to add a new layer to the system below the file entry and above the inode. This new layer was first implemented by Sun Microsystems, which called it the virtual-node, or *vnode*, later. A vnode used by a local filesystem would refer to an inode. A vnode used by a remote filesystem would refer to a protocol control block that described the location and naming information necessary to access the remote file.

### 3.1.2 What Is in the Vnode ?

The vnode is an extensible object-oriented interface. It contains information that is generically useful independent of the underlying filesystem object that it represents. **vnode** structure is defined as,

___________________________________________________ sys/vnode.h

```
86 struct vnode {
87     struct uvm_object  v_uobj;          /* the VM object            */
88 #define v_usecount    v_uobj.uo_refs
89 #define v_interlock   v_uobj.vmobjlock
90     voff_t            v_size;           /* size of file             */
91     int               v_flag;           /* flags                    */
92     int               v_numoutput;      /* number of pending writes */
93     long              v_writecount;     /* reference count of writers */
94     long              v_holdcnt;        /* page & buffer references */
95     u_long            v_id;             /* capability identifier    */
96     struct mount     *v_mount;          /* ptr to vfs we are in     */
97     int      (**v_op) __P((void *));    /* vnode operations vector  */
```

```
 98      TAILQ_ENTRY(vnode) v_freelist;      /* vnode freelist         */
 99      LIST_ENTRY(vnode)  v_mntvnodes;     /* vnodes for mount point */
100      struct buflists    v_cleanblkhd;    /* clean blocklist head   */
101      struct buflists    v_dirtyblkhd;    /* dirty blocklist head   */
102      LIST_ENTRY(vnode)  v_synclist;      /* vnodes with dirty buffers */
103      union {
104        struct mount    *vu_mountedhere;  /* ptr to mounted vfs (VDIR) */
105        struct socket   *vu_socket;       /* unix ipc (VSOCK)       */
106        struct specinfo *vu_specinfo;     /* device (VCHR, VBLK)    */
107        struct fifoinfo *vu_fifoinfo;     /* fifo (VFIFO)           */
108      }                  v_un;
109      struct nqlease    *v_lease;         /* Soft reference to lease */
110      enum vtype         v_type;          /* vnode type             */
111      enum vtagtype      v_tag;           /* type of underlying data */
112      struct lock        v_lock;          /* lock for this vnode    */
113      struct lock       *v_vnlock;        /* pointer to lock        */
114      void              *v_data;          /* private data for fs    */
115      struct klist       v_klist;         /* knotes attached to vnode */
116 #ifdef VERIFIED_EXEC
117      char               fp_status;       /* fingerprint status
118                                             (see below)            */
119 #endif
120 };
121 #define v_mountedhere  v_un.vu_mountedhere
122 #define v_socket       v_un.vu_socket
123 #define v_specinfo     v_un.vu_specinfo
124 #define v_fifoinfo     v_un.vu_fifoinfo
```

———————————————————————————————————————————————— sys/vnode.h

The information stored in a vnode includes the following:

**v_usecount** is the number of file entries that are open for reading
and/or writing that reference the vnode.

**v_numoutput** is the number of buffer write operations in progress. To
speed the flushing of dirty data, the kernel does this operation by
doing asynchronous writes on all the dirty buffers at once. For
local filesystem, this simultaneous push causes all the buffers to be
put into the disk queue, so that they can be sorted into an optimal
order to minimize seeking. System calls that return until the data
are on stable store, such as **fsync** system call, can sleep on the
count of pending output operations, waiting for the count to reach
zero.

**v_writecount** is the number of file entries that are open for writing
that reference the vnode.

**v_holdcnt** is the number of pages and buffers that are associated with
the vnode.

**v_mount** describes the filesystem that contains the object represented
by the vnode.

**v_op** is a pointer to the set of vnode operations defined for the object.

**v_freelist** is a list linking together all the vnodes in the system that
are not being used actively. The free list is used when a filesystem
needs to allocate a new vnode.

**v_mntvnodes** is a list linking together all the vnodes associated with a specific mount point. Thus, when `sync` system call is executed for a filesystem, the kernel can traverse this list to visit all the files active within that filesystem.

**v_cleanblkhd** is the header of vnode clean-buffer list. This list stores all the buffers, about the vnode, that have not been modified, or have been written back since they were last modified.

This list is used to free buffers when a file is deleted. Since the file is never be read again, the kernel can immediately calcel any pending I/O on its dirty buffers, and reclaim all its clean and dirty buffers and place them at the head of the buffer free list, ready for immediate reuse.

**v_dirtyblkhd** is the header of vnode dirty-buffer list. This list stores all the buffers, about the vnode, that have been modified, but not yet been written back.

**v_un** is a reference to state about special devices, sockets, and FIFOs.

**v_lease** is used with NFS. So you need not regard it if you are only interested in local filesystem code.

**v_type** is the type of the underlying object such as regular file, directory, character device, and etc. This type information is not strictly necessary, since a vnode client could always call a vnode operation to get the type of the underlying object. However, because the type often is needed, the type of underlying objects does not change, and it takes time to call through the vnode interface, the object type is cached in the vnode.

This field has a value among `VNON`, `VREG`, `VDIR`, `VBLK`, `VCHR`, `VLNK`, `VSOCK`, `VFIFO`, `VBAD`.

**v_lock** is used for locking the vnode.

**v_data** is a pointer to private information needed for the underlying object. For the local filesystem, this pointer will reference an inode.

### 3.1.3   How to Call Vnode Operations ?

Kernel manipulates vnode by passing requests to the underlying object through a set of defined operations.

As part of the booting process, each filesystem registers the set of vnode operations that is able to support. The kernel then builds a table that lists the union of all operations supported by any filesystem.

Supported operations are filled in with the entry point registered by the filesystem. Filesystems amy opt to have unsupported operations filled in with either a default routine, or a routine that returns the characteristic error.

When a filesystem is mounted on a directory, the previous contents of the directory are hidden; only the contents of the root of the newly mounted filesystem are visible. The `mount` command pushes a new layer onto a vnode stack; an `unmount` command removes a layer.

When a file access such as open, read, stat, or close occurs to a vnode in the stack, that vnode has several options as

- Do the requested operation and resutn a result

- Pass the operation without change to the next-lower vnode on the satck.

- Modify the operations provided with the request, then pass it to the next-lower vnode. When the operation returns from the lower vnode, it may modify the results, or simply return them.

If an operation is passed to the bottom of the stack without any layer taking action on it, then the interface will return the error "operation not supported."

To make pass-operation efficient, the kernel places the vnode operation name and its arguments into an argument structure. This structure is then passed as a single parameter to the vnode operation. Thus all call on a vnode operation will always have exactly one parameter, which is the pointer to the argument structure.

Let's see how this design policy is implemented in NetBSD. When user level `write` system call is executed, kernel executes `vn_write` function of `kern/vfs_syscalls.c` as we have read in the previous chapter. The code of `vn_write` function is again listed here for easy reference.

——————————————————————————————————————— kern/vfs_init.c

```
526 /*
527  * File table vnode write routine.
528  */
529 static int
530 vn_write(fp, offset, uio, cred, flags)
531         struct file *fp;
532         off_t *offset;
533         struct uio *uio;
534         struct ucred *cred;
535         int flags;
536 {
537         struct vnode *vp = (struct vnode *)fp->f_data;
538         int count, error, ioflag = IO_UNIT;
539
540         if (vp->v_type == VREG && (fp->f_flag & O_APPEND))
541                 ioflag |= IO_APPEND;
542         if (fp->f_flag & FNONBLOCK)
543                 ioflag |= IO_NDELAY;
544         if (fp->f_flag & FFSYNC ||
545             (vp->v_mount && (vp->v_mount->mnt_flag & MNT_SYNCHRONOUS)))
546                 ioflag |= IO_SYNC;
547         else if (fp->f_flag & FDSYNC)
548                 ioflag |= IO_DSYNC;
549         if (fp->f_flag & FALTIO)
550                 ioflag |= IO_ALTSEMANTICS;
551         VOP_LEASE(vp, uio->uio_procp, cred, LEASE_WRITE);
552         vn_lock(vp, LK_EXCLUSIVE | LK_RETRY);
553         uio->uio_offset = *offset;
554         count = uio->uio_resid;
555         error = VOP_WRITE(vp, uio, ioflag, cred);
556         if (flags & FOF_UPDATE_OFFSET) {
557                 if (ioflag & IO_APPEND)
558                         *offset = uio->uio_offset;
559                 else
560                         *offset += count - uio->uio_resid;
561         }
562         VOP_UNLOCK(vp, 0);
563         return (error);
```

```
564 }
```

──────────────────────────────────────────────── kern/vfs_init.c

The second virtual file system operation in this function is `VOP_WRITE`.

> `VOP_WRITE(vp, uio, ioflag, cred)` write to a file. The argument `vp`
> is the vnode of the file to write to, `uio` is the location of the data
> to write, `ioflag` is a set of flags and `cred` are the credentials of the
> calling process.
>
> The `ioflag` argument is used to give directives and hints to the
> file system. The low 16 bits are a bit mask which can contain the
> same flags as `VOP_READ()`.
>
> Zero is returned on success, otherwise an error is returned. The
> vnode should be locked on entry and remains locked on exit.

This function is defined in `kern/vnode_if.c` ans `kern/vnode_if.h` twice ! In one
way, it is defined as an inline function, equivalent to macro. And the other way,
it is defined as a general function. Loadable Kernel Module (LKM) used general
function, and normal kernel uses a fast inline function. The code list shown below
is inline function version.

──────────────────────────────────────────────── sys/vnode_if.h

```
374 static __inline int VOP_WRITE(vp, uio, ioflag, cred)
375         struct vnode *vp;
376         struct uio *uio;
377         int ioflag;
378         struct ucred *cred;
379 {
380         struct vop_write_args a;
381         a.a_desc = VDESC(vop_write);
382         a.a_vp = vp;
383         a.a_uio = uio;
384         a.a_ioflag = ioflag;
385         a.a_cred = cred;
386         return (VCALL(vp, VOFFSET(vop_write), &a));
387 }
```

──────────────────────────────────────────────── sys/vnode_if.h

`VCALL` and `VOFFSET` macros are defined as,

──────────────────────────────────────────────── sys/cdefs.h

```
469 /*
470  * VOCALL calls an op given an ops vector.  We break it out because BSD's
471  * vclean changes the ops vector and then wants to call ops with the old
472  * vector.
473  */
474 /*
475  * actually, vclean doesn't use it anymore, but nfs does,
476  * for device specials and fifos.
477  */
478 #define VOCALL(OPSV,OFF,AP) (( *((OPSV)[(OFF)])) (AP))
479
```

```
480 /*
481  * This call works for vnodes in the kernel.
482  */
483 #define VCALL(VP,OFF,AP) VOCALL((VP)->v_op,(OFF),(AP))
484 #define VDESC(OP) (& __CONCAT(OP,_desc))
485 #define VOFFSET(OP) (VDESC(OP)->vdesc_offset)
```

—————————————————————————————————————— sys/cdefs.h

VOFFSET macro used in **line 1476** of **sys/vnode_if.h** is expanded as,

```
        VOFFSET(vop_write)
    -->  (VDESC(vop_write)->vdesc_offset)
    -->  ((& __CONCAT(vop_write))->vdesc_offset,_desc)
    -->  (vop_write_desc->vdesc_offset)
```

Therefore VCALL macro used in **line 1476** is expanded as,

```
        VCALL(vp, VOFFSET(vop_write), &a))
    -->  VCALL(vp, vop_write_desc->vdesc_offset, &a)
    -->  VOCALL((vp)->v_op,(vop_write_desc->vdesc_offset),(&a))
    -->  (( *(((vp)->v_op)[((vop_write_desc->vdesc_offset))]])) (&a))
```

Thus **line 1476** is equivalent to

```
  1476       return (VCALL(vp, VOFFSET(vop_write), &a));

  <====>    return ( *( (vp->v_op) [vop_write_desc->vdesc_offset] ) ) (&a);
                     | |                                          | |
                     | +------------------------------------------+ |
                     |                                              |
                     +----------------------------------------------+
```

vop_write_desc is defined in **kern/vnode_if.c** as,

—————————————————————————————————————— kern/vnode_if.h

```
109 const struct vnodeop_desc vop_bwrite_desc = {
110        2,
111        "vop_bwrite",
112        0,
113        vop_bwrite_vp_offsets,
114        VDESC_NO_OFFSET,
115        VDESC_NO_OFFSET,
116        VDESC_NO_OFFSET,
117        VDESC_NO_OFFSET,
118        NULL,
119 };
```

—————————————————————————————————————— kern/vnode_if.h

where the struct vnodeop_desc is defined in **sys/vnode.h** as,

—————————————————————————————————————— sys/vnode.h

```
389 /*
390  * This structure describes the vnode operation taking place.
391  */
392 struct vnodeop_desc {
393         int            vdesc_offset;   /* offset in vector--first for speed */
394         const char     *vdesc_name;    /* a readable name for debugging */
395         int            vdesc_flags;    /* VDESC_* flags */
396
397         /*
398          * These ops are used by bypass routines to map and locate arguments.
399          * Creds and procs are not needed in bypass routines, but sometimes
400          * they are useful to (for example) transport layers.
401          * Nameidata is useful because it has a cred in it.
402          */
403         const int      *vdesc_vp_offsets;      /* list ended by VDESC_NO_OFFSET */
404         int            vdesc_vpp_offset;       /* return vpp location */
405         int            vdesc_cred_offset;      /* cred location, if any */
406         int            vdesc_proc_offset;      /* proc location, if any */
407         int            vdesc_componentname_offset; /* if any */
408         /*
409          * Finally, we've got a list of private data (about each operation)
410          * for each transport layer.  (Support to manage this list is not
411          * yet part of BSD.)
412          */
413         caddr_t        *vdesc_transports;
414 };
```

——————————————————————————————————————————— sys/vnode.h

Therefore **line 1476** is equivalent to

```
<====>     return ( *( (vp->v_op) [vop_lease_desc->vdesc_offset] ) ) (&a);
<====>     return ( *( (vp->v_op) [2] ) ) (&a);
<====>     return ffs_write (&a);
```

In the next section, we will explain how the vp->v_op pointer is initialized.

## 3.2   Virtual Filesystem Initialization

Virtual filesystem initialization is initiated in `main` function of `kern/init_main.c`
which is practically the first function executed after machine bootstrap. At there,
`vfsinit` function of `kern/vfs_init.c` is called.

That function calls `vfsinit` function of `kern/vfs_init.c`

——————————————————————————————————————————— kern/vfs_init.c

```
320 /*
321  * Initialize the vnode structures and initialize each file system type.
322  */
323 void
324 vfsinit()
325 {
326         extern struct vfsops * const vfs_list_initial[];
327         int i;
328
```

```
329          /*
330           * Initialize the namei pathname buffer pool and cache.
331           */
332          pool_init(&pnbuf_pool, MAXPATHLEN, 0, 0, 0, "pnbufpl",
333                  &pool_allocator_nointr);
334          pool_cache_init(&pnbuf_cache, &pnbuf_pool, NULL, NULL, NULL);
335
336          /*
337           * Initialize the vnode table
338           */
339          vntblinit();
340
341          /*
342           * Initialize the vnode name cache
343           */
344          nchinit();
345
346  #ifdef DEBUG
347          /*
348           * Check the list of vnode operations.
349           */
350          vfs_op_check();
351  #endif
352
353          /*
354           * Initialize the special vnode operations.
355           */
356          vfs_opv_init(vfs_special_vnodeopv_descs);
357
358          /*
359           * Establish each file system which was statically
360           * included in the kernel.
361           */
362          vattr_null(&va_null);
363          for (i = 0; vfs_list_initial[i] != NULL; i++) {
364                  if (vfs_attach(vfs_list_initial[i])) {
365                          printf("multiple '%s' file systems",
366                              vfs_list_initial[i]->vfs_name);
367                          panic("vfsinit");
368                  }
369          }
370  }
```

——————————————————————————————— kern/vfs_init.c

Now, we describes each portion of `vfsinit` function with related source codes.

### 3.2.1   Initializing the namei pathname buffer pool

At **line 332**, `vfsinit` function initializes the namei pathname buffer pool. `MAXPATHLEN` specifies the size of the memory items managed by the pool and is defined to 1024 at `sys/param.h` and `sys/syslimits.h`. The next three zero parameter means there is no alignment constraint in initializing pool and logging facility is not used. When the `top` program executes, we will see "pnbufpl" in the state field when kernel is waiting for allocation of item for pool.

At **line 334**, `vfsinit` function initializes the pool cache for namei pathname buffer pool. This cache, however, is not used anywhere; It is only executed here, but it may be useful for future release of NetBSD operating system.

## 3.2.2 Initializing the vnode table

Vnode table is initialized by calling `vntblinit` function, at **line 339** of `vfs_subr.c`.

———————————————————————————————— kern/vfs_subr.c

```
190 void
191 vntblinit()
192 {
193
194         pool_init(&vnode_pool, sizeof(struct vnode), 0, 0, 0, "vnodepl",
195             &pool_allocator_nointr);
196
197         /*
198          * Initialize the filesystem syncer.
199          */
200         vn_initialize_syncerd();
201 }
```

———————————————————————————————— kern/vfs_subr.c

Just the same way as namei buffer cache pool is initialized, at **line 194** of `vfs_subr.c`, pool for vnode management data structures is initialized.

At **line 200**, `vn_initialize_syncerd` function calls filesystem syncer that flushes cached data to disk at regular intervals.

Surprisingly, filesystem syncer daemon uses a special kind of virtual filesystem called as *syncfs* filesystem ! Syncfs virtual filesystem is implemented in `miscfs/syncfs` directory. When a filesystem is mounted, syncfs is installed, as a virtual filesystem layer, on top of the filesystem by creating a new filesystem syncer vnode for the specified mount point by calling `vfs_allocate_syncvnode` function of `miscfs/syncfs/sync_vnops.c`.

Source code of **vn_initialize_syncerd** function is

———————————————————————————————— miscfs/syncfs_subr.c

```
70 void
71 vn_initialize_syncerd()
72 {
73         int i;
74
75         syncer_last = SYNCER_MAXDELAY + 2;
76
77         syncer_workitem_pending = malloc(
78             syncer_last * sizeof (struct synclist),
79             M_VNODE, M_WAITOK);
79
80         for (i = 0; i < syncer_last; i++)
81                 LIST_INIT(&syncer_workitem_pending[i]);
82
83         lockinit(&syncer_lock, PVFS, "synclk", 0, 0);
84 }
```

———————————————————————————————— miscfs/syncfs_subr.c **line 75** SYNCER_MAXDELAY is maximum delay interval between syncer daemon works
and it is defined to 32 seconds by default.

### 3.2.3   Initializing the Name Cache Buffer

——————————————————————————— kern/vfs_cache.c

```
401 void
402 nchinit(void)
403 {
404
405     TAILQ_INIT(&nclruhead);
406     nchashtbl =
407         hashinit(desiredvnodes, HASH_LIST, M_CACHE, M_WAITOK, &nchash);
408     ncvhashtbl =
409 #ifdef NAMECACHE_ENTER_REVERSE
410         hashinit(desiredvnodes, HASH_LIST, M_CACHE, M_WAITOK, &ncvhash);
411 #else
412         hashinit(desiredvnodes/8, HASH_LIST, M_CACHE, M_WAITOK, &ncvhash);
413 #endif
414     pool_init(&namecache_pool, sizeof(struct namecache), 0, 0, 0,
415             "ncachepl", &pool_allocator_nointr);
416 }
```

——————————————————————————— kern/vfs_cache.c

### 3.2.4   Initialize the Special Vnode Operations

This initialization process is started by **line 356** of kern/vfs_init.c that we have
already listed. We list here again for easy reference.

——————————————————————————— kern/vfs_init.c

```
356         vfs_opv_init(vfs_special_vnodeopv_descs);
```

——————————————————————————— kern/vfs_init.c

The vfs_special_vnodeopv_descs argument used in **line 356** of kern/vfs_init.c
is defined in kern/vfs_init.c as

——————————————————————————— kern/vfs_init.c

```
119 const struct vnodeopv_desc * const vfs_special_vnodeopv_descs[] = {
120         &dead_vnodeop_opv_desc,
121         &fifo_vnodeop_opv_desc,
122         &spec_vnodeop_opv_desc,
123         &sync_vnodeop_opv_desc,
124         NULL,
125 };
```

——————————————————————————— kern/vfs_init.c

where struct vnodeopv_desc is defined in sys/vnode.h as

——————————————————————————— sys/vnode.h

```
449 struct vnodeopv_desc {
450                      /* ptr to the ptr to the vector where op should go */
451         int (***opv_desc_vector_p)(void *);
452         const struct vnodeopv_entry_desc *opv_desc_ops; /* null terminated list */
453 };
```

———————————————————————————————————— sys/vnode.h

For example, the one of the four members contained in `vfs_special_vnodeopv_descs`
array, `sync_vnodeop_opv_desc` is defined in `miscfs/syncfs/sync_vnops.c` as

———————————————————————————— miscfs/syncfs/sync_vnops.c

```
47 int (**sync_vnodeop_p) __P((void *));
48 const struct vnodeopv_entry_desc sync_vnodeop_entries[] = {
49         { &vop_default_desc, vn_default_error },
50         { &vop_close_desc, sync_close },              /* close */
51         { &vop_fsync_desc, sync_fsync },              /* fsync */
52         { &vop_inactive_desc, sync_inactive },        /* inactive */
53         { &vop_reclaim_desc, sync_reclaim },          /* reclaim */
54         { &vop_lock_desc, sync_lock },                /* lock */
55         { &vop_unlock_desc, sync_unlock },            /* unlock */
56         { &vop_print_desc, sync_print },              /* print */
57         { &vop_islocked_desc, sync_islocked },        /* islocked */
58         { &vop_putpages_desc, sync_putpages },        /* islocked */
59         { NULL, NULL }
60 };
61
62 const struct vnodeopv_desc sync_vnodeop_opv_desc =
63         { &sync_vnodeop_p, sync_vnodeop_entries };
```

———————————————————————————— miscfs/syncfs/sync_vnops.c where `struct vnodeopv_entry_desc` is defined in `sys/vnode.h` as,

———————————————————————————————————— sys/vnode.h

```
445 struct vnodeopv_entry_desc {
446         const struct vnodeop_desc *opve_op;     /* which operation this is */
447         int (*opve_impl)(void *);       /* code implementing this operation */
448 };
```

———————————————————————————————————— sys/vnode.h

where `struct vnodeop_desc` is defined in `sys/vnode.h` and we showed this code
in the previous section.

To prevent your confusion, we summarized the relation between `vnodeopv_desc`,
`vnodeopv_entry_desc`, `vnodeop_desc` structures.

```
//////////////////////////////////////////////////////////////
// Vnode Operation Vector Description Table (vfs_init.c)
//
const struct vnodeopv_desc vfs_special_vnodeopv_descs[] = {
        ...
        &sync_vnodeop_opv_desc,
        ...
```

```
    };

    ///////////////////////////////////////////////////////////////
    // Vnode Operation Vector Description (miscfs/syncfs/sync_vnops.c)
    //
    int (**sync_vnodeop_p) __P((void *));
    const struct vnodeopv_desc sync_vnodeop_opv_desc = {
            &sync_vnodeop_p, sync_vnodeop_entries
    };

    ///////////////////////////////////////////////////////////////
    // Vnode Operation Vector Entry Description Table (misc/syncfs/sync_vnops.c)
    //
    const struct vnodeopv_entry_desc sync_vnodeop_entries[] = {
            ...
            { &vop_fsync_desc, sync_fsync },                /* fsync */
            ...
            { NULL, NULL }
    };

    ///////////////////////////////////////////////////////////////
    // Vnode Operation Vector Entry Description (kern/vnode_if.c)
    //
    const int vop_fsync_vp_offsets[] = {
            VOPARG_OFFSETOF(struct vop_fsync_args,a_vp),
            VDESC_NO_OFFSET
    };
    const struct vnodeop_desc vop_fsync_desc = {
            19,
            "vop_fsync",
            0,
            vop_fsync_vp_offsets,
            VDESC_NO_OFFSET,
            VOPARG_OFFSETOF(struct vop_fsync_args, a_cred),
            VOPARG_OFFSETOF(struct vop_fsync_args, a_p),
            VDESC_NO_OFFSET,
            NULL,
    };

    ///////////////////////////////////////////////////////////////
    // Vnode Operation (misc/syncfs/sync_vnops.c)
    //
    int
    sync_fsync(v)
            void *v;
    {
            ...
    }
```

Before going on reading this book, you should clearly understand the relation be-
tween *vnode operation vector description table, vnode operation vector description,
vnode operation vector entry description table, vnode operation vector entry descrip-
tion,* and *vnode operation,* from the above summary.  Only when if you know the relation without confusion, you can clearly understand how `vfs_opv_init` function
work.

`vfs_opv_init` function is defined in `kern/vfs_init.c` as,

————————————————————————————— kern/vfs_init.c

```
240 void
241 vfs_opv_init(vopvdpp)
242     const struct vnodeopv_desc * const *vopvdpp;
243 {
244     int (**opv_desc_vector) __P((void *));
245     int i;
246
247     /*
248      * Allocate the vectors.
249      */
250     for (i = 0; vopvdpp[i] != NULL; i++) {
251             /* XXX - shouldn't be M_VNODE */
252             opv_desc_vector =
253                 malloc(VNODE_OPS_COUNT * sizeof(PFI), M_VNODE, M_WAITOK);
254             memset(opv_desc_vector, 0, VNODE_OPS_COUNT * sizeof(PFI));
255             *(vopvdpp[i]->opv_desc_vector_p) = opv_desc_vector;
256             DODEBUG(printf("vector at %p allocated\n",
257                 opv_desc_vector_p));
258     }
259
260     /*
261      * ...and fill them in.
262      */
263     for (i = 0; vopvdpp[i] != NULL; i++)
264             vfs_opv_init_explicit(vopvdpp[i]);
265
266     /*
267      * Finally, go back and replace unfilled routines
268      * with their default.
269      */
270     for (i = 0; vopvdpp[i] != NULL; i++)
271             vfs_opv_init_default(vopvdpp[i]);
272 }
```

————————————————————————————— kern/vfs_init.c

`for` loop used in **line 250-258** executes 4 times with `vopvdpp` variable set respec-
tively to `dead_vnodeop_opv_desc`, `fifo_vnodeop_opv_desc`, `spec_vnodeop_opv_desc`,
and `sync_vnodeop_opv_desc`. **line 252-255** makes room for storing array of func-
tion pointer indicating each available vnode operation function.

`VNODE_OPS_COUNT` and `PFI` are defined as,

————————————————————————————— kern/vfs_init.c

```
1640 #define VNODE_OPS_COUNT 50
```

————————————————————————————— kern/vfs_init.c

and

———————————————————————————————— kern/vfs_init.c

```
127 /*
128  * This code doesn't work if the defn is **vnodop_defns with cc.
129  * The problem is because of the compiler sometimes putting in an
130  * extra level of indirection for arrays.  It's an interesting
131  * "feature" of C.
132  */
133 typedef int (*PFI) __P((void *));
```

———————————————————————————————— kern/vfs_init.c

After completion of this loop, for example, the value of sync_vnodeop_p used in
**line 63** of miscfs/syncfs/sync_vnops.c changes from NULL to a allocated memory
by **line 252-253** of kern/vfs_init.c.

Now, we will analyze vfs_init_explicit and vfs_init_default functions which
fill the allocated array with function pointers pointing to vnode operation functions.
The code for this function is,

———————————————————————————————— kern/vfs_init.c

```
176 vfs_opv_init_explicit(vfs_opv_desc)
177        const struct vnodeopv_desc *vfs_opv_desc;
178 {
179        int (**opv_desc_vector) __P((void *));
180        const struct vnodeopv_entry_desc *opve_descp;
181
182        opv_desc_vector = *(vfs_opv_desc->opv_desc_vector_p);
183
184        for (opve_descp = vfs_opv_desc->opv_desc_ops;
185             opve_descp->opve_op;
186             opve_descp++) {
187                /*
188                 * Sanity check:  is this operation listed
189                 * in the list of operations?  We check this
190                 * by seeing if its offest is zero.  Since
191                 * the default routine should always be listed
192                 * first, it should be the only one with a zero
193                 * offset.  Any other operation with a zero
194                 * offset is probably not listed in
195                 * vfs_op_descs, and so is probably an error.
196                 *
197                 * A panic here means the layer programmer
198                 * has committed the all-too common bug
199                 * of adding a new operation to the layer's
200                 * list of vnode operations but
201                 * not adding the operation to the system-wide
202                 * list of supported operations.
203                 */
204                if (opve_descp->opve_op->vdesc_offset == 0 &&
205                    opve_descp->opve_op->vdesc_offset != VOFFSET(vop_default)) {
206                        printf("operation %s not listed in %s.\n",
207                            opve_descp->opve_op->vdesc_name, "vfs_op_descs");
208                        panic ("vfs_opv_init: bad operation");
209                }
```

```
210
211                    /*
212                     * Fill in this entry.
213                     */
214                    opv_desc_vector[opve_descp->opve_op->vdesc_offset] =
215                            opve_descp->opve_impl;
216            }
217 }
218
219 static void
220 vfs_opv_init_default(vfs_opv_desc)
221         const struct vnodeopv_desc *vfs_opv_desc;
222 {
223         int j;
224         int (**opv_desc_vector) __P((void *));
225
226         opv_desc_vector = *(vfs_opv_desc->opv_desc_vector_p);
227
228         /*
229          * Force every operations vector to have a default routine.
230          */
231         if (opv_desc_vector[VOFFSET(vop_default)] == NULL)
232                 panic("vfs_opv_init: operation vector without default routine.");
233
234         for (j = 0; j < VNODE_OPS_COUNT; j++)
235                 if (opv_desc_vector[j] == NULL)
236                         opv_desc_vector[j] =
237                                 opv_desc_vector[VOFFSET(vop_default)];
238 }
```

——————————————————————————————— kern/vfs_init.c

If you keep in mind the summary about structures related with vnode operation,
only reading the source code would be sufficient to understand how **vfs_opv_init_explicit**
and **vfs_opv_init_default** function initialize **opv_desc_vector_p** member in *vnode*
*operation vector description* structure.

## 3.3  Attaching Available Static File System

**line 362-369** of **kern/vfs_init.c** attaches available static filesystem.

### 3.3.1  Set vnode attribute to empty

**line 362** of **kern/vfs_init.c** creates **va_null** global variable, defined in **kern/vfs_init.c**,
as a null vnode.

——————————————————————————————— kern/vfs_init.c

```
318 struct vattr va_null;
```

——————————————————————————————— kern/vfs_init.c

This variable is not directly used, but used with **VATTR_NULL** macro used to clear a
vnode. This macro is defined in **sys/vnode.h** as,

———————————————————————————————————————— kern/vfs_init.c

```
281 #define VATTR_NULL(vap) (*(vap) = va_null)      /* initialize a vattr */
```

———————————————————————————————————————— kern/vfs_init.c

Now we list the source code for **vattr_null** function which creates a null vnode. The reason why kernel source uses **VATTR_NULL** macro instead of directly calling **vattr_null** function, is simple since the later is faster.

———————————————————————————————————————— kern/vfs_subr.c

```
372 void
373 vattr_null(vap)
374         struct vattr *vap;
375 {
376
377         vap->va_type = VNON;
378
379         /*
380          * Assign individually so that it is safe even if size and
381          * sign of each member are varied.
382          */
383         vap->va_mode = VNOVAL;
384         vap->va_nlink = VNOVAL;
385         vap->va_uid = VNOVAL;
386         vap->va_gid = VNOVAL;
387         vap->va_fsid = VNOVAL;
388         vap->va_fileid = VNOVAL;
389         vap->va_size = VNOVAL;
390         vap->va_blocksize = VNOVAL;
391         vap->va_atime.tv_sec =
392             vap->va_mtime.tv_sec =
393             vap->va_ctime.tv_sec = VNOVAL;
394         vap->va_atime.tv_nsec =
395             vap->va_mtime.tv_nsec =
396             vap->va_ctime.tv_nsec = VNOVAL;
397         vap->va_gen = VNOVAL;
398         vap->va_flags = VNOVAL;
399         vap->va_rdev = VNOVAL;
400         vap->va_bytes = VNOVAL;
401         vap->va_vaflags = 0;
402 }
```

———————————————————————————————————————— kern/vfs_subr.c

### 3.3.2   How is `vfs_list_initial` initialized ?

**vfs_list_initial** is array of pointer to virtual filesystem operation. The source code for initializing this variable is not included in the kernel source: the needed code is generated automatically when we compile kernel.

Berfore compiling kernel, we executes **config** program. For example, we generates new kenel as

```
# cd /usr/src/syssrc/sys/arch/sparc64/conf
# config MY_KERNEL
# cd ../compile/MY_KERNEL
# make depend; make
```

In the above sample session, `config` program generates `Makefile`, and many header files under `../compile/MY_KERNEL` directory. There is, however, only four C source code is generated: `devsw.c`, `ioconf.c`, `param.c`, `swapnetbsd.c`, `vers.c`

From these automatically generated C source files by `config` program, `ioconf.c` contains the definition of `vfs_list_initial` variable.

For instance, if kernel configuration file contains,

──────────────────────────────────────────── - arch/sparc64/conf/GENERIC32

```
152 ## File systems.  You probably need at least one of FFS or NFS.
153 file-system    FFS           # Berkeley Fast Filesystem
154 file-system    NFS           # Sun NFS-compatible filesystem client
155 file-system    KERNFS        # kernel data-structure filesystem
156 file-system    NULLFS        # NULL layered filesystem
157 file-system    OVERLAY       # overlay file system
158 file-system    MFS           # memory-based filesystem
159 file-system    FDESC         # user file descriptor filesystem
160 file-system    UMAPFS        # uid/gid remapping filesystem
161 file-system    LFS           # Log-based filesystem (still experimental)
162 file-system    PORTAL        # portal filesystem (still experimental)
163 file-system    PROCFS        # /proc
164 file-system    CD9660        # ISO 9660 + Rock Ridge file system
165 file-system    UNION         # union file system
166 file-system    MSDOSFS       # MS-DOS FAT filesystem(s).
```

──────────────────────────────────────────── - arch/sparc64/conf/GENERIC32

then, `ioconf.c` would contain

──────────────────────────────────────────── arch/sparc64/compile/MY_KERNEL/ioconf.c

```
1643 struct vfsops * const vfs_list_initial[] = {
1644        &ffs_vfsops,
1645        &nfs_vfsops,
1646        &kernfs_vfsops,
1647        &nullfs_vfsops,
1648        &overlay_vfsops,
1649        &mfs_vfsops,
1650        &fdesc_vfsops,
1651        &umapfs_vfsops,
1652        &lfs_vfsops,
1653        &portal_vfsops,
1654        &procfs_vfsops,
1655        &cd9660_vfsops,
1656        &union_vfsops,
1657        &msdosfs_vfsops,
1658        NULL,
1659 };
```

———————————————————————— arch/sparc64/compile/MY_KERNEL/ioconf.c where `struct vfsops` is defined in `sys/mount.h` as,

———————————————————————————————————————— sys/mount.h

```
344 struct vfsops {
345         const char *vfs_name;
346         int     (*vfs_mount)    __P((struct mount *mp, const char *path,
347                                     void *data, struct nameidata *ndp,
348                                     struct proc *p));
349         int     (*vfs_start)    __P((struct mount *mp, int flags,
350                                     struct proc *p));
351         int     (*vfs_unmount)  __P((struct mount *mp, int mntflags,
352                                     struct proc *p));
353         int     (*vfs_root)     __P((struct mount *mp, struct vnode **vpp));
354         int     (*vfs_quotactl) __P((struct mount *mp, int cmds, uid_t uid,
355                                     caddr_t arg, struct proc *p));
356         int     (*vfs_statfs)   __P((struct mount *mp, struct statfs *sbp,
357                                     struct proc *p));
358         int     (*vfs_sync)     __P((struct mount *mp, int waitfor,
359                                     struct ucred *cred, struct proc *p));
360         int     (*vfs_vget)     __P((struct mount *mp, ino_t ino,
361                                     struct vnode **vpp));
362         int     (*vfs_fhtovp)   __P((struct mount *mp, struct fid *fhp,
363                                     struct vnode **vpp));
364         int     (*vfs_vptofh)   __P((struct vnode *vp, struct fid *fhp));
365         void    (*vfs_init)     __P((void));
366         void    (*vfs_reinit)   __P((void));
367         void    (*vfs_done)     __P((void));
368         int     (*vfs_sysctl)   __P((int *, u_int, void *, size_t *, void *,
369                                     size_t, struct proc *));
370         int     (*vfs_mountroot) __P((void));
371         int     (*vfs_checkexp) __P((struct mount *mp, struct mbuf *nam,
372                                     int *extflagsp, struct ucred **credanonp));
373         const struct vnodeopv_desc * const *vfs_opv_descs;
374         int     vfs_refcount;
375         LIST_ENTRY(vfsops) vfs_list;
376 };
```

———————————————————————————————————————— sys/mount.h

For example, `ffs_vfsops` variable appeared in **line 1644** of `arch/sparc64/compile/MY_KERNEL/ioconf.c` is initialized in `ufs/ffs/ffs_vfsops.c` as

———————————————————————————————————————— ufs/ffs/ffs_vfsops.c

```
 97 struct vfsops ffs_vfsops = {
 98         MOUNT_FFS,
 99         ffs_mount,
100         ufs_start,
101         ffs_unmount,
102         ufs_root,
103         ufs_quotactl,
104         ffs_statfs,
105         ffs_sync,
```

```
106          ffs_vget,
107          ffs_fhtovp,
108          ffs_vptofh,
109          ffs_init,
110          ffs_reinit,
111          ffs_done,
112          ffs_sysctl,
113          ffs_mountroot,
114          ufs_check_export,
115          ffs_vnodeopv_descs,
116 };
```

———————————————————————————— ufs/ffs/ffs_vfsops.c

You may wonder how `Makefile` for kernel compile know where the filesystem related files are. `Makefile` in `ufs` directory specifies the location of filesystem related kernel sources recursively with the help of system-wide makefile script, `/usr/share/mk/bsd.kinc.mk`.

With this information, we can plant our own filesystem with a different name onto NetBSD kernel !

### 3.3.3   Establish a filesystem and initialize it

**line 363-369** of `vfs_init.c` attaches and initializes all available virtual filesystem layers such as FFS, NFS and LFS, by calling **vfs_attach** function.

———————————————————————————————— kern/vfs_subr.c

```
2634 int
2635 vfs_attach(vfs)
2636          struct vfsops *vfs;
2637 {
2638          struct vfsops *v;
2639          int error = 0;
2640
2641
2642          /*
2643           * Make sure this file system doesn't already exist.
2644           */
2645          LIST_FOREACH(v, &vfs_list, vfs_list) {
2646                  if (strcmp(vfs->vfs_name, v->vfs_name) == 0) {
2647                          error = EEXIST;
2648                          goto out;
2649                  }
2650          }
2651
2652          /*
2653           * Initialize the vnode operations for this file system.
2654           */
2655          vfs_opv_init(vfs->vfs_opv_descs);
2656
2657          /*
2658           * Now initialize the file system itself.
2659           */
2660          (*vfs->vfs_init)();
```

```
2661
2662        /*
2663         * ...and link it into the kernel's list.
2664         */
2665        LIST_INSERT_HEAD(&vfs_list, vfs, vfs_list);
2666
2667        /*
2668         * Sanity: make sure the reference count is 0.
2669         */
2670        vfs->vfs_refcount = 0;
2671
2672 out:
2673        return (error);
2674 }
```

——————————————————————————————————————— kern/vfs_subr.c

In the case of FFS, **line 2660** of **kern/vfs_subr.c** calls **ffs_init** function, since **ffs_vfsops** variable used in **line 1644** of **arch/sparc64/compile/MY_KERNEL/ioconf.c** is initialized so that its **vfs_init** member is set to **ffs_init** by **line 97-116** of **ufs/ffs/vfs_ffsops.c**.

### 3.3.4   Fast Filesystem Initialization

Fast Filesystem as a virtual filesystem layer is initialized by **ffs_init** function. This **ffs_init** function called by **vfs_attach** function that we just described is shown below.

——————————————————————————————————————— kern/vfs_subr.c

```
1368 void
1369 ffs_init()
1370 {
1371        if (ffs_initcount++ > 0)
1372                return;
1373
1374        softdep_initialize();
1375        ufs_init();
1376
1377        pool_init(&ffs_inode_pool, sizeof(struct inode), 0, 0, 0, "ffsinopl",
1378            &pool_allocator_nointr);
1379 }
1380
1381 void
1382 ffs_reinit()
1383 {
1384     softdep_reinitialize();
1385     ufs_reinit();
1386 }
```

——————————————————————————————————————— kern/vfs_subr.c

### 3.3.5   Soft Dependency Module Initialization

To use soft dependency support, 14 additional caches for meta data structure is needed. It is only 5 caches, however, when Fast File System(FFS) without soft de-pendency support is considered. For simplicity, we do not consider caches for soft dependency. For not to use soft dependency support in NetBSD Sparc64, it is sufficient for you to remove a line saying `"options SOFTDEP"` in `arch/sparc64/sparc64/conf/GENERIC32` kernel configuration file.

Be sure to know that even if you turn off the switch, 14 additional caches for soft dependency support is initialized, but they are never used, since every call to soft dependency related functions are avoided by checking `mnt_flag` in `structure mount`. Soft dependency module initialization function, `softdep_initalize` is shown below.

—————————————————————————— ufs/ffs/ffs_softdep.c

```
1050  /*
1051   * Executed during filesystem system initialization before
1052   * mounting any file systems.
1053   */
1054  void
1055  softdep_initialize()
1056  {
1057          int i;
1058
1059          LIST_INIT(&mkdirlisthd);
1060          LIST_INIT(&softdep_workitem_pending);
1061          max_softdeps = desiredvnodes * 4;
1062          pagedep_hashtbl = hashinit(desiredvnodes / 5, HASH_LIST, M_PAGEDEP,
1063              M_WAITOK, &pagedep_hash);
1064          sema_init(&pagedep_in_progress, "pagedep", PRIBIO, 0);
1065          inodedep_hashtbl = hashinit(desiredvnodes, HASH_LIST, M_INODEDEP,
1066              M_WAITOK, &inodedep_hash);
1067          sema_init(&inodedep_in_progress, "inodedep", PRIBIO, 0);
1068          newblk_hashtbl = hashinit(64, HASH_LIST, M_NEWBLK, M_WAITOK,
1069              &newblk_hash);
1070          sema_init(&newblk_in_progress, "newblk", PRIBIO, 0);
1071          pool_init(&sdpcpool, sizeof(struct buf), 0, 0, 0, "sdpcpool",
1072              &pool_allocator_nointr);
1073          for (i = 0; i < PCBPHASHSIZE; i++) {
1074                  LIST_INIT(&pcbphashhead[i]);
1075          }
1076
1077          pool_init(&pagedep_pool, sizeof(struct pagedep), 0, 0, 0,
1078              "pagedeppl", &pool_allocator_nointr);
1079          pool_init(&inodedep_pool, sizeof(struct inodedep), 0, 0, 0,
1080              "inodedeppl", &pool_allocator_nointr);
1081          pool_init(&newblk_pool, sizeof(struct newblk), 0, 0, 0,
1082              "newblkpl", &pool_allocator_nointr);
1083          pool_init(&bmsafemap_pool, sizeof(struct bmsafemap), 0, 0, 0,
1084              "bmsafemappl", &pool_allocator_nointr);
1085          pool_init(&allocdirect_pool, sizeof(struct allocdirect), 0, 0, 0,
1086              "allocdirectpl", &pool_allocator_nointr);
1087          pool_init(&indirdep_pool, sizeof(struct indirdep), 0, 0, 0,
1088              "indirdeppl", &pool_allocator_nointr);
1089          pool_init(&allocindir_pool, sizeof(struct allocindir), 0, 0, 0,
1090              "allocindirpl", &pool_allocator_nointr);
1091          pool_init(&freefrag_pool, sizeof(struct freefrag), 0, 0, 0,
```

```
1092                "freefragpl", &pool_allocator_nointr);
1093            pool_init(&freeblks_pool, sizeof(struct freeblks), 0, 0, 0,
1094                "freeblkspl", &pool_allocator_nointr);
1095            pool_init(&freefile_pool, sizeof(struct freefile), 0, 0, 0,
1096                "freefilepl", &pool_allocator_nointr);
1097            pool_init(&diradd_pool, sizeof(struct diradd), 0, 0, 0,
1098                "diraddpl", &pool_allocator_nointr);
1099            pool_init(&mkdir_pool, sizeof(struct mkdir), 0, 0, 0,
1100                "mkdirpl", &pool_allocator_nointr);
1101            pool_init(&dirrem_pool, sizeof(struct dirrem), 0, 0, 0,
1102                "dirrempl", &pool_allocator_nointr);
1103            pool_init(&newdirblk_pool, sizeof (struct newdirblk), 0, 0, 0,
1104                "newdirblkpl", &pool_allocator_nointr);
1105 }
```

—————————————————————————————————————————— ufs/ffs/ffs_softdep.c

All jumps to the soft dependency code, lives in `ffs_mount`, `ffs_reload`, `ffs_mountfs`, `ffs_unmount`, `ffs_vget` functions of `ufs/ffs/ffs_vfsops.c`.

`ffs_mount` and `ffs_unmount` functions are called respectively when the `mount` and `umount` system call is executed. `ffs_mountfs` is subroutine of `ffs_mount` function and is also used for `ffs_mountroot` function. `ffs_reload` function reloads all incore data for a filesystem after running fsck on the root filesystem and finding things to fix. `ffs_vget` function is called to look up a FFS dinode number to find its incore vnode.

These code are shown in the following list. You do not need to understand it all, since the point is soft dependency functions are not called when kernel configuration is set up so.

—————————————————————————————————————————— ufs/ffs/ffs_vfsops.c

```
177 int
178 ffs_mount(mp, path, data, ndp, p)
179     struct mount *mp;
180     const char *path;
181     void *data;
182     struct nameidata *ndp;
183     struct proc *p;
184 {
...
307             if (mp->mnt_flag & MNT_SOFTDEP)
308                 error = softdep_flushfiles(mp, flags, p);
309             else
310                 error = ffs_flushfiles(mp, flags, p);
...
338         if ((fs->fs_flags & FS_DOSOFTDEP) &&
339             !(mp->mnt_flag & MNT_SOFTDEP) && fs->fs_ronly == 0) {
340 #ifdef notyet
341             flags = WRITECLOSE;
342             if (mp->mnt_flag & MNT_FORCE)
343                 flags |= FORCECLOSE;
344             error = softdep_flushfiles(mp, flags, p);
345             if (error == 0 && ffs_cgupdate(ump, MNT_WAIT) == 0)
346                 fs->fs_flags &= ~FS_DOSOFTDEP;
347                 (void) ffs_sbupdate(ump, MNT_WAIT);
```

```
      348 #elif defined(SOFTDEP)
      349            mp->mnt_flag |= MNT_SOFTDEP;
      350 #endif
      351        }
...
      382            if ((fs->fs_flags & FS_DOSOFTDEP)) {
      383                error = softdep_mount(devvp, mp, fs,
      384                    p->p_ucred);
      385                if (error)
      386                    return (error);
      387            }
...
      427 }
...
...
...
      442 int
      443 ffs_reload(mountp, cred, p)
      444     struct mount *mountp;
      445     struct ucred *cred;
      446     struct proc *p;
      447 {
...
      586     if ((fs->fs_flags & FS_DOSOFTDEP))
      587         softdep_mount(devvp, mountp, fs, cred);
...
      646 }
...
...
...
      651 int
      652 ffs_mountfs(devvp, mp, p)
      653     struct vnode *devvp;
      654     struct mount *mp;
      655     struct proc *p;
      656 {
...
      894     if (ronly == 0 && (fs->fs_flags & FS_DOSOFTDEP)) {
      895         error = softdep_mount(devvp, mp, fs, cred);
      896         if (error) {
      897             free(fs->fs_csp, M_UFSMNT);
      898             goto out;
      899         }
      900     }
...
      916 }
...
...
...
      950 int
      951 ffs_unmount(mp, mntflags, p)
      952     struct mount *mp;
      953     int mntflags;
      954     struct proc *p;
```

```
   955 {
...
   964     if (mp->mnt_flag & MNT_SOFTDEP) {
   965         if ((error = softdep_flushfiles(mp, flags, p)) != 0)
   966             return (error);
   967     } else {
   968         if ((error = ffs_flushfiles(mp, flags, p)) != 0)
   969             return (error);
   970     }
...
  1006 }
...
...
...
  1191 int
  1192 ffs_vget(mp, ino, vpp)
  1193     struct mount *mp;
  1194     ino_t ino;
  1195     struct vnode **vpp;
  1196 {
...
  1286     if (DOINGSOFTDEP(vp))
  1287         softdep_load_inodeblock(ip);
  1288     else
  1289         ip->i_ffs_effnlink = ip->i_ffs_nlink;
  1290     brelse(bp);
...
  1319 }
```

————————————————————————————————— ufs/ffs/ffs_vfsops.c

The `DOINGSOFTDEP()` macro used in the above list is defined in `ufs/inode.h` as

```
#define DOINGSOFTDEP(vp)          ((vp)->v_mount->mnt_flag & MNT_SOFTDEP)
```

As you have seen the codes, there is no need to worry about the operation of soft dependency facility if you removed the kernel option from kernel configuration file, although 14 caches for soft dependency is initialized.

### 3.3.6   UFS Initialization

In **line 1375** of `ufs/ffs/ffs_vfsops.c`, ffs_init function calls **ufs_init**, UFS initialization function that is defined as,

————————————————————————————————— ufs/ufs/ufs_vfsops.c

```
   225 /*
   226  * Initialize UFS filesystems, done only once.
   227  */
   228 void
   229 ufs_init()
   230 {
   231         if (ufs_initcount++ > 0)
   232                 return;
   233
```

```
234          ufs_ihashinit();
235 #ifdef QUOTA
236          dqinit();
237 #endif
238 }
```
——————————————————————————— ufs/ufs/ufs_vfsops.c

where **ufs_ihashint** function that initializes inode hash table is shown below.

——————————————————————————— ufs/ufs/ufs_ihash.c
```
61 /*
62  * Initialize inode hash table.
63  */
64 void
65 ufs_ihashinit()
66 {
67          lockinit(&ufs_hashlock, PINOD, "ufs_hashlock", 0, 0);
68          ihashtbl =
69              hashinit(desiredvnodes, HASH_LIST, M_UFSMNT, M_WAITOK, &ihash);
70          simple_lock_init(&ufs_ihash_slock);
71 }
```
——————————————————————————— ufs/ufs/ufs_ihash.c

Note the **line 69** which creates hash table that can store 'desiredvnodes' elements.
**desiredvnodes** global variable is also defined by **param.c** — autogenerated source
code by **config** program.

——————————————————————————— arch/sparc64/compile/MY_KERNEL/param.c
```
107 int      hz = HZ;
108 int      tick = 1000000 / HZ;
109 int      tickadj = 240000 / (60 * HZ);          /* can adjust 240ms in 60s */
110 int      rtc_offset = RTC_OFFSET;
111 int      maxproc = NPROC;
112 int      desiredvnodes = NVNODE;
113 int      maxfiles = MAXFILES;
114 int      ncallout = 16 + NPROC;  /* size of callwheel (rounded to ^2) */
115 u_long   sb_max = SB_MAX;        /* maximum socket buffer size */
116 int      fscale = FSCALE;        /* kernel uses 'FSCALE', user uses 'fscale' */
```
——————————————————————————— arch/sparc64/compile/MY_KERNEL/param.c where the **NVNODE** macro is defined in **sys/param.h** as,

——————————————————————————— kern/sys/param.h
```
128 #ifndef NPROC
129 #define NPROC   (20 + 16 * MAXUSERS)
130 #endif
131 #ifndef NTEXT
132 #define NTEXT   (80 + NPROC / 8)                /* actually the object cache */
133 #endif
134 #ifndef NVNODE
135 #define NVNODE  (NPROC + NTEXT + 100)
136 #define NVNODE_IMPLICIT
137 #endif
```

—————————————————————————————————— kern/sys/param.h

where the **line 134** means NVNODE parameter can be tuned in *kernel configuration file* using `option` command.  Since default `MAXUSERS` is 64 unless tuned by kernel configuration file,

```
NPROC = 20 + 16 * MAXUSERS
      = 20 + 16 * 64
      = 1044

NTEXT = 80 + NPROC / 8
      = 80 + 1044 / 8
      = 210

NVNODE = NPROC + NTEXT + 100
       = 1044 + 210 + 100
       = 1354
```

So, if you want to change the default value of `desiredvnodes` other than 1354, you can change by tuning `MAXUSERS` parameter in kernel configuration file using `option` command.

Up to now, we showed how virtual file system layer is initialized.  In the next chapter, we will describe a file system is mounted, with exploring the mount process of root file system !

## 3.4  Virtual Filesystem Operations

In a similar fashion to the vnode interface, all operations that are done on a file system are conducted through a single interface that allows the system to carry out operations on a file system without knowing its construction or type.

As we had described earlier, all supported file systems in the kernel have an entry in the **vfs_list_initial table**. This table is generated by `config` program and is a NULL-terminated list of **vfsops** structures.  The **vfsops** structure describes the operations that can be done to a specific file system type.  The **vfsops** structure is shown below.

————————————————————————————————————— sys/mount.h

```
344 struct vfsops {
345     const char *vfs_name;
346     int     (*vfs_mount)    __P((struct mount *mp, const char *path,
347                                 void *data, struct nameidata *ndp,
348                                 struct proc *p));
349     int     (*vfs_start)    __P((struct mount *mp, int flags,
350                                 struct proc *p));
351     int     (*vfs_unmount)  __P((struct mount *mp, int mntflags,
352                                 struct proc *p));
353     int     (*vfs_root)     __P((struct mount *mp, struct vnode **vpp));
354     int     (*vfs_quotactl) __P((struct mount *mp, int cmds, uid_t uid,
355                                 caddr_t arg, struct proc *p));
356     int     (*vfs_statfs)   __P((struct mount *mp, struct statfs *sbp,
357                                 struct proc *p));
358     int     (*vfs_sync)     __P((struct mount *mp, int waitfor,
359                                 struct ucred *cred, struct proc *p));
360     int     (*vfs_vget)     __P((struct mount *mp, ino_t ino,
```

```
361                                     struct vnode **vpp));
362        int     (*vfs_fhtovp)  __P((struct mount *mp, struct fid *fhp,
363                                     struct vnode **vpp));
364        int     (*vfs_vptofh)  __P((struct vnode *vp, struct fid *fhp));
365        void    (*vfs_init)    __P((void));
366        void    (*vfs_reinit)  __P((void));
367        void    (*vfs_done)    __P((void));
368        int     (*vfs_sysctl)  __P((int *, u_int, void *, size_t *, void *,
369                                     size_t, struct proc *));
370        int     (*vfs_mountroot) __P((void));
371        int     (*vfs_checkexp) __P((struct mount *mp, struct mbuf *nam,
372                                     int *extflagsp, struct ucred **credanonp));
373        const struct vnodeopv_desc * const *vfs_opv_descs;
374        int     vfs_refcount;
375        LIST_ENTRY(vfsops) vfs_list;
376 };
```

──────────────────────────────────────────────── sys/mount.h

The following table list the elements of the vfsops vector, the corre- sponding invocation macro, and a description of the element.

```
Vector element          Macro         Description
int (*vfs_mount)()      VFS_MOUNT     Mount a file system
int (*vfs_start)()      VFS_START     Make operational
int (*vfs_unmount)()    VFS_UMOUNT    Unmount a file system
int (*vfs_root)()       VFS_ROOT      Get the file system root vnode
int (*vfs_quotactl)()   VFS_QUOTACTL  Query/modify space quotas
int (*vfs_statfs)()     VFS_STATFS    Get file system statistics
int (*vfs_sync)()       VFS_SYNC      Flush file system buffers
int (*vfs_vget)()       VFS_VGET      Get vnode from file ID
int (*vfs_fhtovp)()     VFS_FHTOVP    NFS file handle to vnode lookup
int (*vfs_vptofh)()     VFS_VPTOFH    Vnode to NFS file handle lookup
void (*vfs_init)()      -             Initialise file system
void (*vfs_reinit)()    -             Reinitialise file system
void (*vfs_done)()      -             Cleanup unmounted file system
int (*vfs_sysctl)()     -             Query/modify kernel state
int (*vfs_mountroot)()  -             Mount the root file system
int (*vfs_checkexp)()   VFS_CHECKEXP  Check if fs is exported
```

Some additional non-function members of the vfsops structure are the file system name **vfs_name** and a reference count **vfs_refcount**. It is not mandatory for a filesystem type to support a particular operation, but it must assign each member function pointer to a suitable function to do the minimum required of it. In most cases, such functions either do nothing or return an error value to the effect that it is not supported.

At system boot, each filesystem with an entry in **vfs_list_initial** is established and initialised. Each initialised file system is recorded by the kernel in the list **vfs_list** and the file system specific initialisation function **vfs_init** in its vfsops vector is invoked. When the filesystem is not longer needed **vfs_done** is invoked to run file system specific cleanups and the file system is removed from the kernel list.

At system boot, the root filesystem is mounted by invoking the file system type specific **vfs_mountroot** function in the **vfsops** vector. All filesystems that can be mounted as a root file system must define this function. It is responsible for initialising to list of **mount** structures for all future mounted file systems.

Kernel state which affects a specific filesystem type can be queried and modified using the `sysctl` interface. The **vfs_sysctl** member of the **vfsops** structure is invoked by filesystem independent code.

## 3.5   References to Source Code

### 3.5.1   `kern/vfs_init.c` - 334 lines, 7 functions

**Gloval Variables**

```
const struct vnodeopv_desc * const vfs_special_vnodeopv_descs[] = {
        &dead_vnodeop_opv_desc,
        &fifo_vnodeop_opv_desc,
        &spec_vnodeop_opv_desc,
        &sync_vnodeop_opv_desc,
        NULL,
};

struct vattr va_null;
```

**Functions**

```
vn_default_error()
vfs_opv_init_explicit()
vfs_opv_init_default()
vfs_opv_init()
vfs_opv_free()
vfs_op_check()
vfsinit()
```

# Chapter 4

# Buffer Cache

*Buffer cache* manages the memory that buffers data being transferred to and from the network or disk, and act as a cache of recently used blocks.

Since we are planning to replace buffer cache, it is essential for us to know the details of buffer cache, and the interaction between vnode operations and buffer cache.

The architecture of buffer cache is best described by [1]. But the details about how the buffer cache is implemented is best described by [2].

The buffer cache is composed of two parts. The first part is the buffer header and the second part is the actual buffer contents.

## 4.1   Buffer Cache Header

The Buffer header of NetBSD release 1.6 is defined in `sys/buf.h` as,

—————————————————————————————— sys/buf.h

```
151 /*
152  * The buffer header describes an I/O operation in the kernel.
153  */
154 struct buf {
155         LIST_ENTRY(buf) b_hash;         /* Hash chain. */
156         LIST_ENTRY(buf) b_vnbufs;       /* Buffer's associated vnode. */
157         TAILQ_ENTRY(buf) b_freelist;    /* Free list position if not active. */
158         TAILQ_ENTRY(buf) b_actq;        /* Device driver queue when active. */
159         struct  proc *b_proc;           /* Associated proc if B_PHYS set. */
160         volatile long   b_flags;        /* B_* flags. */
161         int     b_error;                /* Errno value. */
162         long    b_bufsize;              /* Allocated buffer size. */
163         long    b_bcount;               /* Valid bytes in buffer. */
164         long    b_resid;                /* Remaining I/O. */
165         dev_t   b_dev;                  /* Device associated with buffer. */
166         struct {
167                 caddr_t b_addr;         /* Memory, superblocks, indirect etc. */
168         } b_un;
169         void    *b_saveaddr;            /* Original b_addr for physio. */
170         daddr_t b_lblkno;               /* Logical block number. */
171         daddr_t b_blkno;                /* Underlying physical block number
172                                            (partition relative) */
173         daddr_t b_rawblkno;             /* Raw underlying physical block
```

```
174                                    number (not partition relative) */
175                                    /* Function to call upon completion. */
176      void    (*b_iodone) __P((struct buf *));
177      struct  vnode *b_vp;          /* File vnode. */
178      void    *b_private;           /* Private data for owner */
179      off_t   b_dcookie;            /* Offset cookie if dir block */
180      struct  workhead b_dep;       /* List of filesystem dependencies. */
181 };
```

———————————————————————————————————————————————— sys/buf.h

where

  **b_vnbufs** is a pointer to the vnode whose data the buffer holds.

  **b_flags** tracks status information about the buffer, such as whether
     the buffer contains useful data, whether the buffer is in use, and
     whether the data must be written back to the file before the buffer
     can be reused.

  **b_bufsize** indicates the size of allocated buffer contents, without regard
     to the validity of the data contained.

  **b_bcount** contains the number of valid bytes contained in the buffer.

The possible values of **b_flags** variable are,

———————————————————————————————————————————————— sys/buf.h

```
192 /*
193  * These flags are kept in b_flags.
194  */
195 #define B_AGE       0x00000001  /* Move to age queue when I/O done. */
196 #define B_NEEDCOMMIT 0x00000002 /* Needs committing to stable storage */
197 #define B_ASYNC     0x00000004  /* Start I/O, do not wait. */
198 #define B_BAD       0x00000008  /* Bad block revectoring in progress. */
199 #define B_BUSY      0x00000010  /* I/O in progress. */
200 #define B_SCANNED   0x00000020  /* Block already pushed during sync */
201 #define B_CALL      0x00000040  /* Call b_iodone from biodone. */
202 #define B_DELWRI    0x00000080  /* Delay I/O until buffer reused. */
203 #define B_DIRTY     0x00000100  /* Dirty page to be pushed out async. */
204 #define B_DONE      0x00000200  /* I/O completed. */
205 #define B_EINTR     0x00000400  /* I/O was interrupted */
206 #define B_ERROR     0x00000800  /* I/O error occurred. */
207 #define B_GATHERED  0x00001000  /* LFS: already in a segment. */
208 #define B_INVAL     0x00002000  /* Does not contain valid info. */
209 #define B_LOCKED    0x00004000  /* Locked in core (not reusable). */
210 #define B_NOCACHE   0x00008000  /* Do not cache block after use. */
211 #define B_CACHE     0x00020000  /* Bread found us in the cache. */
212 #define B_PHYS      0x00040000  /* I/O to user memory. */
213 #define B_RAW       0x00080000  /* Set by physio for raw transfers. */
214 #define B_READ      0x00100000  /* Read buffer. */
215 #define B_TAPE      0x00200000  /* Magnetic tape I/O. */
216 #define B_WANTED    0x00800000  /* Process wants this buffer. */
217 #define B_WRITE     0x00000000  /* Write buffer (pseudo flag). */
218 #define B_XXX       0x02000000  /* Debugging flag. */
219 #define B_VFLUSH    0x04000000  /* Buffer is being synced. */
```

———————————————————————————— sys/buf.h

To set and test these flags, convevient macros are provided as,

———————————————————————————— kern/vfs_bio.c

```
70 /* Macros to clear/set/test flags. */
71 #define SET(t, f)        (t) |= (f)
72 #define CLR(t, f)        (t) &= ~(f)
73 #define ISSET(t, f)      ((t) & (f))
```

———————————————————————————— kern/vfs_bio.c

As we analyze buffer cache, we will gradually know every meaning of these flags.

## 4.2 Buffer Cache Contents

Buffer contents are maintaained separately from the header to allow easy manipulation of buffer sizes via the page-mapping hardware.

### 4.2.1 Allocation Virtual Memory to Buffer Cache

Kernel allocates to each buffer `MAXBSIZE` bytes of virtual memory, but the address space is not fully populated with physical memory. Initially, each buffer is assigned 4096 bytes of physical memory. As smaller buffers are allocated, they give up their unused physical memory to buffers that need to hold more than 4096 bytes.

`MAXSIZE` is machine-dependent since it is defined by

———————————————————————————— sys/param.h

```
211 /*
212  * File system parameters and macros.
213  *
214  * The file system is made out of blocks of at most MAXBSIZE units, with
215  * smaller units (fragments) only in the last direct block.  MAXBSIZE
216  * primarily determines the size of buffers in the buffer pool.  It may be
217  * made larger without any effect on existing file systems; however making
218  * it smaller may make some file systems unmountable.
219  */
220 #ifndef MAXBSIZE                                /* XXX */
221 #define MAXBSIZE        MAXPHYS
222 #endif
223 #define MAXFRAG         8
```

———————————————————————————— sys/param.h

and

———————————————————————————— arch/sparc64/include/param.h

```
128 #define DEV_BSIZE       512
129 #define DEV_BSHIFT      9               /* log2(DEV_BSIZE) */
130 #define BLKDEV_IOSIZE   2048
131 #define MAXPHYS         (64 * 1024)
```

———————————————————————————— arch/sparc64/include/param.h

### 4.2.2   Identifying Buffer

How can we identify buffer ? Is there unique ID ?

4.4BSD identify buffers by their logical block number within filesystem by `b_lblkno` member in buffer header.

Since it is difficult to detect aliases for a block belonging to a local file and the same block accessed through the block device disk, kernel prevents this case from occuring: The kernel does not allow the block device from a partition to be opened while that partition is mounted. Nor does the kernel allow a partition to be mounted if the block device from the partition is already open.

## 4.3   Buffer Hash

A buffer with valid contents is contained on exactly one `bufhash` hash chain. The kernel uses the hash chains to determine quickly whether a block is in the buffer pool, and if it is, to locate it.

A buffer is removed from the *buffer hash* only when its contents become invalid or it is reused for different data. Thus, even if the buffer is in use by one process, it can still be found by another process, although `B_BUSY` flag will be set so that it will not be used until the buffer is released.

The *buffer hash* is defined in `kern/vfs_bio.c` as,

——————————————————————————————— kern/vfs_bio.c

```
75 /*
76  * Definitions for the buffer hash lists.
77  */
78 #define BUFHASH(dvp, lbn)          \
79         (&bufhashtbl[(((long)(dvp) >> 8) + (int)(lbn)) & bufhash])
80 LIST_HEAD(bufhashhdr, buf) *bufhashtbl, invalhash;
81 u_long  bufhash;
```

——————————————————————————————— kern/vfs_bio.c

If you as unfamilar with the `LIST_HEAD` macro, review a section describing kernel list data structures in chapter 1. If you know how to use linked list macros, then you would know the above definition of **line 80** is equal to

————————————————————————————————————————

```
struct bufhashhdr {
    struct  buf *lh_first;  /* first element */
} *bufhashtbl, invalhash;
```

————————————————————————————————————————

`bufhashtbl` points to a hash table composed of an array of linked-lists. However, `invalhash` is simply a linked-list, not an array.

## 4.4   Buffer Cache Free Lists

In addition to appearing on the hash list, each unlocked byffer appears on exactly one free list. There are four kinds of free list. They are defined in `vfs_bio.c` as,

——————————————————————————————— kern/vfs_bio.c

```
 92 /*
 93  * Definitions for the buffer free lists.
 94  */
 95 #define BQUEUES         4               /* number of free buffer queues */
 96
 97 #define BQ_LOCKED       0               /* super-blocks &c */
 98 #define BQ_LRU          1               /* lru, useful buffers */
 99 #define BQ_AGE          2               /* rubbish */
100 #define BQ_EMPTY        3               /* buffer headers with no memory */
101
102 TAILQ_HEAD(bqueues, buf) bufqueues[BQUEUES];
103 int needbuffer;
```

——————————————————————————————————————————— kern/vfs_bio.c

If you as unfamilar with the `TAILQ_HEAD` macro, review a section describing
kernel list data structures in chapter 1.

If you had read the section, you would know **line 102** means that four tail
queues are defined, and these tail queues contain elements whose type is `struct`
`buf`. Also, you would know these definition is exactly equivalent to

———————————————————————————————————————————

```
struct bqueues {
        struct buf  *tqh_first;   /* first element */
        struct buf  **tqh_first;  /* addr of last next element */
} bufqueues [BQUEUES];
```

———————————————————————————————————————————

## 4.4.1   LOCKED List

Buffers on this list cannot be flushed from the cache.

## 4.4.2   LRU List

After a buffer is used, the buffer is then returned to the end of the LRU list. When
new buffers are needed, they are taken from the front of the LRU list. As its name
suggests, this list implements a least recently used (LRU) algorithm.

## 4.4.3   AGE List

AGE list holds two kinds of buffers. They are the buffers which are,

> *Blocks of unlinked file:* These buffers are not likely to be reused.  The
> buffers are placed at the *front of the AGE list* where they will be
> reclaimed quickly.

> *Read-ahead block:* These buffers are not proben their usefulness.  The
> buffers are placed at the *end of the AGE list* where they will might
> remain long enough to be used again.

AGE list is used for two purposes.  First, if a block is requested and it is found
on a buffer cache that lives in the AGE list, the buffer is returned to the end of the
LRU list, not the AGE list, because it has proved its usefulness.  Second, when a
new buffer is needed, the front of the AGE list is searched first; only when the AGE
list is empty, the LRU list is used.

### 4.4.4   EMPTY List

The EMPTY list contains buffers that have no physical memory. They are held on this list waiting for another buffer to be reused for a smaller block and thus give up its extra physical memory.

## 4.5   Buffer Cache Initialization

Buffer cache is initialized in the beginning stage of the system bootstrap. Initialization process consists of two stages.

At first, `cpu_startup` function does machine dependent memory allocation. At Second, `bufinit` function called by `cpu_startup` function initializes buffer cache hash and free lists using the memory allocated by previous machine dependent initialization stage.

### 4.5.1   Physical Memory Allocation

`main` function of `kern/init_main.c` is machine independent bootstrap routine, and it calls machine dependent startup routine `cpu_startup` function of `arch/sparc64/sparc64/machdep.c` defined as

——————————————————————————— arch/sparc64/sparc64/machdep.c

```
166 /*
167  * Machine-dependent startup code
168  */
169 void
170 cpu_startup()
171 {
172         caddr_t v;
173         long sz;
174         u_int i, base, residual;
175 #ifdef DEBUG
176         extern int pmapdebug;
177         int opmapdebug = pmapdebug;
178 #endif
179         vaddr_t minaddr, maxaddr;
180         vsize_t size;
181         extern struct user *proc0paddr;
182         char pbuf[9];
183
184 #ifdef DEBUG
185         pmapdebug = 0;
186 #endif
187
188         proc0.p_addr = proc0paddr;
189
190         /*
191          * Good {morning,afternoon,evening,night}.
192          */
193         printf(version);
194         /*identifycpu();*/
195         format_bytes(pbuf, sizeof(pbuf), ctob((u_int64_t)physmem));
196         printf("total memory = %s\n", pbuf);
```

```
197
198         /*
199          * Find out how much space we need, allocate it,
200          * and then give everything true virtual addresses.
201          */
202         sz = (long)allocsys(NULL, NULL);
203         if ((v = (caddr_t)uvm_km_alloc(kernel_map, round_page(sz))) == 0)
204                 panic("startup: no room for %lx bytes of tables", sz);
205         if (allocsys(v, NULL) - v != sz)
206                 panic("startup: table size inconsistency");
207
208         /*
209          * allocate virtual and physical memory for the buffers.
210          */
211         size = MAXBSIZE * nbuf;          /* # bytes for buffers */
212
213         /* allocate VM for buffers... area is not managed by VM system */
214         if (uvm_map(kernel_map, (vaddr_t *) &buffers, round_page(size),
215                     NULL, UVM_UNKNOWN_OFFSET, 0,
216                     UVM_MAPFLAG(UVM_PROT_NONE, UVM_PROT_NONE, UVM_INH_NONE,
217                             UVM_ADV_NORMAL, 0)) != 0)
218                 panic("cpu_startup: cannot allocate VM for buffers");
219
220         minaddr = (vaddr_t) buffers;
221         if ((bufpages / nbuf) >= btoc(MAXBSIZE)) {
222                 bufpages = btoc(MAXBSIZE) * nbuf; /* do not overallocate RAM */
223         }
224         base = bufpages / nbuf;
225         residual = bufpages % nbuf;
226
227         /* now allocate RAM for buffers */
228         for (i = 0 ; i < nbuf ; i++) {
229                 vaddr_t curbuf;
230                 vsize_t curbufsize;
231                 struct vm_page *pg;
232
233                 /*
234                  * each buffer has MAXBSIZE bytes of VM space allocated.  of
235                  * that MAXBSIZE space we allocate and map (base+1) pages
236                  * for the first "residual" buffers, and then we allocate
237                  * "base" pages for the rest.
238                  */
239                 curbuf = (vaddr_t) buffers + (i * MAXBSIZE);
240                 curbufsize = NBPG * ((i < residual) ? (base+1) : base);
241
242                 while (curbufsize) {
243                         pg = uvm_pagealloc(NULL, 0, NULL, 0);
244                         if (pg == NULL)
245                                 panic("cpu_startup: "
246                                     "not enough RAM for buffer cache");
247                         pmap_kenter_pa(curbuf, VM_PAGE_TO_PHYS(pg),
248                             VM_PROT_READ | VM_PROT_WRITE);
249                         curbuf += PAGE_SIZE;
250                         curbufsize -= PAGE_SIZE;
```

```
251                    }
252            }
253            pmap_update(kernel_map->pmap);
254
255            /*
256             * Allocate a submap for exec arguments.  This map effectively
257             * limits the number of processes exec'ing at any time.
258             */
259            exec_map = uvm_km_suballoc(kernel_map, &minaddr, &maxaddr,
260                                    16*NCARGS, VM_MAP_PAGEABLE, FALSE, NULL);
261
262            /*
263             * Finally, allocate mbuf cluster submap.
264             */
265            mb_map = uvm_km_suballoc(kernel_map, &minaddr, &maxaddr,
266                    nmbclusters * mclbytes, VM_MAP_INTRSAFE, FALSE, NULL);
267
268 #ifdef DEBUG
269            pmapdebug = opmapdebug;
270 #endif
271            format_bytes(pbuf, sizeof(pbuf), ptoa(uvmexp.free));
272            printf("avail memory = %s\n", pbuf);
273            format_bytes(pbuf, sizeof(pbuf), bufpages * NBPG);
274            printf("using %u buffers containing %s of memory\n", nbuf, pbuf);
275
276            /*
277             * Set up buffers, so they can be used to read disk labels.
278             */
279            bufinit();
280
281 #if 0
282            pmap_redzone();
283 #endif
284 }
```

―――――――――――――――――――――――― arch/sparc64/sparc64/machdep.c

From the above function, `buffers` global variable is defined, in automatically
compile-time generated code by `config` program, arch/sparc64/compile/MY_KERNEL/param.c,
as

―――――――――――――――――― arch/sparc64/compile/MY_KERNEL/param.c

```
194 /*
195  * These have to be allocated somewhere; allocating
196  * them here forces loader errors if this file is omitted
197  * (if they've been externed everywhere else; hah!).
198  */
199 struct  buf *buf;
200 char    *buffers;
```

―――――――――――――――――― arch/sparc64/compile/MY_KERNEL/param.c and `NBPG` macro is defined, in machine dependent source code, arch/sparc64/include/param.h,
as

─────────────────────────────────────── arch/sparc64/include/param.h

```
298 #define PGSHIFT        13             /* log2(NBPG) */
299 #define NBPG           (1<<PGSHIFT)   /* bytes/page */
300 #define PGOFSET        (NBPG-1)       /* byte offset into page */
```

─────────────────────────────────────── arch/sparc64/include/param.h

Global variables used in `cpu_startup` function, such as `nbuf`, `bufpages`, `bufcache` is defined in `kern/kern_allocsys.c` as

─────────────────────────────────────── arch/sparc64/include/param.h

```
 94 /*
 95  * Declare these as initialized data so we can patch them.
 96  */
 97 #ifndef NBUF
 98 # define NBUF 0
 99 #endif
100
101 #ifndef BUFPAGES
102 # define BUFPAGES 0
103 #endif
104
105 #ifdef BUFCACHE
106 # if (BUFCACHE < 5) || (BUFCACHE > 95)
107 #   error BUFCACHE is not between 5 and 95
108 # endif
109 #else
110    /* Default to 10% of first 2MB and 5% of remaining. */
111 # define BUFCACHE 0
112 #endif
113
114 u_int   nbuf = NBUF;
115 u_int   nswbuf = 0;
116 u_int   bufpages = BUFPAGES;    /* optional hardwired count */
117 u_int   bufcache = BUFCACHE;    /* % of RAM to use for buffer cache */
```

─────────────────────────────────────── arch/sparc64/include/param.h

If you specifies `NBUF`, or `BUFPAGES` macro in your kernel configuration file, then the kernel come to have fixed amount of buffer cache. However, by default, current NetBSD releases automatically calculates the amount of memory allocated for buffer cache by setting the value of `NBUF`, and `BUFPAGES` to zero ! Automatically calculated amount of memory allocated for buffer cache is 0.2 MB of the first system memory plus 5 percent of the remaining system memory.

You can change the 5 percent by setting `BUFCACHE` macro in your kernel configuration file using `option` command.

This calculation is done by `allocsys` function called from **line 202-206** of `cpu_startup` function. The source code of `allocsys` function is in the `kern/kern_allocsys.c` as

─────────────────────────────────────── kern/kern_allocsys.c

```
119 /*
120  * Allocate space for system data structures.  We are given
```

```
121      * a starting virtual address and we return a final virtual
122      * address; along the way we set each data structure pointer.
123      *
124      * We call allocsys() with 0 to find out how much space we want,
125      * allocate that much and fill it with zeroes, and then call
126      * allocsys() again with the correct base virtual address.
127      *
128      */
129
130     caddr_t
131     allocsys(caddr_t v, caddr_t (*mdcallback)(caddr_t))
132     {
133
134             /* Calculate the number of callwheels if necessary. */
135             if (callwheelsize == 0)
136                     callout_setsize();
137
138             ALLOCSYS(v, callwheel, struct callout_queue, callwheelsize);
139     #ifdef CALLWHEEL_STATS
140             ALLOCSYS(v, callwheel_sizes, int, callwheelsize);
141     #endif
142     #ifdef SYSVSHM
143             ALLOCSYS(v, shmsegs, struct shmid_ds, shminfo.shmmni);
144     #endif
145     #ifdef SYSVSEM
146             ALLOCSYS(v, sema, struct semid_ds, seminfo.semmni);
147             ALLOCSYS(v, sem, struct __sem, seminfo.semmns);
148             /* This is pretty disgusting! */
149             ALLOCSYS(v, semu, int, (seminfo.semmnu * seminfo.semusz) / sizeof(int));
150     #endif
151     #ifdef SYSVMSG
152             ALLOCSYS(v, msgpool, char, msginfo.msgmax);
153             ALLOCSYS(v, msgmaps, struct msgmap, msginfo.msgseg);
154             ALLOCSYS(v, msghdrs, struct __msg, msginfo.msgtql);
155             ALLOCSYS(v, msqids, struct msqid_ds, msginfo.msgmni);
156     #endif
157
158             /*
159              * Determine how many buffers to allocate.
160              *
161              *      - If bufcache is specified, use that % of memory
162              *        for the buffer cache.
163              *
164              *      - Otherwise, we default to the traditional BSD
165              *        formula of 10% of the first 2MB and 5% of
166              *        the remaining.
167              */
168             if (bufpages == 0) {
169                     if (bufcache != 0) {
170                             if (bufcache < 5 || bufcache > 95)
171                                     panic("bufcache is out of range (%d)",
172                                         bufcache);
173                             bufpages = physmem / 100 * bufcache;
174                     } else {
```

```
175                              if (physmem < btoc(2 * 1024 * 1024))
176                                      bufpages = physmem / 10;
177                              else
178                                      bufpages = (btoc(2 * 1024 * 1024) + physmem) /
179                                          20;
180                      }
181          }
182
183 #ifdef DIAGNOSTIC
184          if (bufpages == 0)
185                  panic("bufpages = 0");
186 #endif
187
188          /*
189           * Call the mdcallback now; it may need to adjust bufpages.
190           */
191          if (mdcallback != NULL)
192                  v = mdcallback(v);
193
194          /*
195           * Ensure a minimum of 16 buffers.
196           */
197          if (nbuf == 0) {
198                  nbuf = bufpages;
199                  if (nbuf < 16)
200                          nbuf = 16;
201          }
202
203 #ifdef VM_MAX_KERNEL_BUF
204          /*
205           * XXX stopgap measure to prevent wasting too much KVM on
206           * the sparsely filled buffer cache.
207           */
208          if (nbuf > VM_MAX_KERNEL_BUF / MAXBSIZE)
209                  nbuf = VM_MAX_KERNEL_BUF / MAXBSIZE;
210 #endif
211
212          /*
213           * We allocate 1/2 as many swap buffer headers as file I/O buffers.
214           */
215          if (nswbuf == 0) {
216                  nswbuf = (nbuf / 2) &~ 1;       /* force even */
217                  if (nswbuf > 256)
218                          nswbuf = 256;          /* sanity */
219          }
220          ALLOCSYS(v, buf, struct buf, nbuf);
221
222          return (v);
223 }
```

──────────────────────────────────────────────── kern/kern_allocsys.c

where the ALLOCSYS macro is defined in sys/systm.h as

──────────────────────────────────────────────── sys/systm.h

```
325 #define ALLOCSYS(base, name, type, num) \
326                 (name) = (type *)(base); (base) = (caddr_t)ALIGN((name)+(num))
```

———————————————————————————————————— sys/systm.h

where the `ALIGN` macro is defined in machine dependent source, `arch/sparc64/include/param.h`
as

———————————————————————————————— arch/sparc64/include/param.h

```
 95 /*
 96  * Round p (pointer or byte index) up to a correctly-aligned value for
 97  * the machine's strictest data type.  The result is u_int and must be
 98  * cast to any desired pointer type.
 99  *
100  * ALIGNED_POINTER is a boolean macro that checks whether an address
101  * is valid to fetch data elements of type t from on this architecture.
102  * This does not reflect the optimal alignment, just the possibility
103  * (within reasonable limits).
104  *
105  */
106 #define ALIGNBYTES32          0x7
107 #define ALIGNBYTES64          0xf
108 #ifdef __arch64__
109 #define ALIGNBYTES            ALIGNBYTES64
110 #else
111 #define ALIGNBYTES            ALIGNBYTES32
112 #endif
113 #define ALIGN(p)              (((u_long)(p) + ALIGNBYTES) & ~ALIGNBYTES)
114 #define ALIGN32(p)            (((u_long)(p) + ALIGNBYTES32) & ~ALIGNBYTES32)
115 #define ALIGNED_POINTER(p,t)  ((((u_long)(p)) & (sizeof(t)-1)) == 0)
```

———————————————————————————— arch/sparc64/include/param.h

Exactly saying, you may not fully understand `cpu_start` function, until we describe
UVM memory management system. Thing worthy of being remembered is that now
you know

- How the physical memory for buffer cache is allocated ?

- How can I change the amount of buffer cache ?

- After machine dependent, physical memory allocation for buffer
  cache, `nbuf, bufpages` variables are set to relevant values on the
  basis of `BUFCACHE` that is representing how much portion of the
  available physical system memory should be allocated for buffer
  cache.

- `buffer` global variable is a pointer to virtual memory chunk allo-
  cated by UVM for the whole buffer cache.

## 4.5.2   Initialization of Hash and Free List

`bufinit` function is called from **line 279** of `cpu_startup` machine dependent func-
tion. The `bufinit` function initalizes buffer cache hash and its four free lists.

———————————————————————————————————— kern/vfs_bio.c

```
146 /*
147  * Initialize buffers and hash links for buffers.
148  */
149 void
150 bufinit()
151 {
152         struct buf *bp;
153         struct bqueues *dp;
154         u_int i, base, residual;
155
156         /*
157          * Initialize the buffer pool.  This pool is used for buffers
158          * which are strictly I/O control blocks, not buffer cache
159          * buffers.
160          */
161         pool_init(&bufpool, sizeof(struct buf), 0, 0, 0, "bufpl", NULL);
162
163         for (dp = bufqueues; dp < &bufqueues[BQUEUES]; dp++)
164                 TAILQ_INIT(dp);
165         bufhashtbl = hashinit(nbuf, HASH_LIST, M_CACHE, M_WAITOK, &bufhash);
166         base = bufpages / nbuf;
167         residual = bufpages % nbuf;
168         for (i = 0; i < nbuf; i++) {
169                 bp = &buf[i];
170                 memset((char *)bp, 0, sizeof(*bp));
171                 bp->b_dev = NODEV;
172                 bp->b_vnbufs.le_next = NOLIST;
173                 LIST_INIT(&bp->b_dep);
174                 bp->b_data = buffers + i * MAXBSIZE;
175                 if (i < residual)
176                         bp->b_bufsize = (base + 1) * PAGE_SIZE;
177                 else
178                         bp->b_bufsize = base * PAGE_SIZE;
179                 bp->b_flags = B_INVAL;
180                 dp = bp->b_bufsize ? &bufqueues[BQ_AGE] : &bufqueues[BQ_EMPTY];
181                 binsheadfree(bp, dp);
182                 binshash(bp, &invalhash);
183         }
184 }
```

———————————————————————————————————————————— kern/vfs_bio.c

**line 161** initialize the buffer pool. Notice that this buffer pool is completely different thing from buffer cache. Buffer cache holds data block specified by logical file block. Buffer pool, however, is used to transfer data between raw device and user buffers, and bypass the buffer cache.

Buffer pool is used by physical I/O by device driver layers such as SCSI controller. When we describe `ccd` device driver in other chapter, we will explain how the buffer pool is used.

**line 164** Did you review chapter 1 about using linked-list and tail queues ? Then your will know the line is equivalent to

```
dp->tqh_first = NULL;
dp->tqh_last = &dp->tqh_first;
```

This code initializes four buffer cache free lists: *LOCKED*, *LRU*, *AGE*, *EMPTY* lists.

**line 165** initializes buffer cache hash and receives mask value in `bufhash` variable. If are not certain what this line do, review a subsection about description of kernel hash implementatin, in chapter 1.

**line 166-167** `nbuf` is the number of all buffer cache. `bufpages` is the number of all physical memory pages that is available for buffer cache. `nbuf` is equal to `bufpages` unless `NBUF` kernel configuration variable is explicitly set. Therefore, by default, these two line is equivalent to

```
base = 1;
residual = 0;
```

**line 168** Remember that `nbuf` is the total number of buffer cache in kernel. This number is displayed in kernel bootstrap message such as

```
console is keyboard/display
Copyright (c) 1996, 1997, 1998, 1999, 2000, 2001, 2002
    The NetBSD Foundation, Inc.  All rights reserved.
Copyright (c) 1982, 1986, 1989, 1991, 1993
    The Regents of the University of California.  All rights reserved.

NetBSD 1.6K (KERNEL) #0: Sat Nov  9 22:09:36 KST 2002
    cimon@ultra1:/usr/src/syssrc/sys/arch/sparc64/compile/KERNEL
total memory = 128 MB
avail memory = 108 MB
using 832 buffers containing 6656 KB of memory
otpath: /sbus@1f,0/SUNW,fas@e,8800000/sd@0,0
mainbus0 (root): SUNW,Ultra-1
cpu0 at mainbus0: SUNW,UltraSPARC @ 170 MHz, version 0 FPU
cpu0: 32K instruction (32 b/l), 16K data (32 b/l), 512K external (64 b/l)
```

The `nbuf` variable is set to 832, for a system having the above bootstrap message.

**line 169** Do you remember where `buf` variable appeared ? We described, in this section, that `buf` appears in `arch/sparc64/compile/MY_KERNEL/param.c` as

**line 170** clears `i`-th buffer cache header in the system.

———————————————- arch/sparc64/compile/MY_KERNEL/param.c

```
199 struct  buf *buf;
200 char    *buffers;
```

———————————————- arch/sparc64/compile/MY_KERNEL/param.c

`buf` global variables the whole memory chunk that can hold all the buffer cache header.

This is initialized in **line 220** of `kern/kern_allocsys.c`. The line is equivalent to

```
ALLOCSYS(v, buf, struct buf, nbuf);

====>  buf = (struct buf *) v;
       v   = (caddr_t) ALIGN (buf + nbuf);
```

Therefore, `buf` points to a memory chunk that can hold all available buffer cache header in system. Ok ? If you are not certain, review this section. And then you are still not certain, please ask me.

`buffers` global variables the whole memory chunk that can hold all the buffer cache contents. We already explained, in the previous subsection, how `buffers` is initialized.

**line 171** Since the buffer pointed by `bp` is just initialized and empty, this buffer is not associated with any other physical storage device. Therefore set `b_dev` member of the buffer cache header to `NODEV`.

We showed the whole source code of buffer cache header in a previous section. And the definition of `NODEV` macro is in `sys/param.h` as

——————————————————————————————————— sys/param.h

```
203 #define NODEV   (dev_t)(-1)     /* non-existent device */
```

——————————————————————————————————— sys/param.h

**line 172** This is some kinds of ad-hoc approach or hacking to make common linked-list header. `b_vnbufs` member is a link to a linked-list that holds the vnode for the buffer cache. However, the head for the linked-list is not defined. Therefore, instead of using `LIST_INIT` macro requiring head node, this line initializes the virtual link-list !

**line 173** You may disregard it, since it is only used by *Soft Dependency* facilities.

**line 174** set the `b_data` member of buffser cache to point the virtual memory chunk whose size is `MAXBSIZE`.

**line 175-178** by default, only **line 178** is effective.

**line 179** set the status of buffer cache. Because buffer cache is not associated with any *vnode* or valid data, the status is set to `B_INVAL`.

**line 180-181** by default, these two lines are equivalent to

```
binheadfree(bp, &bufqueues[BQ_AGE]);
```

meaning that a buffer cache pointed by `bp` variable is inserted in the head of *AGE* list, since the `binheadfree` is a macro defined as,

——————————————————————————————————— kern/vfs_bio.c

```
110 /*
111  * Insq/Remq for the buffer free lists.
112  */
113 #define binsheadfree(bp, dp)    TAILQ_INSERT_HEAD(dp, bp, b_freelist)
114 #define binstailfree(bp, dp)    TAILQ_INSERT_TAIL(dp, bp, b_freelist)
```

———————————————————————————— kern/vfs_bio.c

**line 182** places a buffer cache pointed by `bp` into *invalid hash list*. It is natural
that the buffer cache does not go to *hash list* since it does not contain any
contents.

———————————————————————————— kern/vfs_bio.c

```
86 /*
87  * Insq/Remq for the buffer hash lists.
88  */
89 #define binshash(bp, dp)        LIST_INSERT_HEAD(dp, bp, b_hash)
90 #define bremhash(bp)            LIST_REMOVE(bp, b_hash)
```

———————————————————————————— kern/vfs_bio.c

## 4.6   Buffer Cache Operation

In this section, we shows list of buffer cache operations that are used by filesystem.
Buffer cache operations are defined in `kern/vfs_bio.c` and declared in `sys/buf.h`
as,

———————————————————————————— sys/buf.h

```
260 void    allocbuf __P((struct buf *, int));
261 void    bawrite __P((struct buf *));
262 void    bdirty __P((struct buf *));
263 void    bdwrite __P((struct buf *));
264 void    biodone __P((struct buf *));
265 int     biowait __P((struct buf *));
266 int     bread __P((struct vnode *, daddr_t, int,
267                   struct ucred *, struct buf **));
268 int     breada __P((struct vnode *, daddr_t, int, daddr_t, int,
269                    struct ucred *, struct buf **));
270 int     breadn __P((struct vnode *, daddr_t, int, daddr_t *, int *, int,
271                    struct ucred *, struct buf **));
272 void    brelse __P((struct buf *));
273 void    bremfree __P((struct buf *));
274 void    bufinit __P((void));
275 int     bwrite __P((struct buf *));
276 void    cluster_callback __P((struct buf *));
277 int     cluster_read __P((struct vnode *, u_quad_t, daddr_t, long,
278                         struct ucred *, struct buf **));
279 void    cluster_write __P((struct buf *, u_quad_t));
280 struct buf *getblk __P((struct vnode *, daddr_t, int, int, int));
281 struct buf *geteblk __P((int));
282 struct buf *getnewbuf __P((int slpflag, int slptimeo));
283 struct buf *incore __P((struct vnode *, daddr_t));
284
285 void    minphys __P((struct buf *bp));
286 int     physio __P((void (*strategy)(struct buf *), struct buf *bp, dev_t dev,
287                    int flags, void (*minphys)(struct buf *), struct uio *uio));
288
```

```
289 void  brelvp __P((struct buf *));
290 void  reassignbuf __P((struct buf *, struct vnode *));
291 void  bgetvp __P((struct vnode *, struct buf *));
```

———————————————————————————————— sys/buf.h

### 4.6.1 Finding a Buffer Cache from Hash: `incore` function

The following `incore` function is used to find a buffer cache related with a vnode that has specific logical file block number. (note that line **596-597**)

———————————————————————————————— kern/vfs_bio.c

```
580 /*
581  * Determine if a block is in the cache.
582  * Just look on what would be its hash chain.  If it's there, return
583  * a pointer to it, unless it's marked invalid.  If it's marked invalid,
584  * we normally don't return the buffer, unless the caller explicitly
585  * wants us to.
586  */
587 struct buf *
588 incore(vp, blkno)
589         struct vnode *vp;
590         daddr_t blkno;
591 {
592         struct buf *bp;
593
594         /* Search hash chain */
595         LIST_FOREACH(bp, BUFHASH(vp, blkno), b_hash) {
596                 if (bp->b_lblkno == blkno && bp->b_vp == vp &&
597                     !ISSET(bp->b_flags, B_INVAL))
598                 return (bp);
599         }
600
601         return (NULL);
602 }
```

———————————————————————————————— kern/vfs_bio.c

**line 595** since buffer cache hash is actually an array of linked-lists, this access is logical. `BUFHASH` chooses an linked-list from the array. If you are not certain this operation, review a section describing usage of data structure in kernel, in chapter 1.

**line 596** shows that buffer cache is identified with its associated *vnode* and *logical file block number.*

From the following sections, we will explain what the functions in the above list do.

## 4.7  Managing Buffer Cache Free Lists

### 4.7.1  Reusing Old Buffer: `bremfree` function

The `bremfree` function is used to remove a buffer cache from a free list.

———————————————————————————————— kern/vfs_bio.c

```
120 void
121 bremfree(bp)
122         struct buf *bp;
123 {
124         int s = splbio();
125
126         struct bqueues *dp = NULL;
127
128         /*
129          * We only calculate the head of the freelist when removing
130          * the last element of the list as that is the only time that
131          * it is needed (e.g. to reset the tail pointer).
132          *
133          * NB: This makes an assumption about how tailq's are implemented.
134          */
135         if (TAILQ_NEXT(bp, b_freelist) == NULL) {
136                 for (dp = bufqueues; dp < &bufqueues[BQUEUES]; dp++)
137                         if (dp->tqh_last == &bp->b_freelist.tqe_next)
138                                 break;
139                 if (dp == &bufqueues[BQUEUES])
140                         panic("bremfree: lost tail");
141         }
142         TAILQ_REMOVE(dp, bp, b_freelist);
143         splx(s);
144 }
```

––––––––––––––––––––––––––––––––––––––––––––––––––––––––––– kern/vfs_bio.c

**124** `splbio` function blocks hardware interrupts from disks and other storage de-
vices so that the buffer cache coherency is not disturbed.

**135** Remind that the definition of `TAILQ_NEXT` and `b_freelist` member in `struct`
`buf` as,

```
#define TAILQ_ENTRY(type)                                          \
struct {                                                           \
        struct type *tqe_next;  /* next element */                 \
        struct type **tqe_prev; /* address of previous next element */  \
 }

...

#define TAILQ_NEXT(elm, field)          ((elm)->field.tqe_next)

and

...
struct buf {
        LIST_ENTRY(buf) b_hash;         /* Hash chain. */
        LIST_ENTRY(buf) b_vnbufs;       /* Buffer's associated vnode. */
        TAILQ_ENTRY(buf) b_freelist;    /* Free list position if not active. */
        TAILQ_ENTRY(buf) b_actq;        /* Device driver queue when active. */

...
```

**line 135** checks whether the buffer cache pointed by `bp` pointer is the last elemenet in any one of four free lists.

**line 136-140** find which free list contains the buffer cache. If the buffer cache to be removed from a free list is not the last element from the free list, there is no need to know the pointer to header.

But, if the buffer cache to be removed from a free list is the last element, there need to know the pointer to header

You may wonder why. As the **line 133** says, we need to know the implementation of tail queues to answer this reason.

**line 142** remove the buffer cache from a free list.

From the below definition of `TAILQ_REMOVE`, we can find the why the pointer to the header of a free list in which the buffer cache pointed by `bp` lives, when the buffer cache is the last element of the free list.

```
#define TAILQ_REMOVE(head, elm, field) do {                      \
        QUEUEDEBUG_TAILQ_PREREMOVE((head), (elm), field)         \
        QUEUEDEBUG_TAILQ_OP((elm), field)                        \
        if (((elm)->field.tqe_next) != NULL)                     \
                (elm)->field.tqe_next->field.tqe_prev =          \
                    (elm)->field.tqe_prev;                       \
        else                                                     \
                (head)->tqh_last = (elm)->field.tqe_prev;        \
        *(elm)->field.tqe_prev = (elm)->field.tqe_next;          \
        QUEUEDEBUG_TAILQ_POSTREMOVE((elm), field);               \
} while (/*CONSTCOND*/0)
```

Ok ?

**line 143** Restore the interrupt process condition changed by **line 124**. For your reference, we show the definition of `splbio` and `splx` function of `arch/sparc64/include/psl.h` as,

—————————————————————————— arch/sparc64/include/psl.h

```
 79 /* Interesting spl()s */
 80 #define PIL_SCSI        3
 81 #define PIL_FDSOFT      4
 82 #define PIL_AUSOFT      4
 83 #define PIL_BIO         5
...
355 #define SPLHOLD(name, newpil) \
356 static __inline int name __P((void)); \
357 static __inline int name() \
358 { \
359         int oldpil; \
360         __asm __volatile("rdpr %%pil,%0" : "=r" (oldpil)); \
361         if (newpil <= oldpil) \
362                 return oldpil; \
363         __asm __volatile("wrpr %%g0,%0,%%pil" : : "n" (newpil)); \
364         return (oldpil); \
365 }
366 #endif
```

```
...
    382 /* Block devices */
    383 SPLHOLD(splbio, PIL_BIO)
...
    448 static __inline void splx(newpil)
    449         int newpil;
    450 #endif
    451 {
    452 #ifdef SPLDEBUG
...
    457 #endif
    458         __asm __volatile("wrpr %%g0,%0,%%pil" : : "rn" (newpil));
    459 }
```

——————————————————————————————— arch/sparc64/include/psl.h

## 4.7.2  Allocating a New Buffer: `getnewbuf` function

If a process wants to read data from a file, the kernel determines which file system
contains the file and which block in the filesystem contains the data. When about
to read data from a particular disk block, the kernel checks the block is in the buffer
cache and, if it is not there, assigns a new free buffer using `getnewbuf` function.

Up to now, we presented elaborated description, and from now on we gives brief
explanation to reduce the size of this report :)   The algorithm of this function is

```
start:
        if (there is a buffer on AGE free list)
        {
                remove the buffer from AGE free list;
        } else if (there is a buffer on LRU free list)
                remove the buffer from LRU free list;
        } else {
                // There is no buffer in any free lists. Oops !
                //
                sleep (event any buffers on free list);
                return NULL;
        }

        if (the buffer is being flushed to storage) {
                // Note that the buffer under flush is
                // just removed from LRU list
                //
                set the buffer go to AGE list when the flush is done;

                // check whether there is another free buffer */
                //
                goto start;
        }

        set the buffer cache as bust;

        if (buffer is marked for delayed write) {
                // Kernel must write the ``delayed write buffer``
```

```
                          // to storage and allocate another buffer !
                          //
                          set the buffer go to AGE list when the flush is done;
                          start asynchronous write of the buffer to disk;
                          return NULL;
                  }

                  // The buffer cache do not have filesystem
                  // logical block number associated with its data.
                  // Since logical block number is the hash key,
                  // the buffer cache no longer exist on hash.
                  //
                  disassociate the buffer cache from related vnode;
                  remove the buffer from old hash entry;

                  return buffer;
```

When the `getnewbuf` function returns NULL pointer, the caller of `getnewbuf` function generally try again calling the `getnewbuf` function.

——————————————————————————————————— kern/vfs_bio.c

```
768 /*
769  * Find a buffer which is available for use.
770  * Select something from a free list.
771  * Preference is to AGE list, then LRU list.
772  */
773 struct buf *
774 getnewbuf(slpflag, slptimeo)
775         int slpflag, slptimeo;
776 {
777         struct buf *bp;
778         int s;
779
780 start:
781         s = splbio();
782         if ((bp = TAILQ_FIRST(&bufqueues[BQ_AGE])) != NULL ||
783             (bp = TAILQ_FIRST(&bufqueues[BQ_LRU])) != NULL) {
784                 bremfree(bp);
785         } else {
786                 /* wait for a free buffer of any kind */
787                 needbuffer = 1;
788                 tsleep(&needbuffer, slpflag|(PRIBIO+1), "getnewbuf", slptimeo);
789                 splx(s);
790                 return (NULL);
791         }
792
793         if (ISSET(bp->b_flags, B_VFLUSH)) {
794                 /*
795                  * This is a delayed write buffer being flushed to disk.  Make
796                  * sure it gets aged out of the queue when it's finished, and
797                  * leave it off the LRU queue.
798                  */
799                 CLR(bp->b_flags, B_VFLUSH);
```

```
800                SET(bp->b_flags, B_AGE);
801                splx(s);
802                goto start;
803        }
804
805        /* Buffer is no longer on free lists. */
806        SET(bp->b_flags, B_BUSY);
807
808        /*
809         * If buffer was a delayed write, start it and return NULL
810         * (since we might sleep while starting the write).
811         */
812        if (ISSET(bp->b_flags, B_DELWRI)) {
813                splx(s);
814                /*
815                 * This buffer has gone through the LRU, so make sure it gets
816                 * reused ASAP.
817                 */
818                SET(bp->b_flags, B_AGE);
819                bawrite(bp);
820                return (NULL);
821        }
822
823        /* disassociate us from our vnode, if we had one... */
824        if (bp->b_vp)
825                brelvp(bp);
826        splx(s);
827
828        if (LIST_FIRST(&bp->b_dep) != NULL && bioops.io_deallocate)
829                (*bioops.io_deallocate)(bp);
830
831        /* clear out various other fields */
832        bp->b_flags = B_BUSY;
833        bp->b_dev = NODEV;
834        bp->b_blkno = bp->b_lblkno = bp->b_rawblkno = 0;
835        bp->b_iodone = 0;
836        bp->b_error = 0;
837        bp->b_resid = 0;
838        bp->b_bcount = 0;
839
840        bremhash(bp);
841        return (bp);
842 }
```

----------------------------------------------------------------- kern/vfs_bio.c

Souce code is exact implementation of the algorithm that we just described. The only exception is **line 828-829** that can be ignored since these two lines is only effective when *Soft Dependency* facility is enabled.

### 4.7.3   Adjusting Buffer Size: `allocbuf` function

The task of `allocbuf` is to ensure that the buffer has enough physical memory allocated to it. The data are for each buffer is allocated `MAXBSIZE` bytes of virtual address space by `bufinit` function.

`allocbuf` compares the size of the intended data block with the amount of physical memory already allocated to the buffer.

- If there is excess physical memory,

  - and there is a buffer available on the *EMPTY* list, the excess memory is put into the empty buffer, and that buffer is then inserted onto the front of the *AGE* list.

  - If there are no buffers on the *EMPTY* lists, the excess physical memory is retained in the original buffer.

- If the buffer has insufficient memory, it takes memory from other buffers. `allocbuf` function does this allocation by calling `getnewbuf` function that we described in the previous subsection, to get a second buffer and transfer the physical memory in the second buffer to the new buffer under construction.

  - If there is memory remaining in the second buffer, the second buffer is released to the front of *AGE* list, otherwise the second buffer is released to the *EMPTY* list.

  - If the new buffer still does not have enough physical memory, the process is repeated.

———————————————————————————— kern/vfs_bio.c

```
677 /*
678  * Expand or contract the actual memory allocated to a buffer.
679  *
680  * If the buffer shrinks, data is lost, so it's up to the
681  * caller to have written it out *first*; this routine will not
682  * start a write.  If the buffer grows, it's the callers
683  * responsibility to fill out the buffer's additional contents.
684  */
685 void
686 allocbuf(bp, size)
687         struct buf *bp;
688         int size;
689 {
690         struct buf *nbp;
691         vsize_t desired_size;
692         int s;
693
694         desired_size = round_page((vsize_t)size);
695         if (desired_size > MAXBSIZE)
696                 panic("allocbuf: buffer larger than MAXBSIZE requested");
697
698         if (bp->b_bufsize == desired_size)
699                 goto out;
700
701         /*
702          * If the buffer is smaller than the desired size, we need to snarf
703          * it from other buffers.  Get buffers (via getnewbuf()), and
704          * steal their pages.
705          */
706         while (bp->b_bufsize < desired_size) {
707                 int amt;
```

```
708
709                 /* find a buffer */
710                 while ((nbp = getnewbuf(0, 0)) == NULL)
711                         ;
712
713                 SET(nbp->b_flags, B_INVAL);
714                 binshash(nbp, &invalhash);
715
716                 /* and steal its pages, up to the amount we need */
717                 amt = min(nbp->b_bufsize, (desired_size - bp->b_bufsize));
718                 pagemove((nbp->b_data + nbp->b_bufsize - amt),
719                         bp->b_data + bp->b_bufsize, amt);
720                 bp->b_bufsize += amt;
721                 nbp->b_bufsize -= amt;
722
723                 /* reduce transfer count if we stole some data */
724                 if (nbp->b_bcount > nbp->b_bufsize)
725                         nbp->b_bcount = nbp->b_bufsize;
726
727 #ifdef DIAGNOSTIC
728                 if (nbp->b_bufsize < 0)
729                         panic("allocbuf: negative bufsize");
730 #endif
731
732                 brelse(nbp);
733         }
734
735         /*
736          * If we want a buffer smaller than the current size,
737          * shrink this buffer.  Grab a buf head from the EMPTY queue,
738          * move a page onto it, and put it on front of the AGE queue.
739          * If there are no free buffer headers, leave the buffer alone.
740          */
741         if (bp->b_bufsize > desired_size) {
742                 s = splbio();
743                 if ((nbp = TAILQ_FIRST(&bufqueues[BQ_EMPTY])) == NULL) {
744                         /* No free buffer head */
745                         splx(s);
746                         goto out;
747                 }
748                 bremfree(nbp);
749                 SET(nbp->b_flags, B_BUSY);
750                 splx(s);
751
752                 /* move the page to it and note this change */
753                 pagemove(bp->b_data + desired_size,
754                     nbp->b_data, bp->b_bufsize - desired_size);
755                 nbp->b_bufsize = bp->b_bufsize - desired_size;
756                 bp->b_bufsize = desired_size;
757                 nbp->b_bcount = 0;
758                 SET(nbp->b_flags, B_INVAL);
759
760                 /* release the newly-filled buffer and leave */
761                 brelse(nbp);
```

```
762          }
763
764 out:
765          bp->b_bcount = size;
766 }
```
———————————————————————————— kern/vfs_bio.c

The only additional information to understand every details of the above code, we think, is

- **b_bcount** member in struct **buf** used in **line 724-724** represents the size of physical memory allocated to that buffer cache

- The reason that **brelse** function is called at **line 761**, instead of directly putting into the **AGE** list, is to awake any possible process for the availability of a new buffer.

- **round_page** macro is defined in **uvm/uvm_param.h** as

———————————————————————————— uvm/uvm_param.h

```
151 /*
152  * Round off or truncate to the nearest page.  These will work
153  * for either addresses or counts (i.e., 1 byte rounds to 1 page).
154  */
155 #define round_page(x)   (((x) + PAGE_MASK) & ~PAGE_MASK)
156 #define trunc_page(x)   ((x) & ~PAGE_MASK)
```
———————————————————————————— uvm/uvm_param.h

where the **PAGE_MASK** is defined as

———————————————————————————— uvm/uvm_param.h

```
 96 /*
 97  *      All references to the size of a page should be done with PAGE_SIZE
 98  *      or PAGE_SHIFT.  The fact they are variables is hidden here so that
 99  *      we can easily make them constant if we so desire.
100  */
101 #define PAGE_SIZE       uvmexp.pagesize        /* size of page */
102 #define PAGE_MASK       uvmexp.pagemask        /* size of page - 1 */
103 #define PAGE_SHIFT      uvmexp.pageshift       /* bits to shift for pages */
```
———————————————————————————— uvm/uvm_param.h

where the **uvmexp.pagesize** is set up in **arch/sparc64/sparc64/pamp.c** as,

———————————————————————————— arch/sparc64/sparc64/pmap.c

```
467 void
468 pmap_bootstrap(kernelstart, kernelend, maxctx)
469          u_long kernelstart, kernelend;
470          u_int maxctx;
```

```
       471 {
...
       491          /*
       492           * set machine page size
       493           */
       494          uvmexp.pagesize = NBPG;
       495          uvmexp.ncolors = pmap_calculate_colors();
       496          uvm_setpagesize();
```

———————————————————————————— arch/sparc64/sparc64/pmap.c

where the `NBPG` is defined to 8192 as

———————————————————————————— arch/sparc64/include/param.h

```
       298 #define PGSHIFT       13              /* log2(NBPG) */
       299 #define NBPG          (1<<PGSHIFT)    /* bytes/page */
       300 #define PGOFSET       (NBPG-1)        /* byte offset into page */
```

———————————————————————————— arch/sparc64/include/param.h

**Consistency of Physical Memory Mapping**

`allocbuf` function ensures that each physical-memory page is mapped into exactly one buffer at all times. So, the kernel maintains the consistency by purging old buffers when files are shortened or removed as follows.

- Whenever a file is *removed*,

    1. the kernel traverses its list of dirty buffers.

    2. For each buffer, the kernel cancels its write requests and

    3. marks the buffer invalid, so that the buffer cannot be found in the buffer pool again.

    4. Each invalid buffer is put at the front of the *AGE* list, so that it will be used before any buffers with potentially useful data.

- For a file being partially *truncated*, only the buffers following the truncation point are invalidated.

## 4.7.4   Getting a Buffer: `getblk` function

This function is the essential part in reading a logical file block into a buffer cache.

The algorithm of `getblk` function is

```
input: logical file block number, and vnode
output: locked buffer that can now be used for reading block,
        but not having read the filesystem yet !

{
start:
        if (block is in hash)
        {
                if (the buffercache is busy ?)
                {
```

```
                    if (UVM is using the block ?)
                    {
                            return NULL;
                    }

                    sleep (until a buffer becomes free);
            }

            mark the buffer cache busy;
            remove the buffer from free list;
    }
    else
    {
            if (try to get a new buffer cache failed ?)
            {
                    goto start;
            }

            place the buffer cache into hash;
            associate the buffer cache with vnode;
    }

    ensure the buffer cache has desired amount of physical memory;
    return the buffer cache;
}
```

The source code for `getblk` function of `kern/kern_bio.c` is

——————————————————————————————— kern/vfs_bio.c

```
604 /*
605  * Get a block of requested size that is associated with
606  * a given vnode and block offset. If it is found in the
607  * block cache, mark it as having been found, make it busy
608  * and return it. Otherwise, return an empty block of the
609  * correct size. It is up to the caller to insure that the
610  * cached blocks be of the correct size.
611  */
612 struct buf *
613 getblk(vp, blkno, size, slpflag, slptimeo)
614         struct vnode *vp;
615         daddr_t blkno;
616         int size, slpflag, slptimeo;
617 {
618         struct buf *bp;
619         int s, err;
620
621 start:
622         bp = incore(vp, blkno);
623         if (bp != NULL) {
624                 s = splbio();
625                 if (ISSET(bp->b_flags, B_BUSY)) {
626                         if (curproc == uvm.pagedaemon_proc) {
627                                 splx(s);
628                                 return NULL;
```

```
629                            }
630                            SET(bp->b_flags, B_WANTED);
631                            err = tsleep(bp, slpflag | (PRIBIO + 1), "getblk",
632                                         slptimeo);
633                            splx(s);
634                            if (err)
635                                    return (NULL);
636                            goto start;
637                    }
638 #ifdef DIAGNOSTIC
639                    if (ISSET(bp->b_flags, B_DONE|B_DELWRI) &&
640                      bp->b_bcount < size && vp->v_type != VBLK)
641                            panic("getblk: block size invariant failed");
642 #endif
643                    SET(bp->b_flags, B_BUSY);
644                    bremfree(bp);
645                    splx(s);
646            } else {
647                    if ((bp = getnewbuf(slpflag, slptimeo)) == NULL)
648                            goto start;
649
650                    binshash(bp, BUFHASH(vp, blkno));
651                    bp->b_blkno = bp->b_lblkno = bp->b_rawblkno = blkno;
652                    s = splbio();
653                    bgetvp(vp, bp);
654                    splx(s);
655            }
656            allocbuf(bp, size);
657            return (bp);
658 }
```

———————————————————————————————— kern/vfs_bio.c

For your reference, we show the code of `bgetvp` function that associate the buffer cache with vnode.

———————————————————————————————— kern/vfs_subr.c

```
135 /*
136  * Insq/Remq for the vnode usage lists.
137  */
138 #define bufinsvn(bp, dp)        LIST_INSERT_HEAD(dp, bp, b_vnbufs)
139 #define bufremvn(bp) {                                          \
140         LIST_REMOVE(bp, b_vnbufs);                              \
141         (bp)->b_vnbufs.le_next = NOLIST;                        \
142 }
...
857 /*
858  * Associate a buffer with a vnode.
859  */
860 void
861 bgetvp(vp, bp)
862         struct vnode *vp;
863         struct buf *bp;
864 {
```

```
865          int s;
866
867          if (bp->b_vp)
868                  panic("bgetvp: not free, bp %p", bp);
869          VHOLD(vp);
870          s = splbio();
871          bp->b_vp = vp;
872          if (vp->v_type == VBLK || vp->v_type == VCHR)
873                  bp->b_dev = vp->v_rdev;
874          else
875                  bp->b_dev = NODEV;
876          /*
877           * Insert onto list for new vnode.
878           */
879          bufinsvn(bp, &vp->v_cleanblkhd);
880          splx(s);
881 }
```

———————————————————————————————————————————— kern/vfs_subr.c

In **line 871-873**, note that **b_vp** and **b_dev** member of the buffer cache is set up to assiciate the vnode with the buffer. The buffer is inserted into the vnode clean list by **line 879**. The VHOLD vnode operation used in **line 869** is defined in **sys/vnode.h** as

———————————————————————————————————————————— sys/vnode.h

```
    260 #define HOLDRELE(vp)       holdrele(vp)
    261 #define VHOLD(vp)          vhold(vp)
    262 #define VREF(vp)           vref(vp)
...
    302 /*
    303  * increase buf or page ref
    304  */
    305 static __inline void
    306 vhold(struct vnode *vp)
    307 {
    308
    309          simple_lock(&vp->v_interlock);
    310          if ((vp->v_freelist.tqe_prev != (struct vnode **)0xdeadb) &&
    311            vp->v_holdcnt == 0 && vp->v_usecount == 0) {
    312                  simple_lock(&vnode_free_list_slock);
    313                  TAILQ_REMOVE(&vnode_free_list, vp, v_freelist);
    314                  TAILQ_INSERT_TAIL(&vnode_hold_list, vp, v_freelist);
    315                  simple_unlock(&vnode_free_list_slock);
    316          }
    317          vp->v_holdcnt++;
    318          simple_unlock(&vp->v_interlock);
    319 }
```

———————————————————————————————————————————— sys/vnode.h

VHOLD vnode operation marks the vnode as active by incrementing **vp->v_holdcnt** and moving the vnode from the freelist to the holdlist. Once on the holdlist, the vnode will not be recycled until it is released with **holdrele** function.

## 4.8    Allocating and Reading Filesystem with Buffer Cache

From this section, the description is presented briefly as possible as we can, so that we improve on analysis speed.

——————————————————————————————— kern/vfs_bio.c

```
186 static __inline struct buf *
187 bio_doread(vp, blkno, size, cred, async)
188         struct vnode *vp;
189         daddr_t blkno;
190         int size;
191         struct ucred *cred;
192         int async;
193 {
194         struct buf *bp;
195         struct proc *p = (curproc != NULL ? curproc : &proc0);  /* XXX */
196
197         bp = getblk(vp, blkno, size, 0, 0);
198
199         /*
200          * If buffer does not have data valid, start a read.
201          * Note that if buffer is B_INVAL, getblk() won't return it.
202          * Therefore, it's valid if it's I/O has completed or been delayed.
203          */
204         if (!ISSET(bp->b_flags, (B_DONE | B_DELWRI))) {
205                 /* Start I/O for the buffer. */
206                 SET(bp->b_flags, B_READ | async);
207                 VOP_STRATEGY(bp);
208
209                 /* Pay for the read. */
210                 p->p_stats->p_ru.ru_inblock++;
211         } else if (async) {
212                 brelse(bp);
213         }
214
215         return (bp);
216 }
```

——————————————————————————————— kern/vfs_bio.c

**line 197** get buffer containing the block or new block from buffer cache hash. This buffer is locked and on hash list, but not on free list.

**line 204** check the block is already containing the desired block.

**line 207** calls filesystem strategy routine. If the target filesystem is Fast Filesystem, then `ufs_strategy` is called. Ths `VOP_STRATEGY` is defined as,

——————————————————————————————— sys/vnode_if.h

```
1606 static __inline int VOP_STRATEGY(bp)
1607         struct buf *bp;
1608 {
```

```
1609            struct vop_strategy_args a;
1610            a.a_desc = VDESC(vop_strategy);
1611            a.a_bp = bp;
1612            return (VCALL(bp->b_vp, VOFFSET(vop_strategy), &a));
1613 }
```

—————————————————————————————————— sys/vnode_if.h

**line 211-212** Why should this buffer be returned to free list ? Since asyncronous read is requested for a block already on buffer cache hash, these lines try to return the buffer to free list.

—————————————————————————————————— ufs/ufs/ufs_vnops.c

```
1655 /*
1656  * Calculate the logical to physical mapping if not done already,
1657  * then call the device strategy routine.
1658  */
1659 int
1660 ufs_strategy(void *v)
1661 {
1662         struct vop_strategy_args /* {
1663                 struct buf *a_bp;
1664         } */ *ap = v;
1665         struct buf      *bp;
1666         struct vnode    *vp;
1667         struct inode    *ip;
1668         int             error;
1669
1670         bp = ap->a_bp;
1671         vp = bp->b_vp;
1672         ip = VTOI(vp);
1673         if (vp->v_type == VBLK || vp->v_type == VCHR)
1674                 panic("ufs_strategy: spec");
1675         KASSERT(bp->b_bcount != 0);
1676         if (bp->b_blkno == bp->b_lblkno) {
1677                 error = VOP_BMAP(vp, bp->b_lblkno, NULL, &bp->b_blkno,
1678                                 NULL);
1679                 if (error) {
1680                         bp->b_error = error;
1681                         bp->b_flags |= B_ERROR;
1682                         biodone(bp);
1683                         return (error);
1684                 }
1685                 if ((long)bp->b_blkno == -1) /* no valid data */
1686                         clrbuf(bp);
1687         }
1688         if ((long)bp->b_blkno < 0) { /* block is not on disk */
1689                 biodone(bp);
1690                 return (0);
1691         }
1692         vp = ip->i_devvp;
1693         bp->b_dev = vp->v_rdev;
1694         VOCALL (vp->v_op, VOFFSET(vop_strategy), ap);
```

```
1695          return (0);
1696 }
```
──────────────────────────────────────────── ufs/ufs/ufs_vnops.c

**line 1672** obtains a pointer to *inode*. The definition of `VTOI` is

──────────────────────────────────────────── ufs/ufs/inode.h

```
191 /* Convert between inode pointers and vnode pointers. */
192 #define VTOI(vp)         ((struct inode *)(vp)->v_data)
193 #define ITOV(ip)         ((ip)->i_vnode)
```
──────────────────────────────────────────── ufs/ufs/inode.h

**line 1677** changes the logical block number of a file relative to the beginning of a
file, to the physical block number of a filesystem relative to the beginning of a
partition. `b_lblkno` member contains the logical block number of a file asso-
ciated with the vnode. `b_blkno` member contains the physical block number
of a filesystem.

**line 1686** clears the buffer's data area. The macro definition is `sys/buf.h`

**line 1692-1694** obtains the vnode for device driver of the filesystem such as CCD
pseudo device driver, or SCSI general layer. And then, the strategy function
of the driver or layer is called via *specfs* virtual filesystem layer !  If the
strategy function of the SCSI general layer is called, the function then calls
the start function of SCSI device driver such as Adaptec Fast-Wide, or LSI
logic controller, using `physio` function of `kern/kern_physio.c`.

`biodone` function is called by a device driver to mark I/O complete on the buffer
that is just read or written.

──────────────────────────────────────────── kern/vfs_bio.c

```
869 /*
870  * Mark I/O complete on a buffer.
871  *
872  * If a callback has been requested, e.g. the pageout
873  * daemon, do so. Otherwise, awaken waiting processes.
874  *
875  * [ Leffler, et al., says on p.247:
876  *      "This routine wakes up the blocked process, frees the buffer
877  *      for an asynchronous write, or, for a request by the pagedaemon
878  *      process, invokes a procedure specified in the buffer structure" ]
879  *
880  * In real life, the pagedaemon (or other system processes) wants
881  * to do async stuff to, and doesn't want the buffer brelse()'d.
882  * (for swap pager, that puts swap buffers on the free lists (!!!),
883  * for the vn device, that puts malloc'd buffers on the free lists!)
884  */
885 void
886 biodone(bp)
887          struct buf *bp;
888 {
```

```
889          int s = splbio();
890
891          if (ISSET(bp->b_flags, B_DONE))
892                  panic("biodone already");
893          SET(bp->b_flags, B_DONE);            /* note that it's done */
894
895          if (LIST_FIRST(&bp->b_dep) != NULL && bioops.io_complete)
896                  (*bioops.io_complete)(bp);
897
898          if (!ISSET(bp->b_flags, B_READ))     /* wake up reader */
899                  vwakeup(bp);
900
901          if (ISSET(bp->b_flags, B_CALL)) {    /* if necessary, call out */
902                  CLR(bp->b_flags, B_CALL);    /* but note callout done */
903                  (*bp->b_iodone)(bp);
904          } else {
905                  if (ISSET(bp->b_flags, B_ASYNC))       /* if async, release */
906                          brelse(bp);
907                  else {                              /* or just wakeup the buffer */
908                          CLR(bp->b_flags, B_WANTED);
909                          wakeup(bp);
910                  }
911          }
912
913          splx(s);
914  }
```

——————————————————————————————— kern/vfs_bio.c

**line 898-899** Recall that the B_READ flag is set by bio_doread function. If this
flag is not set, then biodone is called after write, therefore, decrease number
of pending write in vnode structure. The definition of vwakeup function is

——————————————————————————————— kern/vfs_subr.c

```
630  /*
631   * Update outstanding I/O count and do wakeup if requested.
632   */
633  void
634  vwakeup(bp)
635          struct buf *bp;
636  {
637          struct vnode *vp;
638
639          if ((vp = bp->b_vp) != NULL) {
640                  if (--vp->v_numoutput < 0)
641                          panic("vwakeup: neg numoutput, vp %p", vp);
642                  if ((vp->v_flag & VBWAIT) && vp->v_numoutput <= 0) {
643                          vp->v_flag &= ~VBWAIT;
644                          wakeup((caddr_t)&vp->v_numoutput);
645                  }
646          }
647  }
```

——————————————————————————————— kern/vfs_subr.c

**line 905-906** returns a buffer that just finished asynchronous read to free list, since
it is not immediately used. Remember that `brelse` function clears `B_BUSY` flag
which is set by `getblk` function.

**line 907-910** just wakeup the process waiting from `biowait` function for the com-
pletion of I/O.

### 4.8.1   Just Read: `bread` function

The filesystem allocates and fills buffers by calling the `bread` function. `Bread` func-
tion

- Takes a vnode, a logical block number, and a size, and
- Returns a pointer to a locked buffer.

   It is important to remember that any other process that tries to obtain the
buffer will be put to sleep until the buffer is released.

――――――――――――――――――――――――――――――――――― kern/vfs_bio.c

```
218 /*
219  * Read a disk block.
220  * This algorithm described in Bach (p.54).
221  */
222 int
223 bread(vp, blkno, size, cred, bpp)
224         struct vnode *vp;
225         daddr_t blkno;
226         int size;
227         struct ucred *cred;
228         struct buf **bpp;
229 {
230         struct buf *bp;
231
232         /* Get buffer for block. */
233         bp = *bpp = bio_doread(vp, blkno, size, cred, 0);
234
235         /* Wait for the read to complete, and return result. */
236         return (biowait(bp));
237 }
```

――――――――――――――――――――――――――――――――――― kern/vfs_bio.c

**line 236** Remember that this is synchronous read, not asynchronous. The differ-
ence between them is that synchronous read wait for the completion of read
operation from filesystem, but asynchronous read does not wait. To differen-
tiate this trait, see the `breadn` function.

――――――――――――――――――――――――――――――――――― kern/vfs_bio.c

```
844 /*
845  * Wait for operations on the buffer to complete.
846  * When they do, extract and return the I/O's error value.
847  */
848 int
849 biowait(bp)
```

```
850        struct buf *bp;
851 {
852        int s;
853
854        s = splbio();
855        while (!ISSET(bp->b_flags, B_DONE | B_DELWRI))
856                tsleep(bp, PRIBIO + 1, "biowait", 0);
857        splx(s);
858
859        /* check for interruption of I/O (e.g. via NFS), then errors. */
860        if (ISSET(bp->b_flags, B_EINTR)) {
861                CLR(bp->b_flags, B_EINTR);
862                return (EINTR);
863        } else if (ISSET(bp->b_flags, B_ERROR))
864                return (bp->b_error ? bp->b_error : EIO);
865        else
866                return (0);
867 }
```

——————————————————————————————————— kern/vfs_bio.c

The places where the `biowait` function is called are,

`bread` function to wait for the completion of synchronous read.

`breadn` function to wait for the completion of synchronous read of the only first block: the rest block from the second to the end are asynchoronously read, so that the `biowait` function is not used for the I/O completion of those blocks.

`bwrite` function to wait for the completion of synchronous write.

**line 855-856** sleep until `biodone` function called by the relevant device driver strategy function, awakens this line.

## 4.8.2 Read Ahead Multiple Buffers: `breadn` function

——————————————————————————————————— kern/vfs_bio.c

```
239 /*
240  * Read-ahead multiple disk blocks. The first is sync, the rest async.
241  * Trivial modification to the breada algorithm presented in Bach (p.55).
242  */
243 int
244 breadn(vp, blkno, size, rablks, rasizes, nrablks, cred, bpp)
245        struct vnode *vp;
246        daddr_t blkno; int size;
247        daddr_t rablks[]; int rasizes[];
248        int nrablks;
249        struct ucred *cred;
250        struct buf **bpp;
251 {
252        struct buf *bp;
253        int i;
254
255        bp = *bpp = bio_doread(vp, blkno, size, cred, 0);
256
```

```
257            /*
258             * For each of the read-ahead blocks, start a read, if necessary.
259             */
260            for (i = 0; i < nrablks; i++) {
261                    /* If it's in the cache, just go on to next one. */
262                    if (incore(vp, rablks[i]))
263                            continue;
264
265                    /* Get a buffer for the read-ahead block */
266                    (void) bio_doread(vp, rablks[i], rasizes[i], cred, B_ASYNC);
267            }
268
269            /* Otherwise, we had to start a read for it; wait until it's valid. */
270            return (biowait(bp));
271 }
```

——————————————————————————————————————————— kern/vfs_bio.c

**line 270** only waits for the first block. It does not wait for the other blocks to finish
I/O, since those blocks are read-ahead blocks and processed with asynchronous
read.

### 4.8.3   Read Ahead a Single Buffer: breada function

——————————————————————————————————————————— kern/vfs_bio.c

```
273 /*
274  * Read with single-block read-ahead.  Defined in Bach (p.55), but
275  * implemented as a call to breadn().
276  * XXX for compatibility with old file systems.
277  */
278 int
279 breada(vp, blkno, size, rablkno, rabsize, cred, bpp)
280            struct vnode *vp;
281            daddr_t blkno; int size;
282            daddr_t rablkno; int rabsize;
283            struct ucred *cred;
284            struct buf **bpp;
285 {
286
287            return (breadn(vp, blkno, size, &rablkno, &rabsize, 1, cred, bpp));
288 }
```

——————————————————————————————————————————— kern/vfs_bio.c

## 4.9   Releasing Buffer Cache

A buffer can be relased by four ways: by `brelse`, `bdwrite`, `bawrite`, or `bwrite`
function.  The first one releases clean buffer and the latter three releases dirty
buffer. *dirty buffer* means that a buffer which is modified and not yet written to
storage.

### 4.9.1   Just Release: `brelse` function

`brelse` function releases a buffer when the buffer has NOT BEEN MODIFIED. This function

1. returns the buffer to free list and

2. awakens any process that are awaiting for it.

The essential algorithm of this function[2] is

```
1. wakeup all processes that is waiting for ANY buffer to become free
   :from getnewbuf() function

2. wakeup all processes that is waiting for THIS buffer to become free
   :from getblk() function

3. if (buffer contents is invalid or having no physical memory ?)
   {
           if (buffer is invalid ?)
                   enqueue buffer at beginning of AGE free list
           else
                   enqueue buffer at beginning of EMPTY free list
   }
   else
   {
           if
           enqueue buffer at end of free list
   }

4. unlock buffer
```

—————————————————————————————————————————— kern/vfs_bio.c

```
468 /*
469  * Release a buffer on to the free lists.
470  * Described in Bach (p. 46).
471  */
472 void
473 brelse(bp)
474         struct buf *bp;
475 {
476         struct bqueues *bufq;
477         int s;
478
479         KASSERT(ISSET(bp->b_flags, B_BUSY));
480
481         /* Wake up any processes waiting for any buffer to become free. */
482         if (needbuffer) {
483                 needbuffer = 0;
484                 wakeup(&needbuffer);
485         }
486
487         /* Block disk interrupts. */
488         s = splbio();
489
```

```
490          /* Wake up any proceeses waiting for _this_ buffer to become free. */
491          if (ISSET(bp->b_flags, B_WANTED)) {
492                  CLR(bp->b_flags, B_WANTED|B_AGE);
493                  wakeup(bp);
494          }
495
496          /*
497           * Determine which queue the buffer should be on, then put it there.
498           */
499
500          /* If it's locked, don't report an error; try again later. */
501          if (ISSET(bp->b_flags, (B_LOCKED|B_ERROR)) == (B_LOCKED|B_ERROR))
502                  CLR(bp->b_flags, B_ERROR);
503
504          /* If it's not cacheable, or an error, mark it invalid. */
505          if (ISSET(bp->b_flags, (B_NOCACHE|B_ERROR)))
506                  SET(bp->b_flags, B_INVAL);
507
508          if (ISSET(bp->b_flags, B_VFLUSH)) {
509                  /*
510                   * This is a delayed write buffer that was just flushed to
511                   * disk.  It is still on the LRU queue.  If it's become
512                   * invalid, then we need to move it to a different queue;
513                   * otherwise leave it in its current position.
514                   */
515                  CLR(bp->b_flags, B_VFLUSH);
516                  if (!ISSET(bp->b_flags, B_ERROR|B_INVAL|B_LOCKED|B_AGE))
517                          goto already_queued;
518                  else
519                          bremfree(bp);
520          }
521
522          if ((bp->b_bufsize <= 0) || ISSET(bp->b_flags, B_INVAL)) {
523                  /*
524                   * If it's invalid or empty, dissociate it from its vnode
525                   * and put on the head of the appropriate queue.
526                   */
527                  if (LIST_FIRST(&bp->b_dep) != NULL && bioops.io_deallocate)
528                          (*bioops.io_deallocate)(bp);
529                  CLR(bp->b_flags, B_DONE|B_DELWRI);
530                  if (bp->b_vp) {
531                          reassignbuf(bp, bp->b_vp);
532                          brelvp(bp);
533                  }
534                  if (bp->b_bufsize <= 0)
535                          /* no data */
536                          bufq = &bufqueues[BQ_EMPTY];
537                  else
538                          /* invalid data */
539                          bufq = &bufqueues[BQ_AGE];
540                  binsheadfree(bp, bufq);
541          } else {
542                  /*
543                   * It has valid data.  Put it on the end of the appropriate
```

```
544                     * queue, so that it'll stick around for as long as possible.
545                     * If buf is AGE, but has dependencies, must put it on last
546                     * bufqueue to be scanned, ie LRU. This protects against the
547                     * livelock where BQ_AGE only has buffers with dependencies,
548                     * and we thus never get to the dependent buffers in BQ_LRU.
549                     */
550                    if (ISSET(bp->b_flags, B_LOCKED))
551                            /* locked in core */
552                            bufq = &bufqueues[BQ_LOCKED];
553                    else if (!ISSET(bp->b_flags, B_AGE))
554                            /* valid data */
555                            bufq = &bufqueues[BQ_LRU];
556                    else {
557                            /* stale but valid data */
558                            int has_deps;
559
560                            if (LIST_FIRST(&bp->b_dep) != NULL &&
561                                bioops.io_countdeps)
562                                    has_deps = (*bioops.io_countdeps)(bp, 0);
563                            else
564                                    has_deps = 0;
565                            bufq = has_deps ? &bufqueues[BQ_LRU] :
566                                    &bufqueues[BQ_AGE];
567                    }
568                    binstailfree(bp, bufq);
569            }
570
571 already_queued:
572            /* Unlock the buffer. */
573            CLR(bp->b_flags, B_AGE|B_ASYNC|B_BUSY|B_NOCACHE);
574            SET(bp->b_flags, B_CACHE);
575
576            /* Allow disk interrupts. */
577            splx(s);
578 }
```

———————————————————————————————————————— kern/vfs_bio.c

**line 501-502** might be disregarded, if you do not focus on *LFS*, since only the *LFS* uses the *LOCKED free list*.

**line 505-506** B_NOCACHE flag says that the buffer should not be cached after use. Therefore, it is set up with B_INVAL flag. The buffer with this flag can be on a free list, but the buffer cannot be searched by incore function.

**line 508-520** B_VFLUSH flag says that the buffer is being flushed to disk. Buffers are set with this flag by

- vinvalbuf function of kern/vfs_subr.c that flush out and invalidate all buffers associated with a vnode.

- ffs_full_fsync function of ufs/ffs/ffs_vnops.c to flush out all dirty data associated with a vnode.

- ffs_fsync function of ufs/ffs/ffs_vnops.c to flush out ranged dirty data associated with a vnode.

If otherwise specified, ths buffer with `B_VFLUSH` flag stays in free list longer than other buffers: see **line 793-803** of `getnewbuf` function scheduling a buffer of this kind to move from *LRU* free list to *AGE* free list, instead of immediate reuse.

**line 557-566** If *Soft Dependency* facility is not enabled, `has_deps` variable is set to 1.

The `reassign` function used in **line 531** is used to update the status of vnode associated the buffer cache before calling `brelvp` function in line 532. According to the

———————————————————————————————— kern/vfs_subr.c

```
915 /*
916  * Reassign a buffer from one vnode to another.
917  * Used to assign file specific control information
918  * (indirect blocks) to the vnode to which they belong.
919  *
920  * This function must be called at splbio().
921  */
922 void
923 reassignbuf(bp, newvp)
924         struct buf *bp;
925         struct vnode *newvp;
926 {
927         struct buflists *listheadp;
928         int delay;
929
930         /*
931          * Delete from old vnode list, if on one.
932          */
933         if (LIST_NEXT(bp, b_vnbufs) != NOLIST)
934                 bufremvn(bp);
935         /*
936          * If dirty, put on list of dirty buffers;
937          * otherwise insert onto list of clean buffers.
938          */
939         if ((bp->b_flags & B_DELWRI) == 0) {
940                 listheadp = &newvp->v_cleanblkhd;
941                 if (TAILQ_EMPTY(&newvp->v_uobj.memq) &&
942                     (newvp->v_flag & VONWORKLST) &&
943                     LIST_FIRST(&newvp->v_dirtyblkhd) == NULL) {
944                         newvp->v_flag &= ~VONWORKLST;
945                         LIST_REMOVE(newvp, v_synclist);
946                 }
947         } else {
948                 listheadp = &newvp->v_dirtyblkhd;
949                 if ((newvp->v_flag & VONWORKLST) == 0) {
950                         switch (newvp->v_type) {
951                         case VDIR:
952                                 delay = dirdelay;
953                                 break;
954                         case VBLK:
955                                 if (newvp->v_specmountpoint != NULL) {
```

```
956                                        delay = metadelay;
957                                        break;
958                                }
959                                /* fall through */
960                        default:
961                                delay = filedelay;
962                                break;
963                        }
964                        if (!newvp->v_mount ||
965                            (newvp->v_mount->mnt_flag & MNT_ASYNC) == 0)
966                                vn_syncer_add_to_worklist(newvp, delay);
967                }
968        }
969        bufinsvn(bp, listheadp);
970 }
```

———————————————————————————————— kern/vfs_subr.c

where the definition of `brelvp` function is

———————————————————————————————— kern/vfs_subr.c

```
883 /*
884  * Disassociate a buffer from a vnode.
885  */
886 void
887 brelvp(bp)
888        struct buf *bp;
889 {
890        struct vnode *vp;
891        int s;
892
893        if (bp->b_vp == NULL)
894                panic("brelvp: vp NULL, bp %p", bp);
895
896        s = splbio();
897        vp = bp->b_vp;
898        /*
899         * Delete from old vnode list, if on one.
900         */
901        if (LIST_NEXT(bp, b_vnbufs) != NOLIST)
902                bufremvn(bp);
903
904        if (TAILQ_EMPTY(&vp->v_uobj.memq) && (vp->v_flag & VONWORKLST) &&
905            LIST_FIRST(&vp->v_dirtyblkhd) == NULL) {
906                vp->v_flag &= ~VONWORKLST;
907                LIST_REMOVE(vp, v_synclist);
908        }
909
910        bp->b_vp = NULL;
911        HOLDRELE(vp);
912        splx(s);
913 }
```

———————————————————————————————— kern/vfs_subr.c

### 4.9.2  Delayed Write: `bdwrite` function

`bdwrite` function releases a buffer when the buffer has been MODIFIED and EX-
PECTED to be modified soon again.  This function

1. marks the buffer as dirty with `B_DIRTY` flags, but is not immediately written.
   Instead,

2. returns the buffer to the free list and

3. awakens any processes waiting for it.

—————————————————————————————————————————————— kern/vfs_bio.c

```
380 /*
381  * Delayed write.
382  *
383  * The buffer is marked dirty, but is not queued for I/O.
384  * This routine should be used when the buffer is expected
385  * to be modified again soon, typically a small write that
386  * partially fills a buffer.
387  *
388  * NB: magnetic tapes cannot be delayed; they must be
389  * written in the order that the writes are requested.
390  *
391  * Described in Leffler, et al. (pp. 208-213).
392  */
393 void
394 bdwrite(bp)
395         struct buf *bp;
396 {
397         struct proc *p = (curproc != NULL ? curproc : &proc0);  /* XXX */
398         const struct bdevsw *bdev;
399         int s;
400
401         /* If this is a tape block, write the block now. */
402         /* XXX NOTE: the memory filesystem usurpes major device */
403         /* XXX       number 4095, which is a bad idea.          */
404         if (bp->b_dev != NODEV && major(bp->b_dev) != 4095) {
405                 bdev = bdevsw_lookup(bp->b_dev);
406                 if (bdev != NULL && bdev->d_type == D_TAPE) {
407                         bawrite(bp);
408                         return;
409                 }
410         }
411
412         /*
413          * If the block hasn't been seen before:
414          *      (1) Mark it as having been seen,
415          *      (2) Charge for the write,
416          *      (3) Make sure it's on its vnode's correct block list.
417          */
418         s = splbio();
419
420         if (!ISSET(bp->b_flags, B_DELWRI)) {
421                 SET(bp->b_flags, B_DELWRI);
```

```
422                    p->p_stats->p_ru.ru_oublock++;
423                    reassignbuf(bp, bp->b_vp);
424            }
425
426            /* Otherwise, the "write" is done, so mark and release the buffer. */
427            CLR(bp->b_flags, B_NEEDCOMMIT|B_DONE);
428            splx(s);
429
430            brelse(bp);
431 }
```

——————————————————————————————— kern/vfs_bio.c

### 4.9.3   Asynchronous Write: `bawrite` function

`bawrite` function releases a buffer when the buffer has been MODIFIED and NOT EXPECTED to modified soon again. This function

1. schedules an I/O on the buffer, but

2. allows the caller to continue running while the scheduled I/O completes.

Implementation of `bawrite` is the same as `bwrite` function that does synchronous write except the `bawrite` set `B_ASYNC` flag. Now we will describe the reason with source code.

——————————————————————————————— kern/vfs_bio.c

```
433 /*
434  * Asynchronous block write; just an asynchronous bwrite().
435  */
436 void
437 bawrite(bp)
438        struct buf *bp;
439 {
440
441        SET(bp->b_flags, B_ASYNC);
442        VOP_BWRITE(bp);
443 }
```

——————————————————————————————— kern/vfs_bio.c

where the definition of `VOP_BWRITE` is, unless we are compiling assuming we are not using loadable kernel module (LKM),

——————————————————————————————— sys/vnode_if.h

```
1630 static __inline int VOP_BWRITE(bp)
1631        struct buf *bp;
1632 {
1633        struct vop_bwrite_args a;
1634        a.a_desc = VDESC(vop_bwrite);
1635        a.a_bp = bp;
1636        return (VCALL(bp->b_vp, VOFFSET(vop_bwrite), &a));
1637 }
```

——————————————————————————————— sys/vnode_if.h

where the **line 1636** would call, if we are using FFS, `vn_bwrite` function, since the
*vnode operation vector description table* of FFS is defined in `ufs/ffs/vfs_vnops.c`
as, (notice the **line 121**)

```
71 /* Global vfs data structures for ufs. */
72 int (**ffs_vnodeop_p) __P((void *));
73 const struct vnodeopv_entry_desc ffs_vnodeop_entries[] = {
74         { &vop_default_desc, vn_default_error },
75         { &vop_lookup_desc, ufs_lookup },               /* lookup */
76         { &vop_create_desc, ufs_create },               /* create */
77         { &vop_whiteout_desc, ufs_whiteout },           /* whiteout */
78         { &vop_mknod_desc, ufs_mknod },                 /* mknod */
79         { &vop_open_desc, ufs_open },                   /* open */
80         { &vop_close_desc, ufs_close },                 /* close */
81         { &vop_access_desc, ufs_access },               /* access */
82         { &vop_getattr_desc, ufs_getattr },             /* getattr */
83         { &vop_setattr_desc, ufs_setattr },             /* setattr */
84         { &vop_read_desc, ffs_read },                   /* read */
...
121         { &vop_bwrite_desc, vn_bwrite },               /* bwrite */
...
124         { NULL, NULL }
```

The code of `vn_bwrite` called by FFS `VOP_BWRITE` operation is,

———————————————————————————————————— kern/vfs_bio.c

```
371 int
372 vn_bwrite(v)
373         void *v;
374 {
375         struct vop_bwrite_args *ap = v;
376
377         return (bwrite(ap->a_bp));
378 }
```

———————————————————————————————————— kern/vfs_bio.c

Therefore, for FFS, `bawrite` function calls `bwrite` function after setting `B_ASYNC`
flag on the buffer.

### 4.9.4   Synchronous Write: `bwrite` function

`bwrite` function releases a buffer when the buffer has been MODIFIED. This func-
tion ENSURES that the writing the buffer to storage is complete before proceeding.

———————————————————————————————————— kern/vfs_bio.c

```
290 /*
291  * Block write.  Described in Bach (p.56)
292  */
293 int
```

```
294 bwrite(bp)
295        struct buf *bp;
296 {
297        int rv, sync, wasdelayed, s;
298        struct proc *p = (curproc != NULL ? curproc : &proc0);  /* XXX */
299        struct vnode *vp;
300        struct mount *mp;
301
302        vp = bp->b_vp;
303        if (vp != NULL) {
304                if (vp->v_type == VBLK)
305                        mp = vp->v_specmountpoint;
306                else
307                        mp = vp->v_mount;
308        } else {
309                mp = NULL;
310        }
311
312        /*
313         * Remember buffer type, to switch on it later.  If the write was
314         * synchronous, but the file system was mounted with MNT_ASYNC,
315         * convert it to a delayed write.
316         * XXX note that this relies on delayed tape writes being converted
317         * to async, not sync writes (which is safe, but ugly).
318         */
319        sync = !ISSET(bp->b_flags, B_ASYNC);
320        if (sync && mp != NULL && ISSET(mp->mnt_flag, MNT_ASYNC)) {
321                bdwrite(bp);
322                return (0);
323        }
324
325        /*
326         * Collect statistics on synchronous and asynchronous writes.
327         * Writes to block devices are charged to their associated
328         * filesystem (if any).
329         */
330        if (mp != NULL) {
331                if (sync)
332                        mp->mnt_stat.f_syncwrites++;
333                else
334                        mp->mnt_stat.f_asyncwrites++;
335        }
336
337        wasdelayed = ISSET(bp->b_flags, B_DELWRI);
338
339        s = splbio();
340
341        CLR(bp->b_flags, (B_READ | B_DONE | B_ERROR | B_DELWRI));
342
343        /*
344         * Pay for the I/O operation and make sure the buf is on the correct
345         * vnode queue.
346         */
347        if (wasdelayed)
```

```
348                    reassignbuf(bp, bp->b_vp);
349            else
350                    p->p_stats->p_ru.ru_oublock++;
351
352            /* Initiate disk write.  Make sure the appropriate party is charged. */
353            bp->b_vp->v_numoutput++;
354            splx(s);
355
356            VOP_STRATEGY(bp);
357
358            if (sync) {
359                    /* If I/O was synchronous, wait for it to complete. */
360                    rv = biowait(bp);
361
362                    /* Release the buffer. */
363                    brelse(bp);
364
365                    return (rv);
366            } else {
367                    return (0);
368            }
369 }
```
──────────────────────────────────────────── kern/vfs_bio.c

Buffers that are written using `bawrite` or `bwrite` function are placed on the appropriate output queue.  When the output completes, the `brelse` function is called to return those buffers to the free list and to awaken any processes that are waiting for them.

For asynchronous write, the buffer is returned to free list by **line 906** of `biodone` function called by the relevant device driver strategy function.

For synchronous write, the buffer is returned to free list by **line 363** of `bwrite` function after waiting for the completion of write.

## 4.10   References to Source Code

### 4.10.1   `kern/vfs_bio.c` - 334 lines, 21 functions

**Global Variables**

```
LIST_HEAD(bufhashhdr, buf) *bufhashtbl, invalhash; // buffer cache hash
u_long                     bufhash;                 // buffer cache hash mask
TAILQ_HEAD(bqueues, buf)   bufqueues[BQUEUES];      // buffer cache free lists
int                        needbuffer;              // buffer cache locking
struct pool                bufpool;                 // buffers for physio()
```

**Functions**

```
bremfree()
bufinit()
bio_doread()
bread()
breadn()
breada()
bwrite()
```

```
vn_bwrite()
bdwrite()
bawrite()
bdirty()
brelse()
incore()
getblk()
geteblk()
allocbuf()
getnewbuf()
biowait()
biodone()
count_lock_queue()
vfs_bufstats()
```

# Chapter 5

# Vnode

## 5.1  Introduction

The vnode is the focus of all file activity in NetBSD. There is a unique vnode allocated for each active file, directory, mounted-on file, fifo, domain socket, symbolic link and device. The kernel has no concept of a file's structure and so it relies on the information stored in the vnode to describe the file. Thus, the vnode associated with a file holds all the adminstration information pertaining to it.

When a process requests an operation on a file, the vfs interface passes control to a file system type dependent function to carry out the opera-tion. If the file system type dependent function finds that a vnode rep-resenting the file is not in main memory, it dynamically allocates a new vnode from the system main memory pool. Once allocated, the vnode is at-tached to the data structure pointer associated with the cause of the vn-ode allocation and it remains resident in the main memory until the sys-tem decides that it is no longer needed and can be recycled.

## 5.2  Vnode Management Function

The vnode has the following structure:

```
struct vnode {
        struct uvm_object v_uobj;               /* uvm object */
#define v_usecount      v_uobj.uo_refs
#define v_interlock     v_uobj.vmobjlock
        voff_t          v_size;                 /* size of file */
        int             v_flag;                 /* flags */
        int             v_numoutput;            /* num pending writes */
        long            v_writecount;           /* ref count of writers */
        long            v_holdcnt;              /* page & buffer refs */
        daddr_t         v_lastr;                /* last read */
        u_long          v_id;                   /* capability id */
        struct mount    *v_mount;               /* ptr to vfs we are in */
        int             (**v_op)(void *);       /* vnode ops vector */
        TAILQ_ENTRY(vnode) v_freelist;          /* vnode freelist */
        LIST_ENTRY(vnode) v_mntvnodes;          /* vnodes for mount pt */
        struct buflists v_cleanblkhd;           /* clean blocklist head */
```

```
        struct buflists v_dirtyblkhd;           /* dirty blocklist head */
        LIST_ENTRY(vnode) v_synclist;           /* dirty vnodes */
        union {
                struct mount    *vu_mountedhere;/* ptr to mounted vfs */
                struct socket   *vu_socket;     /* unix ipc (VSOCK) */
                struct specinfo *vu_specinfo;   /* device (VCHR, VBLK) */
                struct fifoinfo *vu_fifoinfo;   /* fifo (VFIFO) */
        } v_un;
#define v_mountedhere   v_un.vu_mountedhere
#define v_socket        v_un.vu_socket
#define v_specinfo      v_un.vu_specinfo
#define v_fifoinfo      v_un.vu_fifoinfo
        struct nqlease  *v_lease;               /* Soft ref to lease */
        enum vtype      v_type;                 /* vnode type */
        enum vtagtype   v_tag;                  /* underlying data type */
        struct lock     v_lock;                 /* lock for this vnode */
        struct lock     *v_vnlock;              /* ptr to vnode lock */
        void            *v_data;                /* private data for fs */
};
```

Most functions discussed in this page that operate on vnodes cannot be called from
interrupt context.  The members v_numoutput, v_holdcnt, v_dirtyblkhd,
v_cleanblkhd, v_freelist, and v_synclist are modified in interrupt context and
must be protected by splbio(9) unless it is certain that there is no chance an
interrupt handler will modify them.  The vnode lock must not be acquired within
interrupt context.

## 5.2.1   Vnode Flag

Vnode flags are recorded by v_flag.   Valid flags are:

|          |                                                              |
|----------|--------------------------------------------------------------|
| VROOT    | This vnode is the root of its file system.                   |
| VTEXT    | This vnode is a pure text prototype                          |
| VEXECMAP | This vnode has executable mappings                           |
| VSYSTEM  | This vnode being used by kernel; only used to skip the vflush() operation quota files. |
| VISTTY   | This vnode represents a tty; used when reading dead vnodes. |
| VXLOCK   | This vnode is currently locked to change underlying type.    |
| VXWANT   | A process is waiting for this vnode.                         |
| VBWAIT   | Waiting for output associated with this vnode to complete.   |
| VALIASED | This vnode has an alias.                                     |
| VDIROP   | This vnode is involved in a directory operation.  This flag is used exclusively by LFS. |
| VLAYER   | This vnode is on a layer filesystem.                         |
| VONWORKLST | This vnode is on syncer work-list.                         |
| VDIRTY   | This vnode possibly has dirty pages.                        |

The VXLOCK flag is used to prevent multiple processes from entering the vnode
reclamation code. It is also used as a flag to indicate that reclamation is in progress.
The VXWANT flag is set by threads that wish to be awaken when reclamation is
finished.  Before v_flag can be modified,

the `v_interlock` simplelock must be acquired. See `lock(9)` for details on the kernel locking API.

`vflush(mp, skipvp, flags)`

> Remove any vnodes in the vnode table belonging to mount point mp. If skipvp is not NULL it is exempt from being flushed. The argument flags is a set of flags modifying the operation of vflush(). If MNT_NOFORCE is specified, there should not be any active vnodes and an error is returned if any are found (this is a user error, not a system error). If MNT_FORCE is specified, active vnodes that are found are detached.

### 5.2.2 Reference Counts

Each vnode has three reference counts: `v_usecount`, `v_writecount` and `v_holdcnt`. The first is the number of active references within the kernel to the vnode. This count is maintained by `vref()`, `vrele()`, and `vput()`. The second is the number of active references within the kernel to the vnode performing write access to the file. It is maintained by the `open(2)` and `close(2)` system calls. The third is the number of refer-ences within the kernel requiring the vnode to remain active and not be recycled. This count is maintained by `vhold()` and `holdrele()`. When both the `v_usecount` and `v_holdcnt` reach zero, the vnode is recycled to the `freelist` and may be reused for another file. The transition to and from the `freelist` is handled by `getnewvnode()`, `ungetnewvnode()` and `vrecycle()`. Access to `v_usecount`, `v_writecount` and `v_holdcnt` is also protected by the `v_interlock` simplelock.

The number of pending synchronous and asynchronous writes on the vnode are recorded in `v_numoutput`. It is used by `fsync(2)` to wait for all writes to complete before returning to the user. Its value must only be modified at splbio. See `spl(9)`. It does not track the number of dirty buffers attached to the vnode.

`vref(vp)`

> Increment v_usecount of the vnode vp. Any kernel thread system which uses a vnode (e.g. during the operation of some algorithm or to store in a data structure) should call vref().

`VREF(vp)`

> This function is an alias for vref().

`vrele(vp)`

> Decrement v_usecount of unlocked vnode vp. Any code in the sys-tem which is using a vnode should call vrele() when it is fin-ished with the vnode. If v_usecount of the vnode reaches zero and v_holdcnt is greater than zero, the vnode is placed on the holdlist. If both v_usecount and v_holdcnt are zero, the vnode is placed on the freelist.

`vput(vp)`

> Unlock vnode vp and decrement its v_usecount. Depending of the reference counts, move the vnode to the holdlist or the freel-ist. This operation is functionally equivalent to calling

        VOP_UNLOCK(9) followed by vrele().

vhold(vp)
        Mark the vnode vp as active by incrementing vp->v_holdcnt and
        moving the vnode from the freelist to the holdlist.  Once on the
        holdlist, the vnode will not be recycled until it is released
        with holdrele().

VHOLD(vp)
        This function is an alias for vhold().

holdrele(vp)
        Mark the vnode vp as inactive by decrementing vp->v_holdcnt and
        moving the vnode from the holdlist to the freelist.

HOLDRELE(vp)
        This function is an alias for holdrele().

getnewvnode(tag, mp, vops, vpp)
        Retrieve the next vnode from the freelist.  getnewvnode() must
        choose whether to allocate a new vnode or recycle an existing
        one.  The criterion for allocating a new one is that the total
        number of vnodes is less than the number desired or there are no
        vnodes on either free list.  Generally only vnodes that have no
        buffers associated with them are recycled and the next vnode
        from the freelist is retrieved.  If the freelist is empty, vn-
        odes on the holdlist are considered.  The new vnode is returned
        in the address specified by vpp.

        The argument mp is the mount point for the file system requested
        the new vnode.  Before retrieving the new vnode, the file system
        is checked if it is busy (such as currently unmounting).  An er-
        ror is returned if the file system is unmounted.

        The argument tag is the vnode tag assigned to *vpp->v_tag.  The
        argument vops is the vnode operations vector of the file system
        requesting the new vnode.  If a vnode is successfully retrieved
        zero is returned, otherwise and appropriate error code is re-
        turned.

ungetnewvnode(vp)
        Undo the operation of getnewvnode().  The argument vp is the vn-
        ode to return to the freelist.  This function is needed for
        VFS_VGET(9) which may need to push back a vnode in case of a
        locking race condition.

vrecycle(vp, inter_lkp, p)
        Recycle the unused vnode vp to the front of the freelist.
        vrecycle() is a null operation if the reference count is greater
        than zero.

vcount(vp)
        Calculate the total number of reference counts to a special de-
        vice with vnode vp.

### 5.2.3 Vnode Identifier

Every time a vnode is reassigned to a new file, the vnode capability identifier `v_id` is changed. It is used to maintain the name lookup cache consistency by providing a unique `<vnode *,v_id>` tuple without requiring the cache to hold a reference. The name lookup cache can later compare the vnode's capability identifier to its copy and see if the `vnode` still points to the same file. See `namecache(9)` for details on the name lookup cache.

### 5.2.4 Links to Virtual File System Information

The link to the file system which owns the vnode is recorded by `v_mount`. See `vfsops(9)` for further information of file system mount status.

The `v_op` pointer points to its vnode operations vector. This vector describes what operations can be done to the file associated with the vn-ode. The system maintains one vnode operations vector for each file system type configured into the kernel. The vnode operations vector con-tains a pointer to a function for each operation supported by the file system. See `vnodeops(9)` for a description of vnode operations.

### 5.2.5 Vnode Cache

When not in use, vnodes are kept on the freelist through `v_freelist`. The vnodes still reference valid files but may be reused to refer to a new file at any time. Often, these vnodes are also held in caches in the system, such as the name lookup cache. When a valid vnode which is on the freelist is used again, the user must call `vget()` to increment the reference count and retrieve it from the freelist. When a user wants a new vnode for another file `getnewvnode()` is invoked to remove a vnode from the freelist and initialise it for the new file.

`vget(vp, lockflags)`

> Reclaim vnode vp from the freelist, increment its reference count and lock it. The argument lockflags specifies the lockmgr(9) flags used to lock the vnode. If the VXLOCK is set in vp's v_flag, vnode vp is being recycled in vgone() and the calling thread sleeps until the transition is complete. When it is awakened, an error is returned to indicate that the vnode is no longer usable (possibly having been recycled to a new file system type).

`vgone(vp)`
> Eliminate all activity associated with the vnode vp in preparation for recycling.

### 5.2.6 Type of Object

The type of object the vnode represents is recorded by `v_type`. It is used by generic code to perform checks to ensure operations are performed on valid file system objects. Valid types are:

> VNON    The vnode has no type.

```
VREG    The vnode represents a regular file.
VDIR    The vnode represents a directory.
VBLK    The vnode represents a block special device.
VCHR    The vnode represents a character special device.
VLNK    The vnode represents a symbolic link.
VSOCK   The vnode represents a socket.
VFIFO   The vnode represents a pipe.
VBAD    The vnode represents a bad file (not currently used).
```

Vnode tag types are used by external programs only (eg pstat(8)), and should never be inspected by the kernel.  Its use is deprecated since new `v_tag` values cannot be defined for loadable file systems.  The `v_tag` member is read-only. Valid tag types are:

```
VT_NON       non file system
VT_UFS       universal file system
VT_NFS       network file system
VT_MFS       memory file system
VT_MSDOSFS   FAT file system
VT_LFS       log-structured file system
VT_LOFS      loopback file system
VT_FDESC     file descriptor file system
VT_PORTAL    portal daemon
VT_NULL      null file system layer
VT_UMAP      sample file system layer
VT_KERNFS    kernel interface file system
VT_PROCFS    process interface file system
VT_AFS       AFS file system
VT_ISOFS     ISO file system(s)
VT_UNION     union file system
VT_ADOSFS    Amiga file system
VT_EXT2FS    Linux's EXT2 file system
VT_CODA      Coda file system
VT_FILECORE  filecore file system
VT_NTFS      Microsoft NT's file system
VT_VFS       virtual file system
VT_OVERLAY   overlay file system
```

## 5.2.7   Vnode Lock

All vnode locking operations use `v_vnlock`.  This lock is acquired by calling `vn_lock(9)` and released by calling `vn_unlock(9)`.  The vnode locking operation is complicated because it is used for many purposes.  Sometimes it is used to bundle a series of vnode operations (see `vnodeops(9)`) into an atomic group.  Many file systems rely on it to prevent race conditions in updating file system type specific data structures rather than using their own private locks.  The vnode lock operates as a multiple-reader (shared-access lock) or single-writer lock (exclusive access lock).  The lock may be held while sleeping.  While the `v_vnlock` is acquired, the holder is guaranteed that the vnode will not be reclaimed or invalidated.  Most file system functions require that you hold the vnode lock on entry.  See `lock(9)` for details on the kernel locking API.

For leaf file systems (such as ffs, lfs, msdosfs, etc), `v_vnlock` will point to `v_lock`. For stacked filesystems, `v_vnlock` will generally point to `v_vlock` of the lowest file system. Additionally, the implementation of the vnode lock is the responsibility of the individual file systems and `v_vnlock` may also be `NULL` indicating that a leaf node does not export a lock for vnode locking. In this case, stacked file systems (such as `nullfs`) must call the underlying file system directly for locking.

```
vwakeup(bp)
        Update outstanding I/O count vp->v_numoutput for the vnode
        bp->b_vp and do wakeup if requested and vp->vflag has VBWAIT
        set.
```

## 5.2.8   Private Area

Files and file systems are inextricably linked with the virtual memory system and `v_uobj` contains the data maintained by the virtual memory system. For compatibility with code written before the integration of `uvm(9)` into NetBSD C-preprocessor directives are used to alias the members of `v_uobj`.

Each file system underlying a vnode allocates its own private area and hangs it from `v_data`. If non-null, this area is freed by `getnewvnode()`.

## 5.2.9   Other Vnode-Manipulating Functions

```
vaccess(type, file_mode, uid, gid, acc_mode, cred)
        Do access checking. The arguments file_mode, uid, and gid are
        from the vnode to check. The arguments acc_mode and cred are
        passed directly to VOP_ACCESS(9).

checkalias(vp, nvp_rdev, mp)
        Check to see if the new vnode vp represents a special device for
        which another vnode represents the same device. If such an
        aliases exists the existing contents and the aliased vnode are
        deallocated. The caller is responsible for filling the new vn-
        ode with its new contents.

bdevvp(dev, vpp)
        Create a vnode for a block device. bdevvp() is used for root
        file systems, swap areas and for memory file system special de-
        vices.

cdevvp(dev, vpp)
        Create a vnode for a character device. cdevvp() is used for the
        console and kernfs special devices.

vfinddev(dev, vtype, vpp)
        Lookup a vnode by device number. The vnode is returned in the
        address specified by vpp.

vdevgone(int maj, int min, int minh, enum vtype type)
        Reclaim all vnodes that correspond to the specified minor number
        range minl to minh (endpoints inclusive) of the specified major
```

maj.

vflushbuf(vp, sync)
          Flush all dirty buffers to disk for the file with the locked vn-
          ode vp.  The argument sync specifies whether the I/O should be
          synchronous and vflushbuf() will sleep until vp->v_numoutput is
          zero and vp->v_dirtyblkhd is empty.

vinvalbuf(vp, flags, cred, p, slpflag, slptimeo)
          Flush out and invalidate all buffers associated with locked vn-
          ode vp.  The argument p and cred specified the calling process
          and its credentials.  The arguments flags, slpflag and slptimeo
          are ignored in the present implementation.  If the operation is
          successful zero is returned, otherwise and appropriate error
          code is returned.

vtruncbuf(vp, lbn, slpflag, slptimeo)
          Destroy any in-core buffers past the file truncation length for
          the locked vnode vp.  The truncation length is specified by lbn.
          vtruncbuf() will sleep while the I/O is performed,  The sleep(9)
          flag and timeout are specified by the arguments slpflag and
          slptimeo respectively.  If the operation is successful zero is
          returned, otherwise and appropriate error code is returned.

vprint(label, vp)
          This function is used by the kernel to dump vnode information
          during a panic.  It is only used if kernel option DIAGNOSTIC is
          compiled into the kernel.  The argument label is a string to
          prefix the information dump of vnode vp.

## 5.3   Vnode Attributes

Vnode attributes describe attributes of a file or directory including file permissions,
owner, group, size, access time and modication time.

A vnode attribute has the following structure:

```
struct vattr {
        enum vtype      va_type;        /* vnode type (for create) */
        mode_t          va_mode;        /* files access mode and type */
        nlink_t         va_nlink;       /* number of references to file */
        uid_t           va_uid;         /* owner user id */
        gid_t           va_gid;         /* owner group id */
        long            va_fsid;        /* file system id (dev for now) */
        long            va_fileid;      /* file id */
        u_quad_t        va_size;        /* file size in bytes */
        long            va_blocksize;   /* blocksize preferred for i/o */
        struct timespec va_atime;       /* time of last access */
        struct timespec va_mtime;       /* time of last modification */
        struct timespec va_ctime;       /* time file changed */
        u_long          va_gen;         /* generation number of file */
        u_long          va_flags;       /* flags defined for file */
        dev_t           va_rdev;        /* device the special file represents */
```

```
        u_quad_t       va_bytes;      /* bytes of disk space held by file */
        u_quad_t       va_filerev;    /* file modification number */
        u_int          va_vaflags;    /* operations flags, see below */
        long           va_spare;      /* remain quad aligned */
};
```

A field value of `VNOVAL` represents a field whose value is unavailable or which is not to be changed. Valid flag values for `va_flags` are:

```
        VA_UTIMES_NULL  utimes argument was NULL
        VA_EXCLUSIVE    exclusive create request
```

Vnode attributes for a file are set by the vnode operation `VOP_SETATTR(9)`. Vnode attributes for a file are retrieved by the vnode operation `VOP_GETATTR(9)`. For more information on vnode operations see `vnodeops(9)`.

## 5.4 Vnode Operation about Filesystem Hierarchy

The vnode operations vector describes what operations can be done to the file associated with the vnode. The system maintains one vnode operations vector for each file system type configured into the kernel. The vnode operations vector contains a pointer to a function for each operation supported by the file system. Many of the functions described in the vnode operations vector are closely related to their corresponding system calls. In most cases, they are called as a result of the system call associated with the operation being invoked.

If the file system type does not support a specific operation, it must nevertheless assign an appropriate function in the vnode operations vector to do the minimum required of it. In most cases, such functions either do nothing or return an error value to the effect that it is not supported.

### 5.4.1 Overview

**Opening a File**

When an applicatin opens a file that does not currently have an in-memory vnode, the client filesystem calls the `getnewvnode` routine to allocate a new vnode.

The `getnewvnode` routine removes the least recently used vnode from the front of the free list and calls the `reclaim` operation to notify the filesystem currently using the vnode that that vnode is about to be reused.

**Closing a File**

When the final file-entry reference to a file is closed, the usage count on the vnode drops to zero and the vnode interface calls the `inactive` vnode operation. The `inactive` call

- notifies the underlying system that the file is no longer being used.

- The filesystem will often use this call to write dirty data back to the file, but will not typically reclaim the buffers.

**Disassociation with Underlying Objects**

The `reclaim` operation

- writes back any dirty data associated with the underlying object such as *inode*,

- removes the underlying object from any lists that it is on (such as hash lists used to find it), and

- frees up any auxiliary storage that was being used by the object.

This ability, combined with the ability to associate new objects with the vnode, provides functionality with usefulness that goes far beyond simply allowing vnodes to be moved from one filesystem to another. By replacing an existing object with an object from the `dead filesystem` — a filesystem in which all operations except *close* fail — the kernel revokes the objects. Internally, this revocation of an object is provided by the `vgone` routine.

The recovation service is used to support forcible unmounting of filesystems. It is also possible to downgrade a mounted filesystem from read-write to read-only. Instead of access being revoked on every active file within the filesystem, only those files with a nonzero number of references for writing have their access revoked. The ability to revoke objects is exported to processes through the `revoke` system call.

**Vnode Locking**

The *lock* and *unlock* operators allow the callers of the vnode interface to provide hints to the code that implement operations on the underlying objects. Stateless filesystem suc has NFS ignore these hints. Stateful filesystems such as FFS, however, can use gints to avoid doing extra work.

For example, an `open` system call requesting that a new file be created requires two major phases: *lookup* and *create*. The details are

1. First, a *lookup* call is done to see if the file already exists.

2. For stateful filesystem, before the lookup is started, a *lock* request is made on the directory being searched.

3. While scanning through the directory checking for the name, the lookup code also identifies a location within the directory that contains enough space to hold the new name.

4. If the name does not already exists, the *open* code verifies that the user has permission to create the file. If the user is not eligible to create the new file, then the *abortop* operator is called to release any resources held in reserve.

5. Otherwise, *create* operation is called.

6. If the filesystem is stateful, then it can simply create the name in the previously identified space.
   However, If the filesystem is stateless, then it cannot lock the directory, so the *create* operator must rescan the directory to find space and to verift that the name has not been created since the lookup.

### 5.4.2   `componentname` structure

Many of the functions in the vnode operations vector take a componentname structure. Is is used to encapsulate many parameters into a singla function argument. It has the following structure:

```
struct componentname {
        /*
         * Arguments to lookup.
```

```
         */
        u_long  cn_nameiop;    /* namei operation */
        u_long  cn_flags;      /* flags to namei */
        struct proc *cn_proc;  /* process requesting lookup */
        struct ucred *cn_cred; /* credentials */
        /*
         * Shared between lookup and commit routines.
         */
        char    *cn_pnbuf;     /* pathname buffer */
        const char *cn_nameptr; /* pointer to looked up name */
        long    cn_namelen;    /* length of looked up component */
        u_long  cn_hash;       /* hash value of looked up name */
        long    cn_consume;    /* chars to consume in lookup() */
};
```

The top half of the structure is used exclusively for the pathname lookups using `VOP_LOOKUP()` and is initialized by the caller. The semantics of the lookup are affected by the operation specified in `cn_nameiop` and the flags specified in `cn_flags`. Valid operations are:

```
        LOOKUP  perform name lookup only
        CREATE  setup for file creation
        DELETE  setup for file deletion
        RENAME  setup for file renaming
        OPMASK  mask for operation
```

Valid values for `cn->cn_flags` are:

```
        LOCKLEAF    lock inode on return
        LOCKPARENT  want parent vnode returned locked
        WANTPARENT  want parent vnode returned unlocked
        NOCACHE     name must not be left in name cache (see namecache(9))
        FOLLOW      follow symbolic links
        NOFOLLOW    do not follow symbolic links (pseudo)
        MODMASK     mask of operational modifiers
```

No vnode operations may be called from interrupt context. Most opera-tions also require the vnode to be locked on entry. To prevent dead-locks, when acquiring locks on multiple vnodes, the lock of parent direc-tory must be acquired before the lock on the child directory.

### 5.4.3  Pathname Searching

```
int (*vop_lookup)()       VOP_LOOKUP      Lookup file name in name cache
```

### 5.4.4  Name Creation

```
int (*vop_create)()       VOP_CREATE      Create a new file
int (*vop_mknod)()        VOP_MKNOD       Make a new device
int (*vop_link)()         VOP_LINK        Link a file
int (*vop_symlink)()      VOP_SYMLINK     Create a symbolic link
int (*vop_mkdir)()        VOP_MKDIR       Make a new directory
```

### 5.4.5   Name Change/Deletion

```
int (*vop_rename)()      VOP_RENAME      Rename a file
int (*vop_remove)()      VOP_REMOVE      Remove a file
int (*vop_rmdir)()       VOP_RMDIR       Remove a directory
```

### 5.4.6   Attribute Manipulation

```
int (*vop_access)()      VOP_ACCESS      Determine file accessibility
int (*vop_getattr)()     VOP_GETATTR     Get file attributes
int (*vop_setattr)()     VOP_SETATTR     Set file attributes
```

### 5.4.7   Object Interpretation

```
int (*vop_open)()        VOP_OPEN        Open a file
int (*vop_readdir)()     VOP_READDIR     Read directory entry
int (*vop_readlink)()    VOP_READLINK    Read contents of a symlink
int (*vop_mmap)()        VOP_MMAP        Map file into user address space
int (*vop_close)()       VOP_CLOSE       Close a file

int (*vop_seek)()        VOP_SEEK        Test if file is seekable
int (*vop_bmap)()        VOP_BMAP        Logical block number conversion
int (*vop_pathconf)()    VOP_PATHCONF    Implement POSIX pathconf support
int (*vop_print)()       VOP_PRINT       Print debugging information
```

### 5.4.8   Process Control

None of these operatos modifies the object in the filestore. They are simply using
the object for naming or directing the desired operation.

```
int (*vop_advlock)()     VOP_ADVLOCK     Advisory record locking
int (*vop_ioctl)()       VOP_IOCTL       Perform device-specific I/O

int (*vop_fcntl)()       VOP_FCNTL       Perform file control
int (*vop_poll)()        VOP_POLL        Test if poll event has occurred
```

### 5.4.9   Object Management

```
int (*vop_lock)()        VOP_LOCK        Sleep until vnode lock is free
int (*vop_unlock)()      VOP_UNLOCK      Wake up process sleeping on lock
int (*vop_inactive)()    VOP_INACTIVE    Release the inactive vnode
int (*vop_reclaim)()     VOP_RECLAIM     Reclaim vnode for another file
int (*vop_abortop)()     VOP_ABORTOP     Abort pending operation

int (*vop_revoke)()      VOP_REVOKE      Eliminate vode activity
int (*vop_islocked)()    VOP_ISLOCKED    Test if vnode is locked
int (*vop_lease)()       VOP_LEASE       Validate vnode credentials
int (*vop_bwrite)()      VOP_BWRITE      Write a file system buffer
int (*vop_whiteout)()    VOP_WHITEOUT    Whiteout vnode
int (*vop_strategy)()    VOP_STRATEGY    Read/write a file system buffer

VOP_INACTIVE(vp, p)
        Release the inactive vnode.  VOP_INACTIVE() is called when the
        kernel is no longer using the vnode.  This may be because the
        reference count reaches zero or it may be that the file system
```

is being forcibly unmounted while there are open files. It can
be used to reclaim space for open but deleted files. The argu-
ment vp is the locked vnode to be released. The argument p is
the calling process. If the operation is successful zero is re-
turned, otherwise an appropriate error code is returned. The
vnode vp must be locked on entry, and will be unlocked on re-
turn.

## 5.5 Vnode Operation about Storage

### 5.5.1 Object Creation and Deletion

```
int (*vop_valloc)()       VOP_VALLOC       Allocate fs-specific data
int (*vop_vfree)()        VOP_VFREE        Release file resources

int (*vop_balloc)()       VOP_BALLOC       Allocate physical blocks
int (*vop_reallocblks)()  VOP_REALLOCBLKS  rearrange blocks as contiguous
```

### 5.5.2 Attribute Update

```
int (*vop_update)()       VOP_UPDATE       Update time on a file
```

### 5.5.3 Object Read and Write

```
int (*vop_blkatoff)()     VOP_BLKATOFF     Retrieve buffer from offset
int (*vop_read)()         VOP_READ         Read from a file
int (*vop_write)()        VOP_WRITE        Write to a file
int (*vop_fsync)()        VOP_FSYNC        Flush pending data to disk

int (*vop_getpages)()     VOP_GETPAGES     Read VM pages from file
int (*vop_putpages)()     VOP_PUTPAGES     Write VM pages to file
```

### 5.5.4 Change in Space Allocation

```
int (*vop_truncate)()     VOP_TRUNCATE     Truncate file and free blocks
```

## 5.6 High-Level Vnode Convenient Function

Vnode operations for a file system type generally should not be called
directly from the kernel, but accessed indirectly through the high-level
convenience functions discussed in vnsubr(9).

```
vn_default_error(v)
```
   A generic "default" routine that just returns error. It is used
   by a file system to specify unsupported operations in the vnode
   operations vector.

### 5.6.1 Filesystem Hierarchy

```
vn_stat(fdata, sb, p)
```
   Common code for a vnode stat operation. The vnode is specified
   by the argument fdata and sb is the buffer to return the stat
   information. The argument p is the calling process. vn_stat()

basically calls the vnode operation VOP_GETATTR(9) and transfer
the contents of a vattr structure into a struct stat.  If the
operation is successful zero is returned, otherwise an appropri-
ate error code is returned.

vn_readdir(fp, buf, segflg, count, done, p, cookies, ncookies) Common code
for reading the contents of a directory.  The argument fp is the
file structure, buf is the buffer for placing the struct dirent
structures.  The arguments cookies and ncookies specify the ad-
dresses for the list and number of directory seek cookies gener-
ated for NFS.  Both cookies and ncookies should be NULL is they
aren't required to be returned by vn_readdir().  If the opera-
tion is successful zero is returned, otherwise an appropriate
error code is returned.

vn_isunder(dvp, rvp, p)
Common code to check if one directory specified by the vnode rvp
can be found inside the directory specified by the vnode dvp.
The argument p is the calling process.  vn_isunder() is intended
to be used in chroot(2), chdir(2), fchdir(2), etc., to ensure
that chroot(2) actually means something.  If the operation is
successful zero is returned, otherwise 1 is returned.

## 5.6.2   General File I/O

vn_open(ndp, fmode, cmode)
Common code for vnode open operations.  The pathname is de-
scribed in the nameidata pointer (see namei(9)).  The arguments
fmode and cmode specify the open(2) file mode and the access
permissions for creation.  vn_open() checks  permissions and in-
vokes the VOP_OPEN(9) or VOP_CREATE(9) vnode operations.  If the
operation is successful zero is returned, otherwise an appropri-
ate error code is returned.

vn_close(vp, flags, cred, p)
Common code for a vnode close.  The argument vp is the locked
vnode of the vnode to close.  vn_close() simply locks the vnode,
invokes the vnode operation VOP_CLOSE(9) and calls vput() to re-
turn the vnode to the freelist or holdlist.  Note that
vn_close() expects an unlocked, referenced vnode and will deref-
erence the vnode prior to returning.  If the operation is suc-
cessful zero is returned, otherwise an appropriate error is re-
turned.

vn_closefile(fp, p)
Common code for a file table vnode close operation.  The file is
described by fp and p is the calling process.  vn_closefile()
simply calls vn_close() with the appropriate arguments.

vn_read(fp, offset, uio, cred, flags)
Common code for a file table vnode read.  The argument fp is the
file structure,  The argument offset is the offset into the
file.  The argument uio is the uio structure describing the mem-
ory to read into.  The caller's credentials are specified in cred.  The flags argument can define FOF_UPDATE_OFFSET to update
the read position in the file.  If the operation is successful
zero is returned, otherwise an appropriate error is returned.

vn_write(fp, offset, uio, cred, flags)
  Common code for a file table vnode write.  The argument fp is
  the file structure,  The argument offset is the offset into the
  file.  The argument uio is the uio structure describing the mem-
  ory to read from.  The caller's credentials are specified in
  cred.  The flags argument can define FOF_UPDATE_OFFSET to update
  the read position in the file.  If the operation is successful
  zero is returned, otherwise an appropriate error is returned.

vn_rdwr(rw, vp, base, len, offset, segflg, ioflg, cred, aresid, p) Com-
  mon code to package up an I/O request on a vnode into a uio and
  then perform the I/O.  The argument rw specifies whether the I/O
  is a read (UIO_READ) or write (UIO_WRITE) operation.  The un-
  locked vnode is specified by vp.  The arguments p and cred are
  the calling process and its credentials.  The remaining argu-
  ments specify the uio parameters.  For further information on
  these parameters see uiomove(9).

vn_bwrite(ap)
  Common code for block write operations.

vn_writechk(vp)
  Common code to check for write permission on the vnode vp.  A
  vnode is read-only if it is in use as a process's text image.
  If the vnode is read-only ETEXTBSY is returned, otherwise zero
  is returned to indicate that the vnode can be written to.

vn_fcntl(fp, com, data, p)
  Common code for a file table vnode fcntl(2) operation.  The file
  is specified by fp.  The argument p is the calling process.
  vn_fcntl() simply locks the vnode and invokes the vnode opera-
  tion VOP_FCNTL(9) with the command com and buffer data.  The vn-
  ode is unlocked on return.  If the operation is successful zero
  is returned, otherwise an appropriate error is returned.

vn_ioctl(fp, com, data, p)
  Common code for a file table vnode ioctl operation.  The file is
  specified by fp.  The argument p is the calling process.
  vn_ioctl() simply locks the vnode and invokes the vnode opera-
  tion VOP_IOCTL(9) with the command com and buffer data.  The vn-
  ode is unlocked on return.  If the operation is successful zero
  is returned, otherwise an appropriate error is returned.

### 5.6.3  Advanced I/O

vn_lock(vp, flags)
  Common code to acquire the lock for vnode vp.  The argument
  flags specifies the lockmgr(9) flags used to lock the vnode.  If
  the operation is successful zero is returned, otherwise an ap-
  propriate error code is returned.  The vnode interlock

v_interlock is releases on return.

vn_lock() must not be called when the vnode's reference count is
zero.   Instead, vget(9) should be used.

vn_poll(fp, events, p)
        Common code for a file table vnode poll operation.   vn_poll()
        simply calls VOP_POLL(9) with the events events and the calling
        process p.   If the operation is success zero is returned, other-
        wise an appropriate error code is returned.

vn_markexec(vp)
        Common code to mark the vnode vp as containing executable code
        of a running process.

vn_setrecurse(vp)
        Common code to enable LK_CANRECURSE on the vnode lock for vnode
        vp.   vn_setrecurse() returns the new lockmgr(9) flags after the
        update.

vn_restorerecurse(vp, flags)
        Common code to restore the vnode lock flags for the vnode vp.
        It is called when done with vn_setrecurse().

## 5.7   References to Source Code

### 5.7.1   vfs_subr.c - 2846 lines, 57 functions

**Gloval Variables**

    iftov_tab
    vttoif_tab
    doforce              Switch to permit forcible unmounting
    prtactive

    [ Global System List ]

    vnode_free_list      Vnode free list
    vnode_hold_list      Vnode hold list
    vnode_pool           Memory pool for vnodes

    mountlist            Mounted filesystem list
    vfs_list             VFS list
    nfs_pub              Publicly exported FS

    [ Root Filesystem and Device Information ]

    rootfs
    rootvnode
    root_device

    [ Locks to Manage Global System Lists ]

```
mountlist_slock
mntid_slock
mntvnode_slock
vnode_free_list_slock
spechash_slock
```

## Functions

[ VFS Management ]

```
vfs_busy            Mark a mount point as busy, to synchronize access and
                       to delay unmount
vfs_unbusy
vfs_getvfs          Lookup a mount point by filesystem identifier
vfs_getnewfsid      Get a new unique fsid
makefstype          Make a 'unique' number from a mount type name
vfs_mountedon       Check to see if a filesystem is mounted on a block device
vfs_getopsbyname    Given a file system name, look up the vfsops
```

[ vnode Management ]

```
vntblinit           Initialize the vnode management data structures
vattr_null          Set vnode attributes to VNOVAL

getnewvnode         Return the next vnode from the free list
ungetnewvnode       used by VFS_VGET functions who may need to push back a vnode
insmntque           Move a vnode from one mount queue to another
vwakeup             Update outstanding I/O count and do wakeup if requested
vinvalbuf           Flush out and invalidate all buffers associated with a vnode
vtruncbuf           Destroy any in core blocks past the truncation length
vflushbuf
bgetvp              Associate a buffer with a vnode
brelvp              Disassociate a buffer with a vnode
reassignbuf         Reassign a buffer from one vnode to another
bdevvp              Create a vnode for a block device
cdevvp              Create a vnode for a character device
getdevvp            Common routine used by bdevvp(), cdevvp()
checkalias          Check to see if the new vnode represents a special device
                    for which we already have a vnode
vget                Grab a particular vnode from the free list
vput                just unlock and vrele()
vrele
vhold
holdrele
vref
vflush              Remove any vnodes in the vnode table belonging to
                       mount point mp
vclean              Disassociate the underlying file system from a vnode
vrecycle            Recycle an unused vnode to the front of the free list
vgone               Eliminate all activity associated with a vnode
vgonel              vgone(), with the vp interlock held
vfinddev            Lookup a vnode by device number
vdevgone            Revoke all the vnodes corresponding to the specified
                       minor number range
```

```
vcount

[ Routines About sysctl support ]

vfs_sysctl
sysctl_vnode

[ Exportable File System ]

vfs_hand_addrlist  Build hash lists of net addresses and hang them off
                       the mount point
vfs_free_netcred
vfs_free_addrlist  Free the net address hash lists that are hanging
                       off the mount points
vfs_export
vfs_setpublicfs    Set the publicly exported filesystem (WebNFS)
vfs_export_lookup
vaccess            Do the usual access checking

[ System Bootstrap and Shutdown ]

vfs_attach
vfs_detach
vfs_reinit
vfs_unmountall     Unmount all file systems
vfs_shutdown       Sync and unmount file systems before shutting down
vfs_mountroot      Mount the root file system
vfs_rootmountalloc Lookup a filesystem type, and allocate and
                       initialize a mount structure

[ Diagnostics ]

vprint
vfs_buf_print
vfs_vnode_print
printlockedvnodes
```

## 5.7.2  vfs_vnops.c - 808 lines, 19 functions

**Gloval Variables**

```
struct  fileops vnops = {
        vn_read, vn_write, vn_ioctl, vn_fcntl, vn_poll,
        vn_statfile, vn_closefile, vn_kqfilter
};
```

**Functions**

```
[ Exported File Operation ]

vn_read            used by vn_rdwr()
vn_write
vn_ioctl
vn_fcntl
```

```
vn_poll
vn_statfile
vn_closefile       File table vnode close routine (just cover function)
vn_kqfilter        [?] File table vnode kqfilter routine

[ High-Level Vnode Convenient Function ]

vn_open            used by sys_open()
vn_writechk        Check for write permissions on the specified vnode.
vn_markexec        Mark a vnode as having executable mappings
vn_marktext        Mark a vnode as being the text of a process
vn_close
vn_rdwr            Package up an I/O request on a vnode into a uio and do it
vn_readdir
vn_stat
vn_lock

[ ? Lock Management ]

vn_setrecurse      [?] Enable LK_CANRECURSE on lock. Return prior status
vn_restorerecurse  [?] Called when done with locksetrecurse
```

### 5.7.3  `vfs_syscalls.c` - 3116 lines, 65 functions

**Gloval Variables**

```
dovfsusermount      When set to 1, any user can mount filesystem
mountcompatnames
nmountcompatnames
```

**Functions**

```
[ System Calls Related with Vnode ! ]

sys_mount
sys_unmount
sys_sync
sys_statfs
sys_fstatfs
sys_getfsstat
sys_fchdir
sys_fchroot        Change this process's notion of the root directory
sys_chdir
sys_chroot         Change notion of root (``/'') directory
sys_open
sys_getfh          Get file handle system call
sys_fhopen         Open a file given a file handle
sys_fhstat
sys_fhstatfs       Returns information about a mounted file system
sys_mknod
sys_mkfifo         Create a named pipe
sys_link
sys_symlink
sys_undelete       Delete a whiteout from the filesystem [WOW ! undelete !]
```

```
sys_unlink
sys_lseek
sys_pread          Positional read system call
sys_preadv
sys_pwrite
sys_pwritev
sys_access
sys___stat13
sys___lstat13
sys_pathconf       Get configurable pathname variables
sys_readlink
sys_chflags
sys_fchflags
sys_lchflags
sys_chmod
sys_fchmod         Change mode of a file given a file descriptor
sys_lchmod         this version does not follow links
sys_chown
sys_fchown
sys_lchown
sys_utime
sys_futime
sys_lutime
sys_truncate
sys_fruncate
sys_fsync          Sync an open file
sys_fdatasync      Sync the data of an open file
sys_rename
sys_mkdir
sys_rmdir
sys_getdents       Read a block of directory in filesystem independent format
sys_umask
sys_revoke

[ POSIX Compatable System Calls ]

sys___posix_chown
sys___posix_fchown
sys___posix_lchown
sys___posix_rename

[ Support Routine ]

checkdirs          Support routine for sys_mount()
dounmount          Actual worker for sys_unmount()
getvnode           Convert a user file descriptor to a kernel file entry

[ Common Routine ]

change_dir         Common routine for chroot and chdir
change_flags       Common routine to change flags of a file
change_mode        Common routine to change mode of a file
change_owner
```

```
change_utimes
rename_files
```

# Chapter 6

# UVM

## 6.1 Introduction

UVM is a virtual memory system of the NetBSD/sparc64 release 1.6. UVM has better performance especially in managing memory-mapped files and copy-on-write memory, than the 4.4BSD VM, which is derived from Mach VM. In UVM, the virtual memory object, fault handling, and pager code is replaced from the 4.4BSD VM. And, a new virtual memory based data movement mechanisms is introduced.

## 6.2 UVM Overview

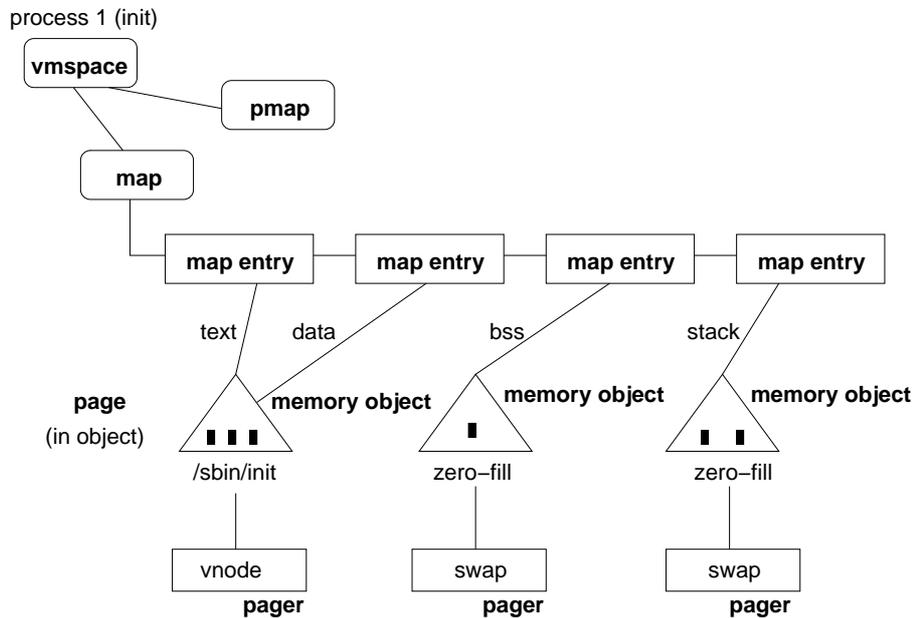

Figure 6.1: The five main machine-independent abstractions in UVM

Both BSD VM and UVM can be divied into two layers: a small mahcine-dependent layer, and a larger machine-independent layer.

The machine-dependent layer used by both BSD VM and UVM is called the *pmap layer*. The pamp layer handles the low level details of programming a processor's MMU. This task conststs of

- managing the mappings of a virtual address.

- managing the mappings of a page of physical memory.

The machine-independent code contains functions that perform the high-level operations of the VM system. Such functions include

- managing a process' file mappings,

- requesting data from backing store,

- paging out memory when it becomes scarce,

- managing the allocation of physical memory, and

- managing copy-on-write memory.

Figure 6.1 shows the five main abstractions that correspond to data structures in both BSD VM and UVM that activities of the machine-independent layer are centered around.

## 6.2.1   Virtual Memory Space

*Virtual memory space* describes both the machine dependent and machine independent parts of process's virtual address space. The `vmspace` structure contains

- pointers to memory `map` structures, and

- statistics on the process's memory usage.

———————————————————————————————— uvm/uvm_extern.h

```
459 /*
460  * Shareable process virtual address space.
461  * May eventually be merged with vm_map.
462  * Several fields are temporary (text, data stuff).
463  */
464 struct vmspace {
465         struct  vm_map vm_map;   /* VM address map */
466         int     vm_refcnt;       /* number of references */
467         caddr_t vm_shm;          /* SYS5 shared memory private data XXX */
468 /* we copy from vm_startcopy to the end of the structure on fork */
469 #define vm_startcopy vm_rssize
470         segsz_t vm_rssize;       /* current resident set size in pages */
471         segsz_t vm_swrss;        /* resident set size before last swap */
472         segsz_t vm_tsize;        /* text size (pages) XXX */
473         segsz_t vm_dsize;        /* data size (pages) XXX */
474         segsz_t vm_ssize;        /* stack size (pages) */
475         caddr_t vm_taddr;        /* user virtual address of text XXX */
476         caddr_t vm_daddr;        /* user virtual address of data XXX */
477         caddr_t vm_maxsaddr;     /* user VA at max stack growth */
478         caddr_t vm_minsaddr;     /* user VA at top of stack */
479 };
```

———————————————————————————————— uvm/uvm_extern.h

## 6.2.2 Memory Map

*Memory map* describes the machine-independent part of the virtual address space of a process or the kernel. Each map structure on the system contains a sorted doubly-linked list of *map entry* structures. Each entry structure contains a record of a mapping in the map's virtual address space. This record includes

- starting and ending virtual address

- a pointer to the memory object mapped into that address range

- the attributes of the mapping

—————————————————————————————————— uvm/uvm_map.h

```
114 /*
115  * Address map entries consist of start and end addresses,
116  * a VM object (or sharing map) and offset into that object,
117  * and user-exported inheritance and protection information.
118  * Also included is control information for virtual copy operations.
119  */
120 struct vm_map_entry {
121         struct vm_map_entry     *prev;          /* previous entry */
122         struct vm_map_entry     *next;          /* next entry */
123         vaddr_t                 start;          /* start address */
124         vaddr_t                 end;            /* end address */
125         union {
126                 struct uvm_object *uvm_obj;     /* uvm object */
127                 struct vm_map   *sub_map;       /* belongs to another map */
128         } object;                               /* object I point to */
129         voff_t                  offset;         /* offset into object */
130         int                     etype;          /* entry type */
131         vm_prot_t               protection;     /* protection code */
132         vm_prot_t               max_protection; /* maximum protection */
133         vm_inherit_t            inheritance;    /* inheritance */
134         int                     wired_count;    /* can be paged if == 0 */
135         struct vm_aref          aref;           /* anonymous overlay */
136         int                     advice;         /* madvise advice */
137 #define uvm_map_entry_stop_copy flags
138         u_int8_t                flags;          /* flags */
139
140 #define UVM_MAP_STATIC          0x01            /* static map entry */
141 #define UVM_MAP_KMEM            0x02            /* from kmem entry pool */
142
143 };
...
199 struct vm_map {
200         struct pmap *           pmap;           /* Physical map */
201         struct lock             lock;           /* Lock for map data */
202         struct vm_map_entry     header;         /* List of entries */
203         int                     nentries;       /* Number of entries */
204         vsize_t                 size;           /* virtual size */
205         int                     ref_count;      /* Reference count */
206         struct simplelock       ref_lock;       /* Lock for ref_count field */
207         struct vm_map_entry *   hint;           /* hint for quick lookups */
208         struct simplelock       hint_lock;      /* lock for hint storage */
```

```
209        struct vm_map_entry *  first_free;   /* First free space hint */
210        int                    flags;        /* flags */
211        struct simplelock      flags_lock;   /* Lock for flags field */
212        unsigned int           timestamp;    /* Version number */
213 #define min_offset            header.start
214 #define max_offset            header.end
215 };
216
217 /* vm_map flags */
218 #define VM_MAP_PAGEABLE     0x01           /* ro: entries are pageable */
219 #define VM_MAP_INTRSAFE     0x02           /* ro: interrupt safe map */
220 #define VM_MAP_WIREFUTURE   0x04           /* rw: wire future mappings */
221 #define VM_MAP_BUSY         0x08           /* rw: map is busy */
222 #define VM_MAP_WANTLOCK     0x10           /* rw: want to write-lock */
223 #define VM_MAP_DYING        0x20           /* rw: map is being destroyed */
```

—————————————————————————————————— uvm/uvm_map.h

## 6.2.3  Memory Object

*Memory object* describes a file, a zero-fill memory area, or a device that can be mapped into a virtual address space. In UVM, a memory object consists of either a **vm_amap** or **uvm_object** structure.

—————————————————————————————————— uvm/uvm_object.h

```
44 /*
45  * uvm_object: all that is left of mach objects.
46  */
47
48 struct uvm_object {
49        struct simplelock      vmobjlock;     /* lock on memq */
50        struct uvm_pagerops    *pgops;        /* pager ops */
51        struct pglist          memq;          /* pages in this object */
52        int                    uo_npages;     /* # of pages in memq */
53        int                    uo_refs;       /* reference count */
54 };
```

—————————————————————————————————— uvm/uvm_object.h

—————————————————————————————————— uvm/uvm_amap.h

```
42 /*
43  * an amap structure contains pointers to a set of anons that are
44  * mapped together in virtual memory (an anon is a single page of
45  * anonymous virtual memory -- see uvm_anon.h).  in uvm we hide the
46  * details of the implementation of amaps behind a general amap
47  * interface.  this allows us to change the amap implementation
48  * without having to touch the rest of the code.  this file is divided
49  * into two parts: the definition of the uvm amap interface and the
50  * amap implementation-specific definitions.
51  */
...
167 struct vm_amap {
168        struct simplelock am_l; /* simple lock [locks all vm_amap fields] */
```

```
169         int am_ref;             /* reference count */
170         int am_flags;           /* flags */
171         int am_maxslot;         /* max # of slots allocated */
172         int am_nslot;           /* # of slots currently in map ( <= maxslot) */
173         int am_nused;           /* # of slots currently in use */
174         int *am_slots;          /* contig array of active slots */
175         int *am_bckptr;         /* back pointer array to am_slots */
176         struct vm_anon **am_anon; /* array of anonymous pages */
177 #ifdef UVM_AMAP_PPREF
178         int *am_ppref;          /* per page reference count (if !NULL) */
179 #endif
180 };
```

———————————————————————————————— uvm/uvm_amap.h

### 6.2.4  Pager

*Pager* describes how backing store can be accessed. Each memory object on the system has a pager that points to a list of functions used by the object to fetch and store pages between physical memory and backing store.

Pages are read in from backing store

- when a process faults on them, or

- in anticipation of a process faulting on them

. Pages are written out to backing store

- at the request of a user (e.g. `msync` system call),

- when physica memory is scarce, or

- when the object that owns the pages is freed.

———————————————————————————————— uvm/uvm_pager.h

```
 90 /*
 91  * pager ops
 92  */
 93
 94 struct uvm_pagerops {
 95
 96         /* init pager */
 97         void    (*pgo_init) __P((void));
 98
 99         /* add reference to obj */
100         void    (*pgo_reference)(struct uvm_object *);
101
102         /* drop reference to obj */
103         void    (*pgo_detach)(struct uvm_object *);
104
105         /* special non-standard fault processing */
106         int     (*pgo_fault)(struct uvm_faultinfo *, vaddr_t, struct vm_page **,
107                         int, int, vm_fault_t, vm_prot_t, int);
108
109         /* get/read pages */
110         int     (*pgo_get)(struct uvm_object *, voff_t, struct vm_page **,
```

```
111                             int *, int, vm_prot_t, int, int);
112
113          /* put/write pages */
114          int     (*pgo_put)(struct uvm_object *, voff_t, voff_t, int);
115 };
```

———————————————————————————————————————————— uvm/uvm_pager.h

## 6.2.5   Page

*Page* describes a page of physical memory.  When the system is booted a `vm_page`
structure is allocated for each page of physical memory that can be used by the VM
system.

———————————————————————————————————————————— uvm/uvm_page.h

```
120 struct vm_page {
121          TAILQ_ENTRY(vm_page)    pageq;          /* queue info for FIFO
122                                                   * queue or free list (P) */
123          TAILQ_ENTRY(vm_page)    hashq;          /* hash table links (O)*/
124          TAILQ_ENTRY(vm_page)    listq;          /* pages in same object (O)*/
125
126          struct vm_anon          *uanon;         /* anon (O,P) */
127          struct uvm_object       *uobject;       /* object (O,P) */
128          voff_t                  offset;         /* offset into object (O,P) */
129          uint16_t                flags;          /* object flags [O] */
130          uint16_t                loan_count;     /* number of active loans
131                                                   * to read: [O or P]
132                                                   * to modify: [O _and_ P] */
133          uint16_t                wire_count;     /* wired down map refs [P] */
134          uint16_t                pqflags;        /* page queue flags [P] */
135          paddr_t                 phys_addr;      /* physical address of page */
136
137 #ifdef __HAVE_VM_PAGE_MD
138          struct vm_page_md       mdpage;         /* pmap-specific data */
139 #endif
140
141 #if defined(UVM_PAGE_TRKOWN)
142          /* debugging fields to track page ownership */
143          pid_t                   owner;          /* proc that set PG_BUSY */
144          char                    *owner_tag;     /* why it was set busy */
145 #endif
146 };
```

———————————————————————————————————————————— uvm/uvm_page.h

**Machine-Dependent Page Structure**

Machine-dependent page structure for sparc64 platform is

———————————————————————————————————————— arch/sparc64/include/vmparam.h

```
152 /*
153  * For each struct vm_page, there is a list of all currently valid virtual
154  * mappings of that page.  An entry is a pv_entry_t.
155  */
```

```
156 struct pmap;
157 typedef struct pv_entry {
158         struct pv_entry *pv_next;      /* next pv_entry */
159         struct pmap     *pv_pmap;      /* pmap where mapping lies */
160         vaddr_t         pv_va;         /* virtual address for mapping */
161 } *pv_entry_t;
162 /* PV flags encoded in the low bits of the VA of the first pv_entry */
163
164 struct vm_page_md {
165         struct pv_entry mdpg_pvh;
166 };
```

———————————————————————- arch/sparc64/include/vmparam.h

where `struct pmap` is defined as

———————————————————————- arch/sparc64/include/pmap.h

```
111 struct pmap {
112         struct uvm_object pm_obj;
113 #define pm_lock pm_obj.vmobjlock
114 #define pm_refs pm_obj.uo_refs
115         LIST_ENTRY(pmap) pm_list;
116         int pm_ctx;                 /* Current context */
117
118         /*
119          * This contains 64-bit pointers to pages that contain
120          * 1024 64-bit pointers to page tables.  All addresses
121          * are physical.
122          *
123          * !!! Only touch this through pseg_get() and pseg_set() !!!
124          */
125         paddr_t pm_physaddr;    /* physical address of pm_segs */
126         int64_t *pm_segs;
127 };
```

———————————————————————- arch/sparc64/include/pmap.h

**Page Fault**

When a process attempts to access an unmapped area of memory a page fault is
generated. In order to find which page should be mapped, the UVM system must
look in the process' map structure for the netry that corresponds to the faulting
address.

- If there is not entry mapping the faulting address, an error signal is generated.

- If an object is mapped at the faulting address,

    - if the requested data is already resident in a page, that page can be
      mapped in.

    - if not, then the fault routine issues a request to the object's pager to
      make the data resident and resolve the fault.

## 6.3   UVM External Interface

We describe parts of UVM external interface **which is essential to understand FFS filesystem source code**. The whole UVM external interfaces can be classified as

- Initialization

- Virtual address space management

- Page fault handling

- Memory mapping files and devices

- Virtual memory I/O

- Management of kernel memory

- Management of physical memory

- Processes

- Page loan

- Miscellaneous functions

From these, we will investigate some functions whose category is *memory mapping files and devices* or *allocation of physical memory*. The first category is mainly related with `open` system call. The second is related with buffer cache and filesystem storage operations.

## 6.4   Memory Mapping Files and Devices

### 6.4.1   Attaching a Memory Object to Vnode: `uvn_attach`

```
struct uvm_object *
uvn_attach(void *arg, vm_prot_t accessprot);

        uvn_attach() attaches a UVM object to vnode arg,
        creating the object if necessary.  The object is returned.

        The values that accessprot maxprot can take are:

        #define VM_PROT_NONE    ((vm_prot_t) 0x00)
        #define VM_PROT_READ    ((vm_prot_t) 0x01) /* read permission */
        #define VM_PROT_WRITE   ((vm_prot_t) 0x02) /* write permission */
        #define VM_PROT_EXECUTE ((vm_prot_t) 0x04) /* execute permission */

        or

        #define UVM_PROT_MASK   0x07    /* protection mask */
        #define UVM_PROT_NONE   0x00    /* protection none */
        #define UVM_PROT_ALL    0x07    /* everything */
        #define UVM_PROT_READ   0x01    /* read */
        #define UVM_PROT_WRITE  0x02    /* write */
        #define UVM_PROT_EXEC   0x04    /* exec */
        #define UVM_PROT_R      0x01    /* read */
```

```
#define UVM_PROT_W      0x02    /* write */
#define UVM_PROT_RW     0x03    /* read-write */
#define UVM_PROT_X      0x04    /* exec */
#define UVM_PROT_RX     0x05    /* read-exec */
#define UVM_PROT_WX     0x06    /* write-exec */
#define UVM_PROT_RWX    0x07    /* read-write-exec */
```

**References to Source Code**

This `uvn_attach` function is used in `vn_open`, vnode high-level operation function which is used to implement `open` system call.

———————————————————————————————————— kern/vfs_vnops.c

```
 87 /*
 88  * Common code for vnode open operations.
 89  * Check permissions, and call the VOP_OPEN or VOP_CREATE routine.
 90  */
 91 int
 92 vn_open(ndp, fmode, cmode)
 93         struct nameidata *ndp;
 94         int fmode, cmode;
 95 {
 96         struct vnode *vp;
 97         struct proc *p = ndp->ni_cnd.cn_proc;
 98         struct ucred *cred = p->p_ucred;
 99         struct vattr va;
100         int error;
...
267         if ((error = VOP_OPEN(vp, fmode, cred, p)) != 0)
268                 goto bad;
269         if (vp->v_type == VREG &&
270             uvn_attach(vp, fmode & FWRITE ? VM_PROT_WRITE : 0) == NULL) {
271                 error = EIO;
272                 goto bad;
273         }
274         if (fmode & FWRITE)
275                 vp->v_writecount++;
276
277         return (0);
278 bad:
279         vput(vp);
280         return (error);
281 }
```

———————————————————————————————————— kern/vfs_vnops.c

## 6.4.2  Setting Vnode Size: `uvn_vnp_setsize`

```
void
uvm_vnp_setsize(struct vnode *vp, voff_t newsize);
```

`uvm_vnp_setsize()` sets the size of vnode vp to newsize.  Caller must hold a reference to the vnode.  If the vnode shrinks, pages no longer used are discarded.

**References to Source Code**

This **uvn_attach** function is used in **ffs_write**, one of the FFS storage function as,

```
    57 #define WRITE                      ffs_write
.....
   184 /*
   185  * Vnode op for writing.
   186  */
   187 int
   188 WRITE(void *v)
   189 {
.....
   312         while (uio->uio_resid > 0) {
   313                 boolean_t extending; /* if we're extending a whole block */
.....
   365                 /*
   366                  * update UVM's notion of the size now that we've
   367                  * copied the data into the vnode's pages.
   368                  *
   369                  * we should update the size even when uiomove failed.
   370                  * otherwise ffs_truncate can't flush soft update states.
   371                  */
   372
   373                 newoff = oldoff + bytelen;
   374                 if (vp->v_size < newoff) {
   375                         uvm_vnp_setsize(vp, newoff);
   376                         extended = 1;
   377                 }
.....
   402         }
.....
   481 }
```

### 6.4.3   Clearing a Vnode: `uvn_vnp_zerorange`

```
/*
 * uvm_vnp_zerorange:  set a range of bytes in a file to zero.
 */

void
uvm_vnp_zerorange(vp, off, len)
        struct vnode *vp;
        off_t off;
        size_t len;
{
        void *win;

        /*
         * XXXUBC invent kzero() and use it
         */

        while (len) {
                vsize_t bytelen = len;
```

```
                    win = ubc_alloc(&vp->v_uobj, off, &bytelen, UBC_WRITE);
                    memset(win, 0, bytelen);
                    ubc_release(win, 0);

                    off += bytelen;
                    len -= bytelen;
            }
    }
```

**References to Source Code**

This `uvn_zerorange` function is used in `ffs_truncate` vnode operation as,

——————————————————————————————————————————————— ffs/ffs/ffs_vnops.c

```
158 /*
159  * Truncate the inode oip to at most length size, freeing the
160  * disk blocks.
161  */
162 int
163 ffs_truncate(v)
164         void *v;
165 {
.....
253         /*
254          * When truncating a regular file down to a non-block-aligned size,
255          * we must zero the part of last block which is past the new EOF.
256          * We must synchronously flush the zeroed pages to disk
257          * since the new pages will be invalidated as soon as we
258          * inform the VM system of the new, smaller size.
259          * We must do this before acquiring the GLOCK, since fetching
260          * the pages will acquire the GLOCK internally.
261          * So there is a window where another thread could see a whole
262          * zeroed page past EOF, but that's life.
263          */
264
265         offset = blkoff(fs, length);
266         if (ovp->v_type == VREG && length < osize && offset != 0) {
267                 voff_t eoz;
268
269                 error = ufs_balloc_range(ovp, length - 1, 1, ap->a_cred,
270                     aflag);
271                 if (error) {
272                         return error;
273                 }
274                 size = blksize(fs, oip, lblkno(fs, length));
275                 eoz = MIN(lblktosize(fs, lblkno(fs, length)) + size, osize);
276                 uvm_vnp_zerorange(ovp, length, eoz - length);
277                 simple_lock(&ovp->v_interlock);
278                 error = VOP_PUTPAGES(ovp, trunc_page(length), round_page(eoz),
279                     PGO_CLEANIT | PGO_DEACTIVATE | PGO_SYNCIO);
280                 if (error) {
281                         return error;
282                 }
283         }
```

```
.....
    456 }
```

─────────────────────────────────────────────── ffs/ffs/ffs_vnops.c

# 6.5   Management of Physical Memory

## 6.5.1   Lock Management for Page Queue: uvm_(un)lock_pageq

```
#define uvm_lock_pageq()        simple_lock(&uvm.pageqlock)
#define uvm_unlock_pageq()      simple_unlock(&uvm.pageqlock)
```

## 6.5.2   Activating Physical Page: uvm_pageactivate

```
/*
 * uvm_pageactivate: activate page
 *
 * => caller must lock page queues
 */

void
uvm_pageactivate(struct vm_page *pg)
```

**References to Source Code**

This function is used in **ufs_balloc_range** data block allocation function as,

─────────────────────────────────────────────── ufs/ufs_inode.c

```
204        /*
205         * read or create pages covering the range of the allocation and
206         * keep them locked until the new block is allocated, so there
207         * will be no window where the old contents of the new block are
208         * visible to racing threads.
209         */
210
211        pagestart = trunc_page(off) & ~(bsize - 1);
212        npages = MIN(ppb, (round_page(neweob) - pagestart) >> PAGE_SHIFT);
213        memset(pgs, 0, npages * sizeof(struct vm_page *));
214        simple_lock(&uobj->vmobjlock);
215        error = VOP_GETPAGES(vp, pagestart, pgs, &npages, 0,
216            VM_PROT_READ, 0, PGO_SYNCIO|PGO_PASTEOF);
217        if (error) {
218                return error;
219        }
220        simple_lock(&uobj->vmobjlock);
221        uvm_lock_pageq();
222        for (i = 0; i < npages; i++) {
223                UVMHIST_LOG(ubchist, "got pgs[%d] %p", i, pgs[i],0,0);
224                KASSERT((pgs[i]->flags & PG_RELEASED) == 0);
225                pgs[i]->flags &= ~PG_CLEAN;
226                uvm_pageactivate(pgs[i]);
227        }
228        uvm_unlock_pageq();
229        simple_unlock(&uobj->vmobjlock);
```

———————————————————————————————— ufs/ufs_inode.c

### 6.5.3   Making Unbusy a Page: `uvm_page_unbusy`

```
/*
 * uvm_page_unbusy: unbusy an array of pages.
 *
 * => pages must either all belong to the same object, or all belong to anons.
 * => if pages are object-owned, object must be locked.
 * => if pages are anon-owned, anons must be locked.
 * => caller must lock page queues if pages may be released.
 */

void
uvm_page_unbusy(struct vm_page **pgs, int npgs);
```

**References to Source Code**

This function is also used in **ufs_balloc_range** data block allocation function as,

———————————————————————————————— ufs/ufs_inode.c

```
247         /*
248          * clear PG_RDONLY on any pages we are holding
249          * (since they now have backing store) and unbusy them.
250          */
251
252         simple_lock(&uobj->vmobjlock);
253         for (i = 0; i < npages; i++) {
254                 pgs[i]->flags &= ~PG_RDONLY;
255                 if (error) {
256                         pgs[i]->flags |= PG_RELEASED;
257                 }
258         }
259         if (error) {
260                 uvm_lock_pageq();
261                 uvm_page_unbusy(pgs, npages);
262                 uvm_unlock_pageq();
263         } else {
264                 uvm_page_unbusy(pgs, npages);
265         }
266         simple_unlock(&uobj->vmobjlock);
```

———————————————————————————————— ufs/ufs_inode.c

### 6.5.4   Looking up a Page: `uvm_pagelookup`

```
/*
 * uvm_pagelookup: look up a page
 *
 * => caller should lock object to keep someone from pulling the page
 *      out from under it
 */

struct vm_page *
uvm_pagelookup(struct uvm_object *obj, voff_t off);
```

**References to Source Code**

Only when the soft dependency facility is used, this function is effective in `ffs_putpages`
as,

———————————————————————————————————— ffs/ffs_vnops.c

```
507 int
508 ffs_putpages(void *v)
509 {
510         struct vop_putpages_args /* {
511                 struct vnode *a_vp;
512                 voff_t a_offlo;
513                 voff_t a_offhi;
514                 int a_flags;
515         } */ *ap = v;
516         struct vnode *vp = ap->a_vp;
517         struct uvm_object *uobj = &vp->v_uobj;
518         struct inode *ip = VTOI(vp);
519         struct fs *fs = ip->i_fs;
520         struct vm_page *pg;
521         off_t off;
522         ufs_lbn_t lbn;
523
524         if (!DOINGSOFTDEP(vp) || (ap->a_flags & PGO_CLEANIT) == 0) {
525                 return genfs_putpages(v);
526         }
527
528         /*
529          * for softdep files, force the pages in a block to be written together.
530          * if we're the pagedaemon and we would have to wait for other pages,
531          * just fail the request.  the pagedaemon will pick a different page.
532          */
533
534         ap->a_offlo &= ~fs->fs_qbmask;
535         lbn = lblkno(fs, ap->a_offhi);
536         ap->a_offhi = blkroundup(fs, ap->a_offhi);
537         if (curproc == uvm.pagedaemon_proc) {
538                 for (off = ap->a_offlo; off < ap->a_offhi; off += PAGE_SIZE) {
539                         pg = uvm_pagelookup(uobj, off);
540
541                         /*
542                          * we only have missing pages here because the
543                          * calculation of offhi above doesn't account for
544                          * fragments.  so once we see one missing page,
545                          * the rest should be missing as well, but we'll
546                          * check for the rest just to be paranoid.
547                          */
548
549                         if (pg == NULL) {
550                                 continue;
551                         }
552                         if (pg->flags & PG_BUSY) {
553                                 simple_unlock(&uobj->vmobjlock);
554                                 return EBUSY;
```

```
555                             }
556                     }
557             }
558             return genfs_putpages(v);
559 }
```

# Chapter 7

# UBC

## 7.1 Introduction

Operating systems allow filesystem data to be accessed using two mechanisms: memory mapping calls such as `mmap`, and I/O system calls such as `read` or `write`. In traditional UNIX, the memory mapping requests are handled by the VM system while I/O system calls are handled by the I/O subsystem. Therefore, the VM subsystem and I/O subsystem each have their own data caching mechanisms that operate semi-independently of each other. This lack of integration leads to degrade in performance and flexibility. The function of Unified Buffer Cache(UBC) is to integrate the two cache mechanisms, to improve system performance.

## 7.2 Traditional Accesses to File

Figure 7.1 shows the flow of data between the disk and the application with a traditional buffer cache and VM page.

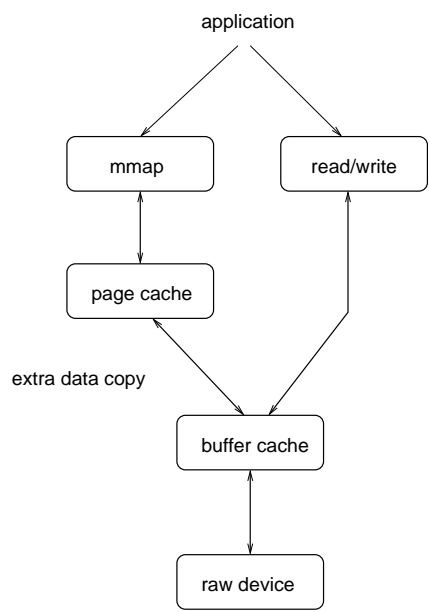

Figure 7.1: NetBSD before UBC

## 7.2.1  I/O Subsystem: `read()` and `write()`

The `read` system call reads data from disk into the kernel's buffer cache, and then copies data from the buffer cache to the application's address space.

The use of the buffer cache for large amounts of data is generally bad, since

- the static sizing of the buffer cache means that the buffer cache is often too small, so that resulting in excessive cache misses for the single large file.

- the excessively high portion of buffer cache about a single large file leaves too little buffer cache for other files.

- or the buffer cache also has the limitation that cached data must always be mapped into kernel vitrual space, since modern hardware can easily have more RAM than kernel virtual memory.

## 7.2.2  Virtual Memory Subsystem: `mmap()`

The `mmap` system call gives the application direct memory-mapped access to the kernel's page cache data. File data is read into the page cache lazily as processes attempt to access the mappings created with `mmap` system call and generate page faults.

To write modified data in page caches back to disk,

1. the new version is copied back to the buffer cache and

2. from the buffer cache, the modified page contents is written to disk.

This double-cacheing of data is a major source of inefficiency, since

- Having two copies of file data means that twice as much memory is used.

- Copying the data back and forth between the buffer cache and the page cache is extra data copy, so that this wastes CPU cycles

- The extra copy also clobbers CPU cache memory and results in performance degrade.

- Having two copies of the dat also allows the possibility that the two sopies will become inconsistent, which can lead to application problems which are difficult to debug

## 7.3    File Access with Unified Buffer Cache

Figure 7.2 shows the changed data flow with UBC. UBC is a new subsystem which solves the problems with the two-cache model.

- File data is read directly into the page cache without going through the buffer cache by creating two new virtual filesystem operations which calls the device driver to read the data from disk if necessary.

- Since the pages of page cache are not always mapped into kernel virtual address space, a new mechanism for providing temporary mappings of page cache pages is provided, to be used by `read` and `write` system call.

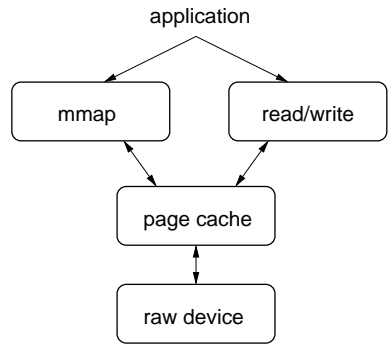

Figure 7.2: NetBSD after UBC

# 7.4 VFS Support for UVM

These new virtual filesystem operations are provided to allow the UVM system to request ranges of pages to be read into memory from disk or written from memory back to disk.

## 7.4.1 `VOP_GETPAGES` Operation

`VOP_GETPAGES` allocate pages from the UVM system for data which is not already cached and then initiate device I/O operations to read all the disk blocks which contain the data for those pages. The functions is defined in `miscfs/genfs/genfs_vnops.c`

```
VOP_GETPAGES(vp, offset, m, count, centeridx, access_type, advice, flags)

        Read VM pages from file.  The argument vp is the locked vnode to
        read the VM pages from.  The argument offset is offset in the
        file to start accessing and m is an array of VM pages.  The ar-
        gument count specifies the number of pages to read.  If the oper-
        ation is successful zero is returned, otherwise an appropriate
        error code is returned.
```

## 7.4.2 `VOP_PUTPAGES` Operation

`VOP_PUTPAGES` initiate device I/Os to write dirty pages back to disk. The functions is defined in `miscfs/genfs/genfs_vnops.c`

```
VOP_PUTPAGES(vp, offset, len, flags)

        Write modified (dirty) VM pages to file.  The argument vp is the
        locked vnode to write the VM pages to and offset and len speci-
        fies the range of VM pages to write.  There seems to be some
        confusion in the code whether offset and len specify the start
        and length of the VM pages for the start and end of the VM
        pages.  The argument flags specifies whether the pages should be
        written asynchronously and also whether they should be marked
        invalid one the write back operation has completed.  If the op-
        eration is successful zero is returned, otherwise an appropriate
        error code is returned.
```

## 7.5   UVM Support for I/O

There are functions that allocate and free temporary mappings of page cache file data.

### 7.5.1   `ubc_alloc` Function

`ubc_alloc` is a page equivalent of the buffer cache funstion, `get_blk`. The functions is defined in `uvm/uvm_bio.c`

```
void *
ubc_alloc(struct uvm_object *uobj, voff_t offset, vsize_t *lenp,
        int flags);

ubc_alloc() creates a kernel mappings of uobj starting at offset offset.
the desired length of the mapping is pointed to by lenp, but the actual
mapping may be smaller than this.  lenp is updated to contain the actual
length mapped.  The flags must be one of

#define UBC_READ      0x01   /* mapping will be accessed for read */
#define UBC_WRITE     0x02   /* mapping will be accessed for write */

Currently, uobj must actually be a vnode object.  Once the mapping is
created, it must be accessed only by methods that can handle faults, such
as uiomove() or kcopy().  Page faults on the mapping will result in the
vnode's VOP_GETPAGES() method being called to resolve the fault.
```

### 7.5.2   `ubc_release` Function

`ubc_release` is a page cache equivalent of the buffer cache funftion, `brelse`. The functions is defined in `uvm/uvm_bio.c`

```
void
ubc_release(void *va, int flags);

ubc_release() frees the mapping at va for reuse.  The mapping may be
cached to speed future accesses to the same region of the object.  The
flags are currently unused.
```

## 7.6   Example

### 7.6.1   Reading from Disk to Buffer with UBC

```
———————————————————————————————————————————— ufs/ufs_readwrite.c
61 /*
62  * Vnode op for reading.
63  */
64 /* ARGSUSED */
65 int
66 ffs_read(void *v)
67 {
68         struct vop_read_args /* {
69                 struct vnode *a_vp;
70                 struct uio *a_uio;
```

```
    71                int a_ioflag;
    72                struct ucred *a_cred;
    73        } */ *ap = v;
    74        struct vnode *vp;
    75        struct inode *ip;
    76        struct uio *uio;
    77        FS *fs;
    78        void *win;
    79        vsize_t bytelen;
    80        struct buf *bp;
    81        ufs_daddr_t lbn, nextlbn;
    82        off_t bytesinfile;
    83        long size, xfersize, blkoffset;
    84        int error;
    85        boolean_t usepc = FALSE;
    86
    87        vp = ap->a_vp;
    88        ip = VTOI(vp);
    89        uio = ap->a_uio;
    90        error = 0;
    91
.....
   104        fs = ip->I_FS;
   105        if ((u_int64_t)uio->uio_offset > fs->fs_maxfilesize)
   106                return (EFBIG);
   107        if (uio->uio_resid == 0)
   108                return (0);
   109        if (uio->uio_offset >= ip->i_ffs_size) {
   110                goto out;
   111        }
   112
.....
   114        usepc = vp->v_type == VREG;
.....
   116        if (usepc) {
   117                while (uio->uio_resid > 0) {
   118                        bytelen = MIN(ip->i_ffs_size - uio->uio_offset,
   119                            uio->uio_resid);
   120                        if (bytelen == 0)
   121                                break;
   122
   123                        win = ubc_alloc(&vp->v_uobj, uio->uio_offset,
   124                                    &bytelen, UBC_READ);
   125                        error = uiomove(win, bytelen, uio);
   126                        ubc_release(win, 0);
   127                        if (error)
   128                                break;
   129                }
   130                goto out;
   131        }
...
   175   out:
   176        if (!(vp->v_mount->mnt_flag & MNT_NOATIME)) {
   177                ip->i_flag |= IN_ACCESS;
```

```
178                 if ((ap->a_ioflag & IO_SYNC) == IO_SYNC)
179                         error = VOP_UPDATE(vp, NULL, NULL, UPDATE_WAIT);
180         }
181         return (error);
182 }
```

———————————————————————————————— ufs/ufs_readwrite.c

## 7.6.2  Writing from Buffer to Disk with UBC

———————————————————————————————— ufs/ufs_readwrite.c

```
184 /*
185  * Vnode op for writing.
186  */
187 int
188 ffs_write(void *v)
189 {
190         struct vop_write_args /* {
191                 struct vnode *a_vp;
192                 struct uio *a_uio;
193                 int a_ioflag;
194                 struct ucred *a_cred;
195         } */ *ap = v;
196         struct vnode *vp;
197         struct uio *uio;
198         struct inode *ip;
199         struct genfs_node *gp;
200         FS *fs;
201         struct buf *bp;
202         struct proc *p;
203         struct ucred *cred;
204         ufs_daddr_t lbn;
205         off_t osize, origoff, oldoff, preallocoff, endallocoff, nsize;
206         int blkoffset, error, flags, ioflag, resid, size, xfersize;
207         int bsize, aflag;
208         int ubc_alloc_flags;
209         int extended=0;
210         void *win;
211         vsize_t bytelen;
212         boolean_t async;
213         boolean_t usepc = FALSE;
214
215         cred = ap->a_cred;
216         ioflag = ap->a_ioflag;
217         uio = ap->a_uio;
218         vp = ap->a_vp;
219         ip = VTOI(vp);
220         gp = VTOG(vp);
221
...
245         fs = ip->I_FS;
...
```

```
270        flags = ioflag & IO_SYNC ? B_SYNC : 0;
271        async = vp->v_mount->mnt_flag & MNT_ASYNC;
272        origoff = uio->uio_offset;
273        resid = uio->uio_resid;
274        osize = ip->i_ffs_size;
275        bsize = fs->fs_bsize;
276        error = 0;
...
311        ubc_alloc_flags = UBC_WRITE;
312        while (uio->uio_resid > 0) {
313                boolean_t extending; /* if we're extending a whole block */
314                off_t newoff;
315
316                oldoff = uio->uio_offset;
317                blkoffset = blkoff(fs, uio->uio_offset);
318                bytelen = MIN(fs->fs_bsize - blkoffset, uio->uio_resid);
319
320                /*
321                 * if we're filling in a hole, allocate the blocks now and
322                 * initialize the pages first.  if we're extending the file,
323                 * we can safely allocate blocks without initializing pages
324                 * since the new blocks will be inaccessible until the write
325                 * is complete.
326                 */
.....
338                        lockmgr(&gp->g_glock, LK_EXCLUSIVE, NULL);
339                        error = GOP_ALLOC(vp, uio->uio_offset, bytelen,
340                            aflag, cred);
341                        lockmgr(&gp->g_glock, LK_RELEASE, NULL);
342                        if (error) {
343                                break;
344                        }
345                        ubc_alloc_flags |= UBC_FAULTBUSY;
.....
347
348                /*
349                 * copy the data.
350                 */
351
352                win = ubc_alloc(&vp->v_uobj, uio->uio_offset, &bytelen,
353                    ubc_alloc_flags);
354                error = uiomove(win, bytelen, uio);
355                if (error && extending) {
356                        /*
357                         * if we haven't initialized the pages yet,
358                         * do it now.  it's safe to use memset here
359                         * because we just mapped the pages above.
360                         */
361                        memset(win, 0, bytelen);
362                }
363                ubc_release(win, 0);
364
.....
383                /*
```

```
384                     * flush what we just wrote if necessary.
385                     * XXXUBC simplistic async flushing.
386                     */
387
388              if (!async && oldoff >> 16 != uio->uio_offset >> 16) {
389                      simple_lock(&vp->v_interlock);
390                      error = VOP_PUTPAGES(vp, (oldoff >> 16) << 16,
391                          (uio->uio_offset >> 16) << 16, PGO_CLEANIT);
392                      if (error) {
393                              break;
394                      }
395              }
396      }
397      if (error == 0 && ioflag & IO_SYNC) {
398              simple_lock(&vp->v_interlock);
399              error = VOP_PUTPAGES(vp, trunc_page(origoff & ~(bsize - 1)),
400                      round_page(blkroundup(fs, uio->uio_offset)),
401                      PGO_CLEANIT | PGO_SYNCIO);
402      }
403      goto out;
.....
466 out:
.....
480      return (error);
481 }
```

——————————————————————————— ufs/ufs_readwrite.c

# Part II

# Analyzing Fast Filesystem

# Chapter 8

# Naming

Filesystem contain files, most of which contain ordinary data. Certain files are distinguished as directories and contain pointers to files that may themselves be directories.

## 8.1  Directories

### 8.1.1  Chunk

Directories are allocated in unites called *chunks* Chunks are broken up into variable-length directory entries to allow filenames to be of nearly arbitrary length. No directory entry can span multiple chunks. The *chunk* is defined as `struct direct` in `ufs/ufs/dir.h` as,

—————————————————————————————— ufs/ufs/dir.h

```
54 /*
55  * A directory consists of some number of blocks of DIRBLKSIZ
56  * bytes, where DIRBLKSIZ is chosen such that it can be transferred
57  * to disk in a single atomic operation (e.g. 512 bytes on most machines).
58  *
59  * Each DIRBLKSIZ byte block contains some number of directory entry
60  * structures, which are of variable length.  Each directory entry has
61  * a struct direct at the front of it, containing its inode number,
62  * the length of the entry, and the length of the name contained in
63  * the entry.  These are followed by the name padded to a 4 byte boundary
64  * with null bytes.  All names are guaranteed null terminated.
65  * The maximum length of a name in a directory is MAXNAMLEN.
66  *
67  * The macro DIRSIZ(fmt, dp) gives the amount of space required to represent
68  * a directory entry.  Free space in a directory is represented by
69  * entries which have dp->d_reclen > DIRSIZ(fmt, dp).  All DIRBLKSIZ bytes
70  * in a directory block are claimed by the directory entries.  This
71  * usually results in the last entry in a directory having a large
72  * dp->d_reclen.  When entries are deleted from a directory, the
73  * space is returned to the previous entry in the same directory
74  * block by increasing its dp->d_reclen.  If the first entry of
75  * a directory block is free, then its dp->d_ino is set to 0.
76  * Entries other than the first in a directory do not normally have
77  * dp->d_ino set to 0.
```

```
78  */
79 #undef  DIRBLKSIZ
80 #define DIRBLKSIZ         DEV_BSIZE
81 #undef  MAXNAMLEN
82 #define MAXNAMLEN       255
83 #define APPLEUFS_DIRBLKSIZ 1024
84
85 struct  direct {
86              u_int32_t d_ino;                /* inode number of entry */
87              u_int16_t d_reclen;             /* length of this record */
88              u_int8_t  d_type;               /* file type, see below */
89              u_int8_t  d_namlen;             /* length of string in d_name */
90              char      d_name[MAXNAMLEN + 1];/* name with length <= MAXNAMLEN */
91 };
92
93 /*
94  * File types
95  */
96 #define DT_UNKNOWN      0
97 #define DT_FIFO         1
98 #define DT_CHR          2
99 #define DT_DIR          4
100 #define DT_BLK          6
101 #define DT_REG          8
102 #define DT_LNK         10
103 #define DT_SOCK        12
104 #define DT_WHT         14
```

──────────────────────────────────────────── ufs/ufs/dir.h

The filesystem records free space in a directory by having entries accumulate the free space in their size fields.

### 8.1.2  Modification of Directory

When an entry is deleted from a directory, the system coalesces the entry's space into the previous entry in the same directory chunk by increasing the size of the previous entry by the size of the deleted entry.

If the filrst entry of a directory chunk is free, then the pointer to the entry's inode is set to zero to show that the entry is unallocated.

## 8.2   Finding of Names in Directories

### 8.2.1   Match Algorithm

First, the length of the sought-after name is compared with the length of the name being checked. If the lengths are identical, a string comparison of the name being sought and the directory entry is made. If they match, the search is complete; if they fail, the search continues with the next entry.

### 8.2.2   Search Performance Improvement

Before starting a directory scan, the kernel looks for the name in the cache. If either a positive or a negative entry is found, the directory scan can be avoided.

## 8.3   Pathname Translation

The translation of a pathname requires a series of interactions between the vnode interface and the underlying filesystems. The pathname-translation process proceeds as follows:

1. The pathname to be translated is copied in from the user process.

2. The starting point if the pathname is determined. The vnode for this directory becomes the *lookup directory* used in the next step.

3. The vnode layer calls the filesystem-specific *lookup* opeartion, and passes the remaining components of the pathname and the current *lookup directory*.

4. Typically, the underlying filesystem will search the *lookup directory* for the next component of the pathname and will return the resulting vnode or an error if the name does not exist.

5. If an error is returned, the top level returns the error. If the pathname has been exhausted, the pathname lookup is done, and the returned vnode is the result of not a directory, then the vnode layer returns "not a directory" error.

6. If there are no errors, the top layer checks to see whether the returned directory is a mount point for another filesystem. If it is, then the *lookup directory* becomes the mounted filesystem; otherwise, the *lookup directory* becomes the vnode returned by the lower layer. The lookup then iterates with step 3.

## 8.4   The Name Cache

Name-cache management is a service that is provided by the vnode management routines. The interface provides a facility

- to add a name and its corresponding vnode,

- to look up a name to get the corresponding vnode,

- to delete a specific name from the cache, and

- to invalidate all names that reference a specific vnode.

### 8.4.1   Vnode's Capability

Each vnode is given a *capability* — a 32-bit number guaranteed to be unique. A vnode's capability is invalidated each time it is reused by `getnewvnode` or, when specifically requested by a client.

When a name is found during a cached lookup, the capability assigned to the name is compared with that of the vnode. If they match, the lookup is successful; if they do not match, the cache entry is freed and failure is returned.

Directory vnodes can have many names that reference them. Using vnode's *capability*, the kernel need not revoke a names for a vnode by scanning the entire name table, thousands of names, looking for references to the vnode in question.

### 8.4.2   Negative Caching

If a name is looked up in a directory and is not found, that name can be entered in the cache, along with a null pointer for its corresponding vnode. When the directory is modified, the kernel must invalidate all the negative names for that directory vnode by assigning the directory a new capability.

### 8.4.3   Special Device Handling

The name and attributes of special devices and FIFOs are maintained by the filesystem in which they reside. However, their operations are maintained by the kernel.

**A Dilemma**

Since a special device is identified solely by its major and minor number, it is possible for two or more instances of the same device to appear within the filesystem name space. Each of these different names has its own vnode and underlying object, Yet all these vnodes must be treated as one from the perspective of identifying blocks in the buffer cache and in other places where the vnode and logical block number are used as a key.

**A Solution**

To ensure that the set of vnodes is treated as a single vnode, the vnode layer provides a routine `checkalias` that is called each time that a new special device vnode comes into existence. This routine looks for other instances of the device, and if it finds them, links them together so that they can act as one.

## 8.5   Links

### 8.5.1   Hard Links

Each file has a single inode, but multiple directory entries in the same filesystem may reference that inode by creating *hard links*.

### 8.5.2   Soft Links

The *symbolic link*, or *soft link* is implemented as a file that contains a pathname. If a symbolic link contains an relative pathname, the contents of the symbolic link are evaluated relative to the location of the link, not relative to the current working directory).

### 8.5.3   The Differences

- A symbolic link can refer to a directory or to a file on a different filsystem; A hard link cannnot

- Since symbolic links may cause loops in the filesystem, the kernel prevents looping by allowing at most eight symbolic link travesal in a single pathname tralslation. If the limit is reached, the kernel produces an `ELOOP` error.

## 8.6   References to Source Code

### 8.6.1   `vfs_cache.c` - 537 lines, 17 functions

**Gloval Variables**

```
    [ Positive Name Cache ]

    nchashtbl          Name cache hash table
    nchash             Magic number to generate hash key
```

```
numcache            Number of cache entries allocated
nclruhead           LRU chain
nchstats            Cache effectiveness statistics

namecache_pool      Pool
doingcache          Switch to enable cache

[ Negative Name Cache ]

ncvhashtbl          Name cache hash table
ncvhash             Magic number to generate hash key
```

### Functions

```
[ Name Cache Management ]

cache_lookup        Look for a the name in the cache
cache_revlookup     Scan cache looking for name of directory entry pointing at vp
cache_enter         Add an entry to the cach
cache_purge         Cache flush, a particular vnode; called when a vnode is renamed
cache_purgevfs      Cache flush, a whole filesystem; when unmounted

[ Name Cache Initialization ]

nchinit
nchreinit

[ Diagnostic ]

namecache_print
```

## 8.6.2  `vfs_lookup.c` - 777 lines, 4 functions

### Gloval Variables

```
pnbuf_pool          Pathname buffer pool
pnbuf_cache         Pathname buffer cache
```

### Functions

```
namei               Convert a pathname into a pointer to a locked inode
namei_hash          Determine the namei hash (for cn_hash) for name
lookup              Search a pathname
relookup            [?] Reacquire a path name component
```

# Chapter 9

# Inode

## 9.1   The Structures of an Inode

To allow files to be allocated concurrently and random access within files, 4.4BSD uses the concept of an *index node*, namely *inode*.

The inode contains information about the contents of the file. Notably missing in the inode is the filename. Chunks are broken up into variable-length directory entries to allow filenames to be of arbitrary length. The fixed parts of a directory entry includes

- An index into a table of on-disk inode structures. This inode structure describes the file.

- The size of the entry in bytes

- The type of the entry.

- The length of the filename contained in the entry in bytes.

The structure definition of the inode is located in `ufs/ufs/inode.h` as,

———————————————————————————————————————————————— ufs/ufs/inode.h

```
67 /*
68  * The inode is used to describe each active (or recently active) file in the
69  * UFS filesystem. It is composed of two types of information. The first part
70  * is the information that is needed only while the file is active (such as
71  * the identity of the file and linkage to speed its lookup). The second part
72  * is the permanent meta-data associated with the file which is read in
73  * from the permanent dinode from long term storage when the file becomes
74  * active, and is put back when the file is no longer being used.
75  */
76 struct inode {
77         struct genfs_node i_gnode;
78         LIST_ENTRY(inode) i_hash;/* Hash chain. */
79         struct  vnode *i_vnode; /* Vnode associated with this inode. */
80         struct  vnode *i_devvp; /* Vnode for block I/O. */
81         u_int32_t i_flag;       /* flags, see below */
82         dev_t     i_dev;        /* Device associated with the inode. */
83         ino_t     i_number;     /* The identity of the inode. */
84
85         union {                 /* Associated filesystem. */
```

```
86               struct  fs *fs;        /* FFS */
87               struct  lfs *lfs;       /* LFS */
88               struct  m_ext2fs *e2fs; /* EXT2FS */
89       } inode_u;
90 #define i_fs    inode_u.fs
91 #define i_lfs   inode_u.lfs
92 #define i_e2fs  inode_u.e2fs
93
94       struct   buflists i_pcbufhd;    /* softdep pagecache buffer head */
95       struct   dquot *i_dquot[MAXQUOTAS]; /* Dquot structures. */
96       u_quad_t i_modrev;      /* Revision level for NFS lease. */
97       struct   lockf *i_lockf;/* Head of byte-level lock list. */
98
99       /*
100       * Side effects; used during directory lookup.
101       */
102      int32_t  i_count;       /* Size of free slot in directory. */
103      doff_t   i_endoff;      /* End of useful stuff in directory. */
104      doff_t   i_diroff;      /* Offset in dir, where we found last entry. */
105      doff_t   i_offset;      /* Offset of free space in directory. */
106      u_int32_t i_reclen;     /* Size of found directory entry. */
107      int      i_ffs_effnlink; /* i_nlink when I/O completes */
108      /*
109       * Inode extensions
110       */
111      union {
112               /* Other extensions could go here... */
113               struct  ext2fs_inode_ext e2fs;
114               struct  lfs_inode_ext lfs;
115      } inode_ext;
116 #define i_e2fs_last_lblk        inode_ext.e2fs.ext2fs_last_lblk
117 #define i_e2fs_last_blk         inode_ext.e2fs.ext2fs_last_blk
118 #define i_lfs_effnblks          inode_ext.lfs.lfs_effnblocks
119 #define i_lfs_fragsize          inode_ext.lfs.lfs_fragsize
120 #define i_lfs_osize             inode_ext.lfs.lfs_osize
121      /*
122       * The on-disk dinode itself.
123       */
124      union {
125               struct  dinode ffs_din; /* 128 bytes of the on-disk dinode. */
126               struct  ext2fs_dinode e2fs_din; /* 128 bytes of the on-disk
127                                        dinode. */
128      } i_din;
130
131 #define i_ffs_atime             i_din.ffs_din.di_atime
132 #define i_ffs_atimensec         i_din.ffs_din.di_atimensec
133 #define i_ffs_blocks            i_din.ffs_din.di_blocks
134 #define i_ffs_ctime             i_din.ffs_din.di_ctime
135 #define i_ffs_ctimensec         i_din.ffs_din.di_ctimensec
136 #define i_ffs_db                i_din.ffs_din.di_db
137 #define i_ffs_flags             i_din.ffs_din.di_flags
138 #define i_ffs_gen               i_din.ffs_din.di_gen
139 #define i_ffs_gid               i_din.ffs_din.di_gid
140 #define i_ffs_ib                i_din.ffs_din.di_ib
```

```
141 #define i_ffs_mode          i_din.ffs_din.di_mode
142 #define i_ffs_mtime         i_din.ffs_din.di_mtime
143 #define i_ffs_mtimensec     i_din.ffs_din.di_mtimensec
144 #define i_ffs_nlink         i_din.ffs_din.di_nlink
145 #define i_ffs_rdev          i_din.ffs_din.di_rdev
146 #define i_ffs_shortlink     i_din.ffs_din.di_shortlink
147 #define i_ffs_size          i_din.ffs_din.di_size
148 #define i_ffs_uid           i_din.ffs_din.di_uid
```

————————————————————————————————————————— ufs/ufs/inode.h

where `struct dinode` of **line 125** is defined as

————————————————————————————————————————— ufs/ufs/dinode.h

```
62 /*
63  * A dinode contains all the meta-data associated with a UFS file.
64  * This structure defines the on-disk format of a dinode. Since
65  * this structure describes an on-disk structure, all its fields
66  * are defined by types with precise widths.
67  */
68
69 typedef int32_t ufs_daddr_t;
70 typedef long ufs_lbn_t;
71
72 #define NDADDR  12                  /* Direct addresses in inode. */
73 #define NIADDR  3                   /* Indirect addresses in inode. */
74
75 struct dinode {
76         u_int16_t       di_mode;        /*  0: IFMT, permissions; see below. */
77         int16_t         di_nlink;       /*  2: File link count. */
78         union {
79                 u_int16_t oldids[2];    /*  4: Ffs: old user and group ids. */
80                 u_int32_t inumber;      /*  4: Lfs: inode number. */
81         } di_u;
82         u_int64_t       di_size;        /*  8: File byte count. */
83         int32_t         di_atime;       /* 16: Last access time. */
84         int32_t         di_atimensec;   /* 20: Last access time. */
85         int32_t         di_mtime;       /* 24: Last modified time. */
86         int32_t         di_mtimensec;   /* 28: Last modified time. */
87         int32_t         di_ctime;       /* 32: Last inode change time. */
88         int32_t         di_ctimensec;   /* 36: Last inode change time. */
89         ufs_daddr_t     di_db[NDADDR];  /* 40: Direct disk blocks. */
90         ufs_daddr_t     di_ib[NIADDR];  /* 88: Indirect disk blocks. */
91         u_int32_t       di_flags;       /* 100: Status flags (chflags). */
92         u_int32_t       di_blocks;      /* 104: Blocks actually held. */
93         int32_t         di_gen;         /* 108: Generation number. */
94         u_int32_t       di_uid;         /* 112: File owner. */
95         u_int32_t       di_gid;         /* 116: File group. */
96         int32_t         di_spare[2];    /* 120: Reserved; currently unused */
97 };
98
99 /*
100  * The di_db fields may be overlaid with other information for
101  * file types that do not have associated disk storage. Block
```

```
102  * and character devices overlay the first data block with their
103  * dev_t value. Short symbolic links place their path in the
104  * di_db area.
105  */
106 #define di_inumber      di_u.inumber
107 #define di_ogid         di_u.oldids[1]
108 #define di_ouid         di_u.oldids[0]
109 #define di_rdev         di_db[0]
110 #define di_shortlink    di_db
111 #define MAXSYMLINKLEN   ((NDADDR + NIADDR) * sizeof(ufs_daddr_t))
112
113 /* NeXT used to keep short symlinks in the inode even when using
114  * FS_42INODEFMT.  In that case fs->fs_maxsymlinklen is probably -1,
115  * but short symlinks were stored in inodes shorter than this:
116  */
117 #define APPLEUFS_MAXSYMLINKLEN 60
118
119 /* File permissions. */
120 #define IEXEC           0000100         /* Executable. */
121 #define IWRITE          0000200         /* Writeable. */
122 #define IREAD           0000400         /* Readable. */
123 #define ISVTX           0001000         /* Sticky bit. */
124 #define ISGID           0002000         /* Set-gid. */
125 #define ISUID           0004000         /* Set-uid. */
126
127 /* File types. */
128 #define IFMT            0170000         /* Mask of file type. */
129 #define IFIFO           0010000         /* Named pipe (fifo). */
130 #define IFCHR           0020000         /* Character device. */
131 #define IFDIR           0040000         /* Directory file. */
132 #define IFBLK           0060000         /* Block device. */
133 #define IFREG           0100000         /* Regular file. */
134 #define IFLNK           0120000         /* Symbolic link. */
135 #define IFSOCK          0140000         /* UNIX domain socket. */
136 #define IFWHT           0160000         /* Whiteout. */
137
138 /* Size of the on-disk inode. */
139 #define DINODE_SIZE     (sizeof(struct dinode))        /* 128 */
```

———————————————————————————————————— ufs/ufs/dinode.h

## 9.1.1   File Flags

4.4BSD added two new system calls, chflags and fchflags, that set a 32-bit flags — di_flags member of dinode structure.

The owner of the file or the superuser can set the low 16 bits. Only the superuser can set the high 16 bits. Once set, the append-only and immutable flags in the top 16 bits cannot be cleared when the system is in *secure mode.*

The flags are defined in sys/stat.h as,

———————————————————————————————————— sys/stat.h

```
232 /*
233  * Definitions of flags stored in file flags word.
234  *
```

```
235    * Super-user and owner changeable flags.
236    */
237 #define UF_SETTABLE     0x0000ffff      /* mask of owner changeable flags */
238 #define UF_NODUMP       0x00000001      /* do not dump file */
239 #define UF_IMMUTABLE    0x00000002      /* file may not be changed */
240 #define UF_APPEND       0x00000004      /* writes to file may only append */
241 #define UF_OPAQUE       0x00000008      /* directory is opaque wrt. union */
242 /*
243    * Super-user changeable flags.
244    */
245 #define SF_SETTABLE     0xffff0000      /* mask of superuser changeable flags */
246 #define SF_ARCHIVED     0x00010000      /* file is archived */
247 #define SF_IMMUTABLE    0x00020000      /* file may not be changed */
248 #define SF_APPEND       0x00040000      /* writes to file may only append */
```

———————————————————————————————— sys/stat.h

Files marked immutable by the superuiser cannot be changed, except by someone with physical access to either the machine or the system console. It is useful in safeguarding the `login` or `su` program from the danger of hacking. The append-only flag is typically used for critical system logs. Although simple in concept, these two features improve the security of a system dramatically.

### 9.1.2 Inode Flags

Unlike *file flags*, *inode flags* is only used to internal purpose by filesystem hierarchy manipulation functions such as `rename`. They are defined in `ufs/ufs/inode.h` as,

———————————————————————————————— ufs/ufs/inode.h

```
170 /* These flags are kept in i_flag. */
171 #define IN_ACCESS       0x0001         /* Access time update request. */
172 #define IN_CHANGE       0x0002         /* Inode change time update request. */
173 #define IN_UPDATE       0x0004         /* Modification time update request. */
174 #define IN_MODIFIED     0x0008         /* Inode has been modified. */
175 #define IN_ACCESSED     0x0010         /* Inode has been accessed. */
176 #define IN_RENAME       0x0020         /* Inode is being renamed. */
177 #define IN_SHLOCK       0x0040         /* File has shared lock. */
178 #define IN_EXLOCK       0x0080         /* File has exclusive lock. */
179 #define IN_CLEANING     0x0100         /* LFS: file is being cleaned */
180 #define IN_ADIROP       0x0200         /* LFS: dirop in progress */
181 #define IN_SPACECOUNTED 0x0400         /* Blocks to be freed in free count. */
```

———————————————————————————————— ufs/ufs/inode.h

### 9.1.3 Inode for Root Directory

Filesystems contain files, most of which contain ordinary data. Certain files are distinguished as directories and contain pointers to files that may themselves be directories. Therefore, an inode can point to a directory as well as to a file.

By convention,

- inode 2 is always reserved for the root directory of a filesystem.

## 9.2   Inode Management

### 9.2.1   Opening a File

Steps in opening a file is

1. Find the file's associated vnode.

   (a) The lookup request is given to the filesystem associated with the directory currently being searched.

   (b) When the local filesystem finds the name in the directory, it gets the inode number of the associated file. If the inode is not in the table, such as the first time a file is opened, the filesystem must request a new vnode. When a new vnode is allocated to the local filesyste, a new structure to hold inode is allocated

   (c) The filesystem searches its collection of inodes to see whether the requested inode is already in memory. To avoid doing a linear scan of all its entries, the system keeps a set of hash chains keyed on *inode number* and *filesystem identifier.*

2. Locate the disk block containing the inode

   (a) When the disk I/O completes, the inode is copied from the disk buffer into the newly allocated inode entry.

   (b) The inode table itself maintains supplement information while the inode is in memory including

      - hash chains managing the inode table
      - flags showing the inode's status
      - reference counts on the inode's use
      - information to manage locks
      - pointers to the superblock.

3. Read the block containg the inode into a buffer in system memory.

### 9.2.2   Closing a File

When the last reference to a file is closed,

1. The local filesystem is notified that the file has become inactive.

2. The inode times will be updated, and the inode may be written to disk.

3. However, it remains on the hash list so that it can be found if it is reopened.

4. After being inactive for a period determined by the vnode layer based on demand for vnodes in all the filesyste, the vnode will be reclaimed.

5. When a vnode for a local file is reclaimed, the inode is removed from the previous filesystem's hash chain and, if the inode is dirty, its contents are written back to disk.

6. Then, the space for the inode is deallocated, so that the vnode will be ready for use by a new filesystem client.

## 9.3 Quotas

The quota mechanism sets limits on both the number of files and the number of disk blocks that a user or members of group may allocate. Quotas connect into the system primarilt as an adjunct to the allocation routines.

### 9.3.1 Soft and Hard Quota

When a process exceeds its soft limit, a warning is printed on the user's terminal; the offending process is not prevented from allocating space unless it exceeds its hard limit. If a user fails to correct the problem for longer than a *grace period*, the soft limit starts to be enforced as the hard limit.

### 9.3.2 Quota Imposing Mechanism

Quota is checked by `chkdq` function. When a new block is requested from the allocation routines, the request is first validated by the quota system with the following steps:

1. If there is a user quota associated with the file, the quota system consults the quota associated with the owner of the file. If the owner has reached or exceeded their limit, the request is denied.

2. The same check is done for the group quota.

3. If the quota tests pass, the request is permitted and is added to the usage statistics for the file.

Quotas are asigned to a filesystem after it has been mounted. For each quota to be imposed, the system opens the appropriate quota file and holds a reference to it in the mount-table nety associated with the mounted filesystem.

### 9.3.3 Quota Records

Quota files are maintained as an array of *quota records* indexed by user or group identifiers. The *Quota record* is defined in **ufs/ufs/quota.h** as,

———————————————————————————————— ufs/ufs/quota.h

```
 94 /*
 95  * The following structure defines the format of the disk quota file
 96  * (as it appears on disk) - the file is an array of these structures
 97  * indexed by user or group number.  The setquota system call establishes
 98  * the vnode for each quota file (a pointer is retained in the ufsmount
 99  * structure).
100  */
101 struct dqblk {
102         u_int32_t dqb_bhardlimit;       /* absolute limit on disk blks alloc */
103         u_int32_t dqb_bsoftlimit;       /* preferred limit on disk blks */
104         u_int32_t dqb_curblocks;        /* current block count */
105         u_int32_t dqb_ihardlimit;       /* maximum # allocated inodes + 1 */
106         u_int32_t dqb_isoftlimit;       /* preferred inode limit */
107         u_int32_t dqb_curinodes;        /* current # allocated inodes */
108         int32_t   dqb_btime;            /* time limit for excessive disk use */
109         int32_t   dqb_itime;            /* time limit for excessive files */
110 };
```

———————————————————————————————— ufs/ufs/quota.h

### 9.3.4    Active Quota Entry: dquot

Active quotas are held in system memory in a dquot structure defined in ufs/ufs/quota.h
as,

──────────────────────────────────────────── ufs/ufs/quota.h

```
115 /*
116  * The following structure records disk usage for a user or group on a
117  * filesystem. There is one allocated for each quota that exists on any
118  * filesystem for the current user or group. A cache is kept of recently
119  * used entries.
120  */
121 struct dquot {
122         LIST_ENTRY(dquot) dq_hash;      /* hash list */
123         TAILQ_ENTRY(dquot) dq_freelist; /* free list */
124         u_int16_t dq_flags;             /* flags, see below */
125         u_int16_t dq_cnt;               /* count of active references */
126         u_int16_t dq_spare;             /* unused spare padding */
127         u_int16_t dq_type;              /* quota type of this dquot */
128         u_int32_t dq_id;                /* identifier this applies to */
129         struct  ufsmount *dq_ump;       /* filesystem that this is taken from */
130         struct  dqblk dq_dqb;           /* actual usage & quotas */
131 };
132 /*
133  * Flag values.
134  */
135 #define DQ_LOCK     0x01                 /* this quota locked (no MODS) */
136 #define DQ_WANT     0x02                 /* wakeup on unlock */
137 #define DQ_MOD      0x04                 /* this quota modified since read */
138 #define DQ_FAKE     0x08                 /* no limits here, just usage */
139 #define DQ_BLKS     0x10                 /* has been warned about blk limit */
140 #define DQ_INODS    0x20                 /* has been warned about inode limit */
```

──────────────────────────────────────────── ufs/ufs/quota.h

The task of finding the dquot structure associated with a file is done when the
file is first opened for writing.  If one or more quotas exist, the inode is set up to
hold a reference to the appropriate dquot, by setting dquot member of the inode
structure.  If a user or a group has multiple files open on the same filesystem, all
inodes describing those files point to the same dquot entry.

**Improvement in Searching a dquot Entry**

To avoid doing a linear scan of all the dquot entries, the system keeps a set of hash
chains keyed on the *filesystem* and on the *user* or *group identifier*.

   If the dquot entry is not resident, such as the first time a file is opened for
writing, the system must reallocate a dquot entry and read in the quota from disk.

   When the reference count on a dquot structure drops to zero, the system puts
that entry onto the end the LRU chain.  The dquot structure is not removed from
its hash chain, so if the structure is needed again soon, it can still be located.

**Dummy dquot Entries**

To prevent cost of going to disk and reading the quota file to discover that a user
has no quota, the system maintains dummy dquot entries.  For a dummy entry,
chkdq routine updates the usage fields, but will not impose any limits.

### 9.3.5 Consistency Maintenance

If the system crashes, leaving the quotas in an inconsistent state, the system administrator must run the quotacheck program to rebuild the usage information in the quota files.

# 9.4 References to Source Code

### 9.4.1 `ufs_bmap.c` - 325 lines, 3 functions

**Gloval Variables**

```
none
```

**Functions**

```
ufs_bmap          converts a the logical block number to its physical block
ufs_bmaparray     does the real conversion for ufs_bmap
ufs_getlbns       (used by ufs_bmaparray)
```

### 9.4.2 `ufs_ihash.c` - 194 lines, 7 functions

**Gloval Variables**

```
ihashtbl          Inode hash table
ihash             Magic number to generate hash key
ufs_hashlock      Lock structure for hash table (used by ffs_vget)
ufs_ihash_slock   Simple lock structure for hash table
```

**Functions**

```
ufs_ihashinit     used by ufs_init()
ufs_ihashreinit   used by ufs_reinit()
ufs_ihashdone     Free inode hash table
ufs_ihashlookup   used only by soft dependency and LFS
ufs_ihashget
ufs_ihashins
ufs_ihashrem
```

### 9.4.3 `ufs_inode.c` - 268 lines, 3 functions

**Gloval Variables**

```
none
```

**Functions**

```
ufs_inactive      Last reference to an inode.  If necessary, write or delete it
ufs_reclaim       Reclaim an inode so that it can be used for other purposes
ufs_balloc_range  Allocate a range of blocks in a file
```

### 9.4.4 `ufs_lookup.c` - 1216 lines, 9 functions

**Gloval Variables**

```
none
```

**Functions**

```
ufs_lookup          Convert a pathname into a pointer to a locked inode
ufs_dirbad
ufs_dirbadentry     Do consistency checking on a directory entry
ufs_makedirentry    Construct a new directory entry after a call to namei
ufs_direnter        Write a directory entry after a call to namei
ufs_dirremove       Remove a directory entry after a call to namei
ufs_dirrewrite      Rewrite an existing directory entry to point at the inode supplied
ufs_dirempty        Check if a directory is empty or not
ufs_checkpath       Check if source directory is in the path of the target directory
```

## 9.4.5  `ufs_quota.c` - 960 lines, 20 functions

**Gloval Variables**

```
none
```

**Functions**

```
getinoquota         Set up the quotas for an inode
chkdq               Update disk usage, and take corrective action
chkdqchg            Check for a valid change to a users allocation
chkiq               Check the inode limit, applying corrective action
chkiqchg            Check for a valid change to a users allocation
chkdquot            Diagnostic

[ Code to process quotactl system calls ]

quotaon             Q_QUOTAON - set up a quota file for a particular file system
quotaoff            Q_QUOTAOFF - turn off disk quotas for a filesystem
getquota            Q_GETQUOTA - return current values in a dqblk structure
setquota            Q_SETQUOTA - assign an entire dqblk structure
setuse              Q_SETUSE - set current inode and block usage
qsync               Q_SYNC - sync quota files to disk

[ Code managing hash table for dquot structures ]

dqinit              Initialize the quota system
dqreinit
dqdone              Free resources held by quota system
dqget               Obtain a dquot structure for the specified identifier and quota fi
dqref               Obtain a reference to a dquot
dqrele              Release a reference to a dquot
dqsync              Update the disk quota in the quota file
dqflush             Flush all entries from the cache for a particular vnode
```

## 9.4.6  `ufs_readwrite.c` - 481 lines, 4 functions

**Gloval Variables**

```
none
```

**Functions**

```
ffs_read
```

```
ffs_write
lfs_read
lfs_write
```

### 9.4.7   `ufs_vfsops.c` - 262 lines, 8 functions

**Gloval Variables**

```
ufs_initcount
```

**Functions**

```
ufs_start
ufs_root          Return the vnode for root of a filesystem
ufs_quotactl      Do operations associated with quotas
ufs_check_export  Verify a remote client has export rights
ufs_fhtovp        generic part of fhtovp
ufs_init
ufs_reinit
ufs_done
```

### 9.4.8   `ufs_vnops.c` - 2074 lines, 30 functions

**Gloval Variables**

```
none
```

**Functions**

```
ufs_vinit         Initialize the vnode associated with a new inode
ufs_makeinode     Allocate a new inode

[ Virtual Filesystem Operations for FFS ]

ufs_access
ufs_advlock
ufs_close
ufs_create
ufs_getattr
ufs_inactive
ufs_link
ufs_lookup
ufs_mkdir
ufs_mknod
ufs_open
ufs_pathconf
ufs_print
ufs_readdir
ufs_readlink
ufs_remove
ufs_rename
ufs_rmdir
ufs_setattr
ufs_strategy
ufs_symlink
```

```
ufs_whiteout        ?

[ Virtual Filesystem Operations for Special File  ]

ufsspec_close
ufsspec_read
ufsspec_write

[ Virtual Filesystem Operations for FIFO ]

ufsfifo_read
ufsfifo_write
ufsfifo_close

[ Generalized Virtual Filesystem Operations ]

#define ufs_lock        genfs_lock
#define ufs_mmap        genfs_mmap
#define ufs_revoke      genfs_revoke
#define ufs_seek        genfs_seek
#define ufs_poll        genfs_poll
#define ufs_unlock      genfs_unlock
#define ufs_abortop     genfs_abortop
#define ufs_fcntl       genfs_fcntl
#define ufs_ioctl       genfs_enoioctl
#define ufs_islocked    genfs_islocked
#define ufs_lease_check genfs_lease_check
```

# Chapter 10

# Berkeley Fast File System

The **FFS fielsore** was designed on the **assumption that buffer caches would be small and thus that files would need to be read often.** It tries to place files likely to be accessed together in the same general location on the disk.

The **LFS filestore** was **designed for fast machines with large buffer caches.** It assumes that writing data to disk is the bottleneck, and it tries to avoid seeking by writing all data together in the order in which they were created. It assumes that active files will remain in the buffer cache, so is little concerned with the time that it takes to retrieve files from the filestore.

## 10.1   Filestore Services

The filestore implementation converts from the user abstraction of a file as an array of bytes to the structure imposed by the underlying physical medium. This operation is called by *Block I/O*.

The *Block I/O* is done by

1. First, the system breaks the user's request into a set of operations to be done on *logical blocks* of the file.

2. The data in each logica lblock are accessed via *physical block* on the disk.

3. A physical disk block is constructed from one or more contiguous *sectors*.

Vnode operations about storage is implemented by underlying filestore based on *block I/O*.

### 10.1.1   Allocating and Freeing Objects

There are four operators for allocating and freeing objects.

```
VOP_VALLOC(pvp, mode, cred, vpp)
        Allocate file system type specific data a new file in the file
        system.  The argument pvp specifies the vnode of the directory
        to create the new file.  The argument mode specifies file system
        type specific flags and cred are the credentials of the calling
        process.  The vnode of the new file is returned in the address
        specified by vpp.

        (implemented as ffs_valloc in ufs/ffs/ffs_alloc.c)
        (used by ufs_mkdir, ufs_makeinode)
```

```
VOP_BALLOC(vp, startoffset, size, cred, flags, bpp)
        Allocate the physical blocks on a device given the vnode vp and
        the offset logical block number startoffset in a file.  The ar-
        gument size specifies the size to be allocated.  The credentials
        of the calling processing are specified by cred.  If the argu-
        ment bpp is not NULL, the buffer is written to the allocated
        blocks.  The argument flags is a set of flags controlling the
        low-level allocation when the buffer is written.  Valid values
        defined in <sys/buf.h> are:

                B_CLRBUF  request allocated buffer be cleared
                B_SYNC    do all allocations synchronously
        If the operation is successful zero is returned, otherwise an
        appropriate error is returned.

        (implemented as ffs_balloc in ufs/ffs/ffs_balloc.c)
        (used by ufs_direnter, ffs_write, ufs_mkdir)

VOP_REALLOCBLKS(vp, buflist)
        Rearrange block in a file to be contiguous.  The argument vp is
        the vnode of the file to manipulate.  The argument buflist is a
        list of buffers to rearrange.  If the operation is successful
        zero is returned, otherwise an appropriate error is returned.

        (implemented as ffs_valloc in ufs/ffs/ffs_alloc.c)
        (used by NONE !)

VOP_VFREE(pvp, ino, mode)
        Release file resources.  This function is used by the file sys-
        tem to release cached file system specific data associated with
        the file when the vnode is recycled.

        (implemented as ffs_vfree in ufs/ffs/ffs_alloc.c)
        (used by ufs_mkdir, ufs_makeinode)
```

## 10.1.2   Updating Inode Attribute

```
VOP_UPDATE(vp, access, modify, flags)
        Update times on file with vnode vp.  The access and modification
        times are specified by the arguments access and modify respec-
        tively.  The change time is always taken from the current time.
        The argument flags is a set of file system type dependent flags
        indicating which times should be updated.

        (implemented as ffs_vfree in ufs/ffs/ffs_inode.c)
        (used by ufs_inactive, ufs_direnter, ufs_setattr, ufs_link,
                ufs_rename, ufs_mkdir, ufs_makeinode,

                ffs_reallocblks, ffs_balloc, ffs_truncate,
                ffs_fsync, ffs_full_fsync, ffs_read, ffs_write)
```

### 10.1.3   Manipulating Existing Objects

The `blkatoff` operator is similar to the `read` operator, except that the `blkatoff` operator simply returns a pointer to a kernel memory buffer with the requested data, instead of copying the data.

```
VOP_READ(vp, uio, ioflag, cred)
        Read the contents of a file.  The argument vp is the vnode of
        the file to read from, uio is the location to read the data in-
        to, ioflag is a set of flags and cred are the credentials of the
        calling process.

        The ioflag argument is used to give directives and hints to the
        file system.  When attempting a read, the high 16 bits are used
        to provide a read-ahead hint (in unit of file system blocks)
        that the file system should attempt.  The low 16 bits are a bit
        mask which can contain the following flags:

                IO_UNIT         do I/O as atomic unit
                IO_APPEND       append write to end
                IO_SYNC         do I/O synchronously
                IO_NODELOCKED   underlying node already locked
                IO_NDELAY       FNDELAY flag set in file table
                IO_VMIO         data already in VMIO space

        Zero is returned on success, otherwise an error is returned.
        The vnode should be locked on entry and remains locked on exit.

VOP_WRITE(vp, uio, ioflag, cred)
        Write to a file.  The argument vp is the vnode of the file to
        write to, uio is the location of the data to write, ioflag is a
        set of flags and cred are the credentials of the calling pro-
        cess.

        The ioflag argument is used to give directives and hints to the
        file system.  The low 16 bits are a bit mask which can contain
        the same flags as VOP_READ().

        Zero is returned on success, otherwise an error is returned.
        The vnode should be locked on entry and remains locked on exit.

VOP_FSYNC(vp, cred, flags, offlo, offhi, p)
        Flush pending data buffers for a file to disk.  The argument vp
        is the locked vnode of the file for flush.  The argument cred is
        the caller's credentials and p the calling process.  The argu-
        ment flags is a set of flags. If FSYNC_WAIT is specified in
        flags, the function should wait for I/O to complete before re-
        turning.  The argument offlo and offhi specify the range of file
        to flush. If the operation is successful zero is returned, oth-
        erwise an appropriate error code is returned.

        This function implements the sync(2) and fsync(2) system calls.

VOP_BLKATOFF(vp, offset, res, bpp)
        Return buffer bpp with the contents of block offset from the be-
```

ginning of directory specified by vnode vp.  If res is non-zero,
fill it in with a pointer to the remaining space in the directo-
ry.

(implemented as ffs_blkatoff in ufs/ffs/ffs_subr.c)
(used by ufs_lookup, ufs_direnter, ufs_dirremove, ufs_dirrewrite)

### 10.1.4   Changing in Space Allocation

Historically, it could be used only to decrease the size of an object.  In 4.4BSD, it
can be used both to increase and to decrease the size of an object.

VOP_TRUNCATE(vp, length, flags, cred, p)
        Truncate the file specified by the vnode vp to at most length
        size and free the unused disk blocks.  The arguments p and cred
        is the calling process and its credentials respectively.  The
        argument flags is a set of I/O flags.  Valid values are:

                IO_UNIT          do I/O as atomic unit
                IO_APPEND        append write to end
                IO_SYNC          sync I/O file integrity completion
                IO_NODELOCKED    underlying node already locked
                IO_NDELAY        FNDELAY flag set in file table
                IO_DSYNC         sync I/O data integrity completion
                IO_ALTSEMANTICS  use alternate i/o semantics

        If the operation is successful zero is returned, otherwise an
        appropriate error is returned.

        (implemented as ffs_truncate in ufs/ffs/ffs_inode.c)
        (used by ufs_inactive, ufs_direnter, ffs_write, ufs_setattr,
         ufs_rename, ufs_rmdir)

### 10.1.5   Virtual Memory System Support

VOP_GETPAGES(vp, offset, m, count, centeridx, access_type, advice, flags)
        Read VM pages from file.  The argument vp is the locked vnode to
        read the VM pages from.  The argument offset is offset in the
        file to start accessing and m is an array of VM pages.  The ar-
        gument count specifies the number of pages to read.  If the op-
        eration is successful zero is returned, otherwise an appropriate
        error code is returned.

        This function is primarily used by the page-fault handing mecha-
        nism.

        (implemented as genfs_getpages in miscfs/genfs/genfs_vnops.c)
        (used by ubc_fault, ubc_alloc, uvn_get)

VOP_PUTPAGES(vp, offset, len, flags)
        Write modified (dirty) VM pages to file.  The argument vp is the
        locked vnode to write the VM pages to and offset and len speci-
        fies the range of VM pages to write.  There seems to be some
        confusion in the code whether offset and len specify the start

```
and length of the VM pages for the start and end of the VM
pages.  The argument flags specifies whether the pages should be
written asynchronously and also whether they should be marked
invalid one the write back operation has completed.  If the op-
eration is successful zero is returned, otherwise an appropriate
error code is returned.

The function is primarily used by the pageout handling mecha-
nism.

(implemented as genfs_putpages in miscfs/genfs/genfs_vnops.c)
(used by uvn_put)
```

## 10.2 Organization of the FFS

To describe the design motivation of the FFS, we describes the problems of tradi-
tional UNIX filesystem before BSD UNIX appeared.

**Long Seek Problem** Traditional UNIX filesystem consists of two area:
inodes area followed by data area. Separation of inode information
from the data resulted a long seek from the file's inode to its data.

**Too Frequent Accesses Problem** The traditional UNIX filesystem
uses a 512-byte physical block size. So seeks between small 512-
byte data transfers are required with long seek frequently.

As a result, the old filesystem was using only about 4 percent of the maximum disk
throughput. The main single reason was that the order of blocks on the free list
quickly became scrambled, occurring too frequeny access to small blocks with long
seek.

### 10.2.1 Superblock

A 4.4BSD filesystem is described by its *superblock*, located at the beginning of
the filesystem's disk partition. The superblock data do not change after filesystem
creation.

The structure of superblock is defined in fs structure of **ffs/ffs/fs.h** as,

———————————————————————————————————————————————— ffs/ffs/fs.h

```
171 /*
172  * Super block for an FFS file system in memory.
173  */
174 struct fs {
175         int32_t  fs_firstfield;       /* historic file system linked list, */
176         int32_t  fs_unused_1;         /*     used for incore super blocks */
177         ufs_daddr_t fs_sblkno;        /* addr of super-block in filesys */
178         ufs_daddr_t fs_cblkno;        /* offset of cyl-block in filesys */
179         ufs_daddr_t fs_iblkno;        /* offset of inode-blocks in filesys */
180         ufs_daddr_t fs_dblkno;        /* offset of first data after cg */
181         int32_t  fs_cgoffset;         /* cylinder group offset in cylinder */
182         int32_t  fs_cgmask;           /* used to calc mod fs_ntrak */
183         int32_t  fs_time;             /* last time written */
184         int32_t  fs_size;             /* number of blocks in fs */
185         int32_t  fs_dsize;            /* number of data blocks in fs */
```

```
186        int32_t  fs_ncg;              /* number of cylinder groups */
187        int32_t  fs_bsize;            /* size of basic blocks in fs */
188        int32_t  fs_fsize;            /* size of frag blocks in fs */
189        int32_t  fs_frag;             /* number of frags in a block in fs */
190 /* these are configuration parameters */
191        int32_t  fs_minfree;          /* minimum percentage of free blocks */
192        int32_t  fs_rotdelay;         /* num of ms for optimal next block */
193        int32_t  fs_rps;              /* disk revolutions per second */
194 /* these fields can be computed from the others */
195        int32_t  fs_bmask;            /* ''blkoff'' calc of blk offsets */
196        int32_t  fs_fmask;            /* ''fragoff'' calc of frag offsets */
197        int32_t  fs_bshift;           /* ''lblkno'' calc of logical blkno */
198        int32_t  fs_fshift;           /* ''numfrags'' calc number of frags */
199 /* these are configuration parameters */
200        int32_t  fs_maxcontig;        /* max number of contiguous blks */
201        int32_t  fs_maxbpg;           /* max number of blks per cyl group */
202 /* these fields can be computed from the others */
203        int32_t  fs_fragshift;        /* block to frag shift */
204        int32_t  fs_fsbtodb;          /* fsbtodb and dbtofsb shift constant */
205        int32_t  fs_sbsize;           /* actual size of super block */
206        int32_t  fs_csmask;           /* csum block offset (now unused) */
207        int32_t  fs_csshift;          /* csum block number (now unused) */
208        int32_t  fs_nindir;           /* value of NINDIR */
209        int32_t  fs_inopb;            /* value of INOPB */
210        int32_t  fs_nspf;             /* value of NSPF */
211 /* yet another configuration parameter */
212        int32_t  fs_optim;            /* optimization preference, see below */
213 /* these fields are derived from the hardware */
214        int32_t  fs_npsect;           /* # sectors/track including spares */
215        int32_t  fs_interleave;       /* hardware sector interleave */
216        int32_t  fs_trackskew;        /* sector 0 skew, per track */
217 /* fs_id takes the space of the unused fs_headswitch and fs_trkseek fields */
218        int32_t  fs_id[2];            /* unique file system id */
219 /* sizes determined by number of cylinder groups and their sizes */
220        ufs_daddr_t  fs_csaddr;       /* blk addr of cyl grp summary area */
221        int32_t  fs_cssize;           /* size of cyl grp summary area */
222        int32_t  fs_cgsize;           /* cylinder group size */
223 /* these fields are derived from the hardware */
224        int32_t  fs_ntrak;            /* tracks per cylinder */
225        int32_t  fs_nsect;            /* sectors per track */
226        int32_t  fs_spc;              /* sectors per cylinder */
227 /* this comes from the disk driver partitioning */
228        int32_t  fs_ncyl;             /* cylinders in file system */
229 /* these fields can be computed from the others */
230        int32_t  fs_cpg;              /* cylinders per group */
231        int32_t  fs_ipg;              /* inodes per group */
232        int32_t  fs_fpg;              /* blocks per group * fs_frag */
233 /* this data must be re-computed after crashes */
234        struct  csum fs_cstotal;      /* cylinder summary information */
235 /* these fields are cleared at mount time */
236        int8_t   fs_fmod;             /* super block modified flag */
237        int8_t   fs_clean;            /* file system is clean flag */
238        int8_t   fs_ronly;            /* mounted read-only flag */
239        int8_t   fs_flags;            /* see FS_ flags below */
```

```
240          u_char   fs_fsmnt[MAXMNTLEN];   /* name mounted on */
241 /* these fields retain the current block allocation info */
242          int32_t  fs_cgrotor;            /* last cg searched (UNUSED) */
243          void     *fs_ocsp[NOCSPTRS];    /* padding; was list of fs_cs buffers */
244          u_int16_t *fs_contigdirs;       /* # of contiguously allocated dirs */
245          struct csum *fs_csp;            /* cg summary info buffer for fs_cs */
246          int32_t *fs_maxcluster;         /* max cluster in each cyl group */
247          int32_t  fs_cpc;                /* cyl per cycle in postbl */
248          int16_t  fs_opostbl[16][8];     /* old rotation block list head */
249          int32_t  fs_snapinum[20];       /* RESERVED for snapshot inode nums */
250          int32_t  fs_avgfilesize;        /* expected average file size */
251          int32_t  fs_avgfpdir;           /* expected # of files per directory */
252          int32_t  fs_sparecon[26];       /* RESERVED for future constants */
253          int32_t  fs_pendingblocks;      /* blocks in process of being freed */
254          int32_t  fs_pendinginodes;      /* inodes in process of being freed */
257          int32_t  fs_inodefmt;           /* format of on-disk inodes */
258          u_int64_t fs_maxfilesize;       /* maximum representable file size */
259          int64_t  fs_qbmask;             /* ~fs_bmask for use with 64-bit size */
260          int64_t  fs_qfmask;             /* ~fs_fmask for use with 64-bit size */
261          int32_t  fs_state;              /* validate fs_clean field (UNUSED) */
262          int32_t  fs_postblformat;       /* format of positional layout tables */
263          int32_t  fs_nrpos;              /* number of rotational positions */
264          int32_t  fs_postbloff;          /* (u_int16) rotation block list head */
265          int32_t  fs_rotbloff;           /* (u_int8) blocks for each rotation */
266          int32_t  fs_magic;              /* magic number */
267          u_int8_t fs_space[1];           /* list of blocks for each rotation */
268 /* actually longer */
269 };
```

———————————————————————————————————————————— ffs/ffs/fs.h

**Block Size**

So that files as large as $2^{32}$ bytes can be created with only two levels of indirection, the minimum size of a filesystem block is 4096 bytes. The block size is recorded in the filesystem's supoerblock, as `fs_bsize` member.

**Filesystem Parameterization**

The goal of parameterizing the processor capabilities and mass-storage characteristics, is to **allocate blocks in an optimum configuration-dependent way**.

Important parameter maintained by filesystem is contained in *superblock* and they includes,

**fs_rotdelay** The expected time in milliseconds to service disk interrupt and to schedule a new disk transfer, depending the speed of the main CPU. It is used to decide how much rotational spacing to place between successive blocks in a file.

    For modern high speed workstation, such as SUN Ultra 1, this parameter should be set to zero, since the the expected time is less than one milliseconds.

**fs_maxcontig** This specifies the maximum number of contiguous blocks that will be laid out before forcing a rotational delay. The default value is one, since most device drivers require an interrupt per disk transfer. Device drivers that can chain sev- eral buffers together in a single transfer should set this to the maximum chain length.

**fs_maxbpg** This indicates the maximum number of blocks any single file can allocate out of a cylinder group before it is forced to begin allocating blocks from another cylinder group. Typically this value is set to about one quarter of the total blocks in a cylinder group. The intent is to prevent any single file from using up all the blocks in a single cylinder group, thus degrading access times for all files subsequently allocated in that cylinder group. The effect of this limit is to cause big files to do long seeks more frequently than if they were allowed to allocate all the blocks in a cylinder group before seeking elsewhere. For file systems with exclusively large files, this parameter should be set higher.

**fs_rps** Number of disk platter revolution per second

**fs_ntrak** Number of tracks per cylinder

**fs_nsect** Number of sectors per track

**fs_npsect** Number of sectors including spares per track

**fs_minfree** This value specifies the percentage of space held back from normal users; the minimum free space threshold. The default value used is 10factor of three in throughput will be lost over the performance obtained at a 10above the current usage level, users will be unable to allocate files until enough files have been deleted to get under the higher threshold.

**fs_interleave** Hardware sector interleave. Used to describe perturbations in the media format to compensate for a slow controller. In- terleave is physical sector interleave on each track, speci- fied as the denominator of the ratio: sectors read/sectors passed over Thus an interleave of 1/1 implies contiguous layout, while 1/2 implies logical sector 0 is separated by one sector from logical sector 1.

**fs_trackskew** This specifies the skew in sectors from one track to the next in a cylinder. The default value is zero, indicating that each track in a cylinder begins at the same rotational position.

**fs_optim** The file system can either try to minimize the time spent allocating blocks, or it can attempt to minimize the space fragmentation on the disk. If the value of minfree (see above) is less than 10running out of full sized blocks. For values of minfree greater than or equal to 10problematical, and the file system can be optimized for time. `fs_optim` can be specified as either space or time.

From `fs_nsect` and `fs_rps`, the allocation routines calculates the number of milliseconds required to skip over a block. With it and processor performance peramater `fs_rotdelay`, the allocation routines calculate the number of blocks to skip over such that the next block in the file will come into position under the disk head in the expected amount of time that it takes to start a new disk-transfer operation. **In fact, for modern SCSI storage device, these parameterization is actually useless, since storage device is internally designed so that it provides optimal performace, without disk interleave regarding disk rotational speed.**

## 10.2.2 Cylinder Group

The FFS filesystem organization divides a disk partition into one or more area, each of which is called a *vylinder group*.

The rationale for using cylinder groups is to create clusters of inodes that are close to the blocks that they reference, instead of them all being located at the beginning of the disk. Then the filesystem attempts to allocate file blocks close to the inodes that describes them to avoid long seeks between getting the inode and getting its associated data.

For each cylinder group, a static number of inodes is allocated at filesystem-creation time. The default policy is to allocate one inode for each 2048 bytes of space in the cylinder group, with the expectation that this amount will be far more than will ever be needed.

Cylinder group contains information including

- a redundant copy of the superblock

- space for inodes

- bitmap describing available blocks in the cylinder group

- summary information describing the usage of data blocks within the cylinder group

To be safe from capastrophic loss, all the bookeeping information about cylinder group is not placed at the beginning of each cylinder group. The offset from each cylinder group is calculated to be about one track father from the beginning than is the preceding cylinder group. In this way, the redundant information spirals down into the pack, so that single track, cylinder, or platter can be lost without all copies of the superblock also being lost.

The structure of cylinder group is defined as `struct cg` of `ffs/ffs/fs.h` as

———————————————————————————————— ffs/ffs/fs.h

```
343 /*
344  * Cylinder group block for a file system.
345  */
346 #define CG_MAGIC        0x090255
347 struct cg {
348         int32_t  cg_firstfield;      /* historic cyl groups linked list */
349         int32_t  cg_magic;           /* magic number */
350         int32_t  cg_time;            /* time last written */
351         int32_t  cg_cgx;             /* we are the cgx'th cylinder group */
352         int16_t  cg_ncyl;            /* number of cyl's this cg */
353         int16_t  cg_niblk;           /* number of inode blocks this cg */
354         int32_t  cg_ndblk;           /* number of data blocks this cg */
355         struct   csum cg_cs;         /* cylinder summary information */
356         int32_t  cg_rotor;           /* position of last used block */
357         int32_t  cg_frotor;          /* position of last used frag */
358         int32_t  cg_irotor;          /* position of last used inode */
359         int32_t  cg_frsum[MAXFRAG];  /* counts of available frags */
360         int32_t  cg_btotoff;         /* (int32) block totals per cylinder */
361         int32_t  cg_boff;            /* (u_int16) free block positions */
362         int32_t  cg_iusedoff;        /* (u_int8) used inode map */
363         int32_t  cg_freeoff;         /* (u_int8) free block map */
364         int32_t  cg_nextfreeoff;     /* (u_int8) next available space */
365         int32_t  cg_clustersumoff;   /* (u_int32) counts of avail clusters */
```

```
366          int32_t  cg_clusteroff;        /* (u_int8) free cluster map */
367          int32_t  cg_nclusterblks;      /* number of clusters this cg */
368          int32_t  cg_sparecon[13];      /* reserved for future use */
369          u_int8_t cg_space[1];          /* space for cylinder group maps */
370 /* actually longer */
```

———————————————————————————————————————————— ffs/ffs/fs.h

where the **struct csum** is defined as

———————————————————————————————————————————— ffs/ffs/fs.h

```
158 /*
159  * Per cylinder group information; summarized in blocks allocated
160  * from first cylinder group data blocks.  These blocks have to be
161  * read in from fs_csaddr (size fs_cssize) in addition to the
162  * super block.
163  */
164 struct csum {
165          int32_t cs_ndir;              /* number of directories */
166          int32_t cs_nbfree;            /* number of free blocks */
167          int32_t cs_nifree;            /* number of free inodes */
168          int32_t cs_nffree;            /* number of free frags */
169 };
```

———————————————————————————————————————————— ffs/ffs/fs.h

### 10.2.3  Fragment

As the block size increases, the amount of space reserved for inodes decreases, but the amount of unused data space at the end of blocks rises quickly to an intolerable level with a minimum allocation of 8192-byte filesystem blocks. To increase space efficiency, the filesystem allow the division of a single filesystem block into one or more *fragments*.

#### Block Map

The block map associated with each cylinder group records the space available in a cylinder group in fragments.

#### Fragmentation Policy

If an 11,000 byte file is to be stored on 4096/1024 filesystem (block/fragment size),

1. this file would use two full-sized blocks and three fragments portion of another block.

2. If no block with three aligned fragments were available at the time, a full-sized block would be split, yielding the necessary three fragments and a single unused fragment.

3. The remaining fragment could be allocated to another file as needed.

## 10.3 Reading a File

Since NetBSD uses *Unified Buffer Cache (UBC)*, the mechanism of reading and writing to a file is different from 4.4BSD. Since UBC integrates filesystem buffer cache and virtual memory caches of file data, *cluster* interface used in 4.4BSD is no longer used.

Key architecture of NetBSD filesystem file read and write is

- The buffer cache functions such as `bread` or `bwrite`, read and write with a device driver strategy routine via `ufs_strategy` function.

- The vnode operation such as `VOP_READ` or `VOP_WRITE`, read and write with UBC interface via `ffs_read` or `ffs_write` function.

### 10.3.1 Regular File Reading Algorithm: using UBC

For 4.4BSD, reading a general file in FFS is processed as, using UBC,

1. `read` — system call from user application program

2. `sys_read` — kernel system call

3. `vn_read` — vnode high-level file operation

4. `VOP_READ` — VFS vnode operation

5. `ffs_read` — FFS vnode operation

6. UBC interaction

### 10.3.2 Non-regular File Reading Algorithm: without UBC

Reading a non-regular file such as directory in FFS is processed as, without using UBC,

1. `read` — system call from user application program

2. `sys_read` — kernel system call

3. `vn_read` — vnode high-level file operation

4. `VOP_READ` — VFS vnode operation

5. `ffs_read` — FFS vnode operation

6. `breadn` — Buffer Cache

7. `VOP_STRATEGY` — VFS vnode operation

8. `ufs_strategy` — FFS vnode operation

9. `VOP_BMAP` — convert logical file block number to physical disk block number

10. `VOP_STRATEGY` — device driver's vnode operation

11. `spec_strategy` — special filesystem vnode operation

12. `*bdev->d_strategy` — device driver strategy function

## 10.3.3  Implementation

—————————————————————————— ufs/ufs/ufs_readwrite.c

```
61 /*
62  * Vnode op for reading.
63  */
64 /* ARGSUSED */
65 int
66 READ(void *v)
67 {
68         struct vop_read_args /* {
69                 struct vnode *a_vp;
70                 struct uio *a_uio;
71                 int a_ioflag;
72                 struct ucred *a_cred;
73         } */ *ap = v;
74         struct vnode *vp;
75         struct inode *ip;
76         struct uio *uio;
77         FS *fs;
78         void *win;
79         vsize_t bytelen;
80         struct buf *bp;
81         ufs_daddr_t lbn, nextlbn;
82         off_t bytesinfile;
83         long size, xfersize, blkoffset;
84         int error;
85         boolean_t usepc = FALSE;
86
87         vp = ap->a_vp;
88         ip = VTOI(vp);
89         uio = ap->a_uio;
90         error = 0;
91
92 #ifdef DIAGNOSTIC
93         if (uio->uio_rw != UIO_READ)
94                 panic("%s: mode", READ_S);
95
96         if (vp->v_type == VLNK) {
97                 if ((int)ip->i_ffs_size < vp->v_mount->mnt_maxsymlinklen ||
98                     (vp->v_mount->mnt_maxsymlinklen == 0 &&
99                     ip->i_ffs_blocks == 0))
100                        panic("%s: short symlink", READ_S);
101        } else if (vp->v_type != VREG && vp->v_type != VDIR)
102                panic("%s: type %d", READ_S, vp->v_type);
103 #endif
104         fs = ip->I_FS;
105         if ((u_int64_t)uio->uio_offset > fs->fs_maxfilesize)
106                 return (EFBIG);
107         if (uio->uio_resid == 0)
108                 return (0);
109         if (uio->uio_offset >= ip->i_ffs_size) {
110                 goto out;
111         }
```

```
112
113 #ifndef LFS_READWRITE
114        usepc = vp->v_type == VREG;
115 #endif
116        if (usepc) {
117                while (uio->uio_resid > 0) {
118                        bytelen = MIN(ip->i_ffs_size - uio->uio_offset,
119                            uio->uio_resid);
120                        if (bytelen == 0)
121                                break;
122
123                        win = ubc_alloc(&vp->v_uobj, uio->uio_offset,
124                                    &bytelen, UBC_READ);
125                        error = uiomove(win, bytelen, uio);
126                        ubc_release(win, 0);
127                        if (error)
128                                break;
129                }
130                goto out;
131        }
132
133        for (error = 0, bp = NULL; uio->uio_resid > 0; bp = NULL) {
134                bytesinfile = ip->i_ffs_size - uio->uio_offset;
135                if (bytesinfile <= 0)
136                        break;
137                lbn = lblkno(fs, uio->uio_offset);
138                nextlbn = lbn + 1;
139                size = BLKSIZE(fs, ip, lbn);
140                blkoffset = blkoff(fs, uio->uio_offset);
141                xfersize = MIN(MIN(fs->fs_bsize - blkoffset, uio->uio_resid),
142                    bytesinfile);
143
144                if (lblktosize(fs, nextlbn) >= ip->i_ffs_size)
145                        error = bread(vp, lbn, size, NOCRED, &bp);
146                else {
147                        int nextsize = BLKSIZE(fs, ip, nextlbn);
148                        error = breadn(vp, lbn,
149                            size, &nextlbn, &nextsize, 1, NOCRED, &bp);
150                }
151                if (error)
152                        break;
153
154                /*
155                 * We should only get non-zero b_resid when an I/O error
156                 * has occurred, which should cause us to break above.
157                 * However, if the short read did not cause an error,
158                 * then we want to ensure that we do not uiomove bad
159                 * or uninitialized data.
160                 */
161                size -= bp->b_resid;
162                if (size < xfersize) {
163                        if (size == 0)
164                                break;
165                        xfersize = size;
```

```
166                     }
167                     error = uiomove((char *)bp->b_data + blkoffset, xfersize, uio);
168                     if (error)
169                             break;
170                     brelse(bp);
171             }
172     if (bp != NULL)
173             brelse(bp);
174
175 out:
176     if (!(vp->v_mount->mnt_flag & MNT_NOATIME)) {
177             ip->i_flag |= IN_ACCESS;
178             if ((ap->a_ioflag & IO_SYNC) == IO_SYNC)
179                     error = VOP_UPDATE(vp, NULL, NULL, UPDATE_WAIT);
180     }
181     return (error);
182 }
```

———————————————————————————————— ufs/ufs/ufs_readwrite.c

## 10.4   Writing a File

### 10.4.1   Regular File Writing Algorithm

Writing a regular file in FFS is processed as, using UBC,

1. **write** — system call from user application program

2. **sys_write** — kernel system call

3. **vn_write** — vnode high-level file operation

4. **VOP_WRITE** — VFS vnode operation

5. **ffs_write** — FFS vnode operation

6. UBC interation

   Writing a non-regular file in FFS such as directory is processed, without using UBC, as shown in the previous subsection.

### 10.4.2   Non-regular File Writing Algorithm

If the file needs to be extended, the request is rounded up to the next fragment size, and only that much space is allocated by **VOP_BALLOC**.

### 10.4.3   Implementation

———————————————————————————————— ufs/ufs/ufs_readwrite.c

```
184 /*
185  * Vnode op for writing.
186  */
187 int
188 WRITE(void *v)
189 {
```

```
190     struct vop_write_args /* {
191         struct vnode *a_vp;
192         struct uio *a_uio;
193         int a_ioflag;
194         struct ucred *a_cred;
195     } */ *ap = v;
196     struct vnode *vp;
197     struct uio *uio;
198     struct inode *ip;
199     struct genfs_node *gp;
200     FS *fs;
201     struct buf *bp;
202     struct proc *p;
203     struct ucred *cred;
204     ufs_daddr_t lbn;
205     off_t osize, origoff, oldoff, preallocoff, endallocoff, nsize;
206     int blkoffset, error, flags, ioflag, resid, size, xfersize;
207     int bsize, aflag;
208     int ubc_alloc_flags;
209     int extended=0;
210     void *win;
211     vsize_t bytelen;
212     boolean_t async;
213     boolean_t usepc = FALSE;
214
215     cred = ap->a_cred;
216     ioflag = ap->a_ioflag;
217     uio = ap->a_uio;
218     vp = ap->a_vp;
219     ip = VTOI(vp);
220     gp = VTOG(vp);
221
222     KASSERT(vp->v_size == ip->i_ffs_size);
223 #ifdef DIAGNOSTIC
224     if (uio->uio_rw != UIO_WRITE)
225         panic("%s: mode", WRITE_S);
226 #endif
227
228     switch (vp->v_type) {
229     case VREG:
230         if (ioflag & IO_APPEND)
231             uio->uio_offset = ip->i_ffs_size;
232         if ((ip->i_ffs_flags & APPEND) && uio->uio_offset != ip->i_ffs_size)
233             return (EPERM);
234         /* FALLTHROUGH */
235     case VLNK:
236         break;
237     case VDIR:
238         if ((ioflag & IO_SYNC) == 0)
239             panic("%s: nonsync dir write", WRITE_S);
240         break;
241     default:
242         panic("%s: type", WRITE_S);
243     }
```

```
244
245      fs = ip->I_FS;
246      if (uio->uio_offset < 0 ||
247          (u_int64_t)uio->uio_offset + uio->uio_resid > fs->fs_maxfilesize)
248          return (EFBIG);
249 #ifdef LFS_READWRITE
250      /* Disallow writes to the Ifile, even if noschg flag is removed */
251      /* XXX can this go away when the Ifile is no longer in the namespace? */
252      if (vp == fs->lfs_ivnode)
253          return (EPERM);
254 #endif
255
256      /*
257       * Maybe this should be above the vnode op call, but so long as
258       * file servers have no limits, I don't think it matters.
259       */
260      p = uio->uio_procp;
261      if (vp->v_type == VREG && p &&
262          uio->uio_offset + uio->uio_resid >
263          p->p_rlimit[RLIMIT_FSIZE].rlim_cur) {
264          psignal(p, SIGXFSZ);
265          return (EFBIG);
266      }
267      if (uio->uio_resid == 0)
268          return (0);
269
270      flags = ioflag & IO_SYNC ? B_SYNC : 0;
271      async = vp->v_mount->mnt_flag & MNT_ASYNC;
272      origoff = uio->uio_offset;
273      resid = uio->uio_resid;
274      osize = ip->i_ffs_size;
275      bsize = fs->fs_bsize;
276      error = 0;
277
278 #ifndef LFS_READWRITE
279      usepc = vp->v_type == VREG;
280 #endif
281      if (!usepc) {
282          goto bcache;
283      }
284
285      prealloccoff = round_page(blkroundup(fs, MAX(osize, uio->uio_offset)));
286      aflag = ioflag & IO_SYNC ? B_SYNC : 0;
287      nsize = MAX(osize, uio->uio_offset + uio->uio_resid);
288      endalloccoff = nsize - blkoff(fs, nsize);
289
290      /*
291       * if we're increasing the file size, deal with expanding
292       * the fragment if there is one.
293       */
294
295      if (nsize > osize && lblkno(fs, osize) < NDADDR &&
296          lblkno(fs, osize) != lblkno(fs, nsize) &&
297          blkroundup(fs, osize) != osize) {
```

```
298          error = ufs_balloc_range(vp, osize, blkroundup(fs, osize) -
299              osize, cred, aflag);
300          if (error) {
301              goto out;
302          }
303          if (flags & B_SYNC) {
304              vp->v_size = blkroundup(fs, osize);
305              simple_lock(&vp->v_interlock);
306              VOP_PUTPAGES(vp, trunc_page(osize & ~(bsize - 1)),
307                  round_page(vp->v_size), PGO_CLEANIT | PGO_SYNCIO);
308          }
309      }
310
311      ubc_alloc_flags = UBC_WRITE;
312      while (uio->uio_resid > 0) {
313          boolean_t extending; /* if we're extending a whole block */
314          off_t newoff;
315
316          oldoff = uio->uio_offset;
317          blkoffset = blkoff(fs, uio->uio_offset);
318          bytelen = MIN(fs->fs_bsize - blkoffset, uio->uio_resid);
319
320          /*
321           * if we're filling in a hole, allocate the blocks now and
322           * initialize the pages first.  if we're extending the file,
323           * we can safely allocate blocks without initializing pages
324           * since the new blocks will be inaccessible until the write
325           * is complete.
326           */
327          extending = uio->uio_offset >= preallocoff &&
328              uio->uio_offset < endallocoff;
329
330          if (!extending) {
331              error = ufs_balloc_range(vp, uio->uio_offset, bytelen,
332                  cred, aflag);
333              if (error) {
334                  break;
335              }
336              ubc_alloc_flags &= ~UBC_FAULTBUSY;
337          } else {
338              lockmgr(&gp->g_glock, LK_EXCLUSIVE, NULL);
339              error = GOP_ALLOC(vp, uio->uio_offset, bytelen,
340                  aflag, cred);
341              lockmgr(&gp->g_glock, LK_RELEASE, NULL);
342              if (error) {
343                  break;
344              }
345              ubc_alloc_flags |= UBC_FAULTBUSY;
346          }
347
348          /*
349           * copy the data.
350           */
351
```

```
352          win = ubc_alloc(&vp->v_uobj, uio->uio_offset, &bytelen,
353              ubc_alloc_flags);
354          error = uiomove(win, bytelen, uio);
355          if (error && extending) {
356              /*
357               * if we haven't initialized the pages yet,
358               * do it now.  it's safe to use memset here
359               * because we just mapped the pages above.
360               */
361              memset(win, 0, bytelen);
362          }
363          ubc_release(win, 0);
364
365          /*
366           * update UVM's notion of the size now that we've
367           * copied the data into the vnode's pages.
368           *
369           * we should update the size even when uiomove failed.
370           * otherwise ffs_truncate can't flush soft update states.
371           */
372
373          newoff = oldoff + bytelen;
374          if (vp->v_size < newoff) {
375              uvm_vnp_setsize(vp, newoff);
376              extended = 1;
377          }
378
379          if (error) {
380              break;
381          }
382
383          /*
384           * flush what we just wrote if necessary.
385           * XXXUBC simplistic async flushing.
386           */
387
388          if (!async && oldoff >> 16 != uio->uio_offset >> 16) {
389              simple_lock(&vp->v_interlock);
390              error = VOP_PUTPAGES(vp, (oldoff >> 16) << 16,
391                  (uio->uio_offset >> 16) << 16, PGO_CLEANIT);
392              if (error) {
393                  break;
394              }
395          }
396      }
397      if (error == 0 && ioflag & IO_SYNC) {
398          simple_lock(&vp->v_interlock);
399          error = VOP_PUTPAGES(vp, trunc_page(origoff & ~(bsize - 1)),
400              round_page(blkroundup(fs, uio->uio_offset)),
401              PGO_CLEANIT | PGO_SYNCIO);
402      }
403      goto out;
404
405  bcache:
```

```
406        simple_lock(&vp->v_interlock);
407        VOP_PUTPAGES(vp, trunc_page(origoff), round_page(origoff + resid),
408            PGO_CLEANIT | PGO_FREE | PGO_SYNCIO);
409        while (uio->uio_resid > 0) {
410            lbn = lblkno(fs, uio->uio_offset);
411            blkoffset = blkoff(fs, uio->uio_offset);
412            xfersize = MIN(fs->fs_bsize - blkoffset, uio->uio_resid);
413            if (fs->fs_bsize > xfersize)
414                flags |= B_CLRBUF;
415            else
416                flags &= ~B_CLRBUF;
417
418            error = VOP_BALLOC(vp, uio->uio_offset, xfersize,
419                ap->a_cred, flags, &bp);
420
421            if (error)
422                break;
423            if (uio->uio_offset + xfersize > ip->i_ffs_size) {
424                ip->i_ffs_size = uio->uio_offset + xfersize;
425                uvm_vnp_setsize(vp, ip->i_ffs_size);
426                extended = 1;
427            }
428            size = BLKSIZE(fs, ip, lbn) - bp->b_resid;
429            if (xfersize > size)
430                xfersize = size;
431
432            error = uiomove((char *)bp->b_data + blkoffset, xfersize, uio);
433
434            /*
435             * if we didn't clear the block and the uiomove failed,
436             * the buf will now contain part of some other file,
437             * so we need to invalidate it.
438             */
439            if (error && (flags & B_CLRBUF) == 0) {
440                bp->b_flags |= B_INVAL;
441                brelse(bp);
442                break;
443            }
444 #ifdef LFS_READWRITE
445            if (!error)
446                error = lfs_reserve(fs, vp, btofsb(fs, (NIADDR + 1) << fs->lfs_bshift));
447            (void)VOP_BWRITE(bp);
448            if (!error)
449                lfs_reserve(fs, vp, -btofsb(fs, (NIADDR + 1) << fs->lfs_bshift));
450 #else
451            if (ioflag & IO_SYNC)
452                (void)bwrite(bp);
453            else if (xfersize + blkoffset == fs->fs_bsize)
454                bawrite(bp);
455            else
456                bdwrite(bp);
457 #endif
458            if (error || xfersize == 0)
459                break;
```

```
460      }
461      /*
462       * If we successfully wrote any data, and we are not the superuser
463       * we clear the setuid and setgid bits as a precaution against
464       * tampering.
465       */
466 out:
467      ip->i_flag |= IN_CHANGE | IN_UPDATE;
468      if (resid > uio->uio_resid && ap->a_cred && ap->a_cred->cr_uid != 0)
469          ip->i_ffs_mode &= ~(ISUID | ISGID);
470      if (resid > uio->uio_resid)
471          VN_KNOTE(vp, NOTE_WRITE | (extended ? NOTE_EXTEND : 0));
472      if (error) {
473          (void) VOP_TRUNCATE(vp, osize, ioflag & IO_SYNC, ap->a_cred,
474              uio->uio_procp);
475          uio->uio_offset -= resid - uio->uio_resid;
476          uio->uio_resid = resid;
477      } else if (resid > uio->uio_resid && (ioflag & IO_SYNC) == IO_SYNC)
478          error = VOP_UPDATE(vp, NULL, NULL, UPDATE_WAIT);
479      KASSERT(vp->v_size == ip->i_ffs_size);
480      return (error);
481 }
```

——————————————————————————————— ufs/ufs/ufs_readwrite.c

## 10.5   Layout Policies

Two methods for improving filesystem performance are to increase locality, and to make larger transfers possible.

Local allocation routine uses a locally optimal scheme to lay out data blocks.

The global layout policies try to improve performance by spreading unrelated data among different cylinder groups. The global policies, using summary information, try to balance the two conflicting goals of localizing data that are concurrently accessed while spreading out unrelated data.

### 10.5.1   Inode Layout Policy

- Try to place all the inodes of files in a directory in the same cylinder group.

- Try to place new directories in cylinder group with a greater-than-average number of free inodes and with the smallest number of directories.

### 10.5.2   Data Block Layout Policy

- Try to place data blocks for a file in the same cylinder group.

- Make the spillover points to force block allocation to be redirected when any file has used about 25 percent of the data blocks in a cylinder group. The newly chosen cylinder group for block allocation is the next cylinder group that has a greater-than-average number of free blocks.

## 10.6   Data Block Allocation Mechanisms

The task of managing block and fragment allocation is done by **ffs_balloc** function.

### 10.6.1  Work Flow

1. The global-policy routines call local-allocation routines with requests for specific blocks, using heuristics based on the partial information that is available.

2. The local-allocation routines will always allocate the requested block if it is free; otherwise, if a requested block is not available, the local allocator uses a four-level allocation strategy:

   (a) Use the next available block rotationally closet to the requested block on the same cylinder.

   (b) If no blocks are available on the same cylinder, choose a block within the same cylinder group.

   (c) If the cylinder group is full, quadratically hash the cylinder group number to choose another cylinder group in which to look for a free block.

   (d) Even so, if the free block is not found, apply an exhaustive search to all cylinder group.

Two conditions when the new block is allocated is

- The file contains no fragmented block, and the final block in the file contains insufficient space to hold the new data.

   1. If the remainder of the new data consists of more than a full block, a full block is allocated. This process is repeated until less than a full block of new data remains.

   2. A block with the necessary number of fragments is located.

- The file contains one or more fragments, but the fragments contain insufficient space to hold the new data.

   1. A new block is allocated

   2. The contents of the gragments are copied to the beginning of the block

   3. The remainder of the block is filled with new data.

   4. The process then continues as in the first condition.

### 10.6.2  Main Function that Does Allocation: `ffs_balloc`

We describe the algorithm of data block allocation with pseudo code shown below. Be sure that the following is simplified algorithm skeleton, and the real source code, `ffs/ffs/ffs_alloc.c` and `ffs/ffs/ffs_balloc.c`, is more complex.

```
/* [A] Balloc defines the structure of file system storage
 *      by allocating the physical blocks on a device given
 *      the inode and the logical block number in a file.
 */
ffs_balloc()
{
    // Select the most desirable block based on the global-policy

    ffs_blkpref();

    // Condition 2: When a fragment has already been allocated.
```

```
    if (fragment is already allocated ?)
    {
        // Try to extend a fragment

        ffs_realloccg();
    }
    else
    // Condition 1: When the file contains no fragmented block

    {
        // Allocate a new block or gragment

        ffs_alloc();
    }
}

/* [H]
 *
 * Select the desired position for the next block in a file.  The file is
 * logically divided into sections. The first section is composed of the
 * direct blocks. Each additional section contains fs_maxbpg blocks.
 *
 * If no blocks have been allocated in the first section, the policy is to
 * request a block in the same cylinder group as the inode that describes
 * the file. If no blocks have been allocated in any other section, the
 * policy is to place the section in a cylinder group with a greater than
 * average number of free blocks.  An appropriate cylinder group is found
 * by using a rotor that sweeps the cylinder groups. When a new group of
 * blocks is needed, the sweep begins in the cylinder group following the
 * cylinder group from which the previous allocation was made. The sweep
 * continues until a cylinder group with greater than the average number
 * of free blocks is found. If the allocation is for the first block in an
 * indirect block, the information on the previous allocation is unavailable;
 * here a best guess is made based upon the logical block number being
 * allocated.
 *
 * If a section is already partially allocated, the policy is to
 * contiguously allocate fs_maxcontig blocks.  The end of one of these
 * contiguous blocks and the beginning of the next is physically separated
 * so that the disk head will be in transit between them for at least
 * fs_rotdelay milliseconds.  This is to allow time for the processor to
 * schedule another I/O transfer.
 */
ufs_daddr_t
ffs_blkpref()
{
    .....

    return (appropriate next block number);   // if not found return zero;
}
```

### 10.6.3 Cylinder Overflow Algorithm: `ffs_hashalloc`

```
/* [C] Implement the cylinder overflow algorithm.
 *
 *     The policy implemented by this algorithm is:
 *        1) allocate the block in its requested cylinder group.
 *        2) quadradically rehash on the cylinder group number.
 *        3) brute force search for a free block.
 */
ffs_hashalloc( *func_ptr_to_allocator )
{
    // Try to find a fragment from preferred cylinder group

    if (*func_ptr_to_allocator() succeeded ?)
        return OK;

    // Quadratic rehash

    for (i = 1; i < fs->fs_ncg; i *= 2)
    {
        intended cylinder group numer += i;
        adjust cylinder group number overflow;
        if (*func_ptr_to_allocator() succeeded ?)
            return OK;
    }

    // Brute force search

    for (i = 2; i < fs->fs_ncg; i++)
    {
        intended cylinder group number = i;
        if (*func_ptr_to_allocator() succeeded ?)
            return OK;
    }

    return FAIL;
}
```

### 10.6.4 Global Policy 1 - Extending an Fragment: `ffs_realloccg`

```
/* [B] Reallocate a fragment to a bigger size
 *
 *     The number and size of the old block is given, and a preference
 *     and new size is also specified. The allocator attempts to extend
 *     the original block. Failing that, the regular block allocator is
 *     invoked to get an appropriate block.
 */
ffs_realloccg()
{
    // Check for extension in the existing location

    if (ffs_fragextend() succeeded ?)
        return OK;
```

```
// Calculate a new disk location from which to allocate a new fragment

switch ((int)fs->fs_optim) {
case FS_OPTSPACE:

    // Allocate an exact sized fragment. Although this makes
    // best use of space, we will waste time relocating it if
    // the file continues to grow. If the fragmentation is
    // less than half of the minimum free reserve, we choose
    // to begin optimizing for time.

    .....

    break;

case FS_OPTTIME:

    // At this point we have discovered a file that is trying to
    // grow a small fragment to a larger fragment. To save time,
    // we allocate a full sized block, then free the unused portion.
    // If the file continues to grow, the 'ffs_fragextend' call
    // above will be able to grow it in place without further
    // copying. If aberrant programs cause disk fragmentation to
    // grow within 2% of the free reserve, we choose to begin
    // optimizing for space.

    .....

    break;
}

// Try to allocate a new fragment honoring the calculated location

if (ffs_hashalloc( ffs_alloccg ) succeeded ?)
    return OK;

return FAIL;
}

/* [E] Determine whether a fragment can be extended.
 *
 *      Check to see if the necessary fragments are available, and
 *      if they are, allocate them.
 */
ffs_fragextend()
{
    ...
}
```

## 10.6.5   Global Policy 2 - Get a New Block: ffs_alloc

```
/* [F] Allocate a block in the file system.
 *
```

```
 *      The size of the requested block is given, which must be some
 *      multiple of fs_fsize and <= fs_bsize.
 *      A preference may be optionally specified. If a preference is given
 *      the following hierarchy is used to allocate a block:
 *        1) allocate the requested block.
 *        2) allocate a rotationally optimal block in the same cylinder.
 *        3) allocate a block in the same cylinder group.
 *        4) quadradically rehash into other cylinder groups, until an
 *           available block is located.
 *      If no block preference is given the following hierarchy is used
 *      to allocate a block:
 *        1) allocate a block in the cylinder group that contains the
 *           inode for the file.
 *        2) quadradically rehash into other cylinder groups, until an
 *           available block is located.
 */
ffs_alloc()
{
    // Set cylinder group number according to the above rules

    cg = .....;

    // Try to allocate a block

    if (ffs_hashalloc( ffs_alloccg ) succeeded ?)
        return OK;

    return FAIL;
}
```

## 10.6.6  Local Policy - Allocate a Block or Fragment: `ffs_alloccg`

```
/* [D] Determine whether a block can be allocated.
 *
 *      From the specified cylinder group and block,
 *      check to see if a block of the appropriate size is available,
 *      and if it is, allocate it.
 */
ffs_alloccg()
{
    //
    // [1] When the called requested a block
    //

        if (requested size == block ?)
        {
            // Allocate a new block and return the block number

            return ffs_alloccgblk();
        }

    //
    // [2] When the caller requested fragments
    //
```

```
// Allocate a new block from which to allocate requested fragments

// Check the cylinder group has free fragments having the requested size
// (allocsiz is the number of fragments which will be allocated)

frags = numfrags(fs, size); // calculate number of requested fragments
for (allocsiz = frags; allocsiz < fs->fs_frag; allocsiz++) {
        if (cgp->cg_frsum[allocsiz] != 0) // if there available frag ?
                break;
}

// When there is no fragments having the requested size

if (allocsiz == fs->fs_frag)
{
    if (if the given cylinder group has the not any free fragment ?)
    {
        // Say that there is no fragment in the given cylinder group

        return 0;
    }

    // Try to allocate a new block

    bno = ffs_alloccgblk();

    // Allocate the requested fragments

    ...

    // Mark the filesystem that fragments are allocated

    ...

    return bno;
}

// Now there are certainly fragments having the requested size,
// in the requested cylinder group

// Find the block number !

bno = ffs_mapresearch( allocsiz );

// Allocate necessary fragments;

...

    return (block number);
}

/* [G] Allocate a block in a cylinder group.
 *
 *      This algorithm implements the following policy:
```

```
 *       1) allocate the requested block.
 *       2) allocate a rotationally optimal block in the same cylinder.
 *       3) allocate the next available block on the block rotor for the
 *          specified cylinder group.
 *    Note that this routine only allocates fs_bsize blocks; these
 *    blocks may be fragmented by the routine that allocates them.
 */
ffs_alloccgblk
{
    //
    // [1] if the requested block is available, use it
    //

        if (ffs_isblock(...) says that there is available block ?)
            goto STEP_YES;

        if (fs->fs_nrpos <= 1 || fs->fs_cpc == 0)
        {
            // Block layout information is not available.
            // Leaving bpref unchanged means we take the
            // next available free block following the one
            // we just allocated. Hopefully this will at
            // least hit a track cache on drives of unknown
            // geometry (e.g. SCSI).

            goto STEP_3;
        }

        // check for a block available on the same cylinder

        if (cg_blktot(...) says that there is no available block ?)
            goto STEP_3;

    //
    // [2] check the summary information to see if a block is
    //     available in the requested cylinder starting at the
    //     requested rotational position and proceeding around.
    //

        // Get the rotational-layout table from superblock

        cylbp = cg_blks(fs, cgp, cylno, needswap);

        // Calculate the intended rotational position

        pos = cbtorpos(fs, bpref);

        // Search for a block to allocate through the summary information for
        // a rotational position with a nonzero block count

        for (i = pos; i < fs->fs_nrpos; i++)
            if (cylbp[i] > 0)
                break;
```

```
        // Search after wrapping

        if (i == fs->fs_nrpos)
        {
            for (i = 0; i < pos; i++)
                if (cylbp[i] > 0)
                    break;
        }

        // When found a rotational position

        if (cylbp[i] > 0)
        {
            // Find the actual block. A panic if none is actually there.

            .....
        }

STEP_3:

    //
    // [3] no blocks in the requested cylinder, so take next
    //     available one in this cylinder group.
    //

        bno = ffs_mapsearch(...);
        if (bno < 0)
            return FAIL;

STEP_YES:

    mark to the filesystem that the block is allocated;
    return block number;
}
```

## 10.6.7   Searching Fragment Descriptor Table: `ffs_mapsearch`

If an appropriate-sized fragment is listed in the fragment summary, then the allocation routine expects to find it in the allocation map.  To speed up the process of scanning the potentially large allocation map, the filesystem uses a table-driven algorithm.  Each byte in the map is treated as an index into a *fragment descriptor table*.  Each entry in the fragment descriptor table describes the fragment that are free for that corresponding map entry.

The fragment descriptor table is defined in `ffs_tables.c` as,

——————————————————————————————————————— ffs/ffs/ffs_tables.c

```
60 /*
61  * Given a block map bit pattern, the frag tables tell whether a
62  * particular size fragment is available.
63  *
64  * used as:
65  * if ((1 << (size - 1)) & fragtbl[fs->fs_frag][map] {
66  *       at least one fragment of the indicated size is available
```

```
67  * }
68  *
69  * These tables are used by the scanc instruction on the VAX to
70  * quickly find an appropriate fragment.
71  */
72  const u_char fragtbl124[256] = {
73          0x00, 0x16, 0x16, 0x2a, 0x16, 0x16, 0x26, 0x4e,
...
104          0x9e, 0x9e, 0x9e, 0xbe, 0xaa, 0xbe, 0xce, 0x8a,
105 };
106
107 const u_char fragtbl8[256] = {
108          0x00, 0x01, 0x01, 0x02, 0x01, 0x01, 0x02, 0x04,
...
139          0x10, 0x11, 0x11, 0x12, 0x20, 0x21, 0x40, 0x80,
140 };
141
142 /*
143  * The actual fragtbl array.
144  */
145 const u_char * const fragtbl[MAXFRAG + 1] = {
146          0, fragtbl124, fragtbl124, 0, fragtbl124, 0, 0, 0, fragtbl8,
147 };
```

——————————————————————————————————— ffs/ffs/ffs_tables.c

The **ffs_mapsearch** function that implements this algorithm is

——————————————————————————————————— ffs/ffs/ffs_alloc.c

```
1708 /*
1709  * Find a block of the specified size in the specified cylinder group.
1710  *
1711  * It is a panic if a request is made to find a block if none are
1712  * available.
1713  */
1714 static ufs_daddr_t
1715 ffs_mapsearch(fs, cgp, bpref, allocsiz)
1716          struct fs *fs;
1717          struct cg *cgp;
1718          ufs_daddr_t bpref;
1719          int allocsiz;
1720 {
1721          ufs_daddr_t bno;
1722          int start, len, loc, i;
1723          int blk, field, subfield, pos;
1724          int ostart, olen;
1725 #ifdef FFS_EI
1726          const int needswap = UFS_FSNEEDSWAP(fs);
1727 #endif
1728
1729          /*
1730           * find the fragment by searching through the free block
1731           * map for an appropriate bit pattern
1732           */
```

```
1733        if (bpref)
1734                start = dtogd(fs, bpref) / NBBY;
1735        else
1736                start = ufs_rw32(cgp->cg_frotor, needswap) / NBBY;
1737        len = howmany(fs->fs_fpg, NBBY) - start;
1738        ostart = start;
1739        olen = len;
1740        loc = scanc((u_int)len,
1741                (const u_char *)&cg_blksfree(cgp, needswap)[start],
1742                (const u_char *)fragtbl[fs->fs_frag],
1743                (1 << (allocsiz - 1 + (fs->fs_frag & (NBBY - 1)))));
1744        if (loc == 0) {
1745                len = start + 1;
1746                start = 0;
1747                loc = scanc((u_int)len,
1748                        (const u_char *)&cg_blksfree(cgp, needswap)[0],
1749                        (const u_char *)fragtbl[fs->fs_frag],
1750                        (1 << (allocsiz - 1 + (fs->fs_frag & (NBBY - 1)))));
1751                if (loc == 0) {
1752                        printf("start = %d, len = %d, fs = %s\n",
1753                                ostart, olen, fs->fs_fsmnt);
1754                        printf("offset=%d %ld\n",
1755                                ufs_rw32(cgp->cg_freeoff, needswap),
1756                                (long)cg_blksfree(cgp, needswap) - (long)cgp);
1757                        panic("ffs_alloccg: map corrupted");
1758                        /* NOTREACHED */
1759                }
1760        }
1761        bno = (start + len - loc) * NBBY;
1762        cgp->cg_frotor = ufs_rw32(bno, needswap);
1763        /*
1764         * found the byte in the map
1765         * sift through the bits to find the selected frag
1766         */
1767        for (i = bno + NBBY; bno < i; bno += fs->fs_frag) {
1768                blk = blkmap(fs, cg_blksfree(cgp, needswap), bno);
1769                blk <<= 1;
1770                field = around[allocsiz];
1771                subfield = inside[allocsiz];
1772                for (pos = 0; pos <= fs->fs_frag - allocsiz; pos++) {
1773                        if ((blk & field) == subfield)
1774                                return (bno + pos);
1775                        field <<= 1;
1776                        subfield <<= 1;
1777                }
1778        }
1779        printf("bno = %d, fs = %s\n", bno, fs->fs_fsmnt);
1780        panic("ffs_alloccg: block not in map");
1781        return (-1);
1782 }
```

——————————————————————————————————————— ffs/ffs/ffs_alloc.c

where the scanc function is defined in lib/libkern/scanc.c as,

———————————————————————————— - lib/libkern/scanc.c

```
41 int
42 scanc(u_int size, const u_char *cp, const u_char table[], int mask)
43 {
44         const u_char *end = &cp[size];
45
46         while (cp < end && (table[*cp] & mask) == 0)
47                 cp++;
48         return (end - cp);
49 }
```

———————————————————————————— - lib/libkern/scanc.c

Notice **line 1740-1760** of `ffs_mapsearch` function. The first part scans from `start` to `len`. The second part scans from `0` to `start`.

### 10.6.8   Rotational Layout Table

## 10.7   Inode Allocation Mechanism

### 10.7.1   Global Policy: `ffs_valloc`

```
/*
 * Allocate an inode in the file system.
 *
 * If allocating a directory, use ffs_dirpref to select the inode.
 * If allocating in a directory, the following hierarchy is followed:
 *   1) allocate the preferred inode.
 *   2) allocate an inode in the same cylinder group.
 *   3) quadradically rehash into other cylinder groups, until an
 *      available inode is located.
 * If no inode preference is given the following hierarchy is used
 * to allocate an inode:
 *   1) allocate an inode in cylinder group 0.
 *   2) quadradically rehash into other cylinder groups, until an
 *      available inode is located.
 */
ffs_valloc(v)
{
        // If allocating a directory, use ffs_dirpref to select the inode.

        if ((mode & IFMT) == IFDIR)
                ipref = ffs_dirpref(pip);

        // Set preferred inode and cylinder group according to the above rules.

        ipref = .....;
        cg = .....;

        // Try to allocate

        ffs_hashalloc( ffs_nodealloccg );
}
```

### 10.7.2   Local Policy 1: `ffs_dirpref`

```
/*
 * Find a cylinder group in which to place a directory.
 *
 * The policy implemented by this algorithm is to allocate a
 * directory inode in the same cylinder group as its parent
 * directory, but also to reserve space for its files inodes
 * and data. Restrict the number of directories which may be
 * allocated one after another in the same cylinder group
 * without intervening allocation of files.
 *
 * If we allocate a first level directory then force allocation
 * in another cylinder group.
 */
static ino_t
ffs_dirpref(pip)
{
     .....
}
```

### 10.7.3   Local Policy 2: `ffs_nodealloccg`

```
/*
 * Determine whether an inode can be allocated.
 *
 * Check to see if an inode is available, and if it is,
 * allocate it using the following policy:
 *   1) allocate the requested inode.
 *   2) allocate the next available inode after the requested
 *      inode in the specified cylinder group.
 */
static ufs_daddr_t
ffs_nodealloccg(ip, cg, ipref, mode)
{
     .....
}
```

## 10.8   Synchronous Operations

To ensure that the on-disk state of the filesystem can always be returned to a
consistent state, the system must do three operations synchronously.

- Write a newly allocated inode to disk before its name is entered into a directory
  containing the file indicated by the inode.

- Remove a directory name before the inode is deallocated.

- Write a deallocated inode to disk before its blocks are placed into the cylinder
  group free list.

# 10.9 Filesystem Semantics

## 10.9.1 Large File Sizes

4.4BSD FFS support 64-bit file size. However, the interface to the filesystem is still limited to 31-bit sizes !

This is not quite enough for modern enormous storage systems. Therefore if we want to extend it to 128-bit file size, we shold redefine **off_t** type and make many changes to FFS functions that implement system calls including, but not limited to,

- lseek

- stat, fstat, lstat

- truncate

- mmap

- getrlimit, setrlimit

# 10.10 References to Source Code

## 10.10.1 `fs.h` - 574 lines

**Type Definitions**

```
struct fs                  Superblock
struct csum                Cylinder group summary information
struct cg                  Cylinder hroup
struct ocg                 Old cylinder hroup
struct appleufslabel       Apple UFS superblock
```

**Macro Functions**

```
fs_postbl(fs, cylno)       Get the base address of positional layout table
fs_rotbl(fs)               Get blocks for each rotation

CGSIZE(fs)                 Get the size of a cylinder group

fs_cs(fs, indx)            Get the base address of cylinder group summary info

[ Macros for access to cylinder group without regard to superblock version ]

cg_blktot(cgp, ns)         Get the block totals per cylinder
cg_blks(fs, cgp, cylno, ns) Get the base address of free block positions
cg_inosused(cgp, ns)       Get the base address of used inode map
cg_blksfree(cgp, ns)       Get the base address of free block map
cg_chkmagic(cgp, ns)       Get the filesystem version magic number
cg_clustersfree(cgp, ns)   Get base address of free cluster map
cg_clustersum(cgp, ns)     Get the counts of available clusters

[ Converter between from filesystem block number and disk block address ]

fsbtodb(fs, b)             From the filesystem block to disk block
dbtofsb(fs, b)             Reverse
```

```
[ Macros to locate things in cylinder groups ]

cgbase(fs, c)                Locate the cylinder group base
cgdmin(fs, c)                Locate the data block
cgimin(fs, c)                Locate the inode block
cgsblock(fs, c)              Locate the superblock
cgtod(fs, c)                 Locate the cylinder group block itself
cgstart(fs, c)               Locate the cylinder start some after base

[ Convert inode number to some other things ]

ino_to_cg(fs, x)             Convert to cylinder group number
ino_to_fsba(fs, x)           Convert to ????????
ino_to_fsbo(fs, x)           Convert to ????????

[ Convert filesystem block number to some other things ]

dtog(fs, d)                  Convert to cylinder group number
dtogd(fs, d)                 Convert to cylinder group block number

[ ? ]

blkmap(fs, map, loc)         ?
cbtocylno(fs, bno)           ?
cbtorpos(fs, bno)            ?

[ Determine the number of available frags given a percent to hold in reserve ]

freespace(fs, percentreserv) ?

[ Determining the size of a file block in the file system ]

blksize(fs, ip, lbn)         ?
dblksize(fs, dip, lbn)       ?

[ Convert something into the number of sectors ]

NSPB(fs)                     Convert a block size
NSPF(fs)                     Convert a sector size

[ Convert something into the number of inodes ]

INOPB(fs)                    Convert a block size
INOPF(fs)                    Convert a fragment size

[ Misc. ]

blkoff(fs, loc)              /* calculates (loc % fs->fs_bsize) */ \
fragoff(fs, loc)             /* calculates (loc % fs->fs_fsize) */ \
lblktosize(fs, blk)          /* calculates ((off_t)blk * fs->fs_bsize) */ \
lblkno(fs, loc)              /* calculates (loc / fs->fs_bsize) */ \
numfrags(fs, loc)            /* calculates (loc / fs->fs_fsize) */ \
blkroundup(fs, size)         /* calculates roundup(size, fs->fs_bsize) */ \
```

```
fragroundup(fs, size)          /* calculates roundup(size, fs->fs_fsize) */ \
fragstoblks(fs, frags)         /* calculates (frags / fs->fs_frag) */ \
blkstofrags(fs, blks)          /* calculates (blks * fs->fs_frag) */ \
fragnum(fs, fsb)               /* calculates (fsb % fs->fs_frag) */ \
blknum(fs, fsb)                /* calculates rounddown(fsb, fs->fs_frag) */ \

NINDIR(fs)                     [?] Number of indirects in a file system block
```

## 10.10.2  `ffs_vfsops.c` - 1518 lines, 18 functions

**Gloval Variables**

```
ffs_vnodeopv_descs
ffs_vfsops
ffs_genfsops
ffs_inode_pool
```

**Functions**

```
ffs_mountroot
ffs_mount
ffs_reload
ffs_mountfs
ffs_oldfscompat
ffs_unmount
ffs_flushfiles
ffs_statfs
ffs_sync
ffs_vget            Read a FFS dinode using inode cache
ffs_fhtovp
ffs_vptofh
ffs_init
ffs_reinit
ffs_done
ffs_sysctl
ffs_sbupdate
ffs_cgupdate
```

## 10.10.3  `ffs_vnops.c` - 580 lines, 6 functions

**Gloval Variables**

```
ffs_vnodeop_opv_desc
ffs_specop_opv_desc
ffs_fifoop_opv_desc
```

**Functions**

```
ffs_fsync
ffs_full_fsync
ffs_reclaim
ffs_getpages
ffs_putpages
ffs_gop_size        Return the last logical file offset that should be written
```

### 10.10.4   `ffs_alloc.c` - 1895 lines, 18 functions

**Gloval Variables**

    none

**Functions**

    [ Block Allocation ]

    ffs_alloc          Provide a global policy for getting a new block
    ffs_realloccg      Provide a global policy for extending an fragment
    ffs_alloccg        Allocate a block or fragment with a local policy

    ffs_fragextend     Extend an fragment
    ffs_alloccgblk     Allocate a block in the specified cylinder group

    ffs_blkpref        Guess a proper location of a new block
    ffs_hashalloc      Implement  cylinder overflow algorithm

    ffs_reallocblks    Gather blocks together [Disabled Now]
    ffs_clusteralloc   Determine whether a cluster can be allocated [Disabled Now]

    [ Inode Allocation ]

    ffs_valloc         Provide a global policy for allocating a new inode
    ffs_dirpref        Guess a proper cylinder group for a new directory
    ffs_nodealloccg    Allocate a new inode

    [ Block & Inode De-allocation ]

    ffs_blkfree        Free a block or fragment
    ffs_vfree          Cover function for freeing an inode
    ffs_freefile       Do the actual free operation for an inode

    [ Misc. ]

    ffs_mapsearch      Find a block or fragments of the specified size from a cyl.
    ffs_clusteracct    Update the cluster map because of an allocation of free

    [ Debug ]

    ffs_fserr          Prints the name of a filesystem with an error diagnostic

### 10.10.5   `ffs_balloc.c` - 552 lines, 2 functions

**Gloval Variables**

    none

**Functions**

    ffs_balloc         Non-UBC: Allocate a requested block or fragments
    ffs_gop_alloc      UBC: Allocate blocks

### 10.10.6   `ffs_inode.c` - 582 lines, 3 functions

**Gloval Variables**

    none

**Functions**

| | |
|---|---|
| `ffs_update` | Update the access, modified, and inode change time |
| `ffs_truncate` | Force the length of the inode to a specified value |
| `ffs_indirtrunc` | Called by ffs_truncate to deal indirect pointer |

### 10.10.7   `ffs_subr.c` - 295 lines, 7 functions

**Gloval Variables**

    none

**Functions**

| | |
|---|---|
| `ffs_blkatoff` | Return buffer with the contents of block |
| `ffs_fragacct` | Update Fragment Summary |
| `ffs_isblock` | Check if a block is available |
| `ffs_isfreeblock` | Check if a block is free |
| `ffs_clrblock` | Take a block out of the map |
| `ffs_setblock` | Put a block into the map |
| `ffs_checkoverlap` | Diagnostic |

### 10.10.8   `ffs_tables.c` - 147 lines, 0 functions

**Gloval Variables**

| | |
|---|---|
| `fragtbl` | Fragment table for fast free fragment searching |

**Functions**

    none

### 10.10.9   `ffs_swap.c` - 158 lines, 3 functions

**Gloval Variables**

    none

**Functions**

| | |
|---|---|
| `ffs_sb_swap` | Adjust superblock for machine indenpendent meta-data |
| `ffs_dinode_swap` | Adjust inode for machine indenpendent meta-data |
| `ffs_csum_swap` | Adjust cyl group summary information for the same reason |

# Chapter 11

# Mounting Root File System

In this chapter, the procedure involved in mounting root filesystem is described.

## 11.1 System Bootstrapping

After bootstrap, the system initialization is started in `main` function of `kern/init_main.c` as follows.

    `main` function initialize the world, create process 0, mount root filesystem, and fork to create init and pagedaemon. Most of the hard work is done in the lower-level initialization routines including `start` function of `arch/sparc64/sparc64/locore.s`, which does memory initialization and autoconfiguration.

———————————————————————————————————————— kern/init_main.c

```
171 void
172 main(void)
173 {
...
188         /*
189          * Initialize the current process pointer (curproc) before
190          * any possible traps/probes to simplify trap processing.
191          */
...
219         /* Initialize kqueues. */
...
235         /* Initialize the sysctl subsystem. */
236         sysctl_init();
237
...
248         /*
249          * Create process 0 (the swapper).
250          */
...
344         /* Configure virtual memory system, set vm rlimits. */
345         uvm_init_limits(p);
346
347         /* Initialize the file systems. */
...
361         vfsinit();
...
```

——————————————————————————— kern/init_main.c

At **line 361**, virtual filesystem layer initialization is started by calling `vfsinit` function.

Note that before virtual filesystem layer is initialized, various system initializing procedure including creation of swapper and initialization of UVM[3] virtual memory system.

As we described at previous section, Virtual filesystem initialization is initiated in `main` function of `kern/init_main.c` which is practically the first function executed after machine bootstrap.

That function calls `vfsinit` function of `kern/vfs_init.c` which we studied in the previous chapter.

## 11.2   Before Mounting

Up to now, we followed execution of `main` function in `kern/init_main.c` just after kernel bootstrap. Now let keep going on analyzing `kern/init_main.c` just after we have traced.

We will show the whole source code that will be described in this section and then describe each part of it. To minimize the list, all optional parts (ifdef block) are removed.

——————————————————————————— kern/init_main.c

```
363     /* Configure the system hardware.  This will enable interrupts.  */
364     configure();
365
366     ubc_init();              /* must be after autoconfig */
367
368     /* Lock the kernel on behalf of proc0. */
369     KERNEL_PROC_LOCK(p);
...
386     /* Attach pseudo-devices. */
387     for (pdev = pdevinit; pdev->pdev_attach != NULL; pdev++)
388         (*pdev->pdev_attach)(pdev->pdev_count);
389
390     /*
391      * Initialize protocols.  Block reception of incoming packets
392      * until everything is ready.
393      */
394     s = splnet();
395     ifinit();
396     domaininit();
397     if_attachdomain();
398     splx(s);
...
405     /* Initialize system accouting. */
406     acct_init();
...
411     /*
412      * Initialize signal-related data structures, and signal state
413      * for proc0.
414      */
415     signal_init();
```

```
416        p->p_sigacts = &sigacts0;
417        siginit(p);
418
419        /* Kick off timeout driven events by calling first time. */
420        schedcpu(NULL);
421
422        /*
423         * Create process 1 (init(8)).  We do this now, as Unix has
424         * historically had init be process 1, and changing this would
425         * probably upset a lot of people.
426         *
427         * Note that process 1 won't immediately exec init(8), but will
428         * wait for us to inform it that the root file system has been
429         * mounted.
430         */
431        if (fork1(p, 0, SIGCHLD, NULL, 0, start_init, NULL, NULL, &initproc))
432            panic("fork init");
433
434        /*
435         * Create any kernel threads who's creation was deferred because
436         * initproc had not yet been created.
437         */
438        kthread_run_deferred_queue();
439
440        /*
441         * Now that device driver threads have been created, wait for
442         * them to finish any deferred autoconfiguration.  Note we don't
443         * need to lock this semaphore, since we haven't booted any
444         * secondary processors, yet.
445         */
446        while (config_pending)
447            (void) tsleep((void *)&config_pending, PWAIT, "cfpend", 0);
448
449        /*
450         * Finalize configuration now that all real devices have been
451         * found.  This needs to be done before the root device is
452         * selected, since finalization may create the root device.
453         */
454        config_finalize();
455
456        /*
457         * Now that autoconfiguration has completed, we can determine
458         * the root and dump devices.
459         */
460        cpu_rootconf();
461        cpu_dumpconf();
462
463        /* Mount the root file system. */
464        do {
465            domountroothook();
466            if ((error = vfs_mountroot())) {
467                printf("cannot mount root, error = %d\n", error);
468                boothowto |= RB_ASKNAME;
469                setroot(root_device,
```

```
470                        (rootdev != NODEV) ? DISKPART(rootdev) : 0);
471        }
472   } while (error != 0);
473   mountroothook_destroy();
474
475   CIRCLEQ_FIRST(&mountlist)->mnt_flag |= MNT_ROOTFS;
476   CIRCLEQ_FIRST(&mountlist)->mnt_op->vfs_refcount++;
477
478   /*
479    * Get the vnode for '/'.  Set filedesc0.fd_fd.fd_cdir to
480    * reference it.
481    */
482   if (VFS_ROOT(CIRCLEQ_FIRST(&mountlist), &rootvnode))
483           panic("cannot find root vnode");
484   cwdi0.cwdi_cdir = rootvnode;
485   VREF(cwdi0.cwdi_cdir);
486   VOP_UNLOCK(rootvnode, 0);
487   cwdi0.cwdi_rdir = NULL;
488
489   /*
490    * Now that root is mounted, we can fixup initproc's CWD
491    * info.  All other processes are kthreads, which merely
492    * share proc0's CWD info.
493    */
494   initproc->p_cwdi->cwdi_cdir = rootvnode;
495   VREF(initproc->p_cwdi->cwdi_cdir);
496   initproc->p_cwdi->cwdi_rdir = NULL;
497
498   /*
499    * Now can look at time, having had a chance to verify the time
500    * from the file system.  Reset p->p_rtime as it may have been
501    * munched in mi_switch() after the time got set.
502    */
503   proclist_lock_read();
504   s = splsched();
505   for (p = LIST_FIRST(&allproc); p != NULL;
506       p = LIST_NEXT(p, p_list)) {
507           p->p_stats->p_start = mono_time = boottime = time;
508           if (p->p_cpu != NULL)
509                   p->p_cpu->ci_schedstate.spc_runtime = time;
510           p->p_rtime.tv_sec = p->p_rtime.tv_usec = 0;
511   }
512   splx(s);
513   proclist_unlock_read();
514
515   /* Create the pageout daemon kernel thread. */
516   uvm_swap_init();
517   if (kthread_create1(uvm_pageout, NULL, NULL, "pagedaemon"))
518           panic("fork pagedaemon");
519
520   /* Create the process reaper kernel thread. */
521   if (kthread_create1(reaper, NULL, NULL, "reaper"))
522           panic("fork reaper");
523
```

```
524     /* Create the filesystem syncer kernel thread. */
525     if (kthread_create1(sched_sync, NULL, NULL, "ioflush"))
526             panic("fork syncer");
527
528     /* Create the aiodone daemon kernel thread. */
529     if (kthread_create1(uvm_aiodone_daemon, NULL, NULL, "aiodoned"))
530             panic("fork aiodoned");
531
...
536
537     /* Initialize exec structures */
538     exec_init(1);
539
...
544
545     /*
546      * Okay, now we can let init(8) exec!  It's off to userland!
547      */
548     start_init_exec = 1;
549     wakeup((void *)&start_init_exec);
550
551     /* The scheduler is an infinite loop. */
552     uvm_scheduler();
553     /* NOTREACHED */
554 }
```

——————————————————————————————————————— kern/init_main.c

## 11.2.1  Creating stopped `init` process

**line 366-421** is out of our concern, since it is not directly related with filesystem code. Where the `init` process is created is **line 422-432** kernel level `fork1` function creates a new process out of `start_init` function of `init_main.c`, which is assumed to be the current process.

——————————————————————————————————————— kern/init_main.c

```
586 static void
587 start_init(void *arg)
588 {
...
604     /*
605      * Now in process 1.
606      */
607     strncpy(p->p_comm, "init", MAXCOMLEN);
608
609     /*
610      * Wait for main() to tell us that it's safe to exec.
611      */
612     while (start_init_exec == 0)
613         (void) tsleep((void *)&start_init_exec, PWAIT, "initexec", 0);
```

——————————————————————————————————————— kern/init_main.c

After `main` function of `kern/init_main.c` mounts root filesystem, the temporally stopped execution of `start_init` function at **line 612-613** is continued.

## 11.2.2   Finding Where is the Root File System

**line 460-461** in main function of `kern/init_main.c` finds where the root filesystem are. It also finds where the dump device is.

`cpu_rootconf` function does the work and this function is machine dependent.

—————————————————————————————— arch/sparc64/sparc64/autoconf.c

```
491 void
492 cpu_rootconf()
493 {
494         struct bootpath *bp;
495         struct device *bootdv;
496         int bootpartition;
497
498         bp = nbootpath == 0 ? NULL : &bootpath[nbootpath-1];
499         bootdv = bp == NULL ? NULL : bp->dev;
500         bootpartition = bootdv == NULL ? 0 : bp->val[2];
501
502         setroot(bootdv, bootpartition);
503 }
```

—————————————————————————————— arch/sparc64/sparc64/autoconf.c

`setroot` function of `kern/kern_subr.c` set global variable `struct device *root_device` to proper value.

You may wonder how the root device specified in `bootpath` entry of Sparc PROM is delivered to `nbootpath` variable in **line 498** of `arch/sparc64/sparc64/autoconf.c` The answer is at `configure` function at the **line 364** of `kern/init_main.c`. This `configure` function is,

—————————————————————————————— kern/subr_autoconf.c

```
226 /*
227  * Configure the system's hardware.
228  */
229 void
230 configure(void)
231 {
232
233     /* Initialize data structures. */
234     config_init();
235
236 #ifdef USERCONF
237     if (boothowto & RB_USERCONF)
238         user_config();
239 #endif
240
241     /*
242      * Do the machine-dependent portion of autoconfiguration.  This
243      * sets the configuration machinery here in motion by "finding"
244      * the root bus.  When this function returns, we expect interrupts
245      * to be enabled.
246      */
247     cpu_configure();
248
```

```
249    /*
250     * Now that we've found all the hardware, start the real time
251     * and statistics clocks.
252     */
253    initclocks();
254
255    cold = 0;   /* clocks are running, we're warm now! */
256
257    /*
258     * Now callback to finish configuration for devices which want
259     * to do this once interrupts are enabled.
260     */
261    config_process_deferred(&interrupt_config_queue, NULL);
262 }
```

——————————————————————————————— kern/subr_autoconf.c

This `configure` function calls machine dependent `cpu_configure` function of
`arch/sparc64/sparc64/autoconf.c`,

—————————————————————————————- arch/sparc64/sparc64/autoconf.c

```
458 /*
459  * Determine mass storage and memory configuration for a machine.
460  * We get the PROM's root device and make sure we understand it, then
461  * attach it as 'mainbus0'.  We also set up to handle the PROM 'sync'
462  * command.
463  */
464 void
465 cpu_configure()
466 {
467
468         /* build the bootpath */
469         bootpath_build();
470
471 #if notyet
472         /* FIXME FIXME FIXME  This is probably *WRONG!!!**/
473         OF_set_callback(sync_crash);
474 #endif
475
476         /* block clock interrupts and anything below */
477         splclock();
478         /* Enable device interrupts */
479         setpstate(getpstate()|PSTATE_IE);
480
481         if (config_rootfound("mainbus", NULL) == NULL)
482                 panic("mainbus not configured");
483
484         /* Enable device interrupts */
485         setpstate(getpstate()|PSTATE_IE);
486
487         (void)spl0();
488 }
```

—————————————————————————————- arch/sparc64/sparc64/autoconf.c

This `cpu_configure` function calls again `bootpath_build` function and this function reads `bootpath` entry of SPARC PROM !

—————————————————————————————— arch/sparc64/sparc64/autoconf.c

```
274 static void
275 bootpath_build()
276 {
277     register char *cp, *pp;
278     register struct bootpath *bp;
279     register long chosen;
280     char buf[128];
281
282     bzero((void*)bootpath, sizeof(bootpath));
283     bp = bootpath;
284
285     /*
286      * Grab boot path from PROM
287      */
288     chosen = OF_finddevice("/chosen");
289     OF_getprop(chosen, "bootpath", buf, sizeof(buf));
290     cp = buf;
291     while (cp != NULL && *cp == '/') {
292         /* Step over '/' */
293         ++cp;
294         /* Extract name */
295         pp = bp->name;
296         while (*cp != '@' && *cp != '/' && *cp != '\0')
297             *pp++ = *cp++;
298         *pp = '\0';
299         if (*cp == '@') {
300             cp = str2hex(++cp, &bp->val[0]);
301             if (*cp == ',')
302                 cp = str2hex(++cp, &bp->val[1]);
303             if (*cp == ':')
304                 /* XXX - we handle just one char */
305                 bp->val[2] = *++cp - 'a', ++cp;
306         } else {
307             bp->val[0] = -1; /* no #'s: assume unit 0, no
308                             sbus offset/adddress */
309         }
310         ++bp;
311         ++nbootpath;
312     }
313     bp->name[0] = 0;
314
315     bootpath_print(bootpath);
316
317     /* Setup pointer to boot flags */
318     OF_getprop(chosen, "bootargs", buf, sizeof(buf));
319     cp = buf;
320
321     /* Find start of boot flags */
322     while (*cp) {
323         while(*cp == ' ' || *cp == '\t') cp++;
```

```
324          if (*cp == '-' || *cp == '\0')
325               break;
326          while(*cp != ' ' && *cp != '\t' && *cp != '\0') cp++;
327
328      }
329      if (*cp != '-')
330          return;
331
332      for (;*++cp;) {
333          int fl;
334
335          fl = 0;
336          BOOT_FLAG(*cp, fl);
337          if (!fl) {
338              printf("unknown option '%c'\n", *cp);
339              continue;
340          }
341          boothowto |= fl;
342
343          /* specialties */
344          if (*cp == 'd') {
345 #if defined(KGDB)
346              kgdb_debug_panic = 1;
347              kgdb_connect(1);
348 #elif defined(DDB)
349              Debugger();
350 #else
351              printf("kernel has no debugger\n");
352 #endif
353          } else if (*cp == 't') {
354              /* turn on traptrace w/o breaking into kdb */
355              extern int trap_trace_dis;
356
357              trap_trace_dis = 0;
358          }
359      }
```

———————————————————————————- arch/sparc64/sparc64/autoconf.c

This function reads bootpath, from SPARC PROM, such as

> bootpath: /sbus@1f,0/SUNW,fas@e,8800000/sd@0,0

and stores this machine dependent `struct bootpath` structure.

For the time being, it is sufficient for us only to know that after execution of `cpu_rootconf` function at **line 460** of `kern/init_main.c`, we obtains the location of the root device in global variable `struct device *root_device`. There is no need for us to analyze the very detail of this function, since we are only interested in filesystem related code ! Instead, we summarize how kernel founds root filesystem.

```
main() ........................... (kern/init_main.c)
|
+-> configure() ................... (kern/subr_autoconf.c)
     |
|    +-> cpu_configure() ........... (arch/sparc64/sparc64/autoconf.c)
```

```
|         |
|         +-> bootpath_build() ...... (arch/sparc64/sparc64/autoconf.c)
|
|                 o set 'bootpath' variable from SPARC PROM 'bootpath' entry
|                 o set 'boothowto' variable from SPARC PROM 'bootargs' entry
|
+-> cpu_rootconf() ................ (arch/sparc64/sparc64/autoconf.c)
    |
    |   o set 'bootdv' variable from 'bootpath' variable
    |   o set 'bootpartition' variable from 'bootpath' variable
    |
    +-> setroot() ............. (kern/kern_subr.c)

            o set 'rootdev', 'root_device', 'mountroot' variable
```

Pay attention the three variables that is set by `setroot` function.

1. **`mountroot`** pointer to function indicates the function to mount root filesystem. For example, if root filesystem is FFS, this function pointer is set to **`ffs_mountroot`** function of **`ufs/ffs/ffs_vfsops.c`**.

   **`mountroot`** variable is defined in automatically generated **`swapnetbsd.c`** by **`config`** program as,

   ――――――――――――――――――――

   arch/sparc64/compile/MY_KERNEL/swapnetbsd.c

   ```
    1 #include <sys/param.h>
    2 #include <sys/conf.h>
    3
    4 const char *rootspec = NULL;
    5 dev_t   rootdev = NODEV;        /* wildcarded */
    6
    7 const char *dumpspec = NULL;
    8 dev_t   dumpdev = NODEV;        /* unspecified */
    9
   10 int (*mountroot)(void) = NULL;
   ```

   ――――――――――――――――――――

   arch/sparc64/compile/MY_KERNEL/swapnetbsd.c

   where **`rootspec`**, **`dumpdev`** variable is set to other value, if you specified root device or dump device in kernel configuration file. If kernel configuration file set this variable to other values, it overrides the value of SPARC PROM's bootpath entry !

2. **`rootdev`** is major number of device driver controlling a storage device containing root file system. **`dev_t`** type is simply defined as **`u_int32_t`** in /usr/include/sys/types.h

3. **`root_device`** is defined as **`structure device`**. It contains device information corresponding to **`rootdev`** variable. **`structure device`** is defined in **`sys/device.h`** as,

   ――――――――――――――――――――――――――――――――― sys/device.h

   ```
   107 struct device {
   108         enum    devclass dv_class;      /* this device's classification *
   ```

```
109          TAILQ_ENTRY(device) dv_list;   /* entry on list of all devices */
110          struct  cfdata *dv_cfdata;     /* config data that found us
111                                            (NULL if pseudo-device) */
112          struct  cfdriver *dv_cfdriver; /* our cfdriver */
113          struct  cfattach *dv_cfattach; /* our cfattach */
114          int     dv_unit;               /* device unit number */
115          char    dv_xname[16];          /* external name (name + unit) */
116          struct  device *dv_parent;     /* pointer to parent device
117                                            (NULL if pesudo- or root node) */
118          int     dv_flags;              /* misc. flags; see below */
119 };
```

———————————————————————————————— sys/device.h

### 11.2.3 Executing Root Mount Hook

After somewhat long preparation after system bootstrap, the system is now ready
to mount the root filesystem. **line 463-476** of `kern/init_main.c` does the work.

*Root mount hook* is a list of functions that should be called before root filesystem
mount as a preparation process.

> **mountroothook_establish** registers a function that should be executed
> before root filesystem, we can register with **mountrootbook_establish**.

> **domountroothook** executes all registered *root mount hook*.

For reference, we listed the source code below.

———————————————————————————————— kern/kern_subr.c

```
504 /*
505  * "Mountroot hook" types, functions, and variables.
506  */
507
508 hook_list_t mountroothook_list;
509
510 void *
511 mountroothook_establish(fn, dev)
512          void (*fn) __P((struct device *));
513          struct device *dev;
514 {
515          return hook_establish(&mountroothook_list, (void (*)__P((void *)))fn,
516              dev);
517 }
518
519 void
520 mountroothook_disestablish(vhook)
521          void *vhook;
522 {
523          hook_disestablish(&mountroothook_list, vhook);
524 }
525
526 void
527 mountroothook_destroy()
528 {
529          hook_destroy(&mountroothook_list);
```

```
530 }
531
532 void
533 domountroothook()
534 {
535         struct hook_desc *hd;
536
537         LIST_FOREACH(hd, &mountroothook_list, hk_list) {
538                 if (hd->hk_arg == (void *)root_device) {
539                         (*hd->hk_fn)(hd->hk_arg);
540                         return;
541                 }
542         }
543 }
```

————————————————————————————————— kern/kern_subr.c where `hook_establish`, `hook_disestablish`, and `hook_destroy` functions are simply implemented in `kern/kern_subr.c` as,

————————————————————————————————— kern/kern_subr.c

```
383 static void *
384 hook_establish(list, fn, arg)
385         hook_list_t *list;
386         void (*fn) __P((void *));
387         void *arg;
388 {
389         struct hook_desc *hd;
390
391         hd = malloc(sizeof(*hd), M_DEVBUF, M_NOWAIT);
392         if (hd == NULL)
393                 return (NULL);
394
395         hd->hk_fn = fn;
396         hd->hk_arg = arg;
397         LIST_INSERT_HEAD(list, hd, hk_list);
398
399         return (hd);
400 }
401
402 static void
403 hook_disestablish(list, vhook)
404         hook_list_t *list;
405         void *vhook;
406 {
407 #ifdef DIAGNOSTIC
408         struct hook_desc *hd;
409
410         LIST_FOREACH(hd, list, hk_list) {
411                 if (hd == vhook)
412                         break;
413         }
414
415         if (hd == NULL)
```

```
416                 panic("hook_disestablish: hook %p not established", vhook);
417 #endif
418         LIST_REMOVE((struct hook_desc *)vhook, hk_list);
419         free(vhook, M_DEVBUF);
420 }
421
422 static void
423 hook_destroy(list)
424         hook_list_t *list;
425 {
426         struct hook_desc *hd;
427
428         while ((hd = LIST_FIRST(list)) != NULL) {
429                 LIST_REMOVE(hd, hk_list);
430                 free(hd, M_DEVBUF);
431         }
432 }
```

———————————————————————————————— kern/kern_subr.c

Those functions can be used by software RAID, since before mounting root filesystem, kernel should get information to mount root file system from the unconfigured RAID. That information from unconfigured RAID, can be obtained by mount root hook.

For more important, if you want to newly design any RAID-like filesystem, and want to mount it as a root filesystem, you may need to use *root mount hook*.

# 11.3   Let's Mount the Root File System !

## 11.3.1   Telling the VFS to Mount the Root Filesystem

In this section, we will show how VFS calls a FFS function for mounting root filesystem. For you reference, we again shows the code mounting root file system: the **line 464-476** of `kern/init_main.c`.

———————————————————————————————— kern/init_main.c

```
463         /* Mount the root file system. */
464         do {
465                 domountroothook();
466                 if ((error = vfs_mountroot())) {
467                         printf("cannot mount root, error = %d\n", error);
468                         boothowto |= RB_ASKNAME;
469                         setroot(root_device,
470                             (rootdev != NODEV) ? DISKPART(rootdev) : 0);
471                 }
472         } while (error != 0);
473         mountroothook_destroy();
```

———————————————————————————————— kern/init_main.c

`vfs_mountroot` function of `kern/vfs_subr.c` does the work and its source code is

———————————————————————————————— kern/vfs_subr.c

```
2549 /*
2550  * Mount the root file system.  If the operator didn't specify a
2551  * file system to use, try all possible file systems until one
2552  * succeeds.
2553  */
2554 int
2555 vfs_mountroot()
2556 {
2557         struct vfsops *v;
2558
2559         if (root_device == NULL)
2560                 panic("vfs_mountroot: root device unknown");
2561
2562         switch (root_device->dv_class) {
2563         case DV_IFNET:
2564                 if (rootdev != NODEV)
2565                         panic("vfs_mountroot: rootdev set for DV_IFNET "
2566                             "(0x%08x -> %d,%d)", rootdev,
2567                             major(rootdev), minor(rootdev));
2568                 break;
2569
2570         case DV_DISK:
2571                 if (rootdev == NODEV)
2572                         panic("vfs_mountroot: rootdev not set for DV_DISK");
2573                 break;
2574
2575         default:
2576                 printf("%s: inappropriate for root file system\n",
2577                     root_device->dv_xname);
2578                 return (ENODEV);
2579         }
2580
2581         /*
2582          * If user specified a file system, use it.
2583          */
2584         if (mountroot != NULL)
2585                 return ((*mountroot)());
2586
2587         /*
2588          * Try each file system currently configured into the kernel.
2589          */
2590         for (v = LIST_FIRST(&vfs_list); v != NULL; v = LIST_NEXT(v, vfs_list)) {
2591                 if (v->vfs_mountroot == NULL)
2592                         continue;
2593 #ifdef DEBUG
2594                 printf("mountroot: trying %s...\n", v->vfs_name);
2595 #endif
2596                 if ((*v->vfs_mountroot)() == 0) {
2597                         printf("root file system type: %s\n", v->vfs_name);
2598                         break;
2599                 }
2600         }
2601
2602         if (v == NULL) {
```

```
2603                printf("no file system for %s", root_device->dv_xname);
2604                if (root_device->dv_class == DV_DISK)
2605                        printf(" (dev 0x%x)", rootdev);
2606                printf("\n");
2607                return (EFTYPE);
2608        }
2609        return (0);
2610 }
```
───────────────────────────────────────────────────── kern/vfs_subr.c

Suppose that the SPARC PROM `bootpath` entry is set to indicate a partition
containing FFS filesystem.  Then `mountroot` variable in **line 2585** of `kern/vfs_subr.c`
is already set to `ffs_mountroot` function by `setroot` function.  We described how
`mountroot` variable is set, in the previous subsection.

Now we turn to `ffs_mountroot` function of `ufs/ffs/ffs_vfsops.c`.

───────────────────────────────────────────────────── ufs/ffs/ffs_vfsops.c

```
130 int
131 ffs_mountroot()
132 {
133        struct fs *fs;
134        struct mount *mp;
135        struct proc *p = curproc;        /* XXX */
136        struct ufsmount *ump;
137        int error;
138
139        if (root_device->dv_class != DV_DISK)
140                return (ENODEV);
141
142        /*
143         * Get vnodes for rootdev.
144         */
145        if (bdevvp(rootdev, &rootvp))
146                panic("ffs_mountroot: can't setup bdevvp's");
147
148        if ((error = vfs_rootmountalloc(MOUNT_FFS, "root_device", &mp))) {
149                vrele(rootvp);
150                return (error);
151        }
152        if ((error = ffs_mountfs(rootvp, mp, p)) != 0) {
153                mp->mnt_op->vfs_refcount--;
154                vfs_unbusy(mp);
155                free(mp, M_MOUNT);
156                vrele(rootvp);
157                return (error);
158        }
159        simple_lock(&mountlist_slock);
160        CIRCLEQ_INSERT_TAIL(&mountlist, mp, mnt_list);
161        simple_unlock(&mountlist_slock);
162        ump = VFSTOUFS(mp);
163        fs = ump->um_fs;
164        memset(fs->fs_fsmnt, 0, sizeof(fs->fs_fsmnt));
165        (void)copystr(mp->mnt_stat.f_mntonname, fs->fs_fsmnt, MNAMELEN - 1, 0);
```

```
166          (void)ffs_statfs(mp, &mp->mnt_stat, p);
167          vfs_unbusy(mp);
168          inittodr(fs->fs_time);
169          return (0);
170 }
```

————————————————————————————————————— ufs/ffs/ffs_vfsops.c

### 11.3.2   Getting Vnode for Root Device

### 11.3.3   Allocating `Mount` Structure

```
126 struct mount {
127          CIRCLEQ_ENTRY(mount) mnt_list;        /* mount list */
128          struct vfsops   *mnt_op;              /* operations on fs */
129          struct vnode    *mnt_vnodecovered;    /* vnode we mounted on */
130          struct vnode    *mnt_syncer;          /* syncer vnode */
131          struct vnodelst mnt_vnodelist;        /* list of vnodes this mount */
132          struct lock     mnt_lock;             /* mount structure lock */
133          int             mnt_flag;             /* flags */
134          int             mnt_maxsymlinklen;    /* max size of short symlink */
135          int             mnt_fs_bshift;        /* offset shift for lblkno */
136          int             mnt_dev_bshift;       /* shift for device sectors */
137          struct statfs   mnt_stat;             /* cache of filesystem stats */
138          void            *mnt_data;            /* private data */
139          int             mnt_wcnt;             /* count of vfs_busy waiters */
140          struct proc     *mnt_unmounter;       /* who is unmounting */
141 };
```

### 11.3.4   Reading Superblock

```
171 /*
172  * Super block for an FFS file system in memory.
173  */
174 struct fs {
175          int32_t  fs_firstfield;        /* historic file system linked list, */
176          int32_t  fs_unused_1;          /*     used for incore super blocks */
177          ufs_daddr_t fs_sblkno;         /* addr of super-block in filesys */
178          ufs_daddr_t fs_cblkno;         /* offset of cyl-block in filesys */
179          ufs_daddr_t fs_iblkno;         /* offset of inode-blocks in filesys */
180          ufs_daddr_t fs_dblkno;         /* offset of first data after cg */
181          int32_t  fs_cgoffset;          /* cylinder group offset in cylinder */
182          int32_t  fs_cgmask;            /* used to calc mod fs_ntrak */
183          int32_t  fs_time;              /* last time written */
184          int32_t  fs_size;              /* number of blocks in fs */
185          int32_t  fs_dsize;             /* number of data blocks in fs */
186          int32_t  fs_ncg;               /* number of cylinder groups */
187          int32_t  fs_bsize;             /* size of basic blocks in fs */
188          int32_t  fs_fsize;             /* size of frag blocks in fs */
189          int32_t  fs_frag;              /* number of frags in a block in fs */
190 /* these are configuration parameters */
```

```
191        int32_t  fs_minfree;        /* minimum percentage of free blocks */
192        int32_t  fs_rotdelay;       /* num of ms for optimal next block */
193        int32_t  fs_rps;            /* disk revolutions per second */
194 /* these fields can be computed from the others */
195        int32_t  fs_bmask;          /* ``blkoff'' calc of blk offsets */
196        int32_t  fs_fmask;          /* ``fragoff'' calc of frag offsets */
197        int32_t  fs_bshift;         /* ``lblkno'' calc of logical blkno */
198        int32_t  fs_fshift;         /* ``numfrags'' calc number of frags */
199 /* these are configuration parameters */
200        int32_t  fs_maxcontig;      /* max number of contiguous blks */
201        int32_t  fs_maxbpg;         /* max number of blks per cyl group */
202 /* these fields can be computed from the others */
203        int32_t  fs_fragshift;      /* block to frag shift */
204        int32_t  fs_fsbtodb;        /* fsbtodb and dbtofsb shift constant */
205        int32_t  fs_sbsize;         /* actual size of super block */
206        int32_t  fs_csmask;         /* csum block offset (now unused) */
207        int32_t  fs_csshift;        /* csum block number (now unused) */
208        int32_t  fs_nindir;         /* value of NINDIR */
209        int32_t  fs_inopb;          /* value of INOPB */
210        int32_t  fs_nspf;           /* value of NSPF */
211 /* yet another configuration parameter */
212        int32_t  fs_optim;          /* optimization preference, see below */
213 /* these fields are derived from the hardware */
214        int32_t  fs_npsect;         /* # sectors/track including spares */
215        int32_t  fs_interleave;     /* hardware sector interleave */
216        int32_t  fs_trackskew;      /* sector 0 skew, per track */
217 /* fs_id takes the space of the unused fs_headswitch and fs_trkseek fields */
218        int32_t  fs_id[2];          /* unique file system id */
219 /* sizes determined by number of cylinder groups and their sizes */
220        ufs_daddr_t fs_csaddr;      /* blk addr of cyl grp summary area */
221        int32_t  fs_cssize;         /* size of cyl grp summary area */
222        int32_t  fs_cgsize;         /* cylinder group size */
223 /* these fields are derived from the hardware */
224        int32_t  fs_ntrak;          /* tracks per cylinder */
225        int32_t  fs_nsect;          /* sectors per track */
226        int32_t  fs_spc;            /* sectors per cylinder */
227 /* this comes from the disk driver partitioning */
228        int32_t  fs_ncyl;           /* cylinders in file system */
229 /* these fields can be computed from the others */
230        int32_t  fs_cpg;            /* cylinders per group */
231        int32_t  fs_ipg;            /* inodes per group */
232        int32_t  fs_fpg;            /* blocks per group * fs_frag */
233 /* this data must be re-computed after crashes */
234        struct  csum fs_cstotal;    /* cylinder summary information */
235 /* these fields are cleared at mount time */
236        int8_t   fs_fmod;           /* super block modified flag */
237        int8_t   fs_clean;          /* file system is clean flag */
238        int8_t   fs_ronly;          /* mounted read-only flag */
239        int8_t   fs_flags;          /* see FS_ flags below */
240        u_char   fs_fsmnt[MAXMNTLEN]; /* name mounted on */
241 /* these fields retain the current block allocation info */
242        int32_t  fs_cgrotor;        /* last cg searched (UNUSED) */
243        void    *fs_ocsp[NOCSPTRS]; /* padding; was list of fs_cs buffers */
244        u_int16_t *fs_contigdirs;   /* # of contiguously allocated dirs */
```

```
245        struct csum *fs_csp;          /* cg summary info buffer for fs_cs */
246        int32_t *fs_maxcluster;       /* max cluster in each cyl group */
247        int32_t  fs_cpc;              /* cyl per cycle in postbl */
248        int16_t  fs_opostbl[16][8];   /* old rotation block list head */
249        int32_t  fs_snapinum[20];     /* RESERVED for snapshot inode nums */
250        int32_t  fs_avgfilesize;      /* expected average file size */
251        int32_t  fs_avgfpdir;         /* expected # of files per directory */
252        int32_t  fs_sparecon[26];     /* RESERVED for future constants */
253        int32_t  fs_pendingblocks;    /* blocks in process of being freed */
254        int32_t  fs_pendinginodes;    /* inodes in process of being freed */
255        int32_t  fs_contigsumsize;    /* size of cluster summary array */
256        int32_t  fs_maxsymlinklen;    /* max length of an internal symlink */
257        int32_t  fs_inodefmt;         /* format of on-disk inodes */
258        u_int64_t fs_maxfilesize;     /* maximum representable file size */
259        int64_t  fs_qbmask;           /* ~fs_bmask for use with 64-bit size */
260        int64_t  fs_qfmask;           /* ~fs_fmask for use with 64-bit size */
261        int32_t  fs_state;            /* validate fs_clean field (UNUSED) */
262        int32_t  fs_postblformat;     /* format of positional layout tables */
263        int32_t  fs_nrpos;            /* number of rotational positions */
264        int32_t  fs_postbloff;        /* (u_int16) rotation block list head */
265        int32_t  fs_rotbloff;         /* (u_int8) blocks for each rotation */
266        int32_t  fs_magic;            /* magic number */
267        u_int8_t fs_space[1];         /* list of blocks for each rotation */
268 /* actually longer */
--- ufs/ffs/fs.h
```

```
71 /* This structure describes the UFS specific mount structure data. */
72 struct ufsmount {
73        struct  mount *um_mountp;          /* filesystem vfs structure */
74        dev_t   um_dev;                    /* device mounted */
75        struct  vnode *um_devvp;           /* block device mounted vnode */
76        u_int32_t um_flags;                /* UFS-specific flags - see below
77        union {                            /* pointer to superblock */
78                struct  fs *fs;            /* FFS */
79                struct  lfs *lfs;          /* LFS */
80                struct  m_ext2fs *e2fs;    /* EXT2FS */
81        } ufsmount_u;
82 #define um_fs    ufsmount_u.fs
83 #define um_lfs   ufsmount_u.lfs
84 #define um_e2fs  ufsmount_u.e2fs
85 #define um_e2fsb ufsmount_u.e2fs->s_es
86
87        struct  vnode *um_quotas[MAXQUOTAS];   /* pointer to quota files */
88        struct  ucred *um_cred[MAXQUOTAS];     /* quota file access cred */
89        u_long  um_nindir;                     /* indirect ptrs per block */
90        u_long  um_lognindir;                  /* log2 of um_nindir */
91        u_long  um_bptrtodb;                   /* indir ptr to disk block */
92        u_long  um_seqinc;                     /* inc between seq blocks */
93        time_t  um_btime[MAXQUOTAS];           /* block quota time limit */
94        time_t  um_itime[MAXQUOTAS];           /* inode quota time limit */
95        char    um_qflags[MAXQUOTAS];          /* quota specific flags */
96        struct  netexport um_export;           /* export information */
97        u_int64_t um_savedmaxfilesize;         /* XXX - limit maxfilesize */
```

```
    98 };
--- ufs/ufsmount.h
```

### 11.3.5   Mount !

## 11.4   What Must Be Done after Mount ?

### 11.4.1   Find vnode for '/' — root directory

### 11.4.2   Set current working directory of `init` process

### 11.4.3   Check File System Time

### 11.4.4   Create Kernel Threads about File System

### 11.4.5   Start Up `init` processor

# Part III

# Storage Systems

# Chapter 12

# Storage Device

In this chapter, we describes how we can manages to storage devices such as SCSI hard disk drives.

## 12.1 Generic Disk Framework

The NetBSD generic disk framework is designed to provide flexible, scalable, and consistent handling of disk state and metrics information.

### 12.1.1 `disk` Structure

The fundamental component of this framework is the `disk` structure, which is defined in as follows:

———————————————————————————————— sys/disk.h

```
100 struct disk {
101         TAILQ_ENTRY(disk) dk_link;      /* link in global disklist */
102         char            *dk_name;       /* disk name */
103         int             dk_bopenmask;   /* block devices open */
104         int             dk_copenmask;   /* character devices open */
105         int             dk_openmask;    /* composite (bopen|copen) */
106         int             dk_state;       /* label state   ### */
107         int             dk_blkshift;    /* shift to convert DEV_BSIZE to blks */
108         int             dk_byteshift;   /* shift to convert bytes to blks */
109
110         /*
111          * Metrics data; note that some metrics may have no meaning
112          * on certain types of disks.
113          */
114         int             dk_busy;        /* busy counter */
115         u_int64_t       dk_rxfer;       /* total number of read transfers */
116         u_int64_t       dk_wxfer;       /* total number of write transfers */
117         u_int64_t       dk_seek;        /* total independent seek operations */
118         u_int64_t       dk_rbytes;      /* total bytes read */
119         u_int64_t       dk_wbytes;      /* total bytes written */
120         struct timeval  dk_attachtime;  /* time disk was attached */
121         struct timeval  dk_timestamp;   /* timestamp of last unbusy */
122         struct timeval  dk_time;        /* total time spent busy */
123
```

```
124        struct  dkdriver *dk_driver;    /* pointer to driver */
125
126        /*
127         * Disk label information.  Storage for the in-core disk label
128         * must be dynamically allocated, otherwise the size of this
129         * structure becomes machine-dependent.
130         */
131        daddr_t          dk_labelsector;        /* sector containing label */
132        struct disklabel *dk_label;     /* label */
133        struct cpu_disklabel *dk_cpulabel;
134 };
```

—————————————————————————————————— sys/disk.h

The system maintains a global linked-list of all disks attached to the system. This list, called `disklist`, may grow or shrink over time as disks are dynamically added and removed from the system. Drivers which currently make use of the detachment capability of the framework are the *ccd* and *vnd* pseudo-device drivers.

## 12.1.2   Disk Interfaces

disk_init()        Initialize the disklist and other data structures used
                   by the framework.  Called by main() before autoconfigu-
                   ration.

disk_attach()      Attach a disk; allocate storage for the disklabel, set
                   the ''attached time'' timestamp, insert the disk into
                   the disklist, and increment the system disk count.

disk_detach()      Detach a disk; free storage for the disklabel, remove
                   the disk from the disklist, and decrement the system
                   disk count.  If the count drops below zero, panic.

disk_busy()        Increment the disk's ''busy counter''.  If this counter
                   goes from 0 to 1, set the timestamp corresponding to
                   this transfer.

disk_unbusy()      Decrement a disk's busy counter.  If the count drops
                   below zero, panic.  Get the current time, subtract it
                   from the disk's timestamp, and add the difference to
                   the disk's running total.  Set the disk's timestamp to
                   the current time.  If the provided byte count is
                   greater than 0, add it to the disk's running total and
                   increment the number of transfers performed by the
                   disk.

disk_resetstat()   Reset the running byte, transfer, and time totals.

disk_find()        Return a pointer to the disk structure corresponding to
                   the name provided, or NULL if the disk does not exist.

**disk_attach function**

—————————————————————————————————— kern/disk_subr.c

```
206 /*
207  * Attach a disk.
208  */
209 void
210 disk_attach(struct disk *diskp)
211 {
212     int s;
213
214     /*
215      * Allocate and initialize the disklabel structures.  Note that
216      * it's not safe to sleep here, since we're probably going to be
217      * called during autoconfiguration.
218      */
219     diskp->dk_label = malloc(sizeof(struct disklabel), M_DEVBUF, M_NOWAIT);
220     diskp->dk_cpulabel = malloc(sizeof(struct cpu_disklabel), M_DEVBUF,
221         M_NOWAIT);
222     if ((diskp->dk_label == NULL) || (diskp->dk_cpulabel == NULL))
223         panic("disk_attach: can't allocate storage for disklabel");
224
225     memset(diskp->dk_label, 0, sizeof(struct disklabel));
226     memset(diskp->dk_cpulabel, 0, sizeof(struct cpu_disklabel));
227
228     /*
229      * Set the attached timestamp.
230      */
231     s = splclock();
232     diskp->dk_attachtime = mono_time;
233     splx(s);
234
235     /*
236      * Link into the disklist.
237      */
238     simple_lock(&disklist_slock);
239     TAILQ_INSERT_TAIL(&disklist, diskp, dk_link);
240     simple_unlock(&disklist_slock);
241     ++disk_count;
242 }
```

──────────────────────────────────── kern/disk_subr.c

**disk_busy function**

───────────────────────────────────────

```
267 /*
268  * Increment a disk's busy counter.  If the counter is going from
269  * 0 to 1, set the timestamp.
270  */
271 void
272 disk_busy(struct disk *diskp)
273 {
274     int s;
275
276     /*
277      * XXX We'd like to use something as accurate as microtime(),
```

```
278        * but that doesn't depend on the system TOD clock.
279        */
280       if (diskp->dk_busy++ == 0) {
281            s = splclock();
282            diskp->dk_timestamp = mono_time;
283            splx(s);
284       }
285 }
```

---

**disk_unbusy function**

---

```
287 /*
288  * Decrement a disk's busy counter, increment the byte count, total busy
289  * time, and reset the timestamp.
290  */
291 void
292 disk_unbusy(struct disk *diskp, long bcount, int read)
293 {
294      int s;
295      struct timeval dv_time, diff_time;
296
297      if (diskp->dk_busy-- == 0) {
298           printf("%s: dk_busy < 0\n", diskp->dk_name);
299           panic("disk_unbusy");
300      }
301
302      s = splclock();
303      dv_time = mono_time;
304      splx(s);
305
306      timersub(&dv_time, &diskp->dk_timestamp, &diff_time);
307      timeradd(&diskp->dk_time, &diff_time, &diskp->dk_time);
308
309      diskp->dk_timestamp = dv_time;
310      if (bcount > 0) {
311           if (read) {
312                diskp->dk_rbytes += bcount;
313                diskp->dk_rxfer++;
314           } else {
315                diskp->dk_wbytes += bcount;
316                diskp->dk_wxfer++;
317           }
318      }
319 }
```

---

**disk_unbusy function**

---

```
244 /*
245  * Detach a disk.
246  */
247 void
248 disk_detach(struct disk *diskp)
249 {
250
251          /*
252           * Remove from the disklist.
253           */
254          if (--disk_count < 0)
255                  panic("disk_detach: disk_count < 0");
256          simple_lock(&disklist_slock);
257          TAILQ_REMOVE(&disklist, diskp, dk_link);
258          simple_unlock(&disklist_slock);
259
260          /*
261           * Free the space used by the disklabel structures.
262           */
263          free(diskp->dk_label, M_DEVBUF);
264          free(diskp->dk_cpulabel, M_DEVBUF);
265 }
```

### 12.1.3   Using the Framework

This section includes a description on basic use of the framework and example usage of its functions. Actual implementation of a device driver which utilizes the framework may vary.

A special routine, `disk_init`, is provided to perform basic initialization of data structures used by the framework. It is called exactly once by the system, in `main` function, before device autoconfiguration.

**Attaching**

Each device in the system uses a "softc" structure which contains autoconfiguration and state information for that device. In the case of disks, the softc should also contain one instance of the disk structure, e.g.:

```
struct foo_softc {
        struct  device sc_dev;          /* generic device information */
        struct  disk sc_dk;             /* generic disk information */
        [ . . . more . . . ]
};
```

In order for the system to gather metrics data about a disk, the disk must be registered with the system. The `disk_attach` routine performs all of the functions currently required to register a disk with the system including allocation of disklabel storage space, recording of the time since boot that the disk was attached, and insertion into the disklist. Note that since this function allocates storage space for the disklabel, it must be called before the disklabel is read from the media or used in any other way. Before `disk_attach` is called, a portions of the disk structure must be initialized with data specific to that disk. For example, in the "foo" disk driver, the following would be performed in the autoconfiguration "attach" routine:

```
void
fooattach(parent, self, aux)
        struct device *parent, *self;
        void *aux;
{
        struct foo_softc *sc = (struct foo_softc *)self;
        [ . . . ]

        /* Initialize and attach the disk structure. */
        sc->sc_dk.dk_driver = &foodkdriver;
        sc->sc_dk.dk_name = sc->sc_dev.dv_xname;
        disk_attach(&sc->sc_dk);

        /* Read geometry and fill in pertinent parts of disklabel. */
        [ . . . ]
}
```

The foodkdriver above is the disk's "driver" switch. This switch currently includes
a pointer to the disk's "strategy" routine. This switch needs to have global scope
and should be initialized as follows:

```
void    foostrategy(struct buf *);
struct  dkdriver foodkdriver = { foostrategy };
```

**Gathering Metrics during Disk Operations**

Once the disk is attached, metrics may be gathered on that disk. In order to
gather metrics data, the driver must tell the framework when the disk starts and
stops operations. This functionality is provided by the disk_busy and disk_unbusy
routines. The disk_busy routine should be called immediately before a command
to the disk is sent, e.g.:

```
void
foostart(sc)
        struct foo_softc *sc;
{
        [ . . . ]

        /* Get buffer from drive's transfer queue. */
        [ . . . ]

        /* Build command to send to drive. */
        [ . . . ]

        /* Tell the disk framework we're going busy. */
        disk_busy(&sc->sc_dk);

        /* Send command to the drive. */
        [ . . . ]
}
```

When disk_busy is called, a timestamp is taken if the disk's busy counter moves
from 0 to 1, indicating the disk has gone from an idle to non-idle state. Note that
disk_busy must be called at splbio(). At the end of a transaction, the disk_unbusy
routine should be called. This routine performs some consistency checks, such as
ensuring that the calls to disk_busy and disk_unbusy are balanced. This routine also performs the actual metrics calculation. A timestamp is taken, and the difference from the timestamp taken in disk_busy is added to the disk's total running time. The disk's timestamp is then updated in case there is more than one pending transfer on the disk. A byte count is also added to the disk's running total, and if greater than zero, the number of transfers the disk has performed is incremented.

```
void
foodone(xfer)
        struct foo_xfer *xfer;
{
        struct foo_softc = (struct foo_softc *)xfer->xf_softc;
        struct buf *bp = xfer->xf_buf;
        long nbytes;
        [ . . . ]

        /*
         * Get number of bytes transfered.  If there is no buf
         * associated with the xfer, we are being called at the
         * end of a non-I/O command.
         */
        if (bp == NULL)
                nbytes = 0;
        else
                nbytes = bp->b_bcount - bp->b_resid;

        [ . . . ]

        /* Notify the disk framework that we've completed the transfer. */
        disk_unbusy(&sc->sc_dk, nbytes);

        [ . . . ]
}
```

Like disk_busy, disk_unbusy must be called at splbio().

At some point a driver may wish to reset the metrics data gathered on a particular disk. For this function, the disk_resetstat routine is provided.

## 12.2  Disk Label

### 12.2.1  What does it have ?

Each disk or disk pack on a system may contain a disk label which provides detailed information about

- the geometry of the disk and

- the partitions into which the disk is divided.

It should be initialized when the disk is formatted, and may be changed later with the disklabel(8) pro- gram.

This information is used by

- the system disk driver and

- by the bootstrap program to determine how to program the drive and where to find the filesystems on the disk partitions.

- Additional information is used by the filesystem in order to use the disk most efficiently and to locate important filesystem information.

The description of each partition contains an identifier for the partition type (standard filesystem, swap area, etc.). The filesystem updates the in-core copy of the label if it contains incomplete information about the filesystem.

## 12.2.2  disklabel structure

————————————————————————————————————— sys/disklabel.h

```
 98 struct disklabel {
 99        u_int32_t d_magic;              /* the magic number */
100        u_int16_t d_type;               /* drive type */
101        u_int16_t d_subtype;            /* controller/d_type specific */
102        char      d_typename[16];       /* type name, e.g. "eagle" */
103
104        /*
105         * d_packname contains the pack identifier and is returned when
106         * the disklabel is read off the disk or in-core copy.
107         * d_boot0 and d_boot1 are the (optional) names of the
108         * primary (block 0) and secondary (block 1-15) bootstraps
109         * as found in /usr/mdec.  These are returned when using
110         * getdiskbyname(3) to retrieve the values from /etc/disktab.
111         */
112        union {
113                char    un_d_packname[16];    /* pack identifier */
114                struct {
115                        char *un_d_boot0;     /* primary bootstrap name */
116                        char *un_d_boot1;     /* secondary bootstrap name */
117                } un_b;
118        } d_un;
119 #define d_packname      d_un.un_d_packname
120 #define d_boot0         d_un.un_b.un_d_boot0
121 #define d_boot1         d_un.un_b.un_d_boot1
122
123                        /* disk geometry: */
124        u_int32_t d_secsize;            /* # of bytes per sector */
125        u_int32_t d_nsectors;           /* # of data sectors per track */
126        u_int32_t d_ntracks;            /* # of tracks per cylinder */
127        u_int32_t d_ncylinders;         /* # of data cylinders per unit */
128        u_int32_t d_secpercyl;          /* # of data sectors per cylinder */
129        u_int32_t d_secperunit;         /* # of data sectors per unit */
130
131        /*
132         * Spares (bad sector replacements) below are not counted in
133         * d_nsectors or d_secpercyl.  Spare sectors are assumed to
134         * be physical sectors which occupy space at the end of each
135         * track and/or cylinder.
136         */
137        u_int16_t d_sparespertrack;     /* # of spare sectors per track */
138        u_int16_t d_sparespercyl;       /* # of spare sectors per cylinder */
139        /*
140         * Alternative cylinders include maintenance, replacement,
141         * configuration description areas, etc.
```

```
142           */
143           u_int32_t d_acylinders;          /* # of alt. cylinders per unit */
144
145                              /* hardware characteristics: */
146           /*
147            * d_interleave, d_trackskew and d_cylskew describe perturbations
148            * in the media format used to compensate for a slow controller.
149            * Interleave is physical sector interleave, set up by the
150            * formatter or controller when formatting.  When interleaving is
151            * in use, logically adjacent sectors are not physically
152            * contiguous, but instead are separated by some number of
153            * sectors.  It is specified as the ratio of physical sectors
154            * traversed per logical sector.  Thus an interleave of 1:1
155            * implies contiguous layout, while 2:1 implies that logical
156            * sector 0 is separated by one sector from logical sector 1.
157            * d_trackskew is the offset of sector 0 on track N relative to
158            * sector 0 on track N-1 on the same cylinder.  Finally, d_cylskew
159            * is the offset of sector 0 on cylinder N relative to sector 0
160            * on cylinder N-1.
161            */
162           u_int16_t d_rpm;                 /* rotational speed */
163           u_int16_t d_interleave;          /* hardware sector interleave */
164           u_int16_t d_trackskew;           /* sector 0 skew, per track */
165           u_int16_t d_cylskew;             /* sector 0 skew, per cylinder */
166           u_int32_t d_headswitch;          /* head switch time, usec */
167           u_int32_t d_trkseek;             /* track-to-track seek, usec */
168           u_int32_t d_flags;               /* generic flags */
169 #define NDDATA 5
170           u_int32_t d_drivedata[NDDATA];   /* drive-type specific information */
171 #define NSPARE 5
172           u_int32_t d_spare[NSPARE];       /* reserved for future use */
173           u_int32_t d_magic2;              /* the magic number (again) */
174           u_int16_t d_checksum;            /* xor of data incl. partitions */
175
176                              /* filesystem and partition information: */
177           u_int16_t d_npartitions;         /* number of partitions in following */
178           u_int32_t d_bbsize;              /* size of boot area at sn0, bytes */
179           u_int32_t d_sbsize;              /* max size of fs superblock, bytes */
180           struct  partition {              /* the partition table */
181                   u_int32_t p_size;        /* number of sectors in partition */
182                   u_int32_t p_offset;      /* starting sector */
183                   union {
184                           u_int32_t fsize; /* FFS, ADOS:
185                                               filesystem basic fragment size */
186                           u_int32_t cdsession; /* ISO9660: session offset */
187                   } __partition_u2;
188 #define p_fsize          __partition_u2.fsize
189 #define p_cdsession      __partition_u2.cdsession
190                   u_int8_t p_fstype;       /* filesystem type, see below */
191                   u_int8_t p_frag;         /* filesystem fragments per block */
192                   union {
193                           u_int16_t cpg;   /* UFS: FS cylinders per group */
194                           u_int16_t sgs;   /* LFS: FS segment shift */
195                   } __partition_u1;
```

```
196 #define p_cpg    __partition_u1.cpg
197 #define p_sgs    __partition_u1.sgs
198          } d_partitions[MAXPARTITIONS];  /* actually may be more */
199 };
```
――――――――――――――――――――――――――― sys/disklabel.h

and machine dependent definition in sparc64 architecture is,

## 12.2.3   Where is the Disk Label ?

The label is located in sector number LABELSECTOR of the drive, usually sector 0 where it may be found without any information about the disk ge- ometry. It is at an offset LABELOFFSET from the beginning of the sector, to allow room for the initial bootstrap. The disk sector containing the label is normally made read-only so that it is not accidentally overwrit- ten by pack-to-pack copies or swap operations; the DIOCWLABEL ioctl(2), which is done as needed by the disklabel(8) program.

LABELSECTOR and LABELOFFSET macros are machine-dependent parameter and defined in `arch/sparc64/include/disklabel.h` as

――――――――――――――――――――――― - arch/sparc64/include/disklabel.h

```
1 /*      $NetBSD: disklabel.h,v 1.2 2002/07/20 11:52:21 mrg Exp $      */
2
3 #include <sparc/disklabel.h>
```

――――――――――――――――――――――― - arch/sparc64/include/disklabel.h

where `arch/sparc/include/disklabel.h` is

――――――――――――――――――――――― arch/sparc/include/disklabel.h

```
36 #define LABELSECTOR     0                    /* sector containing label */
37 #define LABELOFFSET     128                  /* offset of label in sector */
38 #define MAXPARTITIONS   8                    /* number of partitions */
39 #define RAW_PART        2                    /* raw partition: xx?c */
40
41 struct cpu_disklabel {
42          char    cd_block[512];
43 };
```

――――――――――――――――――――――― arch/sparc/include/disklabel.h

## 12.2.4   General Disk Label Interfaces

```
                  /* get and set disklabel; DIOCGPART used internally */

DIOCGDINFO      _IOR('d', 101, struct disklabel) /* get */
DIOCSDINFO      _IOW('d', 102, struct disklabel) /* set */
DIOCWDINFO      _IOW('d', 103, struct disklabel) /* set, update disk */
DIOCGPART       _IOW('d', 104, struct partinfo)  /* get partition */

                  /* do format operation, read or write */

DIOCRFORMAT     _IOWR('d', 105, struct format_op)
DIOCWFORMAT     _IOWR('d', 106, struct format_op)
```

```
DIOCSSTEP       _IOW('d', 107, int)              /* set step rate */
DIOCSRETRIES    _IOW('d', 108, int)              /* set # of retries */
DIOCKLABEL      _IOW('d', 119, int)              /* keep/drop label on close? */
DIOCWLABEL      _IOW('d', 109, int)              /* write en/disable label */

DIOCSBAD        _IOW('d', 110, struct dkbad)     /* set kernel dkbad */
DIOCEJECT       _IOW('d', 112, int)              /* eject removable disk */
ODIOCEJECT      _IO('d', 112)                     /* eject removable disk */
DIOCLOCK        _IOW('d', 113, int)              /* lock/unlock pack */

                /* get default label, clear label */

DIOCGDEFLABEL   _IOR('d', 114, struct disklabel)
DIOCCLRLABEL    _IO('d', 115)

                /* disk cache enable/disable */

DIOCGCACHE      _IOR('d', 116, int)              /* get cache enables */
DIOCSCACHE      _IOW('d', 117, int)              /* set cache enables */

DKCACHE_READ    0x000001                          /* read cache enabled */
DKCACHE_WRITE   0x000002                          /* write(back) cache enabled */
DKCACHE_RCHANGE 0x000100                          /* read enable is changeable */
DKCACHE_WCHANGE 0x000200                          /* write enable is changeable */
DKCACHE_SAVE    0x010000               /* cache parameters are savable/save them */

                /* sync disk cache */

#define DIOCCACHESYNC   _IOW('d', 118, int)       /* sync cache (force?) */
```

**Terms**

**in-core label:** The content of disklabel which is read in memory.

**on-disk label:** The actual disklabel stored in storage device.

## 12.2.5   Reading Diak Label: `DIOCGDINFO`

A copy of the in-core label for a disk can be obtained with the DIOCGDINFO ioctl(2); this works with a file descriptor for a block or character ("raw") device for any partition of the disk.

Reading in-core label means that the contents are not directly from storage device, but from the kernel structure which is filled when the storage device drive attaches the disk at the initial configuration stage.

———————————————————————————————— dev/scsipi/sd.c

```
896 /*
897  * Perform special action on behalf of the user
898  * Knows about the internals of this device
899  */
900 int
901 sdioctl(dev, cmd, addr, flag, p)
902     dev_t dev;
```

```
903      u_long cmd;
904      caddr_t addr;
905      int flag;
906      struct proc *p;
907 {
908      struct sd_softc *sd = sd_cd.cd_devs[SDUNIT(dev)];
909      struct scsipi_periph *periph = sd->sc_periph;
910      int part = SDPART(dev);
911      int error;
.....
946      switch (cmd) {
947      case DIOCGDINFO:
948          *(struct disklabel *)addr = *(sd->sc_dk.dk_label);
949          return (0);
.....
```

———————————————————————————————————— dev/scsipi/sd.c

### 12.2.6   Writing In-Core Disk Label: `DIOCSDINFO`

The in-core copy of the label is set by the DIOCSDINFO ioctl(2).

The kernel device drivers will not allow the size of a disk partition to be decreased or the offset of a partition to be changed while it is open. Some device drivers create a label containing only a single large parti- tion if a disk is unlabeled; thus, the label must be written to the "a" partition of the disk while it is open. This sometimes requires the de- sired label to be set in two steps, the first one creating at least one other partition, and the second setting the label on the new partition while shrinking the "a" partition.

### 12.2.7   Writing On-Disk Disk Label: `DIOCWDINFO`

Finally, the DIOCWDINFO ioctl(2) operation sets the in-core label and then up- dates the on-disk label; there must be an existing label on the disk for this operation to succeed. Thus, the initial label for a disk or disk pack must be installed by writing to the raw disk. All of these operations are normally done using disklabel(8).

———————————————————————————————————————

```
900 int
901 sdioctl(dev, cmd, addr, flag, p)
902      dev_t dev;
903      u_long cmd;
904      caddr_t addr;
905      int flag;
906      struct proc *p;
907 {
908      struct sd_softc *sd = sd_cd.cd_devs[SDUNIT(dev)];
909      struct scsipi_periph *periph = sd->sc_periph;
910      int part = SDPART(dev);
911      int error;
.....
946      switch (cmd) {
.....
966      case DIOCWDINFO:
967      case DIOCSDINFO:
```

```
.....
   972     {
   973         struct disklabel *lp;
.....
   982         lp = (struct disklabel *)addr;
   983
   984         if ((flag & FWRITE) == 0)
   985             return (EBADF);
   986
   987         if ((error = sdlock(sd)) != 0)
   988             return (error);
   989         sd->flags |= SDF_LABELLING;
   990
   991         error = setdisklabel(sd->sc_dk.dk_label,
   992             lp, /*sd->sc_dk.dk_openmask : */0,
   993             sd->sc_dk.dk_cpulabel);
   994         if (error == 0) {
   995             if (cmd == DIOCWDINFO)
.....
  1000                 error = writedisklabel(SDLABELDEV(dev),
  1001                     sdstrategy, sd->sc_dk.dk_label,
  1002                     sd->sc_dk.dk_cpulabel);
  1003         }
  1005         sd->flags &= ~SDF_LABELLING;
  1006         sdunlock(sd);
  1007         return (error);
  1008     }
.....
```

where `sdlock` and `sdunlock` is defined as

```
   322 /*
   323  * Wait interruptibly for an exclusive lock.
   324  *
   325  * XXX
   326  * Several drivers do this; it should be abstracted and made MP-safe.
   327  */
   328 int
   329 sdlock(sd)
   330     struct sd_softc *sd;
   331 {
   332     int error;
   333
   334     while ((sd->flags & SDF_LOCKED) != 0) {
   335         sd->flags |= SDF_WANTED;
   336         if ((error = tsleep(sd, PRIBIO | PCATCH, "sdlck", 0)) != 0)
   337             return (error);
   338     }
   339     sd->flags |= SDF_LOCKED;
   340     return (0);
   341 }
```

```
342
343 /*
344  * Unlock and wake up any waiters.
345  */
346 void
347 sdunlock(sd)
348     struct sd_softc *sd;
349 {
350
351     sd->flags &= ~SDF_LOCKED;
352     if ((sd->flags & SDF_WANTED) != 0) {
353         sd->flags &= ~SDF_WANTED;
354         wakeup(sd);
355     }
356 }
```

## 12.2.8   Restrictions of Disk Label in sparc64

On the sparc, sparc64, sun2 and sun3 NetBSD systems, the size of each partition must be a multiple of the number of sectors per cylinder (i.e. each partition must be an integer number of cylinders), or the boot ROMs will declare the label invalid and fail to boot the system.

If the disk partition is not specified in the disk name (i.e. "xy0" instead of "/dev/rxy0c"), disklabel will construct the full pathname of the disk and use the "a" partition on the tahoe, the "d" partition on i386 or hpcmips or arc, and the "c" partition on all others including sparc64.

On some machines the bootstrap code may not fit entirely in the area al- located for it by some filesystems. As a result, it may not be possible to have filesystems on some partitions of a "bootable" disk. When in- stalling bootstrap code, disklabel checks for these cases. If the in- stalled boot code would overlap a partition of type FS_UNUSED it is marked as type FS_BOOT. The newfs(8) utility will disallow creation of filesystems on FS_BOOT partitions. Conversely, if a partition has a type other than FS_UNUSED or FS_BOOT, disklabel will not install bootstrap code that overlaps it.

# 12.3   Concatenated Disk Driver

## 12.3.1   Strcture

```
struct ccdbuf {
        struct buf      cb_buf;          /* new I/O buf */
        struct buf      *cb_obp;         /* ptr. to original I/O buf */
        struct ccd_softc *cb_sc;         /* pointer to ccd softc */
        int             cb_comp;         /* target component */
        SIMPLEQ_ENTRY(ccdbuf) cb_q;      /* fifo of component buffers */
};

/*
 * This structure is used to configure a ccd via ioctl(2).
 */
struct ccd_ioctl {
        char    **ccio_disks;            /* pointer to component paths */
        u_int   ccio_ndisks;             /* number of disks to concatenate */
```

```
            int     ccio_ileave;            /* interleave (DEV_BSIZE blocks) */
            int     ccio_flags;             /* see sc_flags below */
            int     ccio_unit;              /* unit number: use varies */
            size_t  ccio_size;              /* (returned) size of ccd */
};

/*
 * Component info table.
 * Describes a single component of a concatenated disk.
 */
struct ccdcinfo {
            struct vnode    *ci_vp;                 /* device's vnode */
            dev_t           ci_dev;                 /* XXX: device's dev_t */
            size_t          ci_size;                /* size */
            char            *ci_path;               /* path to component */
            size_t          ci_pathlen;             /* length of component path */
};

/*
 * Interleave description table.
 * Computed at boot time to speed irregular-interleave lookups.
 * The idea is that we interleave in "groups".  First we interleave
 * evenly over all component disks up to the size of the smallest
 * component (the first group), then we interleave evenly over all
 * remaining disks up to the size of the next-smallest (second group),
 * and so on.
 *
 * Each table entry describes the interleave characteristics of one
 * of these groups.  For example if a concatenated disk consisted of
 * three components of 5, 3, and 7 DEV_BSIZE blocks interleaved at
 * DEV_BSIZE (1), the table would have three entries:
 *
 *      ndisk   startblk        startoff        dev
 *      3       0               0               0, 1, 2
 *      2       9               3               0, 2
 *      1       13              5               2
 *      0       -               -               -
 *
 * which says that the first nine blocks (0-8) are interleaved over
 * 3 disks (0, 1, 2) starting at block offset 0 on any component disk,
 * the next 4 blocks (9-12) are interleaved over 2 disks (0, 2) starting
 * at component block 3, and the remaining blocks (13-14) are on disk
 * 2 starting at offset 5.
 */
struct ccdiinfo {
            int     ii_ndisk;       /* # of disks range is interleaved over */
            daddr_t ii_startblk;    /* starting scaled block # for range */
            daddr_t ii_startoff;    /* starting component offset (block #) */
            int     *ii_index;      /* ordered list of components in range */
};

/*
 * Concatenated disk pseudo-geometry information.
 */
```

```
struct ccdgeom {
        u_int32_t       ccg_secsize;    /* # bytes per sector */
        u_int32_t       ccg_nsectors;   /* # data sectors per track */
        u_int32_t       ccg_ntracks;    /* # tracks per cylinder */
        u_int32_t       ccg_ncylinders; /* # cylinders per unit */
};

struct ccdbuf;

/*
 * A concatenated disk is described after initialization by this structure.
 */
struct ccd_softc {
        int             sc_flags;               /* flags */
        size_t          sc_size;                /* size of ccd */
        int             sc_ileave;              /* interleave */
        u_int           sc_nccdisks;            /* number of components */
#define CCD_MAXNDISKS   65536
        struct ccdinfo  *sc_cinfo;              /* component info */
        struct ccdiinfo *sc_itable;             /* interleave table */
        struct ccdgeom  sc_geom;                /* pseudo geometry info */
        char            sc_xname[8];            /* XXX external name */
        struct disk     sc_dkdev;               /* generic disk device info */
        struct lock     sc_lock;                /* lock on this structure */
};
```

## 12.3.2   Gloval Variables

```
struct pool ccd_cbufpool;
const struct bdevsw ccd_bdevsw = {
        ccdopen, ccdclose, ccdstrategy, ccdioctl, ccddump, ccdsize, D_DISK
};

const struct cdevsw ccd_cdevsw = {
        ccdopen, ccdclose, ccdread, ccdwrite, ccdioctl,
        nostop, notty, nopoll, nommap, nokqfilter, D_DISK
};
struct  ccd_softc *ccd_softc;
int     numccd = 0;
```

## 12.3.3   Functions

```
[ Common Device Driver Entry ]

int   ccdopen  (dev_t dev, int flags, int fmt, struct proc *p);
int   ccdclose (dev_t dev, int flags, int fmt, struct proc *p);
int   ccdioctl (dev_t dev, u_long cmd, caddr_t data, int flag, struct proc *p);

[ Block Device Driver Entry ]

void  ccdstrategy (struct buf *bp);
int   ccdsize     (dev_t dev);
```

```
int    ccddump    (dev_t dev, daddr_t blkno, caddr_t va, size_t size);

[ Character Device Driver Entry ]

int    ccdread    (dev_t dev, struct uio *uio, int flags);
int    ccdwrite   (dev_t dev, struct uio *uio, int flags);

[ Device Driver Autoconfiguration ]

void   ccdattach   (int num);

[ Sub-function ]

ccdinit                used by ccdioctl() - CCDIOCSET
ccdinterleave          used by ccdinit()
ccdstart               used by ccdstrategy()
ccdbuffer              used by ccdstart()
ccdintr                used by ccdiodone()
ccdiodone              used by biodone() which is called by ccdintr()
ccdlookup              used by ccdioctl() - CCDIOCSET
ccdgetdefaultlabel     used by ccdgetdisklabel(), ccdioctl() - DIOCGDEFLABEL
ccdgetdisklabel        used by ccdopen(), ccdioctl() - DIOCSET
ccdmakedisklabel       used by ccdgetdisklabel()
```

# Chapter 13

# Logical Volume Manager

RAIDframe is a kind of Logical Volume Manager not included in NetBSD/sparc64. In this chapter, we describes other logical volume manager, such as VERITAS Volume Manager 3.1 under HP-UX 11i, and LVM under HP-UX 10.

## 13.1  RAIDframe

### 13.1.1  Introduction

The raid driver provides RAID 0, 1, 4, and 5 (and more!) capabilities to NetBSD. This document assumes that the reader has at least some famil-iarity with RAID and RAID concepts. The reader is also assumed to know how to configure disks and pseudo-devices into kernels, how to generate kernels, and how to partition disks.

RAIDframe provides a number of different RAID levels including:

```
RAID 0  provides simple data striping across the components.

RAID 1  provides mirroring.

RAID 4  provides data striping across the components, with parity stored
        on a dedicated drive (in this case, the last component).

RAID 5  provides data striping across the components, with parity dis-
        tributed across all the components.
```

There are a wide variety of other RAID levels supported by RAIDframe, in-cluding Even-Odd parity, RAID level 5 with rotated sparing, Chained declustering, and Interleaved declustering. The reader is referred to the RAIDframe documentation mentioned in the HISTORY section for more detail on these various RAID configurations.

Depending on the parity level configured, the device driver can support the failure of component drives. The number of failures allowed depends on the parity level selected. If the driver is able to handle drive failures, and a drive does fail, then the system is operating in "degraded mode". In this mode, all missing data must be reconstructed from the data and parity present on the other components. This results in much slower data accesses, but does mean that a failure need not bring the system to a complete halt.

## 13.1.2   Component Labels

The RAID driver supports and enforces the use of 'component labels'. A 'component label' contains important information about the component, in-cluding a user-specified serial number, the row and column of that compo-nent in the RAID set, and whether the data (and parity) on the component is 'clean'. If the driver determines that the labels are very inconsis-tent with respect to each other (e.g. two or more serial numbers do not match) or that the component label is not consistent with it's assigned place in the set (e.g. the component label claims the component should be the 3rd one a 6-disk set, but the RAID set has it as the 3rd component in a 5-disk set) then the device will fail to configure. If the driver de-termines that exactly one component label seems to be incorrect, and the RAID set is being configured as a set that supports a single failure, then the RAID set will be allowed to configure, but the incorrectly la-beled component will be marked as 'failed', and the RAID set will begin operation in degraded mode. If all of the components are consistent among themselves, the RAID set will configure normally.

Component labels are also used to support the auto-detection and auto-configuration of RAID sets. A RAID set can be flagged as auto-config-urable, in which case it will be configured automatically during the ker-nel boot process. RAID filesystems which are automatically configured are also eligible to be the root filesystem. There is currently only limited support (alpha and pmax architectures) for booting a kernel di-rectly from a RAID 1 set, and no support for booting from any other RAID sets. To use a RAID set as the root filesystem, a kernel is usually ob-tained from a small non-RAID partition, after which any auto-configuring RAID set can be used for the root filesystem.

## 13.1.3   Hot Spares

The driver supports 'hot spares', disks which are on-line, but are not actively used in an existing filesystem. Should a disk fail, the driver is capable of reconstructing the failed disk onto a hot spare or back on-to a replacement drive. If the components are hot swapable, the failed disk can then be removed, a new disk put in its place, and a copyback op-eration performed. The copyback operation, as its name indicates, will copy the reconstructed data from the hot spare to the previously failed (and now replaced) disk. Hot spares can also be hot-added.

## 13.1.4   Hierarchical Organization

If a component cannot be detected when the RAID device is configured, that component will be simply marked as 'failed'. The user-land utility for doing all raid configuration and other opera-tions is `raidctl` command. Most importantly, `raidctl` command must be used with the `-i` option to initialize all RAID sets. In particular, this initialization includes re-building the parity data. This rebuilding of parity data is also required when either a) a new RAID device is brought up for the first time or b) after an unclean shutdown of a RAID device. By using the `-P` option to `raidctl` command, and performing this on-demand recomputation of all parity before doing a `fsck` command or a `newfs` command, filesystem integrity and parity integrity can be ensured. It bears repeating again that parity recomputation is required before any filesystems are created or used on the RAID device. If the parity is not correct, then missing data cannot be correctly recovered.

RAID levels may be combined in a hierarchical fashion. For example, a RAID 0 device can be constructed out of a number of RAID 5 devices (which, in turn, may be constructed out of the physical disks, or of other RAID devices).

## 13.1.5 Kernel Configuration

It is important that drives be hard-coded at their respective addresses (i.e. not left free-floating, where a drive with SCSI ID of 4 can end up as `/dev/sd0c`) for well-behaved functioning of the RAID device. This is true for all types of drives, including IDE, HP-IB, etc. For normal SCSI drives, for example, the following can be used to fix the device address-es:

```
sd0     at scsibus0 target 0 lun ?      \# SCSI disk drives
sd1     at scsibus0 target 1 lun ?      \# SCSI disk drives
sd2     at scsibus0 target 2 lun ?      \# SCSI disk drives
sd3     at scsibus0 target 3 lun ?      \# SCSI disk drives
sd4     at scsibus0 target 4 lun ?      \# SCSI disk drives
sd5     at scsibus0 target 5 lun ?      \# SCSI disk drives
sd6     at scsibus0 target 6 lun ?      \# SCSI disk drives
```

The rationale for fixing the device addresses is as follows: Consider a system with three SCSI drives at SCSI ID's 4, 5, and 6, and which map to components `/dev/sd0e`, `/dev/sd1e`, and `/dev/sd2e` of a RAID 5 set. If the drive with SCSI ID 5 fails, and the system reboots, the old `/dev/sd2e` will show up as `/dev/sd1e`. The RAID driver is able to detect that component positions have changed, and will not allow normal `configuration`. If the device addresses are hard coded, however, the RAID driver would detect that the middle component is unavailable, and bring the RAID 5 set up in degraded mode. Note that the auto-detection and auto-configuration code does not care about where the components live. The auto-configuration code will correctly configure a device even after any number of the components have been rearranged.

The first step to using the raid driver is to ensure that it is suitably configured in the kernel. This is done by adding a line similar to:

```
pseudo-device   raid   4       \# RAIDframe disk device
```

to the kernel configuration file. The 'count' argument ( '4', in this case), specifies the number of RAIDframe drivers to configure. To turn on component auto-detection and auto-configuration of RAID sets, simply add:

```
        options    RAID_AUTOCONFIG
```

to the kernel configuration file.

All component partitions must be of the type `FS_BSDFFS` (e.g. `4.2BSD`) or `FS_RAID`. The use of the latter is strongly encouraged, and is required if auto-configuration of the RAID set is desired. Since RAIDframe leaves room for disklabels, RAID components can be simply raw disks, or partitions which use an entire disk.

It is highly recommended that the steps to reconstruct, copyback, and re-compute parity are well understood by the system admin-istrators before a component failure. Doing the wrong thing when a component fails may result in data loss.

Additional internal consistency checking can be enabled by specifying:

```
        options    RAID_DIAGNOSTIC
```

These assertions are disabled by default in order to improve performance.

## 13.2  VERITAS Volume Manager

This section describes what VERITAS Volume Manager is, how it works, how you can communicate with it through the user interfaces, and Volume Manager concepts.

### 13.2.1  Introduction

Volume Manager provides easy-to-use online disk storage management for computing environments. Traditional disk storage management often requires that machines be taken off-line at a major inconvenience to users. In the distributed client/server environment, databases and other resources must maintain high availability, be easy to access, and be Volume Manager provides the tools to improve performance and ensure data availability and integrity. Volume Manager also dynamically configures disk storage while the system is active.

### 13.2.2  Volume Manager Overview

The Volume Manager uses objects to do storage management. The two types of objects used by Volume Manager are *physical objects* and *virtual objects.*

**physical objects** Volume Manager uses two physical objects: physical disks and partitions. Partitions are created on the physical disks

**virtual objects** Volume Manager creates virtual objects, called volumes. Each volume records and retrieves data from one or more physical disks. Volumes are accessed by a file system, a database, or other applications in the same way that physical disks are accessed. Volumes are also composed of other virtual objects that are used to change the volume configuration. Volumes and their virtual components are called virtual objects.

### 13.2.3 Physical Objects

A *physical disk* is the basic storage device (media) where the data is ultimately stored. You can access the data on a physical disk by using a device name (devname) to locate the disk. The physical disk device name varies with the computer system you use. Not all parameters are used on all systems. Typical device names can include: c#t#d#, where:

```
c\# is the controller
t\# is the target ID
d\# is the disk number
```

On some computer systems, a physical disk can be divided into one or more *partitions*. The partition number, or s#, is added at the end of the device name. Note that a partition can be an entire physical disk.

### 13.2.4 Volumes and Virtual Objects

Volume Manager creates virtual objects and makes logical connections between the objects. The *virtual objects* are then used by Volume Manager to do storage management tasks.

A *volume* is a virtual disk device that appears to applications, databases, and file systems as a physical disk. However, a volume does not have the limitations of a physical disk. When you use Volume Manager, applications access volumes created on Volume Manager disks (VM Disks) rather than physical disks.

**Volume Manager Disks**

When you place a physical disk under Volume Manager control, a *Volume Manager disk (or VM Disk)* is assigned to the physical disk. A VM Disk is under Volume Manager control and is usually in a disk group. Each VM disk corresponds to at least one physical disk. Volume Manager allocates storage from a contiguous area of Volume Manager disk space.

A VM disk typically includes a *public region* (allocated storage) and a *private region* where Volume Manager internal configuration information is stored.

Each VM Disk has a unique disk *media name* (a virtual disk name). You can supply the disk name or allow Volume Manager to assign a default name that typically takes the form disk##.

**Disk Groups**

A *disk group* is a collection of VM disks that share a common configuration. A disk group configuration is a set of records with detailed information about related Volume Manager objects, their attributes, and their connections. The default disk group is rootdg (the root disk group).

You can create additional disk groups as necessary. Disk groups allow the administrator to group disks into logical collections. A disk group and its components can be moved as a unit from one host machine to another.

Volumes are created within a disk group. A given volume must be configured from disks in the same disk group.

**Subdisks**

A *subdisk* is a set of contiguous disk blocks. A block is a unit of space on the disk. Volume Manager allocates disk space using subdisks. A VM disk can be divided into one or more subdisks. Each subdisk represents a specific portion of a VM disk, which is mapped to a specific region of a physical disk.

The default name for a VM disk is disk## (such as disk01) and the default name for a subdisk is disk##-##.

A VM disk can contain multiple subdisks, but subdisks cannot overlap or share the same portions of a VM disk.

Any VM disk space that is not part of a subdisk is free space. You can use free space to create new subdisks.

Volume Manager release 3.0 or higher supports the concept of layered volumes in which subdisk objects can contain volumes. For more information, see "Layered Volumes".

**Plexes**

The Volume Manager uses subdisks to build virtual objects called plexes. A plex consists of one or more subdisks located on one or more physical disks.

You can organize data on the subdisks to form a plex by using these methods:

- concatenation
- striping (RAID-0)
- striping with parity (RAID-5)
- mirroring (RAID-1)

**Volumes**

A volume is a virtual disk device that appears to applications, databases, and file systems like a physical disk device, but does not have the physical limitations of a physical disk device. A volume consists of one or more plexes, each holding a copy of the selected data in the volume. Due to its virtual nature, a volume is not restricted to a particular disk or a specific area of a disk. The configuration of a volume can be changed by using the Volume Manager user interfaces. Configuration changes can be done without causing disruption to applications or file systems that are using the volume. For example, a volume can be mirrored on separate disks or moved to use different disk storage.

The Volume Manager uses the default naming conventions of vol## for volumes and vol##-## for plexes in a volume. Administrators must select meaningful names for their volumes.

A volume can consist of up to 32 plexes, each of which contains one or more subdisks. A volume must have at least one associated plex that has a complete set of the data in the volume with at least one associated subdisk. Note that all subdisks within a volume must belong to the same disk group.

# Appendix

## A. References to NetBSD Kernel Sources

```
src/syssrc/sys
               /kern ......................... 73,735 lines
                   /vfs_??? (VFS)............ 10,822 lines <<<<<
               /sys ......................... 31,066 lines
               /ufs ......................... 38,381 lines
                   /ufs ..................... 6,975 lines <<<<<
                   /ffs ..................... 12,970 lines
                   /ffs (without softdep) .... 6,639 lines <<<<<
                   /lfs ..................... 10,529 lines
                   /ext2fs .................. 6,994 lines
                   /mfs ..................... 912 lines
               /msdosfs ..................... 7,832 lines
               /ntfs ........................ 5,358 lines
               /arch ........................ 1,284,122 lines
                   /sparc64 ................. 54,785 lines
               /uvm ......................... 24,905 lines
               /net ......................... 56,506 lines
               /netinet (IPv4) .............. 47,558 lines
               /netinet6 (IPv6) ............. 38,958 lines
               /dev ......................... 938,890 lines
               /nfs ......................... 23,235 lines

ultra1: {205} ls kern
CVS                kern_event.c         makesyscalls.sh    tty_conf.c
Make.tags.inc      kern_exec.c          subr_autoconf.c    tty_pty.c
Makefile           kern_exit.c          subr_devsw.c       tty_subr.c
cnmagic.c          kern_fork.c          subr_disk.c        tty_tb.c
core_elf32.c       kern_kthread.c       subr_extent.c      tty_tty.c
core_elf64.c       kern_ktrace.c        subr_log.c         uipc_domain.c
core_netbsd.c      kern_lkm.c           subr_pool.c        uipc_mbuf.c
exec_aout.c        kern_lock.c          subr_prf.c         uipc_mbuf2.c
exec_conf.c        kern_malloc.c        subr_prof.c        uipc_proto.c
exec_ecoff.c       kern_malloc_debug.c  subr_prop.c        uipc_socket.c
exec_elf32.c       kern_ntptime.c       subr_userconf.c    uipc_socket2.c
exec_elf64.c       kern_physio.c        subr_xxx.c         uipc_syscalls.c
exec_elf_common.c  kern_proc.c          sys_generic.c      uipc_usrreq.c
exec_macho.c       kern_prot.c          sys_pipe.c         vfs_bio.c
exec_script.c      kern_ras.c           sys_pmc.c          vfs_cache.c
exec_subr.c        kern_resource.c      sys_process.c      vfs_getcwd.c
```

```
genassym.awk        kern_sig.c            sys_socket.c      vfs_init.c
genassym.sh         kern_subr.c           syscalls.c        vfs_lockf.c
genlintstub.awk     kern_synch.c          syscalls.conf     vfs_lookup.c
init_main.c         kern_sysctl.c         syscalls.master   vfs_subr.c
init_sysent.c       kern_systrace.c       sysv_ipc.c        vfs_syscalls.c
kern_acct.c         kern_time.c           sysv_msg.c        vfs_vnops.c
kern_allocsys.c     kern_verifiedexec.c   sysv_sem.c        vnode_if.c
kern_clock.c        kern_xxx.c            sysv_shm.c        vnode_if.sh
kern_descrip.c      kgdb_stub.c           tty.c             vnode_if.src
```

# Bibliography

[1] Marshall Kirk McKusick, Keith Bostic, Michael J. Karels, and John S. Quarterman, *The design and implementation of the 4.4BSD Operating System,* pp. 193-196, Addision Wesley, 1996.

[2] Maurice J. Bach, *Design of the Unix Operating System*, pp. 46-56, Prentice Hall, 1986.

[3] Chales D. Cranor, "The Design and Implementation of the UVM Virtual Mem-ory System", *Ph.D. Dissertation*, Department of Computer Science, Washington University, 1998.

[4] Uresh Vahalia, *UNIX internals: the new frontiers*, Prentice Hall, 1996.

[5] Chuck Silvers, UBC: An Efficient Unified I/O and Memory Caching Subsystem *for NetBSD,* i n *Proceedings of USENIX Annual Technical Conference*, June 2000, pp. 285-290.